\documentclass{optica-article}

\newcommand{\vect}[1]{\vec{\textbf{#1}}}
\newcommand{\uvect}[1]{\boldsymbol{\hat{\textbf{#1}}}}
\journal{opticajournal} 

\articletype{Research Article}

\usepackage{lineno}

\begin{document}

\title{Temporal dissipative solitons and optical frequency combs in coherently driven Kerr resonators}

\author{Stuart G. Murdoch,\authormark{1,2,*} Francois Leo,\authormark{3}  Xiaoxiao Xue,\authormark{4} St\'ephane Coen,\authormark{1,2} and Miro Erkintalo\authormark{1,2}}

\address{\authormark{1} Department of Physics, University of Auckland, Auckland 1010, New Zealand\\
\authormark{2} The Dodd-Walls Centre for Photonic and Quantum Technologies, Auckland, New Zealand\\
\authormark{3} OPERA-photonics, Universit\'e libre de Bruxelles, 50 Avenue F. D. Roosevelt, B-1050 Brussels, Belgium\\
\authormark{4} Department of Electronic Engineering, Beijing National Research Center for Information Science and Technology, Tsinghua University, Beijing, China}

\email{\authormark{*}s.murdoch@auckland.ac.nz} 


\begin{abstract*} 
Kerr frequency combs have recently emerged as an exciting new photonic technology, with applications across science and engineering. Their formation within driven optical resonators that possess a Kerr nonlinearity is enabled through the rich landscape of localized nonlinear dissipative structures intrinsic to these systems. This article offers a comprehensive review of the physics that underpins these nonlinear comb-generating structures. Particular attention is placed on bright temporal cavity solitons and nonlinear switching waves -- the canonical stable comb-generating states in the anomalous and normal dispersion regimes, respectively. Written as both a review and tutorial, the article also includes an in-depth treatment of the numerical methods required to simulate driven Kerr resonators, alongside a comprehensive discussion of the laboratory techniques used to experimentally realize and characterize Kerr combs.

\end{abstract*}

\tableofcontents

\section{Introduction}

A continuous wave (cw) laser beam injected into a dispersive resonator with a Kerr-type optical nonlinearity can drive the generation of temporal localized structures within the resonator~\cite{wabnitz_suppression_1993, leo_temporal_2010, herr_temporal_2014,xue_mode-locked_2015}. These structures manifest as persistent (bright or dark) pulses of light that maintain their shape and energy over time, constituting robust steady-state configurations of the resonator system. Because some portion of light within the resonator is coupled out each round trip, a single localized structure delivers a periodic train of pulses at the resonator output. In the frequency domain, this dynamic results in the transformation of the single-frequency input into a regularly spaced array of spectral lines -- an optical frequency comb~\cite{delhaye_optical_2007,kippenberg_microresonator-based_2011,kippenberg_dissipative_2018}.

Kerr resonators can host several fundamentally different types of localized structures, accessed through judicious choice of system parameters and operating regimes. Arguably the most widely studied is the bright temporal cavity soliton (CS)~\cite{leo_temporal_2010}, also known as a dissipative Kerr soliton (DKS)~\cite{kippenberg_dissipative_2018}. These canonical structures emerge when a pure Kerr resonator is driven by a monochromatic pump field in the regime of anomalous group-velocity dispersion (GVD) and can be seen as an extension of the celebrated soliton solution of the nonlinear Schr\"odinger equation (NLSE)~\cite{zakharov_exact_1972,hasegawa_transmission_1973-1} into the dissipative resonator context. 

As with the fundamental soliton solution of the NLSE, CSs exhibit a hyperbolic secant temporal profile. However, in stark contrast to the NLSE soliton, CSs sit atop a cw background corresponding to (one of) the cw steady-state solutions of the driven resonator system. Moreover, whilst standard NLSE solitons form a continuous family of solutions with different peak amplitudes and temporal widths, CSs are unique attractors: except for trivial time translations, all the CSs that exist for given system parameters are identical. This fundamental difference is a manifestation of the ubiquitous dichotomy between \emph{conservative} and \emph{dissipative} solitons~\cite{akhmediev_dissipative_2005, firth_temporal_2010, grelu_dissipative_2012}, firmly placing CSs within the latter category.

Although temporal cavity solitons (CSs) were first theoretically proposed in 1993~\cite{wabnitz_suppression_1993}, it was not until 2010 that they were conclusively observed and investigated in laboratory experiments~\cite{leo_temporal_2010}. In these first pioneering experiments, CSs manifested themselves as picosecond pulses of light persisting in a 380-m-long resonator made from standard single-mode telecommunications optical fiber. The initial application envisaged for temporal CSs was to use them as bits in an all-optical buffer, leveraging the demonstrated fact that a sequence of multiple CSs (representing a bit stream) could be controllably excited and persist within the resonator \cite{leo_temporal_2010, jang_all-optical_2016}. However, as history shows, the CSs’ true application potential lay elsewhere.

In addition to enabling the first experimental observation of temporal CSs~\cite{leo_temporal_2010}, macroscopic optical fiber resonators have played an important role in subsequent CS research, providing a highly controllable experimental platform in which to explore many aspects of CS physics including their stability~\cite{leo_dynamics_2013, anderson_observations_2016}, manipulation and excitation~\cite{jang_temporal_2015, luo_spontaneous_2015, jang_writing_2015, wang_addressing_2018}, and interactions~\cite{jang_ultraweak_2013, erkintalo_bunching_2015, jang_controlled_2016, wang_universal_2017}. A key breakthrough occurred in 2014, when CSs were conclusively observed in a radically different resonator configuration: a monolithic magnesium fluoride microresonator with a round-trip length of just 6.1 mm~\cite{herr_temporal_2014}. It established CSs as powerful new mechanism for generating coherent optical frequency combs with properties unobtainable through other approaches.


The potential of Kerr microresonators for optical comb generation had first been demonstrated in~2007 \cite{delhaye_optical_2007}, drawing substantial interest as a potential route to translate the Nobel prize winning frequency comb technology~\cite{hansch_nobel_2006, hall_nobel_2006} to platforms with small footprint and potential for mass manufacture. However, it was arguably not until the realization of CS combs in a magnesium fluoride resonator in~2014 that an explicit path to the systematic generation of coherent combs with a smooth spectral envelope was revealed~\cite{herr_temporal_2014}. 

Following their first observations in a microresonator platform, temporal CS frequency combs were rapidly observed in many different resonator geometries and materials (see, e.g., \cite{yi_soliton_2015,webb_experimental_2016,brasch_photonic_2016,yu_mode-locked_2016}). Of particular interest has been the ability to generate soliton combs in resonator platforms compatible with microelectronics‑derived fabrication techniques (e.g. silicon~\cite{yu_mode-locked_2016}, silicon nitride~\cite{brasch_photonic_2016,joshi_thermally_2016,wang_intracavity_2016,li_stably_2017,guo_universal_2017}, aluminum nitride~\cite{gong_high-fidelity_2018,liu_aluminum_2021}, and lithium niobate~\cite{he_self-starting_2019,gong_near-octave_2020}). This has raised the tantalizing possibility of fully integrating coherent and broadband optical frequency comb generation on a single photonic chip~\cite{gaeta_photonic-chip-based_2019}. Compounded by the unique characteristics associated with microresonator-based frequency combs (such as large comb line spacing and high power per line), CSs frequency combs have now been demonstrated in many applications that can benefit from numerous equally-spaced coherent frequency components, including spectroscopy~\cite{suh_microresonator_2016,dutt_-chip_2018, yu_silicon-chip-based_2018}, telecommunications~\cite{marin-palomo_microresonator-based_2017-1,corcoran_ultra-dense_2020}, optical distance measurements and ranging~\cite{trocha_ultrafast_2018, suh_soliton_2018, riemensberger_massively_2020, jang_nanometric_2021}, calibration of astronomical spectrographs~\cite{obrzud_microphotonic_2019,suh_searching_2019}, low-noise microwave generation~\cite{liang_high_2015,liu_photonic_2020,lucas_ultralow-noise_2020}, and frequency metrology~\cite{spencer_optical-frequency_2018,newman_architecture_2019}. Beyond microresonator and fiber systems, temporal CSs have also recently been reported in free-space enhancement cavities~\cite{lilienfein2019temporal}, opening promising new avenues for applications including the temporal compression of high peak-power pulses and cavity-enhanced nonlinear frequency conversion.

Whilst CS generation is restricted to spectral regions that exhibit anomalous GVD, a fundamentally different kind of temporal localized structure can manifest itself in the opposite regime of normal GVD~\cite{xue_mode-locked_2015}. In this case, the structure within the resonator is a \emph{dark} pulse, i.e., an intensity dip on a cw background. Spectral signatures of such dark pulse Kerr combs were first observed in experiments using crystalline Kerr microresonators~\cite{coillet_azimuthal_2013,liang_generation_2014}, followed by more unequivocal observations in integrated silicon nitride resonators that combined spectral measurements with line-by-line pulse shaping to reveal the combs' dark pulse temporal characteristics~\cite{xue_mode-locked_2015}. For clarity, at this point we note that, a bright pulse within the resonator will result in a dark pulse at the resonator output and vice versa. This is because the output field consists of a superposition of the field extracted from the resonator and the portion of the cw driving field that is not coupled into the resonator. When referring to bright or dark pulses, we exclusively refer to the field within the resonator. 

Dark pulse Kerr combs were found to result from the interlocking of switching waves (SWs) between different cw states~\cite{parra-rivas_origin_2016, parra-rivas_dark_2016}. In contrast to bright CSs, they often feature a more intricate spectral envelope and are generally more difficult to excite. Counterbalancing these deficiencies, dark pulse Kerr combs can exhibit a much larger conversion efficiency compared to bright CSs~\cite{xue_microresonator_2017}, making them an attractive alternative for selected applications. For this reason, substantial research efforts have been dedicated to understanding the physics that underpin dark pulse Kerr combs in Kerr resonators, and to design system configurations that enable their turn-key generation~\cite{xue_normal-dispersion_2015, kim_turn-key_2019, helgason_dissipative_2021}.

In this Article, we review the core physics that underpin the bright and dark temporal dissipative structures that manifest themselves in externally driven, dispersive Kerr resonators. We begin by synthesizing the key theoretical frameworks that govern the salient systems, and we use these frameworks to describe the basic properties of externally-driven Kerr resonators (Chapter~2). We then leverage the synthesized models to discuss the physics and dynamics of the bright temporal Kerr CSs in resonators with anomalous GVD (Chapter~3), and of the dark pulses (and related switching wave structures) in the regime of normal GVD (Chapter~4). In Chapters 5 and~6, we will provide a brief overview of the salient experimental and theoretical techniques that are relevant to dissipative soliton (and optical comb) generation in driven Kerr resonators. Chapter 7 concludes the article and discusses the remaining challenges as well as the future outlook.

\subsection{Scope of the Article}

Before proceeding, we wish to briefly delineate the scope of our Article. As described above, our primary focus is on temporal localized structures (and corresponding frequency combs) that can be generated in \emph{passive} externally-driven \emph{Kerr} resonators. Temporal dissipative structures that can be generated in other non-equilibrium optical systems --- including mode-locked lasers~\cite{grelu_dissipative_2012,marconi_how_2014}, systems with delayed-feedback~\cite{garbin_topological_2015,marconi_vectorial_2015}, and quadratically nonlinear $\chi^{(2)}$ resonators~\cite{jankowski_temporal_2018,ricciardi_optical_2020,roy_temporal_2022} --- are beyond the scope of our Article. Similarly, our focus is on \emph{temporal} solitons in \emph{dispersive} resonators, and we will therefore not discuss the substantial body of literature that exists on \emph{spatial} dissipative solitons in \emph{diffractive} Kerr resonators (for excellent and accessible reviews, see, e.g., refs.~\cite{firth_cavity_2002,firth_theory_2001,ackemann_dissipative_2005,ackemann_chapter_2009}). We nonetheless take this opportunity to highlight that such spatial (cavity) solitons are intimately related to the temporal CSs at the focus of our Article: in fact, many of the fundamental properties and behaviors of temporal CSs were first observed and studied in the context of analogous spatial solitons in diffractive resonators~\cite{firth_optical_1996,maggipinto_cavity_2000,barland_cavity_2002,pedaci_positioning_2006,pedaci_all-optical_2008,caboche_microresonator_2009,caboche_cavity-soliton_2009}. 

Thematically, our primary focus is to describe the physics and (nonlinear) dynamics that govern temporal CSs and related structures. As such, we do not intend to present a comprehensive review of the extensive body of applied research that has been reported on the topic. We refer the reader to several review articles that have been published over the past few years that more broadly provide excellent overviews of the different applications of CSs and microresonator frequency combs~\cite{kippenberg_dissipative_2018,gaeta_photonic-chip-based_2019,yang_efficient_2024,zhang_advances_2024}. 

Lastly, whilst the context of our Article is situated within the field of \emph{optics}, we briefly acknowledge here the fundamental inter-disciplinarity associated with some of the salient topics. Pulses of light (bright or dark) that persist in driven resonators belong to the universal class of dissipative solitons (and structures) that are ubiquitous in nature~\cite{akhmediev_dissipative_2008, purwins_dissipative_2010}, from physiology and chemistry, to fluid dynamics, granular matter, plama physics, and vegetation patterns~\cite{hodgkin_quantitative_1952, lee_pattern_1993, wu_observation_1984, clerc_soliton_2009, umbanhowar_localized_1996, lioubashevski_dissipative_1996, astrov_plasma_2001, cramer_are_2013, fernandez-oto_strong_2014, ruiz-reynes_fairy_2017}. Likewise, the dynamics that underpin the formation of such pulses are intimately tied to general processes such as pattern formation~\cite{turing_chemical_1952} and front dynamics~\cite{pomeau_front_1986}. More explicitly, as will be discussed below, the canonical equation that models a driven, dispersive Kerr resonator (and thus bright and dark localized structures within) is the AC-driven nonlinear Schr\"odinger equation (often referred to as the Lugiato-Lefever equation in optics)~\cite{lugiato_spatial_1987, coen_modeling_2013, chembo_spatiotemporal_2013}, which is also to be found in studies of long Josephson junctions, rf-driven plasmas, ferromagnets in microwave fields, etc~\cite{barashenkov_existence_1996}. In this context, it is an interesting anecdote that the bright soliton solution of that equation (which mathematically corresponds to the temporal CS at the focus of our Article) was, to the best of our knowledge, first identified and studied in the context of one-dimensional condensates~\cite{kaup_theory_1978} and then in plasma physics~\cite{nozaki_chaotic_1985}.



\section{Modeling and properties of dispersive Kerr resonators}
\label{sec:Modeling}

We begin by describing the theoretical models that govern the dynamics and properties of light circulating within externally, coherently, driven dispersive resonators with a Kerr optical nonlinearity. We will describe several different modeling approaches used in the literature, emphasizing the relationships, advantages, and disadvantages of the different theoretical paradigms. Subsequently, we will use the models to elucidate the basic properties of driven Kerr resonators, focusing primarily on homogeneous, continuous wave (cw) states; non-homogeneous (e.g., solitonic) states will be discussed in subsequent Sections. Note that our focus is on temporal and spectral dynamics of the electromagnetic field, and so we shall not extensively review the spatial part of the problem that yields the linear eigenmodes of the system (for thorough treatments of the spatial problem, see~\cite{matsko_optical_2006, strekalov_nonlinear_2016, chembo_modal_2010,pasquazi_micro-combs_2018,agrawal_nonlinear_2013}). 





\begin{figure*}[!b]
	\centering
	\includegraphics[width = 0.6\textwidth, clip=true]{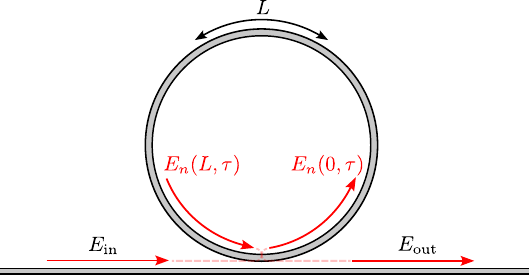}
	\caption{Schematic illustration. A resonator with round-trip length $L$ is coherently driven with a field with complex amplitude $E_\mathrm{in}$, giving rise to an intracavity field whose complex amplitude at the $n^\mathrm{th}$ transit around the resonator is denoted as $E_n(z,t)$. The cavity output field $E_\mathrm{out}$ is a superposition of the field coupled out from the resonator and the part of the driving field that is reflected at the coupler.} 
	\label{fig:resonator_schematic}
\end{figure*}

We consider a dispersive resonator as a segment of optical waveguide (of length $L$) that is looped on itself with a point coupler with power transmission coefficient $\theta$ (see Fig.~\ref{fig:resonator_schematic}). This model is intuitively appropriate for a macroscopic resonator made, e.g., from optical fiber, but is also applicable to monolithic microresonators, including whispering gallery mode resonators and chip-integrated resonators. We assume that the electromagnetic wave within the resonator (i.e., the ``intracavity'' field) propagates in a single (polarization or spatial) mode of the waveguide, and we write the electric field of the wave during the $n^\mathrm{th}$ transit around the resonator at position $\vect{r}=(x,y,z)$ and time~$t$ as
\begin{equation}
    \label{carrier_envelope}
    \vec{\mathcal{E}}_n(\vect{r},t) = \uvect{p}\left[F(x,y)E_n(z,t)e^{i\beta_0 z - i\omega_0 t} + \text{c.c.}\right],
\end{equation}
where $\uvect{p}$, $F(x,y)$, and $\beta_0=\omega_0 n_\mathrm{eff}/c$ are respectively the polarization unit vector, transverse profile, and propagation constant of the mode, with $n_\mathrm{eff}$~the effective refractive index of the waveguide, each evaluated at the carrier angular frequency $\omega_0$ (c.c. denotes complex conjugation). $E_n(z,t)$ is a slowly-varying temporal envelope that evolves with propagation along the waveguide that forms the resonator (i.e., along the $z$ coordinate), normalized such that $|E_n(z,t)|^2$ is the instantaneous power in Watts~\cite{agrawal_nonlinear_2013}. The Fourier transform~(FT) of $E_n(z,t)$ is defined as 
\begin{equation}
    \tilde{E}_n(z,\Omega) = \int_{-\infty}^\infty E_n(z,t)e^{i\Omega t}\, \mathrm{d}t\label{FFT} = \mathrm{FT} \left[E_n(z,t)\right]\,,
\end{equation}
and yields the optical intensity spectrum as $|\tilde{E}_n(z,\Omega)|^2$, with $\Omega = \omega-\omega_0$.  

The slowly-varying envelope~$E_n(z,t)$ encapsulates all information pertaining to the temporal and spectral intensity profiles of the intracavity field, and is thus the central quantity of interest for understanding the nonlinear dynamics of light in a resonator. In what follows, we discuss the theoretical models that will allow us to describe how $E_n(z,t)$ behaves as the field circulates around the driven resonator multiple times.

\subsection{Generalized nonlinear Schr\"odinger equation}

\begin{figure*}[!t]
    \centering
    \includegraphics[width = \textwidth, clip=true]{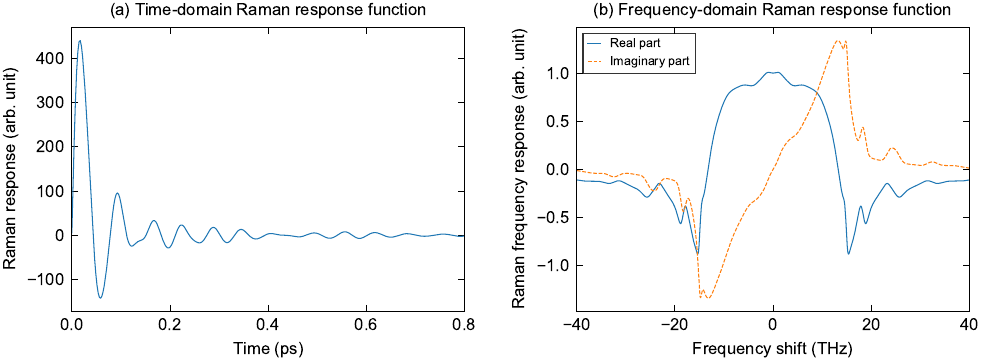}
    \caption{(a) Temporal- and (b) frequency-domain representation of the Raman response function of fused silica (multi-vibrational mode model~\cite{hollenbeck_multiple-vibrational-mode_2002}).}
    \label{fig:SRS}
\end{figure*}

Over a single pass through the Kerr nonlinear waveguide that forms the resonator (i.e., from ${z=0}$ to ${z = L}$), the evolution of the slowly-varying envelope $E_n(z,t)$ is described by the generalized nonlinear Schr\"odinger equation (GNLSE). The GNLSE is well-known to permit accurate modeling of the complex interplay between dispersive and nonlinear effects, allowing, e.g., to quantitatively understand the generation of octave-spanning supercontinuum generation in photonic crystal fibers and integrated waveguides \cite{dudley_supercontinuum_2006}. We write the GNLSE equation in a reference frame that moves at the group velocity~$v_\mathrm{g}$ of light at the carrier frequency~$\omega_0$. Specifically, we define $\tau = t-\beta_1 z$, where $\beta_1 = 1/v_\mathrm{g}$, to obtain
\begin{equation}
    \frac{\partial E_n(z,\tau)}{\partial z} = i\hat{D}\left(i\frac{\partial}{\partial\tau}\right)E_n +i\gamma \left[(1-f_\mathrm{R})|E_n|^2 + f_\mathrm{R}h_\mathrm{R}(\tau)\ast|E_n|^2 \right]E_n.
    \label{GNLSE}
\end{equation}
Here, the terms on the right-hand-side describe chromatic group-velocity dispersion, instantaneous Kerr nonlinearity, and (for the sake of completeness) stimulated Raman scattering (SRS), respectively. $\hat{D}$~is the linear dispersion operator defined as
\begin{equation}
    \hat{D}\left(i\frac{\partial}{\partial \tau}\right) = \sum_{k\geq2} \frac{\beta_k}{k!}\left(i \frac{\partial}{\partial \tau}\right)^k,
    \label{Dhat}
\end{equation}
with $\beta_k = d^k\beta/d\omega^k|_{\omega=\omega_0}$ the Taylor series expansion coefficients of the mode propagation constant $\beta(\omega)$ around~$\omega_0$.  The dispersion operator can equivalently be written in the frequency domain as $\hat{D}(\Omega) = \beta(\Omega) - \beta_0 -\beta_1 \Omega$. The parameter $\gamma = n_2\omega_0/(c A_\mathrm{eff})$ is the Kerr nonlinearity coefficient, with $n_2$~the nonlinear refractive index of the waveguide material and $A_\mathrm{eff}$ the effective mode area, while $f_\mathrm{R}$ is the Raman fraction of the nonlinearity~\cite{agrawal_nonlinear_2013}. $h_\mathrm{R}(\tau)$ is the time-domain Raman response function; the real and imaginary parts of the Fourier transform of $h_\mathrm{R}(\tau)$ are proportional respectively, to the phase shift and gain due to SRS. For fused silica, these are both plotted in Figure~\ref{fig:SRS} (based on the multi-vibrational mode model of~\cite{hollenbeck_multiple-vibrational-mode_2002}). The nonlinear contribution of the Raman term then takes the form of a convolution (denoted by the `$\ast$' operator) of this Raman reponse function with the temporal intensity of the intracavity field. It is also worth noting that SRS does not play a significant role under all experimental conditions (e.g., some resonator materials or geometries intrinsically feature a negligible Raman response). When appropriate, SRS can be ignored from the modeling simply by setting~${f_\mathrm{R} = 0}$. 

The GNLSE, as written in Eq.~\eqref{GNLSE}, omits a number of higher-order effects that typically play a minor role, but that can become important under some conditions. For example, the nonlinearity coefficient~$\gamma$ is in general frequency-dependent. This can be accounted for by expanding it as a Taylor series in the frequency domain \cite{kibler_supercontinuum_2005}, with these Taylor coefficients then used to introduce a temporal operator similar to the dispersion operator~$\hat{D}$. When considering this expansion to only first-order, the effect is often referred to as ``self-steepening''~\cite{dudley_supercontinuum_2006, lamont_route_2013}). Linear mode coupling between multiple waveguide modes is another higher-order effect that is often encountered. This, likewise, can be accounted for by generalizing the dispersion operator~$\hat{D}$, as will be discussed in Section~\ref{sec:mode_coupling}.

\subsection{Ikeda map}
\label{sec:Ikeda}

To describe the evolution of the intracavity field over multiple transits of the cavity, the GNLSE must be used in conjunction with a discrete map that describes how the field transforms from the end of one round trip (at ${z=L}$) to the beginning of the next round trip at the point coupler (${z=0}$). The discrete map can be written as
\begin{equation}
    \label{map}
    E_{n+1}(z=0,\tau) = \sqrt{\theta}\,E_\mathrm{in}(\tau)+\sqrt{1-2\alpha}\,E_n(z=L,\tau)e^{i\phi_0},
\end{equation}
where $E_\mathrm{in}(\tau)$ is the complex amplitude of the coherent field driving the resonator (same units as~$E_n$), $\alpha$ describes (half) the total power losses experienced by the intracavity field over one round trip, and $\phi_0$ is the linear phase accumulated over one round trip by a frequency component of the intracavity field at~$\omega_0$ with respect to the driving field [see Eq.~\eqref{carrier_envelope}],
\begin{equation}
  \phi_0 = \beta(\omega_0) L = \beta_0 L = \frac{\omega_0 n_\mathrm{eff} L}{c}\,.
  \label{eq:phi0}
\end{equation}
We emphasize that Eqs.~\eqref{GNLSE} and~\eqref{map} lump all the losses in the system at the point coupler; the coefficient $\alpha$ not only accounts for external losses due to out-coupling, but also (propagation) losses due to absorption, scattering, and other loss mechanisms. If the losses are only due to coupling and propagation (with intrinsic power loss coefficient per length~$\alpha_\mathrm{i}$), the relationship $\sqrt{1-2\alpha}=\sqrt{1-\theta}\exp(-\alpha_\mathrm{i}L/2)$ holds, with $\alpha\approx (\theta + \alpha_\mathrm{i}L)/2$ in the limit of low losses. Of course, the assumption that all losses occur at the coupler is in general an approximation, but it holds very well for the low-loss resonators which are the focus of this Article. In this context, it is also worth pointing out that Eq.~\eqref{map} approximates the coupler power transmission coefficient~$\theta$, and the resonator losses~$\alpha$, to be independent of frequency. The frequency-dependence of both coefficients can be readily included in the analysis by expressing the map   Eq.~\eqref{map} in the Fourier domain~\cite{hansson_single_2016}.

The iterative use of Eqs.~\eqref{GNLSE} and~\eqref{map} allows us to calculate the evolution of the slowly-varying envelope~$E_n(z,\tau)$ within the resonator. This modeling approach is often referred to as the ``Ikeda~map,'' acknowledging the pioneering work by Ikeda who used this formulation to study the (chaotic) dynamics of driven Kerr resonators in the cw regime~\cite{ikeda_multiple-valued_1979}; the map described by Eqs.~\eqref{GNLSE} and~\eqref{map} reduces to the specific model studied by Ikeda when ignoring dispersion and SRS. Finally, before proceeding, it is worth emphasizing that $E_\mathrm{n}(z,\tau)$ describes the field inside the resonator; the field at the cavity output,~$E_\mathrm{out}$, consists of a superposition of the portion of the intracavity field that is coupled out, and the portion of the driving field that is reflected off from the input coupler. Mathematically, the output field can be written as
\begin{equation}
    E_{\mathrm{out},n}(\tau) = \sqrt{\theta}\,\exp(-\alpha_\mathrm{i}L/2)\,E_n(L,\tau)\,e^{i\phi_0} - \sqrt{1-\theta}\,E_\mathrm{in}. 
    \label{eq:Eout_map}
\end{equation}
Note that the subtraction of the two fields arises from the presence of a $\pi$~phase shift in the coupler, as required to satisfy energy conservation.

\subsubsection{Linear resonator characteristics}

\begin{figure}[b]
    \centering
    \includegraphics{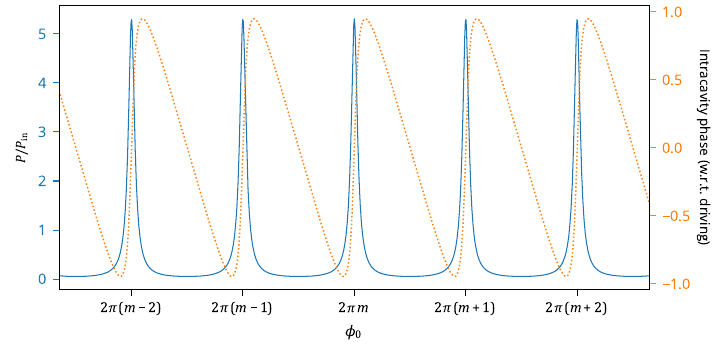}
    \caption{Steady-state intracavity power with respect to driving power, $P/P_\mathrm{in}$ (blue, left axis), and phase response (dotted orange, right axis) of a passive resonator in the linear regime versus linear round-trip phase shift, $\phi_0$. The power curve corresponds to the well known Airy distribution of linear Fabry-Perot resonators, Eq.~\eqref{eq:Airy}, and illustrates the characteristic equidistant resonance peaks and the cavity power enhancement. The phase is that of the intracavity field~$E$ with respect to the driving field~$E_\mathrm{in}$, as obtained from Eq.~\eqref{eq:Es_lin}. Finesse set at~15, and assuming critical coupling}.
    \label{fig:airy}
\end{figure}

We now derive the linear characteristics of a resonator (i.e., we neglect the Kerr nonlinearity,~${\gamma = 0}$). To proceed, we restrict our initial analysis to cw fields with $E_\mathrm{in}(\tau)$ and $E_n(z,\tau)$ both set to be independent of~$\tau$. In this case, Eq.~\eqref{GNLSE} shows that the slowly varying envelope~$E_n$ only depends on the round-trip index~$n$, i.e., $E_n(z,\tau)\rightarrow E_n$. We seek the steady-state intracavity field amplitude by requiring ${E_{n+1} = E_n=E}$ in Eq.~\eqref{map}, leading to, 
\begin{equation}
    \label{eq:Es_lin}
    E = \frac{\sqrt{\theta}}{1-\sqrt{T}\,e^{i\phi_0}}\,E_\mathrm{in}.
\end{equation}
where $T=1-2\alpha$. The corresponding steady-state intracavity power level $P =|E|^2$ reads
\begin{equation}
    \label{eq:Airy}
    P = \frac{\theta P_{\mathrm{in}}}{\bigl(1-\sqrt{T}\bigr)^2}\frac{1}{1+ F \sin^2(\phi_0/2)},
\end{equation}
where $F = 4\sqrt{T}/\bigl(1-\sqrt{T}\bigr)^2$. Equation~\eqref{eq:Airy} corresponds to the well known Airy distribution of linear Fabry-Perot resonators, expressed as a function of the linear round-trip phase shift~$\phi_0$. Periodic maxima --- resonances --- occur when $\phi_0 = 2\pi m$ with $m$ an integer (see Fig.~\ref{fig:airy}). Denoting the angular frequencies that would drive two adjacent resonances as $\omega_0'$ and $\omega_1'$ (note that apostrophes highlight resonance frequencies throughout the text), the corresponding round-trip phase shifts are $\phi_0'=\beta(\omega_0') L = 2\pi m_0$ and $\phi_1' = \beta(\omega_1')L = 2\pi(m_0+1)$ and we have $[\beta(\omega_1')-\beta(\omega_0')]L = 2\pi$. Expanding the $\beta$'s at first order in this expression, we find the frequency spacing between adjacent resonances, also known as the free-spectral range (FSR), as  
\begin{equation}
    \mathrm{FSR} = \frac{\omega_1'-\omega_0'}{2\pi} = \frac{1}{\beta_1 L}=\frac{1}{t_\mathrm{R}},
    \label{eq:FSR}
\end{equation}
where we have used $\beta_1=1/v_\mathrm{g}$ to relate the FSR to the cavity round-trip time~$t_\mathrm{R}$. 

Eq.~\eqref{eq:Airy} also reveals that, on resonance, the intracavity power~$P$ can be much larger than the driving power~$P_\mathrm{in}$ (see Fig.~\ref{fig:airy}). Maximum on-resonance intracavity power is obtained when $\sqrt{T} = 1-\theta$ (or $\alpha \simeq \theta$), a condition known as critical coupling. In this case, $P=P_\mathrm{in}/\theta$. This effect, referred to as cavity power enhancement, is the key to harness effectively nonlinear optical effects in resonators and explain why small losses are particularly important for such applications. Note that when on resonance at critical coupling, $E_\mathrm{out}=0$ [see Eq.~\eqref{eq:Eout_map}], which means that all the driving power is effectively coupled into the resonator.

There are several mutually related metrics that are commonly used to quantify the overall losses of a resonator. The finesse parameter~$\mathcal{F}$ is defined as the ratio between the resonances' separation and their width and is (approximately) inversely proportional to the loss parameter $\alpha$ in Eq.~\eqref{map}:
\begin{equation}
    \label{eq:finesse}
    \mathcal{F} = \frac{\mathrm{FSR}}{\Delta f} = \frac{2\pi}{\Delta\phi_0} \approx \frac{\pi\sqrt{F}}{2} \approx \frac{\pi}{\alpha},
\end{equation}
where $\Delta f$ and $\Delta\phi_0$ are the full-width at half maximum (FWHM) of the resonances in terms of driving frequency and round-trip phase shift, respectively. The approximations above apply to the low-loss limit ($\alpha\ll 1$), for which the individual resonances predicted by Eq.~\eqref{eq:Es_lin} are narrow and have a Lorentzian shape. In this case, the finesse only depends on the losses, and we have $\Delta f = \alpha\cdot\mathrm{FSR}/\pi$ and $\Delta\phi_0 = 2\alpha$. The equation above also makes clear that, at critical coupling, the cavity power enhancement factor is simply given by $\mathcal{F}/\pi$, hence is directly proportional  to the finesse (see Fig.~\ref{fig:airy}).

Another important metric is the quality factor~$Q$, defined as the ratio between the frequency of a resonance and its width (FWHM), 
\begin{equation}
    \label{eq:Qfactor}
    Q = \frac{\omega_0'/(2\pi)}{\Delta f} = \omega_0'\, t_\mathrm{R}\,\frac{\mathcal{F}}{2\pi}.
\end{equation}
Yet a third measure of cavity loss is the photon lifetime, $t_\mathrm{ph}$, which corresponds to the mean lifetime of a photon in the resonator. Equivalently, it can be expressed as the reciprocal of the decay constant describing the exponential decay of the intracavity resonator photon number without driving. The photon lifetime is related to the round-trip time~$t_\mathrm{R}$, the cavity losses~$\alpha$, the finesse parameter~$\mathcal{F}$ and the $Q$-factor as
\begin{equation}
    \label{eq:photon_lifetime}
    t_\mathrm{ph} = \frac{t_\mathrm{R}}{2\alpha} = \frac{\mathcal{F}}{2\pi}\,t_\mathrm{R} = \frac{Q}{\omega'_0}.
\end{equation}
 
\subsubsection{Kerr tilt and bistability}
\label{Kerrtilt}

In the presence of the Kerr nonlinearity, the intracavity field accumulates a nonlinear phase shift due to self-phase modulation (SPM)~\cite{agrawal_nonlinear_2013}, which qualitatively changes the profile of the resonances. Neglecting Raman scattering ($f_\mathrm{R}=0$), and again assuming cw operation, such that $E_n(z,\tau)\rightarrow E_n(z)$, the GNLSE Eq.~\eqref{GNLSE} can be analytically solved to yield 
\begin{equation}
    E_n(L) = E_n(0)\,e^{i\gamma L |E_n|^2}. 
\end{equation}
(Note that $|E_n|^2$ is independent of~$z$ in this case.) The total accumulated phase per round trip thus becomes
\begin{equation}
    \label{eq:nlphase}
    \phi = \phi_0 + \gamma L P\,.
\end{equation}
In this case, Eqs.~\eqref{eq:Es_lin} and~\eqref{eq:Airy} for the steady state intracavity field amplitude~$E$ and power~$P$ still hold, provided that $\phi_0$ is replaced by~$\phi$ in the expressions, leading to
\begin{align}
    \label{eq:Es_NL}
    E &= \frac{\sqrt{\theta}}{1-\sqrt{T}\,e^{i(\phi_0+\gamma L P)}}\,E_\mathrm{in},\\
    \label{eq:AiryNL}
    P &= \frac{\theta P_{\mathrm{in}}}{\bigl(1-\sqrt{T}\bigr)^2} \frac{1}{1 + F \sin^2 \left[ (\phi_0 + \gamma L P)/2\right]}.
\end{align}

\begin{figure}
    \centerline{\includegraphics{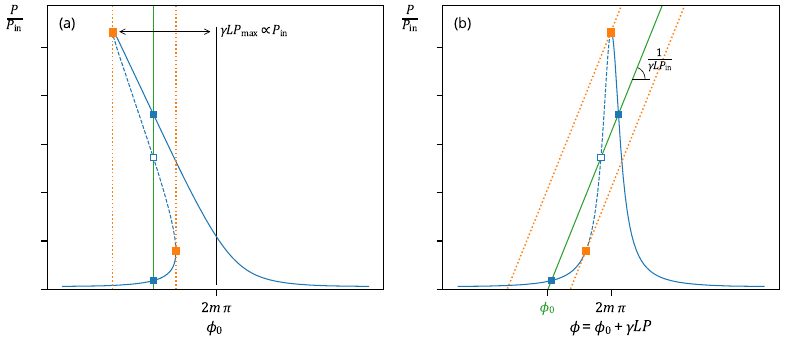}}
    \caption{In blue, steady-state response of a Kerr resonator around a resonance in the nonlinear regime versus (a) the linear round-trip phase shift~$\phi_0$ and (b) the total accumulated phase $\phi=\phi_0+\gamma L P$ for a driving power~$P_\mathrm{in}$ high enough to give rise to optical bistability. The green line is plotted for a particular value of $\phi_0$ in the bistability region and in (b) corresponds to Eq.~\eqref{eq:PPin_phi_lin}. The orange dotted lines, with the same slope as the green lines, are tangent to the blue curves at the orange points and delineate the bistability region. The blue markers show intersections between the blue resonance curves and the green lines. The dashed part of the blue curves and the open blue markers indicate unstable states. Both panels are plotted for the same parameter values.}
    \label{fig:KerrTilt}
\end{figure}

SPM makes the total phase~$\phi$ accumulated by the intracavity field dependent upon the field power~$P$, changing the resonance condition from $\phi_0 = 2\pi m$ in the linear regime to $\phi_0 = 2\pi m - \gamma L P_\mathrm{max}$, where $P_\mathrm{max} = \theta P_\mathrm{in}/(1-\sqrt{T})^2$ is the (maximal) on-resonance cw intracavity power. In contrast, the tails of the resonances, where $P$~is small, are unaffected. Assuming a positive nonlinearity coefficient,~$\gamma>0$, the net effect is that the resonances become tilted towards smaller values of $\phi$ (i.e. red-detuned driving frequencies) [blue curve in Fig.~\ref{fig:KerrTilt}(a)]. If the Kerr shift of the peak of the resonance ($\gamma L P_\mathrm{max}$) is sufficiently large, the resonances can fold upon themselves, as illustrated in Fig.~\ref{fig:KerrTilt}(a). In this condition, there exists a range of values of linear phase shifts~$\phi_0$ (and hence driving frequencies~$\omega_0$) where the resonator response is multivalued, i.e., it exhibits three different steady-states with different power levels. These are highlighted by the blue markers in Fig.~\ref{fig:KerrTilt}(a) for a particular value of~$\phi_0$. This leads to Kerr-induced bistability and hysteresis~\cite{gibbs_differential_1976, kaplan_directionally_1982, smith_optical_1984}. 

To provide a better intuitive understanding of Kerr bistability, we next introduce a geometric framework for the nonlinear response, enabling a simple visualization of the underlying physics \cite{fraile-pelaez_transmission_1991}. To this end, we plot the resonance curve~$P/P_\mathrm{in}$,  Eq.~\eqref{eq:AiryNL}, versus the \emph{total} accumulated phase~$\phi$, as illustrated by the blue curve in Fig.~\ref{fig:KerrTilt}(b). Note how this curve effectively matches with the \emph{linear} resonance (as in Fig.~\ref{fig:airy}). Then, from the expression of the total accumulated phase, Eq.\eqref{eq:nlphase}, we can derive a second relation between $P/P_\mathrm{in}$ and~$\phi$,  
\begin{equation}
    \label{eq:PPin_phi_lin}
    \frac{P}{P_{\mathrm{in}}} = \frac{\phi-\phi_0}{\gamma L P_{\mathrm{in}}}.
\end{equation}
In the geometric representation of Fig.~~\ref{fig:KerrTilt}(b), the above relation corresponds to a straight line with an $x$-intercept of~$\phi_0$ and a slope of~$1/(\gamma L P_\mathrm{in})$, which is inversely proportional to the driving power~$P_\mathrm{in}$ [green line in Fig.~\ref{fig:KerrTilt}(b)]. The steady state intracavity power levels~$P$ must satisfy both relations, which geometrically correspond to the intersections of the green line with the blue resonance curve (blue markers). Note how, in this representation, the green line slides along the horizontal axis and gets a different inclination with a change of $\phi_0$ and~$P_\mathrm{in}$, respectively. 

Figure~\ref{fig:KerrTilt}(b) illustrates the construct described above for a driving power sufficiently high for the green line to have three intersections with the blue resonance curve, leading to a multivalued response. Clearly, this only occurs for a certain range of $\phi_0$ values: the green line must lie in between the two tangents to the resonant curves with the same slope (orange dotted lines, with tangent points indicated by orange markers); the response is univalued outside that range. These conclusions match with the properties of the nonlinear response curve plotted in terms of~$\phi_0$ in Fig.~\ref{fig:KerrTilt}(a). We note that the same parameters are used in both panels of Fig.~\ref{fig:KerrTilt}, and all lines and markers match. 

For low driving powers, we recover the linear resonance with the green line in~(b) nearly vertical and only one intersection for all values of~$\phi_0$. The resonance is thus univalued for all pump phase detunings. In the representation of panel~(b), the threshold of driving power for a multivalued response corresponds to a green line with a slope \emph{lower} than that at the inflexion point of the (linear) resonance curve (with positive slope), or $1/(\gamma L) < \mathrm{max}(d P/d\phi)$. Conversely, if the driving power is high enough, the green line can be inclined sufficiently away from the vertical so as to intersect adjacent neighboring resonances (not shown), leading to more than three intersections (each extra intercepted resonance contributes two extra intersections). This corresponds to a maximum Kerr phase shift $\gamma L P_\mathrm{max}$ exceeding~$2\pi$, but such large nonlinear shifts are not commonly encountered (see, e.g., Fig.~\ref{Fig:Lorentzian_vs_Airy}) \cite{hansson_frequency_2015, anderson_coexistence_2017}.  

The representation of Fig.~\ref{fig:KerrTilt}(b) also provides insights into the cw stability of the nonlinear response of a Kerr resonator. To demonstrate this, we consider a perturbative increase to the intracavity power from a steady-state value. Such increase is accompanied with a larger total accumulated phase~$\phi$, corresponding to an upward displacement along the green line in panel~(b). From the middle intersection point, indicated by the open blue marker, such change in~$\phi$ brings the system closer to resonance and to a further increase in power. This is because the slope of the blue curve is larger than that of the green curve. This positive feedback pushes the system away from the middle point, and towards the other side of the resonance, corresponding to the upper branch in panel~(a). Conversely, a decrease in power from the middle point pushes the system towards the lower branch. The instability condition is thus $dP/d\phi > 1/(\gamma L)$, indicated by the dashed part of the blue curves in Fig.~\ref{fig:KerrTilt}(a) and~(b), delimited by the orange tangency points. As a consequence, only the lower and upper steady-states are stable, and the Kerr tilt thus gives rise to optical \emph{bistability}. By extension, intersections with the next resonance at higher driving power leads to optical \emph{tristability} and so on. 

\subsection{Mean-field model}
\label{sec:mean_field_model}

The discrete map described by Eqs.~\eqref{GNLSE} and~\eqref{map} forms an accurate description of a coherently-driven Kerr cavity --- subject of course to the approximations described above. However, the map model suffers from the fact that (i) analytical analyses (e.g., stability of steady-states) tends to be cumbersome, and (ii) numerical integration tends to be computationally expensive, requiring at least one integration step per round trip. In high-finesse (low-loss) cavities, where the intracavity field undergoes very small changes each round trip, a key simplification can be applied. Here, all the physical effects described by Eqs.~\eqref{GNLSE} and~\eqref{map} can be averaged over one round trip, yielding a mean-field model that describes the intracavity evolution. The derivation of the equation is well-known~\cite{haelterman_dissipative_1992}, so we simply quote the final result: 
\begin{equation}
    t_\mathrm{R}\frac{\partial E(t,\tau)}{\partial t} =
        \left[ -\alpha-i\delta_0 + i L \hat{D} \left( i\frac{\partial}{\partial\tau} \right) \right] E 
	+ i \gamma L \left[ (1-f_\mathrm{R})|E|^2 + f_\mathrm{R} h_\mathrm{R}(\tau)\ast|E|^2 \right] E
        + \sqrt{\theta}\,E_\mathrm{in}.
    \label{GLLE}
\end{equation}
Here, the time $t$ is interpreted as a ``slow time'' variable that describes the evolution of the intracavity field~$E(t,\tau)$ over consecutive round trips, and is related to the round-trip index~$n$ in the discrete map model via $n = t/t_\mathrm{R}$. With that terminology, $\tau$, which describes the field’s temporal structure at
shorter timescales, is referred to as ``fast time.''

The cavity linear round-trip phase shift~$\phi_0$ appears in the mean-field model through the phase detuning parameter $\delta_0 = 2\pi m_0 - \phi_0 = 2\pi m_0 -\beta(\omega_0)L$ that describes the offset of the driving field from the closest cavity resonance (with order~$m_0$). Denoting $\omega'_0$~as the resonance (angular) frequency closest to the driving frequency $\omega_0$, we have $\beta(\omega'_0)L = 2\pi m_0$. Expanding $\beta(\omega'_0)$ as a Taylor series around~$\omega_0$, and retaining only the linear term, we can derive the relationship
\begin{equation}
    \delta_0 = \left[\beta(\omega'_0)-\beta(\omega_0)\right]L \approx \frac{\omega'_0-\omega_0}{\text{FSR}},
\end{equation}
where we used $\beta_1 L = t_\mathrm{R} = \text{FSR}^{-1}$. Thus, the phase detuning $\delta_0$ can be seen to be linearly proportional to the frequency detuning~$\Delta\omega_0 = \omega'_0-\omega_0$ between the driving field and the closest resonance.

 \begin{figure*}[!t]
    \centering
    \includegraphics[width = \textwidth, clip=true]{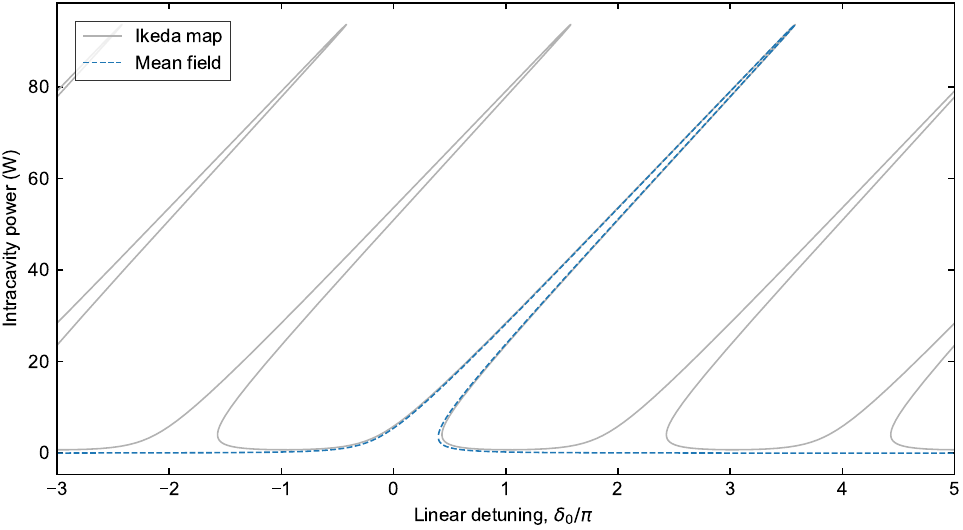}
    \caption{Intracavity power as a function of linear cavity detuning as predicted by the Ikeda map (solid gray curve) and the mean field model (dashed blue curve). Both calculations use $\theta = 0.1$, $P_\mathrm{in} = 25$~W, $\gamma = 1.2~\mathrm{W^{-1}\,km^{-1}}$, and $L = 100$~m. The Ikeda map calculation uses $\alpha = 0.150$ whilst the mean field calculation uses $\alpha = 0.163$. These different $\alpha$ values were chosen so as to match the peak intracavity powers across the models, which would otherwise not agree due to the comparatively low finesse ($\mathcal{F}\approx 20$) of the resonator considered here. }
    \label{Fig:Lorentzian_vs_Airy}
\end{figure*}

The cw ($\partial E/\partial \tau = 0$) steady-state ($\partial E/\partial t = 0$) solutions of Eq.~\eqref{GLLE} are given by
\begin{equation}
    \label{ssL}
    E = \frac{\sqrt{\theta}\,E_\mathrm{in}}{\alpha+i(\delta_0-\gamma L |E|^2)},
\end{equation}
and the corresponding steady-state power levels $P = |E|^2$ then satisfy the cubic polynomial
\begin{equation}
    (\gamma L)^2 P^3 - 2\delta_0 \gamma L P^2 + (\alpha^2 + \delta_0^2)P - \theta P_\mathrm{in} = 0,
    \label{eq:cwcubic_dim}
\end{equation}
where $P_\mathrm{in} = |E_\mathrm{in}|^2$. Notably, Eq.~\eqref{ssL} describes a \emph{single} tilted Lorentzian resonance, which is in stark contrast to the periodic resonances predicted by Eq.~\eqref{eq:Es_NL} of the discrete map [see Fig.~\ref{Fig:Lorentzian_vs_Airy} for a comparison]. It is worth noting also that Eq.~\eqref{ssL} can be derived from Eq.~\eqref{eq:Es_NL} under the assumption that $\alpha$, $\delta_0$, $\gamma L P \ll 1$. This means that, in addition to the requirement that the resonator exhibits low-losses, for Eq.~\eqref{GLLE} to be valid, also the nonlinear (as well as the dispersive) phase-shifts should be small over one round trip. These latter requirements ensure that no neighboring resonances are excited; such excitation can give rise to resonant Kelly-like sidebands [see Section~\ref{sec:Kelly}] and coexistence of multiple nonlinear states~\cite{hansson_frequency_2015,anderson_coexistence_2017} that are not captured by the single-resonance mean-field model.

The mean-field model of Eq.~\eqref{GLLE} provides a highly accurate description of most of the physics at play in a high finesse coherently-driven Kerr resonator. We have written the equation in a particular form that is commonly used in the literature when modeling macroscopic resonators. Before discussing additional physical phenomena, and how such phenomena can be included within this mean-field formalism, in the next Section we discuss the different forms of Eq.~\eqref{GLLE} that can be found in the literature, and the relationships between these. For completeness, we also again emphasize that within the mean-field framework $E(t,\tau)$ explicitly refers to the intracavity field, with the field at the cavity output set by~\cite{leo_temporal_2010},
\begin{equation}
    E_\mathrm{out}(t,\tau) = \sqrt{\theta}\,E-E_\mathrm{in}.
\end{equation}

\subsection{Coupled mode formalism and the integrated dispersion}
\label{sec:CM}

Equation~\eqref{GLLE} can be seen as a time-domain model of the intracavity dynamics. We can derive an equivalent frequency-domain description by expressing the intracavity field $E(t,\tau)$ as a Fourier series of the form
\begin{equation}
\label{FS}
    E(t,\tau) = \sum_\mu a_\mu(t)e^{-i\mu D_1\tau},
\end{equation}
where $D_1 = 2\pi\cdot\text{FSR}$. The coefficients $a_\mu(t)$ represent the modal amplitudes associated with different cavity resonances. 
We note that the validity of the Fourier series representation is commensurate with the fact that the intracavity envelope $E(t,\tau)$ must be periodic in fast time~$\tau$ with periodicity~$t_\mathrm{R}$. 

Substituting Eq.~\eqref{FS} into Eq.~\eqref{GLLE}, and ignoring SRS for simplicity, we can derive the following coupled rate equations for the mode amplitudes $a_\mu(t)$,
\begin{equation}
\label{cme}
    \frac{da_\mu}{dt} = \left[-\frac{\kappa}{2} - i\Delta\omega_0 + i \frac{L}{t_\mathrm{R}}\hat{D}(\mu D_1) \right]a_\mu + i\frac{\gamma L}{t_\mathrm{R}} \sum_{\mu = p+q-r} a_p a_q a_r^\ast + \frac{\sqrt{\theta}}{t_\mathrm{R}}\,E_\mathrm{in}\,\delta_{0\mu}.
\end{equation}
Here $\kappa = 2\pi\Delta f=2\alpha/t_\mathrm{R}$ is the resonance FWHM in angular frequency, $\Delta\omega_0 = \omega'_0-\omega_0 \approx\delta_0/t_\mathrm{R}$ is the frequency detuning of the driving field, and $\delta_{nm}$ is the Kronecker delta.

Before proceeding, we recall that there are two distinct approaches to describing the dispersion of a resonator~\cite{pasquazi_micro-combs_2018}. The first approach (and the one adopted above) describes dispersion using the Taylor series expansion of the frequency-dependent  propagation constant and uses the reduced dispersion  $\hat{D}(\omega-\omega_0)=\beta(\omega)-\beta_0-\beta_1(\omega-\omega_0)$; this formalism is particularly popular in the description of macroscopic fiber resonators, where the mode density is large. On the other hand, in the context of small resonators with large mode spacing, it is common to describe dispersion using a series expansion of the resonance frequencies $\omega'_\mu$ around the driven mode $\omega'_0$. In the latter formalism, the salient effects are captured by the ``integrated dispersion'' defined as
\begin{equation}
    \label{Dint}
    D_\mathrm{int}(\mu) = \omega'_\mu - \omega'_0- \mu D_1 = \sum_{k\geq 2}\frac{D_k}{k!}\mu^k,
\end{equation}
where $D_1 = 2\pi \cdot \text{FSR}$ and $D_k$ with $k>1$ account for deviations of the resonance frequencies~$\omega'_\mu$ from an equidistant, FSR-spaced, grid defined by $\omega'_0 + \mu D_1$. In this context, $\mu = m-m_0$ can be interpreted as a relative mode index, with $m$ and~$m_0$ the actual mode indices of $\omega'_\mu$ and~$\omega'_0$, respectively. 

The two dispersion formalisms are naturally linked via the resonance condition, $\beta(\omega'_\mu)L~=~2\pi m$. Specifically, by subtracting $\beta(\omega'_0)L = 2\pi m_0$, i.e., the resonance condition for the mode closest to the driving, and expanding the $\beta$'s, we have 
\begin{equation}
    \label{eq:beta_D_mu}
    \left[ \beta_0 + \beta_1 (\omega'_\mu-\omega_0) + \hat{D}(\omega'_\mu-\omega_0) \right] L
    -
    \left[ \beta_0 + \beta_1 (\omega'_0-\omega_0) + \hat{D}(\omega'_0-\omega_0) \right] L = 2\pi\mu\,.
\end{equation}
From Eq.~\eqref{Dint}, we can express $\omega'_\mu -\omega_0$ as $\mu D_1 + D_\mathrm{int}(\mu) + (\omega'_0-\omega_0)$, which can be substituted in the first order terms above. Recalling that $\beta_1 L=1/\mathrm{FSR}$ and noting that $\beta_1 L D_1=2\pi$, a number of simplification occurs, leading to the relation
\begin{equation}
    D_\mathrm{int}(\mu) = -L \cdot \mathrm{FSR} \left[ 
    \hat{D}(\omega'_\mu-\omega_0) -
    \hat{D}(\omega'_0-\omega_0)
    \right].
    \label{eq:DintDhat_exact}
\end{equation}
In this expression, the second term (which stems from expanding $\beta$ around $\omega_0$, and not $\omega'_0$) takes into account that GVD at the driving frequency ($\omega_0$) is not quite the same than at the adjacent resonance ($\omega'_0$). However, when operating close to resonance in high finesse resonators, this contribution is very small and can be neglected. In fact, the way we have defined the FSR, Eq.~\eqref{eq:FSR}, also neglects the difference between group velocities at~$\omega_0$ vs that at $\omega'_0$ such that strictly speaking $\beta_1 L D_1$ is only approximately equal to~$2\pi$. Furthermore, we can approximate $\hat{D}(\omega'_\mu-\omega_0)$ as $\hat{D}(\mu D_1)$ provided that $|D_\mathrm{int}(\mu) + (\omega_0'-\omega_0)| \ll |\mu D_1|$ [see Eq.~\eqref{Dint}]. This amounts to assuming that the resonance frequencies only deviate slightly from an equidistant grid, which is generally valid. With these assumptions, Eq.~\eqref{eq:DintDhat_exact} becomes
\begin{equation}
    \label{Dintbeta}
    D_\mathrm{int}(\mu) \approx -L \cdot \text{FSR} \cdot \hat{D}(\mu D_1).
\end{equation}
From this result, we obtain the following approximate relation between the coefficients~$D_k$ appearing in the series expansion of the resonance frequencies and the expansion coefficients~$\beta_k$ associated with the propagation constant~\cite{pasquazi_micro-combs_2018}
\begin{equation}
  D_k \approx -L\cdot\text{FSR}\cdot D_1^k\cdot \beta_k.
  \label{eq:Dkbetak}  
\end{equation}

The above relation is exact for $k=1$ and~$k=2$. For $k>2$, an exact relation can be obtained by inserting the series expansion of Eq.~\eqref{Dint} into Eq.~\eqref{eq:beta_D_mu} and, using the multinomial expansion, identifying the coefficients of each power of~$\mu$, leading to (again neglecting the difference between $\omega_0$ and $\omega_0'$)
\begin{equation}
  D_k = -L\cdot\text{FSR} \cdot k! \,\sum_{l=2}^{k}\ \frac{\beta_l}{l!}\ \sum_{\substack{l_1, l_2, l_3, \ldots \ge 0\\[0.5ex]\sum l_p=l,\ \sum p l_p=k}} \left[ \left(\frac{l!}{l_1!\,l_2!\,l_3!\, \ldots}\right) \prod_{p\ge1} \left(\frac{D_p}{p!}\right)^{l_p} \right]\,.
\end{equation}
A more compact but formal expression can also be obtained from the Lagrange inversion theorem, 
\begin{equation}
  \label{eq:LagInvTheo}
  D_k = \left( \frac{2\pi}{L} \right)^k \, \frac{d^{k-1}}{d\omega^{k-1}} \left( \left[ \frac{\omega-\omega_0'}{\beta(\omega)-\beta(\omega_0')} \right]^k\right)_{\omega \rightarrow \omega_0'}\,.
\end{equation}
Equation~\eqref{eq:LagInvTheo} was first presented in \cite{pasquazi_micro-combs_2018}, unfortunately with a typo, which is fixed here.

To obtain a commonly used form of coupled mode equations, we now substitute Eq.~\eqref{Dintbeta} into Eq.~\eqref{cme} and renormalize the mode amplitudes by defining $b_\mu(t) = a_\mu(t) \sqrt{t_\mathrm{R}/(\hbar\omega_0)}$, such that $|b_\mu(t)|^2$ represents the number of photons per mode. This yields
\begin{equation}
\label{cme2}
    \frac{db_\mu}{dt} = \left[-\frac{\kappa}{2} - i\Delta\omega_0 - i D_\mathrm{int}(\mu) \right]b_\mu + ig \sum_{\mu = p+q-r} b_pb_qb_r^\ast + \sqrt{\kappa_\mathrm{ext}}\,s_\mathrm{in}\,\delta_{0\mu},
\end{equation}
where $g = \hbar\omega_0L\gamma/t_\mathrm{R}^2$ represents the Kerr-induced resonance frequency shift due to one photon, $\kappa_\mathrm{ext} = \theta/t_\mathrm{R}$ is the coupling rate and $s_\mathrm{in}=\sqrt{P_\mathrm{in}/(\hbar\omega_0)}$. 

Dispersion parameters for both formalisms can be obtained from polynomial fit to an experimental measurement of a cavity's resonant frequencies: directly as a function of mode index in the case of the integrated dispersion $D_\mathrm{int}$, and from a polynomial fit to Eq.~\eqref{eq:beta_D_mu} for the propagation constant $\beta$. Typically, for small resonators whose resonance frequencies can be accurately measured the $D_\mathrm{int}$ formalism has been favored in the literature, but both formalisms are equally valid. In addition to the coupled-mode equations, the integrated dispersion can also be readily implemented in envelope models similar to Eq.~\eqref{GLLE}. For example, a notable formulation can be obtained by replacing  $\beta_k \approx -D_k/(L\cdot\text{FSR}\cdot D_1^k)$ in Eq.~\eqref{GLLE} whilst simultaneously introducing {the change of variables $\tau = \Phi/D_1$ and $B(t,\Phi) = E(t,\Phi)\sqrt{t_\mathrm{R}/(\hbar\omega_0)}$, thus yielding
\begin{equation}
    \frac{\partial B(t,\Phi)}{\partial t} = \left[-\frac{\kappa}{2}+i(g|B|^2-\Delta\omega_0) - iD_\mathrm{int}\left(i\frac{\partial}{\partial\Phi}\right) \right]B + \sqrt{\kappa_\mathrm{ext}}\,s_\mathrm{in}. \label{GLLE2}
\end{equation}

In the above model, the variable $\Phi$ corresponds to an azimuthal angle around the circumference of the resonator ($\Phi\in [0,2\pi]$), which is why the model~\eqref{GLLE2} is sometimes referred to as a ``spatiotemporal'' model~\cite{chembo_spatiotemporal_2013} to differentiate from the ``two-time'' models~\cite{matsko_mode-locked_2011,coen_modeling_2013} that invoke the concepts of slow and fast time [e.g., Eq.~\eqref{GLLE}]. However, as described above, all of the different formulations are closely related. In what follows, we will primarily use the two-time model given by Eq.~\eqref{GLLE} or its variants or extensions.

\subsection{Polychromatic, pulsed, and partially coherent driving}
\label{sec:polychromatic}
The models described above have assumed that the resonator is driven with a purely monochromatic cw field, such that the driving amplitude $E_\mathrm{in}$ is a complex scalar. Whilst such simple driving is applicable to many systems and experiments, there is increasing appreciation that polychromatic driving can provide new advantageous functionalities~\cite{jang_temporal_2015,obrzud_temporal_2017-1,hendry_spontaneous_2018,li_efficiency_2022,zhang_spectral_2020,moille_ultra-broadband_2021,qureshi_soliton_2022}. In general, to model more complex (and realistic) driving configurations, the term $E_\mathrm{in}$ must be allowed to vary both in slow and fast time, i.e., $E_\mathrm{in}\rightarrow E_\mathrm{in}(t,\tau)$. In what follows, we will describe the modeling of three important driving paradigms that go beyond purely monochromatic cw fields.

\subsubsection{Polychromatic driving}

As will be discussed in Sections~\ref{sec:synchronization} and ~\ref{sec:extension}, the injection of two (or more) cw fields into a resonator can be used to extend the frequency comb spectrum generated within the resonator, or to lock and control the spacing of that comb. Such polychromatic injection can be modeled with a driving field of the form
\begin{equation}
\label{polychromatic}
    E_\mathrm{in}(t,\tau) = \sum_{q} E_{\mathrm{in},q}e^{-i\mu_q D_1\tau + i[\delta_q-\delta_0 + L\hat{D}(\mu_q D_1)]\frac{t}{t_\mathrm{R}}},
\end{equation}
where the summation is over all the different cw fields launched into the resonator ($q$~is an integer), with the index~$\mu_q$ designating the relative mode number that the field with complex amplitude~$E_{\mathrm{in}, q}$ is exciting~\cite{zhang_spectral_2020,taheri_optical_2017,taheri_all-optical_2022,qureshi_soliton_2022}. The slow time dependence of the terms in Eq.~\eqref{polychromatic} is such that $\delta_q$~is the linear phase detuning of the $q^\mathrm{th}$ field from the resonance it is exciting (with relative mode number~$\mu_q$). It is straightforward to show that 
\begin{equation}
\frac{\delta_q - \delta_0 + L\hat{D}(\mu_q D_1)}{t_\mathrm{R}} = \omega_0 + \mu_q D_1 - \omega_{q},
\end{equation}
where $\omega_q$ is the angular frequency of the $q^\mathrm{th}$ driving field. Thus, the slow time dependence of the terms in Eq.~\eqref{polychromatic} describes the frequency deviation of the $q^\mathrm{th}$ driving field from an equidistant (e.g., frequency comb) grid defined by $\omega_0 + \mu D_1$.

Bichromatic driving, where one of the fields oscillates at the reference frequency $\omega_0$, is arguably the simplest form of polychromatic driving. 
Such driving can be modeled as
\begin{equation}
\label{eq:bichromatic}
    E_\mathrm{in}(t,\tau) = E_{\mathrm{in}, 0} + E_{\mathrm{in}, 1}e^{-i\mu_1D_1\tau + i[\delta_1-\delta_0 + L\hat{D}(qD_1)]\frac{t}{t_\mathrm{R}}}.
\end{equation}
The above expression is obviously a special case of Eq.~\eqref{polychromatic}, and has been written out explicitly as an example.

\subsubsection{Pulsed driving}
\label{sec:pulsed}

In many experiments, a Kerr resonator is driven with a train of pulses whose periodicity, $t_\mathrm{P}$, is approximately equal to the resonator round-trip time $t_\mathrm{R}$ (or a harmonic multiple or division of the round-trip time)~\cite{obrzud_temporal_2017-1,xu_harmonic_2020}. As will be discussed in subsequent sections, such driving can improve the efficiency and fidelity of soliton frequency comb generation and provide control over the comb's characteristics. 

Pulsed driving is a special case of polychromatic driving, where the optical frequencies of the driving field are equally-spaced as $\omega_q = \omega_0 + 2\pi q/t_\mathrm{P}$. Focusing on the specific scenario of near-synchronous driving, the excited modes are equally spaced with a spacing of one FSR, such that $\mu_q = q$, and the detunings~$\delta_q$ can be shown to satisfy
\begin{equation}
    \label{pulsed_deltas}
    \delta_q \simeq \delta_0 - L\hat{D}(qD_1) - q D_1 \Delta \tau,
\end{equation}
where $\Delta \tau = t_\mathrm{R}-t_\mathrm{P}$. Substituting $\mu_q = q$ and $\delta_q$ from Eq.~\eqref{pulsed_deltas} into Eq.~\eqref{polychromatic} reveals a Fourier series representation, where the left-hand-side can be identified as
\begin{equation}
    \label{pulsed_drive}
    E_\mathrm{in}(t,\tau) \rightarrow E_\mathrm{in}\left(\tau + \Delta \tau\frac{t}{t_\mathrm{R}}\right).
\end{equation}
Thus, pulsed driving can be modeled in a temporal envelope equation, such as Eq.~\eqref{GLLE}, by simply using a fast-time dependent pulse profile with appropriate characteristics (e.g., peak power and duration) as the driving term $E_\mathrm{in}$. Non-zero desynchronization ($\Delta \tau \neq 0$) can be modelled by shifting the driving envelope in fast time by $\Delta \tau$ per round trip. Physically, this shifting occurs because the envelope equation is written in a reference frame that propagates with the group velocity of light at $\omega_0$, where one round trip takes $t_\mathrm{R}$ to complete. In some instances, it is useful to redefine the reference frame to be such that the pulse driving is stationary. This can be achieved through a simple change of variable $\tau' = \tau + \Delta \tau\cdot t/t_\mathrm{R}$. As an example, using this change of variable in Eq.~\eqref{GLLE} and assuming pulsed driving, we obtain the equation~\cite{coen_convection_1999,li_ultrashort_2024}
\begin{align}
    t_\mathrm{R}\frac{\partial E}{\partial t} = & \left[-\alpha-i\delta_0-\Delta \tau \frac{\partial}{\partial \tau'} + i L \hat{D}\left(i\frac{\partial}{\partial\tau'}\right) \right]E + \sqrt{\theta} \, E_\mathrm{in}(\tau') \nonumber \\
    & +i\gamma L \left[(1-f_\mathrm{R})|E|^2 + f_\mathrm{R}h_\mathrm{R}(\tau')\ast|E|^2 \right]E. 
    \label{GLLEpulsed}
\end{align}
As can be seen, the driving term is now independent of the slow time~$t$, but a convective drift term proportional to~$\Delta \tau$ has appeared to reflect the change in reference frame. We note that for practical purposes (e.g., for numerical implementation) the convective term can be simply incorporated into the linear dispersion operator $\hat{D}$ by extending the sum therein to $k=1$ [see Eq.~\eqref{Dhat}] and defining an effective $\beta_1$~coefficient such that $\beta_1 L = \Delta \tau$. This further highlights the fact that the convective drift term added to the equation above simply corresponds to a change in group velocity, making the intracavity field slip (or advance) with respect to the driving field at the prescribed rate. 

\subsubsection{Continuous wave pump with phase noise}
All realistic driving fields carry noise. For example, a realistic monochromatic cw injection will exhibit amplitude and phase jitter that can influence the intracavity dynamics. Noise can in general be modeled by including appropriate statistical fluctuations in the driving term $E_\mathrm{in}(t,\tau)$ along the fast and slow times. In most cases, however, the noise bandwidth is much smaller than the resonator FSR, in which case fluctuations along the fast time can be ignored. In what follows we discuss, as a particular example, the modeling of a cw driving field subject only to phase noise. 

A cw driving field subject to phase noise can be modeled as
\begin{equation}
    E_\mathrm{in}(t) = A_\mathrm{in} e^{i\delta\phi(t)},
\end{equation}
where $A_\mathrm{in}$ is the scalar amplitude of the field and $\delta\phi(t)$ is a random phase fluctuation. The phase fluctuation is related to the corresponding fluctuation in the driving field's instantaneous frequency viz.~\cite{frosz_soliton_2006, anderson_photonic_2021}
\begin{equation}
\delta f(t) = \frac{1}{2\pi}\frac{d(\delta\phi)}{dt},
\end{equation}
which allows us to write
\begin{equation}
\label{deltaphi}
\delta\phi(t) = 2\pi \int_0^t \delta f(t')\,dt'.
\end{equation}

 \begin{figure*}[!b]
    \centering
    \includegraphics[width = \textwidth, clip=true]{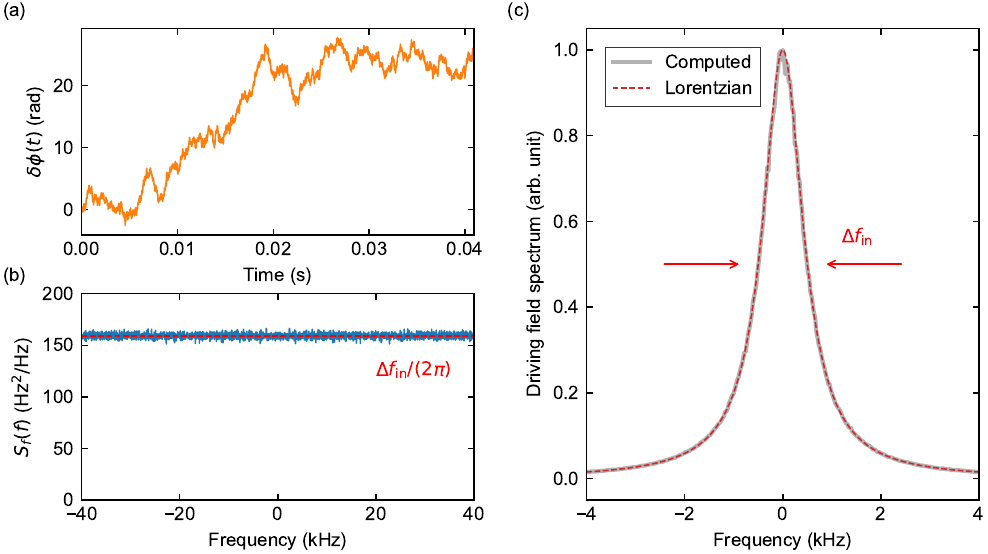}
    \caption{Illustrative modeling of driving laser phase noise. (a) One realization of the slow time dependent phase fluctuation (Brownian motion) arising from frequency noise with constant frequency noise PSD. (b) Blue curve depicts the ensemble average of 5000 frequency noise PSD realizations, showing white noise with constant $S_f(f) = S_0 = \Delta f_\mathrm{in}/(2\pi)$ as highlighted by the red dashed line. (c) Gray curve shows the ensemble averaged optical spectrum arising from 5000 phase fluctuation realizations similar to that shown in panel (a). The red dashed curve shows a Lorentzian profile with FWHM~$\Delta f_\mathrm{in}$.}
    \label{Fig:phasenoise}
\end{figure*}

The precise form of the phase fluctuations depends upon the random process that governs the frequency fluctuation $\delta f(t)$. The details are encapsulated in the frequency and phase noise power spectral densities (PSDs) $S_f(f)$ and $S_\phi(f)$ which correspond to the Fourier transforms of the autocorrelation functions of $\delta f(t)$ and $\delta \phi(t)$, respectively: 
\begin{align}
S_{\phi}(f)&=\int^{\infty}_{-\infty}\langle\delta \phi(t)\delta \phi(t+t')\rangle e^{-i2\pi f t'}dt'\nonumber\\
&=\int_{-\infty}^{\infty} \left( \lim_{T \to \infty} \frac{1}{T} \int_{-T/2}^{T/2} \delta \phi(t) \delta \phi(t + t') \, dt \right)e^{-i2\pi f t'}dt'\\
S_{f}(f)&=\int_{-\infty}^{\infty} \left( \lim_{T \to \infty} \frac{1}{T} \int_{-T/2}^{T/2} \delta f(t) \delta f(t + t') \, dt \right)e^{-i2\pi f t'}dt'
\end{align}

Given Eq.~\eqref{deltaphi}, the two PSDs are related through $S_\phi(f)=S_f(f)/f^2$~\cite{goodman2015statistical,kim2016ultralow}. For a driving field with constant frequency PSD $S_f(f) = S_0$ (and thus a phase noise PSD that decreases as $S_\phi(f)=S_0/f^2$), the frequency fluctuation $\delta f(t)$ is Gaussian white noise with zero mean and variance $\sigma_{\delta f}^2$ that is proportional to the optical linewidth ($\Delta f_\mathrm{in} =2\pi S_0$) of the driving field. Fundamental quantum limited phase noise is an example of this process, with $\Delta f_\mathrm{in}$ in this case the Schawlow-Townes linewidth.

To implement the phase noise model described above in simulations, the slow time $t$ is first discretized with a sampling time $t_\mathrm{s}$ (i.e., $t_{n+1}-t_n = t_\mathrm{s}$) that ideally coincides with the integration step size used in the simulations. The variance of the frequency fluctuation is then computed explicitly as $\sigma_{\delta f}^2 = \Delta f_\mathrm{in}/(2\pi t_\mathrm{s})$~\cite{frosz_soliton_2006}, which allows one to construct the discretized frequency fluctuation $\delta f(t_n)$, the phase fluctuation $\delta\phi(t_n)$, and finally the driving field $E_\mathrm{in}(t_n)$. When implemented correctly, an ensemble average of this process yields a uniform frequency noise PSD ($S_f(f) = S_0$) and a Lorentzian profile for the spectrum (in slow time) of the driving field, with linewidth $\Delta f_\mathrm{in} = 2\pi S_0$ [see Fig.~\ref{Fig:phasenoise}]. Lastly, we note that the above process can also be readily used to model the timing jitter of a pulsed driving field by including appropriate slow-time dependent phase fluctuations $\delta\phi_q(t)$ in Eq.~\eqref{polychromatic}. In the special (but important) case where the amplitude of phase fluctuations increases linearly as a function of relative mode number, viz. $\delta\phi_q(t) = qD_1 j(t)$, where $j(t)$ is an appropriate stochastic process, the pulsed driving Eq.~\eqref{pulsed_drive} can be written in the time domain as
\begin{equation}
E_\mathrm{in}(t,\tau)\rightarrow E_\mathrm{in}\left(\tau+\Delta\tau\frac{t}{t_\mathrm{R}}+j(t)\right).
\end{equation}
This form identifies $j(t)$ as the timing jitter of the pulsed driving source.

\subsection{Mode coupling}
\label{sec:mode_coupling}
In general, resonators can sustain more than one (spatial or polarization) mode family~\cite{matsko_optical_2006, strekalov_nonlinear_2016}. Whilst theoretical mode analyses often show that the different mode families are orthogonal, this result is strictly valid only in idealized resonators. Pragmatically, perturbations such as fabrication imperfections and material inhomogeneities can break the mode orthogonality, resulting in linear coupling between different mode families~\cite{strekalov_nonlinear_2016, carmon_static_2008, herr_mode_2014,ramelow_strong_2014, liu_investigation_2014}. Analogous linear coupling also occurs when multiple resonators are physically coupled to one another~\cite{xue_normal-dispersion_2015,helgason_dissipative_2021, liao_photonic_2020}, with such systems sometimes referred to as photonic molecules to highlight this coupling. Moreover, even in idealised resonators with purely orthogonal modes, nonlinear effects can cause the intracavity field in one mode to affect the intracavity field in another mode~\cite{xu_spontaneous_2021, coen_nonlinear_2024}. Mode coupling, whether linear or nonlinear, and whether occurring within a single resonator or across multiple coupled resonators, can profoundly influence resonator dynamics. Below we describe the physics and modeling of two particularly important coupling mechanisms.

\subsubsection{Linear mode coupling and avoided crossings}
\label{avoided_crossings}

As described above, unavoidable perturbations in real resonators break the orthogonality between different mode families, resulting in linear coupling that can convert energy from one mode to another. In the context of dissipative solitons and optical frequency combs, one of the most important consequences of linear mode coupling is the resultant splitting and shifting of cavity resonances, which results in spectrally localized perturbations to the resonator dispersion~\cite{strekalov_nonlinear_2016, carmon_static_2008, herr_mode_2014,ramelow_strong_2014, liu_investigation_2014}. To understand this phenomenon, we begin by neglecting any nonlinearity and consider cw fields associated with two modes with complex amplitudes $E_1(t)$ and $E_2(t)$ and that are driven at the same frequency~$\omega_0$. We generalize Eq.~\eqref{cme} to account for the two linearly-coupled modes and write the resultant equations in the following form,
\begin{align}
    \frac{d E_1(t)}{d t} &= \left[-\frac{\kappa_1}{2}-i\Delta\omega_{1} \right]E_1 - i\zeta_{12}E_2 + \sqrt{\theta_1}\,E_\mathrm{in,1}, \label{Eq:AMX1}\\
    \frac{d E_2(t)}{d t} &= \left[-\frac{\kappa_2}{2}-i\Delta\omega_{2} \right]E_2 - i\zeta_{21}E_1 + \sqrt{\theta_2}\,E_\mathrm{in,2},
    \label{Eq:AMX2}
\end{align}
where $\Delta\omega_{1,2}$ represent frequency detuning of the two intracavity fields with their respective resonances and $\zeta_{12} = \zeta_{21}^\ast = \zeta$ are the mode coupling coefficients. All other variables, previously defined, are distinguished by subscripts corresponding to the two modes. 

Because we are interested in the shift in resonance frequencies, we ignore losses and driving in Eqs.~\eqref{Eq:AMX1} and~\eqref{Eq:AMX2} to derive the resonances of the coupled system. Moreover, we peel away the carrier-envelope decomposition by writing $E_{1,2}(t) = \mathcal{E}_{1,2}(t)\exp(i\omega_0 t)$ to obtain
\begin{align}
\frac{d \mathcal{E}_1(t)}{d t} &= -i\omega'_{01} \mathcal{E}_1 - i\zeta\mathcal{E}_2 \\
\frac{d \mathcal{E}_2(t)}{d t} &= -i\omega'_{02} \mathcal{E}_2 - i\zeta^\ast\mathcal{E}_1,
\end{align}
where $\omega'_{01}$ and $\omega'_{02}$ are the resonance frequencies of the two modes in the absence of coupling. The dynamics of this linear coupled system is governed by the matrix \begin{equation}
\text{M} = \begin{bmatrix}
\omega'_{01} & \zeta  \\
\zeta^\ast & \omega'_{02},
\end{bmatrix}
\end{equation}
whose eigenvalues are the new resonance frequencies. These eigenvalues can be readily calculated as,
\begin{equation}
    \omega'_\pm = \frac{\omega'_{01} + \omega'_{02}}{2} \pm \sqrt{\frac{(\omega'_{01} - \omega'_{02})^2}{4} + |\zeta|^2}. \label{ompn}
\end{equation}

\begin{figure*}[!t]
    \centering
    \includegraphics[width = \textwidth, clip=true]{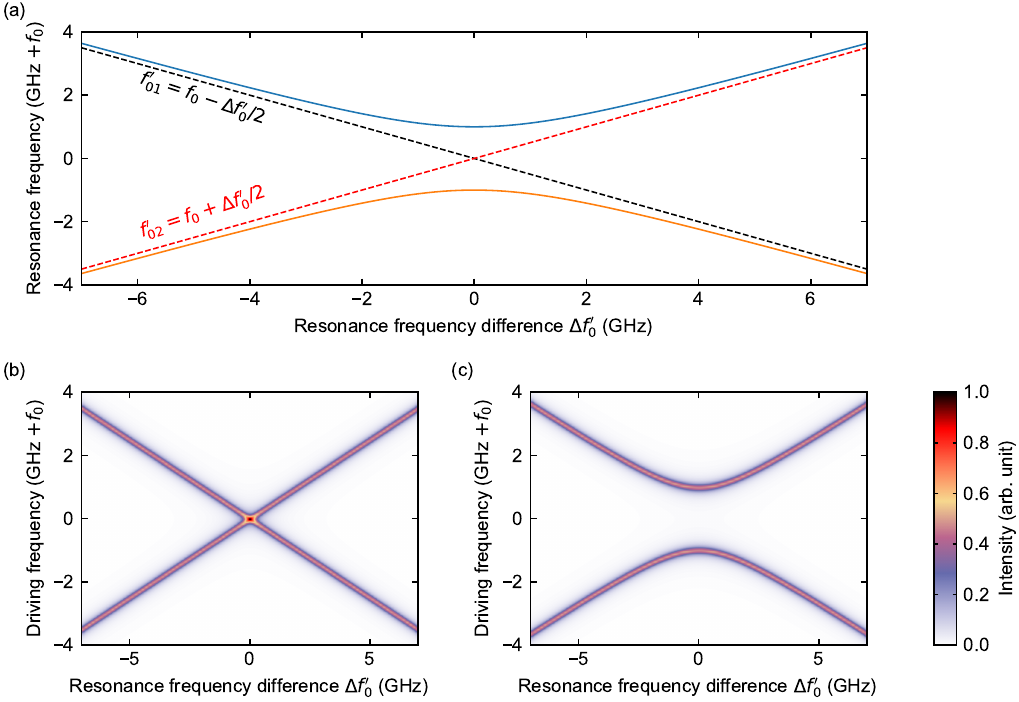}
    \caption{(a) Solid curves show the hybridized resonance frequencies $f'_\pm = \omega'_\pm/(2\pi)$ that arise due to linear mode coupling (with $\zeta = 2\pi\times 1~\mathrm{GHz}$) of two cavity modes with resonance frequencies $f'_{01} = f_0 -\Delta f'_0/2$ and $f'_{02}= f_0 + \Delta f'_0/2$, plotted as a function of the frequency difference~$\Delta f'_0$. Dashed curves show the resonance frequencies $f'_{01}$ and $f'_{02}$ without mode coupling ($\zeta = 0$). (b) and (c) show the normalized steady-state intracavity intensity obtained by solving for the roots of Eqs.~\eqref{Eq:AMX1} and~\eqref{Eq:AMX2} when (b) neglecting and (c) including mode coupling. Both calculations use $\kappa_{1} = \kappa_2 = 2\pi\times 0.2~\mathrm{GHz}$, with $\zeta = 0$ in panel (b) and $\zeta = 2\pi\times 1~\mathrm{GHz}$ in panel (c). In the presence of mode coupling, the intracavity intensities depend sensitively on the relative phase of the driving fields $E_\mathrm{in,2}$ and $E_\mathrm{in,1}$; the calculations in (b) and (c) used $E_\mathrm{in,1} = E_\mathrm{in,2}\exp(i\pi/2)$ to yield a symmetric intensity distribution in (c).}
    \label{fig:AMX}
\end{figure*}

The eigenstates associated with the resonance frequencies~$\omega'_\pm$ are mixtures of the modes $E_1$ and~$E_2$, often referred to as hybrid modes or supermodes \cite{helgason_dissipative_2021,liao_photonic_2020}. Figure~\ref{fig:AMX}(a) schematically illustrates the hybridized resonance frequencies $f'_\pm = \omega'_\pm/(2\pi)$ as a function of the frequency difference between the original resonances, $\Delta f'_0=f'_{02}-f'_{01}$. As can be seen, for $|\Delta f'_0|~\gg~|\zeta|/(2\pi)$, the hybridized resonance frequencies (eigenstates) tend towards the original resonances $f'_{01}$ and~$f'_{02}$ of modes $E_1$ and~$E_2$ respectively. Strong coupling and mode hybridization only occurs when the resonance frequencies are sufficiently close to each other. In particular, whilst the uncoupled resonances would cross as $\Delta f_0' \rightarrow 0$,  coupling causes the hybrid modes to avoid each other, resulting in an ``avoided mode crossing.''

\begin{figure*}[!t]
    \centering
    \includegraphics[width = \textwidth, clip=true]{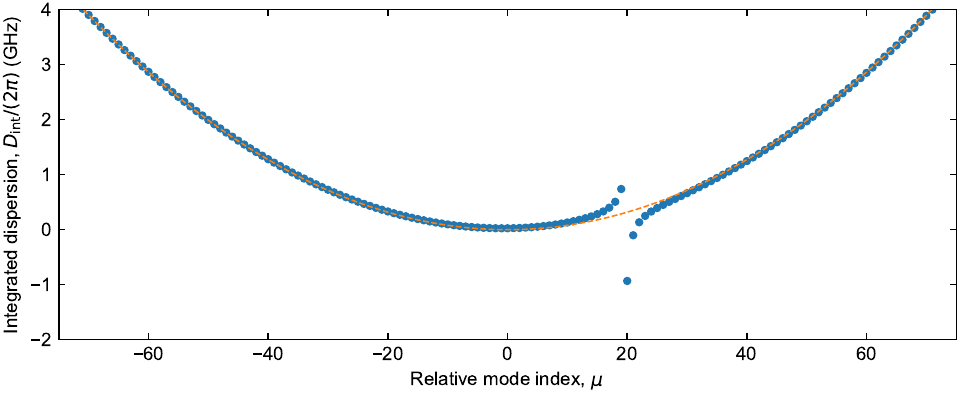}
    \caption{Blue solid circles show the integrated dispersion of the $\omega'_+$ supermode in the presence of linear mode coupling between two modes with $\zeta = 2\pi\times2~\mathrm{GHz}$. The two modes have an FSR difference of 3~GHz and their resonances overlap at~$\mu = 20$, giving rise to a strong, localized perturbation on the integrated dispersion (dashed curves shows the integrated dispersion of the modes without mode coupling).  }
    \label{fig:AMX_dispersion}
\end{figure*}

Figures~\ref{fig:AMX}(b) and~(c) illustrates the avoided mode crossing phenomenon by explicitly depicting the steady-state solutions to Eqs.~\eqref{Eq:AMX1} and~\eqref{Eq:AMX2} in the absence [Fig.~\ref{fig:AMX}(b)] and presence [Fig.~\ref{fig:AMX}(c)] of mode coupling. Here, the pseudocolor plot shows the total steady-state intracavity intensity $|E_1|^2 + |E_2|^2$ as a function of the driving laser frequency (y-axis) for the same range of resonance frequencies and $\Delta f'_0$ as in Fig.~\ref{fig:AMX}(a). We see clearly the avoided mode crossing in the presence of coupling, with the observed intensity distributions following closely the hybridized resonance frequencies derived from Eq.~\eqref{ompn}. 

Cavity dynamics in the presence of linear mode coupling can be rigorously modeled using coupled equations such as Eqs.~\eqref{Eq:AMX1} and~\eqref{Eq:AMX2} when generalized to account for all the different physical phenomena of interest as in Eq.~\eqref{GLLE} [including, e.g., nonlinearity and dispersion, as well as a relative walk-off term analogous to desynchronization as in Eq.~\eqref{GLLEpulsed} to model modes with different FSRs~\cite{xue_super-efficient_2019} or complex driving configurations as in Section~\ref{sec:polychromatic}]. A simpler modeling alternative is to use a single mean-field equation [e.g., Eq.~\eqref{GLLE}], but modify the dispersion of the driven mode family to account for the shifting of the hybridized resonance around the mode crossing(s). In this context, because the interacting modes often have a large FSR difference, mode coupling typically results in dispersive perturbations that are localized in frequency (see Fig.~\ref{fig:AMX_dispersion}). Such localized perturbations cannot easily be modeled using a polynomial expansion of the dispersion operator~$\hat{D}$ (or equivalently the integrated dispersion~$D_\mathrm{int})$. Instead, one typically uses the full dispersion profile (without polynomial approximations) in the frequency domain when solving the relevant evolution equation. The dispersion profile can be straightforwardly synthesized based upon the actual (measured or simulated) resonance frequencies of the system to construct the integrated dispersion $D_\mathrm{int}$ [see Eq.~\eqref{Dint}], which can then be mapped to the dispersion operator~$\hat{D}$ [see Eq.~\eqref{eq:DintDhat_exact}] if desired. As a final point regarding linear mode coupling, we emphasize that, whilst our discussion has revolved around the coupling between two different mode families in a single resonator, the formalism also applies to systems comprised of multiple physically coupled resonators; in this case, $E_1$ and $E_2$ in Eqs.~\eqref{Eq:AMX1} and~\eqref{Eq:AMX2} would refer to intracavity fields within two distinct resonators~\cite{xue_normal-dispersion_2015,helgason_dissipative_2021, liao_photonic_2020}. In ring-type resonators, the formalism also applies to the two possible counter-rotating directions that can be coupled, e.g., via Rayleigh backscattering~\cite{weiss_splitting_1995,gorodetsky_rayleigh_2000,yang_counter-propagating_2017} or via a photonic-crystal topology fabricated on the resonator~\cite{yu_spontaneous_2021,yu_continuum_2022}.

\subsubsection{Nonlinear mode coupling and spontaneous symmetry breaking}
\label{sec_ssb}

Even when two modes are completely orthogonal, i.e., not linearly coupled, they can still interact through the Kerr effect. In particular, the Kerr nonlinearity gives rise to intensity-dependence of the refractive index, which can cause two modes to modify each other's phase through the process of cross-phase-modulation (XPM)~\cite{agrawal_nonlinear_2013}. Assuming cw operation (such that dispersion can be neglected), and likewise ignoring SRS, we can generalize Eq.~\eqref{GLLE} to model two modes coupled via Kerr XPM:  
\begin{align}
t_\mathrm{R}\frac{d E_1(t)}{d t} &= \left[-\alpha_1-i\delta_{01} + i\gamma L(|E_1|^2 + B|E_2|^2) \right]E_1 + \sqrt{\theta_1}\,E_\mathrm{in,1}, \label{Eq:SSB1}
\\
t_\mathrm{R}\frac{d E_2(t)}{d t} &= \left[-\alpha_2 - i\delta_{02}+i\gamma L(|E_2|^2 + B|E_1|^2) \right]E_2 + \sqrt{\theta_2}\,E_\mathrm{in,2},
\label{Eq:SSB2}
\end{align}
where $B$ is the XPM coefficient. 

A particularly interesting aspect of the model described by Eqs.~\eqref{Eq:SSB1} and~\eqref{Eq:SSB2} is that it exhibits spontaneous symmetry breaking (SSB)~\cite{kaplan_directionally_1982, haelterman_polarization_1994}. Specifically, even if both modes have identical characteristics (same losses, detunings, and driving amplitudes), such that the system is symmetric with respect to an interchange of the two modes ($E_1\rightleftharpoons E_2$), the steady-state intracavity power levels can be different, i.e., $|E_1|^2\neq|E_2|^2$. This is illustrated in Fig.~\ref{Fig:SSB}, which shows the cw steady-state intracavity power levels as a function of the common detuning $\delta_0 = \delta_{01} = \delta_{02}$ with all other parameters listed in the caption. As can be seen, the symmetric steady-state solution loses its stability at some detuning threshold, through a pitchfork bifurcation, making way to \emph{two} asymmetric, mirror-like (interchanging $E_1$ and~$E_2$ is also a solution by symmetry) steady-states that are stable. As the detuning is increased further, the asymmetric states disappear through a reverse pitchfork bifurcation, resulting in the re-emergence of a stable symmetric state.

\begin{figure}[t]
  \centerline{\includegraphics{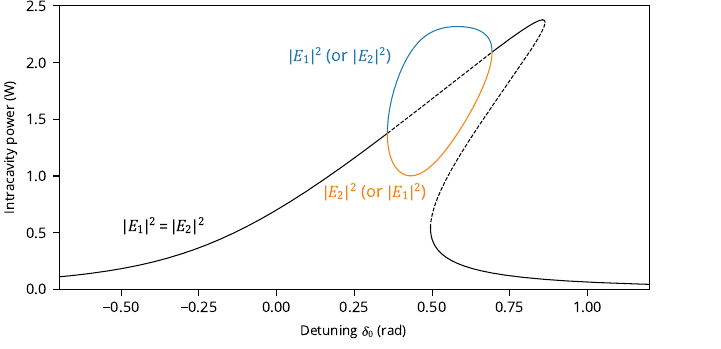}}
  \caption{Intracavity power levels of cw states in a two-mode resonator illustrating the occurrence of spontaneous symmetry breaking in some parameter ranges. Black curves represent symmetric solutions ($|E_1|^2 = |E_2|^2$) while blue and orange curves highlight asymmetric states ($|E_1|^2\neq|E_2|^2$). Dashed lines correspond to unstable solutions. $B=2$, $P_\mathrm{in}=1.26$~W (split equally between the two modes). All other parameters are the same as in Fig.~\ref{Fig:Lorentzian_vs_Airy}.}
  \label{Fig:SSB}
\end{figure}

The emergence of the asymmetric states described above is a characteristic feature of SSB. In the context of driven Kerr resonators, such symmetry breaking dynamics have been experimentally observed using two counter-propagating fields~\cite{cao_experimental_2017, del_bino_symmetry_2017} as well as two co-propagating fields with orthogonal polarizations~\cite{garbin_asymmetric_2020}. In this latter work, SSB has also been shown to be possible under asymmetric conditions, provided that two different asymmetries are properly balanced. Additionally, a topological symmetry protection mechanism has also been identified, involving a periodic roundtrip-to-roundtrip alternation of the two modes, which leads to an average of all asymmetries~\cite{coen_nonlinear_2024}. Whilst the majority of these studies have focused on the SSB of cw states, experiments have also shown that SSB also affects localized temporal structures, giving rise to asymmetric two-component bright CSs~\cite{xu_spontaneous_2021, xu_breathing_2022, coen_nonlinear_2024} as well as optical domain walls~\cite{garbin_dissipative_2021, coen_nonlinear_2024}.

\subsection{Thermal nonlinearity}
\label{sec:thermal}

In addition to the Kerr and Raman nonlinearities, externally driven resonators can also experience strong thermal nonlinearity.  Fundamentally, the resonator’s temperature is governed by its average intracavity power, a portion of which is absorbed and transformed into heat. This thermal buildup alters the cavity’s resonance frequencies via the thermorefractive effect and thermal expansion. Thermal effects are particularly significant for very small resonators (e.g., monolithic microresonators), which struggle to efficiently dissipate heat to the environment. 

To understand the effect of the thermal nonlinearity, we recount the phenomenological analysis in~\cite{carmon_dynamical_2004}. We start from the resonance condition for the driven mode with frequency $\omega_0'$, assumed to be associated with mode index~$m_0$
\begin{equation}
\label{Eq:thermal1}
    \beta(\omega_0')L = 2\pi m_0,
\end{equation}
Here, $\beta(\omega_0') = \omega_0' n_\mathrm{eff} /c$ defines the propagation constant of the mode, with $n_\mathrm{eff}$ the effective refractive index. Thermal expansion modifies the physical resonator length according to $L = L_0(1 + \alpha_\mathrm{L} \Delta T)$, with $\alpha_\mathrm{L}$ representing the linear thermal expansion coefficient, $\Delta T$ the deviation from ambient temperature, and $L_0$ the ``cold'' resonator length (i.e., for $\Delta T = 0$). Additionally, the thermo-refractive effect alters the refractive index such that, to first order, $n_\mathrm{eff} = n_{\mathrm{eff},0} + \Delta T dn_\mathrm{eff}/dT$, where $n_{\mathrm{eff},0}$ is the cold-cavity effective index. Incorporating these thermal dependencies into Eq.~\eqref{Eq:thermal1} yields:
\begin{equation}
\label{Eq:thermal2}
    \omega_0'\approx \omega_{0,\mathrm{c}}'(1-a\Delta T),
\end{equation}
where $\omega_{0,\mathrm{c}}'$ is the cold cavity resonance frequency and $a = \alpha_\mathrm{L} + n_\mathrm{eff,0}^{-1} dn_\mathrm{eff}/dT$. The resonator heats (cools) via absorption of intracavity light (heat dissipation), allowing to write an expression for the rate of change of the resonator temperature as
\begin{equation}
\label{Eq:thermal3}
    \frac{d\Delta T}{dt} = \frac{\alpha_\mathrm{abs}}{C_\mathrm{p}}P- \frac{\Delta T}{\tau_0},
\end{equation}
where $\alpha_\mathrm{abs}$ is the proportion of light absorbed by the resonator per round trip, $C_\mathrm{p}$ is the heat capacity (with units of~J\,K$^{-1}$), and $\tau_0 = C_\mathrm{P}/K$ is the thermal response time with $K$ the  thermal conductance (with units of~W\,K$^{-1}$). For small resonators, which are most susceptible to thermal effects, typical thermal response times are of the order of microseconds or less, corresponding to multiple resonator round-trip times. Thus, the intracavity power $P$ in Eq.~\eqref{Eq:thermal3} should be understood as the average, rather than the instantaneous, intracavity power, i.e.,
\begin{equation}
    P(t) = \frac{1}{t_\mathrm{R}}\int_{-t_\mathrm{R}/2}^{t_\mathrm{R}/2} |E(t,\tau)|^2\, \mathrm{d}\tau.
\end{equation}

In steady-state, $d\Delta T/dt = 0$, yielding $\Delta T = \tau_0\alpha_\mathrm{abs}P/C_\mathrm{P}$. Substituting this value into Eq.~\eqref{Eq:thermal2}, we find that, due to thermal effects, the resonance frequencies shift in linear proportion to the average intracavity power as
\begin{equation}
    \omega_0' = \omega_{0,\mathrm{c}}'\left(1 - \frac{a\tau_0\alpha_\mathrm{abs}}{C_\mathrm{P}} P\right)\,.
\end{equation}
Accordingly, the frequency detuning between the external driving source and the cavity resonance becomes similarly power-dependent,
\begin{equation}
\label{Eq:thermDet}
    \Delta \omega = \Delta\omega_{\mathrm{c}} - \zeta P,
\end{equation}  
where $\Delta\omega_{\mathrm{c}} = \omega_{0,\mathrm{c}}' - \omega_0$ is the cold cavity detuning and $\zeta = a\omega_{0,\mathrm{c}}' \tau_0\alpha_\mathrm{abs}/C_\mathrm{P}$.

Recalling that $\delta_0 = \Delta\omega_0 t_\mathrm{R}$, and that for cw fields the average power $P = |E|^2$, the cw steady-state solutions of the cavity [see Eq.~\eqref{ssL}] in the presence of thermal effects become
\begin{equation}
      E = \frac{\sqrt{\theta}E_\mathrm{in}}{\alpha+i(\delta_{0,\mathrm{c}} - t_\mathrm{R}\zeta|E|^2-\gamma L |E|^2)}.
\end{equation}
This expression shows that, just like the Kerr nonlinearity, the thermal effect adds a power dependent phase shift to the intracavity field, resulting in a ``thermal tilt'' of the cavity resonances. Most resonators of interest have a positive thermo-refractive coefficient ($dn_\mathrm{eff}/dT > 0$) such that $\zeta>0$; in this case, the thermal nonlinearity tilts the resonances from blue to red (just like the focussing Kerr nonlinearity with $\gamma > 0$). We note that the theoretical framework established in Section~\ref{Kerrtilt} for the Kerr nonlinearity can be trivially extended to include the thermal nonlinearity, with the same methodology. Accordingly, it should be clear that a sufficiently large thermal nonlinearity gives rise to three steady-state solutions, only two of which are stable (i.e., bistability). 
 
When both Kerr and thermal nonlinearities are included, the total accumulated phase shift per round trip can be written as,
\begin{equation}
    \phi = \phi_0 + \gamma L P\left(1+ \frac{t_\mathrm{R}\zeta}{\gamma L} \right).
\end{equation}
This expression highlights how the thermal nonlinearity contributes to the slope of the straight lines defined by Eq.~\eqref{eq:PPin_phi_lin}, and that are visualised in Fig.~\ref{fig:KerrTilt}(b)~\cite{li_stably_2017}. We note that in physically large resonators with efficient thermal dissipation, such as macroscopic fiber ring resonators, the Kerr nonlinearity typically dominates over thermal effects. By contrast, physically small systems such as microresonators often exhibit pronounced thermal nonlinearities, with thermally induced phase shifts often exceeding those arising from Kerr effects. Considering, for example, a typical silicon nitride microresonator~\cite{li_stably_2017} with $dn/dT=2.5\times10^{-5}~{\mathrm{K}^{-1}}$, $t_\mathrm{R}= 1.0~\mathrm{ps}$, $L = 2\pi\times 23~\mathrm{\mu m}$, $\gamma = 1.0~\mathrm{W^{-1}m^{-1}}$, $K = C_\mathrm{P}/\tau_0 = 2.8\times 10^{-4}~\mathrm{W K^{-1}}$ and $\alpha_\mathrm{abs} = 4.1\times 10^{-5}$ (corresponding to an absorption $Q_\mathrm{abs} = \omega_0 t_\mathrm{R}/\alpha_\mathrm{abs}$ which in turn corresponds to $K_\mathrm{eff}^{-1} = 0.6$ as defined in ref.~\cite{li_stably_2017}), we obtain a thermal phaseshift: $\phi = \phi_0 + \gamma L P\times 16$. This example shows that the thermal nonlinearity is more than an order of magnitude stronger than the instantaneous Kerr nonlinearity for these typical microresonator parameters. Finally, we note that the thermo-refractive coefficient varies significantly with temperature in many materials. As a result, operating microresonators within a cryogenic environment can suppress thermal effects sufficiently to recover pure Kerr-induced bistability, even in devices with very small physical dimensions~\cite{moille_Kerr-microresonator_2019}.



The descriptions above have focused on the steady-state configurations. Outside of steady-state, the impact of thermal effects can be modeled (see, e.g., ref.~\cite{herr_temporal_2014}) by simultaneously solving the differential equations for the intracavity optical field [e.g., Eq.~\eqref{GLLE}], with detuning $\delta_0 = \delta_{0,\mathrm{c}} - \omega_{0,\mathrm{c}}'t_\mathrm{R} a\Delta T$, and the temperature change [Eq.~\eqref{Eq:thermal3}]. The salient thermal parameters ($\tau_0$ and $\alpha_\mathrm{abs}/C_\mathrm{P}$) can be estimated from experimental measurements. Due to the very different timescales of the near-instantaneous Kerr nonlinearity and the microsecond to millisecond thermal nonlinearity, the interplay between the two effects can give rise to rich dynamical behaviours~\cite{diallo_giant_2015, lauro_parametric_2017}.

\subsection{Normalization and the Lugiato-Lefever equation}
\label{sec:norm}

The discussions above provide the theoretical background to model and understand the dynamics of temporal localized structures and corresponding frequency combs in dispersive, externally driven Kerr resonators. In the following Sections, we will discuss the physics and dynamics of the structures that manifest themselves under different dispersion conditions, focusing on the minimal conditions that are required to describe each structure. To this end, we will typically neglect the thermal nonlinearity and only invoke it phenomenologically when necessary. To conclude this section, we introduce a normalized form of the Lugiato-Lefever equation. This requires that we introduce the following variable transformations:
\begin{align}
    t\rightarrow \frac{\alpha t}{t_\mathrm{R}}, \qquad \tau\ (\text{or}\ \tau') \rightarrow \frac{\tau}{\tau_\mathrm{s}}, \qquad E\rightarrow E\sqrt{\frac{\gamma L}{\alpha}},
\end{align}
where the fast time normalization timescale $\tau_\mathrm{s} = \sqrt{|\beta_2|L/(2\alpha)}$. Applying these transformations to Eqs.~\eqref{GLLE} and~\eqref{GLLEpulsed} (the latter including the drift term describing desynchronization) yields
\begin{align}
    \frac{\partial E}{\partial t} = 
        \left[ -1 -i\Delta -d \frac{\partial}{\partial\tau} + i \hat{D} \left(i\frac{\partial}{\partial\tau}\right) \right] E 
        + i \left[ (1-f_\mathrm{R})|E|^2 + f_\mathrm{R}\Gamma(\tau,\tau_\mathrm{s})\ast|E|^2 \right] E 
        + S.
    \label{GLLEN}
\end{align}
Here, the normalized detuning is defined as $\Delta = \delta_0/\alpha$, while the normalized Raman response function takes the form $\Gamma(\tau,\tau_\mathrm{s}) = \tau_\mathrm{s}h_\mathrm{\tau}(\tau\tau_\mathrm{s})$~\cite{wang_stimulated_2018}, and the normalised driving strength is expressed as $S = E_\mathrm{in}\sqrt{\gamma L \theta/\alpha^3}$. Additionally, the dispersion operator is normalized as
\begin{equation}
    \hat{D}\left(i\frac{\partial}{\partial\tau}\right) \rightarrow \sum_{k\geq 2} d_k \left(i \frac{\partial}{\partial \tau}\right)^k,
    \label{eq:Dhat_normalized}
\end{equation}
with the normalized dispersion coefficients $d_k = \beta_k L/(\alpha\, \tau_\mathrm{s}^k\, k!)$. Again the sum can be extended to $k=1$ to incorporate the desynchronization term, with $d_1$ defined as the other $d_k$~coefficients but using the effective group-velocity coefficient introduced in Section~\ref{sec:pulsed} through the relation $\beta_1 L = \Delta \tau$, with $\Delta \tau$ the desynchronization time per round trip. This yields
\begin{equation}
    d_1 = d = \frac{\Delta \tau}{\alpha\,\tau_\mathrm{s}}.
    \label{d_coefficient_normalized}
\end{equation}

When the spectral width of the intracavity field is sufficiently narrow, group-velocity dispersion (GVD) can be truncated at second order, such that $\beta_k = 0$ for all $k > 2$. Additionally, in the absence of desynchronization ($\Delta\tau = 0$) and assuming stimulated Raman scattering is negligible ($f_\mathrm{R} = 0$) Eq.~\eqref{GLLEN} reduces to:

\begin{equation}
\label{LLE}
    \frac{\partial E}{\partial t} = \left[ -1+i(|E|^2-\Delta) - i\eta \frac{\partial^2}{\partial\tau^2} \right] E + S,
\end{equation}
where $\eta = d_2 = \text{sign}[\beta_2]$. 

Equation~\eqref{LLE} is the ``canonical form'' of a dispersive, coherently-driven Kerr resonator. The equation is equivalent with that derived by Lugiato and Lefever in~1987 to describe pattern formation in spatially-diffractive Kerr resonators \cite{lugiato_spatial_1987}, and for this reason, the equation above is often referred to as the Lugiato-Lefever equation (LLE). In the context of dispersive resonators, the LLE was first derived and applied by Haelterman and Wabnitz in 1992 to describe the dynamics of an optical fiber ring resonator \cite{haelterman_dissipative_1992}. Often in today's literature, Eq.~\eqref{LLE} as well as its generalizations [such as Eq.~\eqref{GLLE}] are simply referred to as the LLE; we adopt the same relaxed convention in this Article, adding further specifications only where necessary. 

Before proceeding, we note for the sake of completeness that, in normalized form, the cw steady-state solutions of Eq.~\eqref{LLE} satisfy
\begin{equation}
    E = \frac{S}{1+i(\Delta-Y)},
    \label{eq:normss}
\end{equation}
where $Y = |E|^2$ is the normalized intracavity power level. This power level in turn satisfies the well known cubic relation of optical bistability~\cite{haelterman_additive-modulation-instability_1992},
\begin{equation}
    Y^3 - 2\Delta Y^2 + (\Delta^2+1)Y = X,
    \label{eq:cwcubic_norm}
\end{equation}
where $X=|S|^2$ is the normalized driving power. Of course, relations given by Eqs.~\eqref{eq:normss} and~\eqref{eq:cwcubic_norm} are simply the normalized forms of Eqs.~\eqref{ssL} and~\eqref{eq:cwcubic_dim}, respectively. Operating in normalized units, however, results in a particularly elegant formulation of the bistability equations. For example, the peak of the tilted resonance is  intuitively located at normalised detuning $\Delta=X$. For this reason, the normalised values of detuning $\Delta=\delta_0/\alpha$, intracavity power $Y=|E|^2$, and drive power $X=|E_\mathrm{in}|^2\gamma L \theta/\alpha^3$ will be used extensively throughout the remainder of the manuscript.

In the next Section, we will begin by using Eq.~\eqref{LLE} to explain the origins and canonical dynamics of arguably the most important localized structure in Kerr resonators -- the dissipative temporal Kerr cavity soliton. 



\section{Temporal cavity solitons in the anomalous dispersion regime}
\label{sec:CSanomalous}

The preceding Chapter has introduced the salient theoretical frameworks that are required to model and explain the generation, dynamics, and characteristics of dissipative solitons in coherently driven Kerr resonators. In this Section, we will apply these theoretical frameworks to describe the specific physics of temporal Kerr cavity solitons (CSs\cite{leo_temporal_2010}; also known as dissipative Kerr solitons, DKS\cite{herr_temporal_2014}) --- bright pulses of light that can manifest themselves in the regime of anomalous group-velocity dispersion. However, before discussing CSs themselves, we begin by describing another universal nonlinear wave phenomenon that also plays an important role in driven Kerr resonators, namely modulation instability (MI -- also known as the Benjamin-Feir instability~\cite{benjamin_disintegration_1967}). As we shall see, MI is a key precursor for the emergence of CSs.

\subsection{Modulation instability} \label{sec_MI}
Modulation instability describes the growth and evolution of periodic perturbations on a plane wave (i.e., cw) background~\cite{zakharov_modulation_2009}. It is a universal phenomenon that manifests itself in many different physical contexts, including single-pass optical fiber propagation~\cite{hasegawa_tunable_1980,van_simaeys_experimental_2001,kibler_peregrine_2010}, hydrodynamics~\cite{benjamin_disintegration_1967, benjamin_bifurcation_1978}, cold atoms~\cite{nguyen_formation_2017} and plasma physics~\cite{dewar_modulational_1972}. In the context of driven Kerr resonators, MI corresponds to a sideband instability of the cw steady-states of the system, which results in the emergence of a periodic pattern that fills the entire resonator~\cite{haelterman_dissipative_1992}. We note that there is a close analogy between MI in driven cavities and ubiquitous pattern formation dynamics in systems driven far from equilibrium~\cite{cross_pattern_1993, purwins_dissipative_2010}.

MI can be theoretically examined through a linear stability analysis of the canonical LLE, Eq.~\eqref{LLE}. Substituting an ansatz of the form
\begin{equation}
    E(t,\tau) = E_\mathrm{s} + a(t)e^{-i\Omega\tau} + b(t)e^{i\Omega\tau}
\end{equation}
into Eq.~\eqref{LLE}, where $E_\mathrm{s}$ is the cw steady-state of the system, and linearizing with respect to the small-signal sideband amplitudes $a(t)$ and $b(t)$, one can derive the linear system $d\vec{u}/dt = M \vec{u}$, where $\vec{u} = [a(t),b^\ast(t)]^\mathrm{T}$ and the matrix
\begin{equation}\label{MI_matrix}
M = \begin{bmatrix}
-1+i\kappa & iE_\mathrm{s}^2  \\
-i(E_\mathrm{s}^2)^\ast & -1-i\kappa
\end{bmatrix},
\end{equation}
with $\kappa = 2Y-\Delta+\eta\Omega^2$ and $Y = |E_\mathrm{s}|^2$. The eigenvalues $\lambda_\pm$ of the matrix $M$ describe the exponential growth rate of the sideband amplitudes, and can readily be derived as
\begin{align}\label{eq:MIgain}
    \lambda_\pm &= -1 \pm \sqrt{Y^2 - \kappa^2}, \\
    & = -1 \pm \sqrt{4Y(\Delta - \eta \Omega^2) - (\Delta - \eta\Omega^2)^2 - 3Y^2}.
\end{align}

MI occurs when $\lambda_+$ is real and positive, which strictly requires that the intracavity intensity $Y > 1$. The frequency that experiences maximum gain can be found from the condition $\kappa = 0$ to be $\Omega_\mathrm{M} = \sqrt{\eta (\Delta - 2Y)}$. The requirement that $\Omega_\mathrm{M}$ is real sets a second condition for the occurrence of MI (in addition to $Y>1$). In the anomalous dispersion regime, we have $\eta = -1$ and MI can therefore occur only when $Y > \Delta/2$; this corresponds to the upper branch of the bistable cavity response [see Fig.~\ref{fig:MIgain}(a)]. Figure~\ref{fig:MIgain}(b) shows typical MI gain spectra as a function of the detuning $\Delta$ for the case of anomalous dispersion ($\eta = -1$), considering a normalized driving power $X = 10$. As can be seen, the frequencies of maximum gain (dashed red curves) shift further from the driving frequency as the detuning increases, and the bandwidth of the gain spectrum increases. We note that this analysis also predicts that, unlike MI in conservative NLSE systems, driven cavities can support MI in the normal dispersion regime ($\eta = +1$). A detailed analysis of this problem finds that the resultant MI occurs for a small range of detunings close to the up-switching point on the lower branch of the cavity's bistability curve~\cite{coen_modulational_1997, coen_bistable_1999}. Because of this, scalar intracavity MI in the normal dispersion regime is often irrelevant and will not be considered further here. 

\begin{figure}[t]
  \centerline{\includegraphics[width = \textwidth, clip=true]{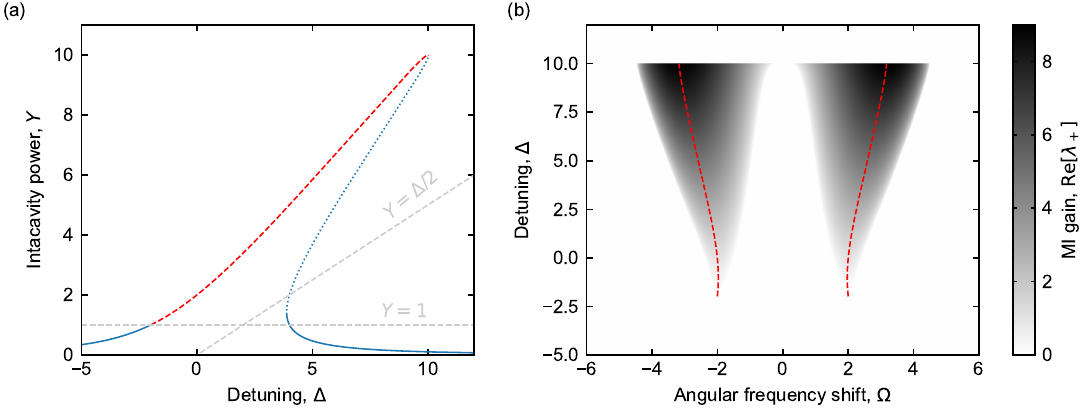}}
  \caption{(a) Steady-state intracavity power levels computed using Eq.~\eqref{eq:cwcubic_norm} with normalized driving power $X = 10$. In the anomalous dispersion regime ($\eta = - 1$), MI occurs for intracavity power levels $Y > 1$ and $Y>\Delta/2$, which can occur in the upper branch of the bistability curve (red dashed curve). The middle branch with negative slope is homogeneously unstable (blue dotted curve). (b) MI gain spectra computed for~$X = 10$ as a function of~$\Delta$. Red dashed curve highlights the peak MI gain frequency, $\Omega_\mathrm{M} = \sqrt{\eta(\Delta-2Y)}$.}
  \label{fig:MIgain}
\end{figure}

In contrast to single-pass fiber propagation systems, where MI gives rise to periodic recurrence akin to the celebrated Fermi-Pasta-Ulam-Tsingou recurrence~\cite{fermi_studies_1955,akhmediev_deja_2001,van_simaeys_experimental_2001}, in dissipative cavities MI results in the irreversible transformation of the cw background into a train of ultrashort pulses. Figure~\ref{fig:MIdynamics} presents numerical results [obtained via direct integration of Eq.~\eqref{LLE} with an initial condition corresponding to white noise] that show typical temporal and spectral characteristics of resonator MI in the anomalous dispersion regime for two different values of the cavity detuning~$\Delta$ (both calculations use ${X = 10}$ as in Fig.~\ref{fig:MIgain}). Figures~\ref{fig:MIdynamics}(a)--(d) consider a low detuning $\Delta = -1.5$ that is just above the MI threshold of $\Delta = 1-\sqrt{X-1} = -2$ (see Fig.~\ref{fig:MIgain}) [the MI threshold of detuning is obtained by setting $Y=1$ into Eq.~\eqref{eq:cwcubic_norm}]. We see how spectral sidebands are amplified which results in the formation of a stable periodic pattern (often referred to as a Turing pattern) that reaches steady state. Figures~\ref{fig:MIdynamics}(e)--(h) shows MI dynamics for a larger detuning of~$\Delta = 3$. Here, in stark contrast, we observe that the emergent pattern never reaches steady-state but instead exhibits spatiotemporally chaotic dynamics~\cite{liu_characterization_2017}. 

The Turing patterns emerging from MI [c.f.\ Fig.~\ref{fig:MIdynamics}(c)] correspond to stable steady-state solutions of the LLE. Careful analysis shows that such patterned solutions exist over a broad range of detunings (and for a broad range of temporal periods), but that the solutions become unstable as the detuning increases~\cite{coen_universal_2013}. The spatiotemporally complex dynamics observed in Fig.~\ref{fig:MIdynamics}(e) and~(f) can be viewed as a manifestation of the instability of an underlyling steady-state pattern. As we shall see in the following Section, the existence of patterned states is closely linked to the existence of bright CSs.

\begin{figure}[t]
  \centerline{\includegraphics[width = \textwidth, clip=true]{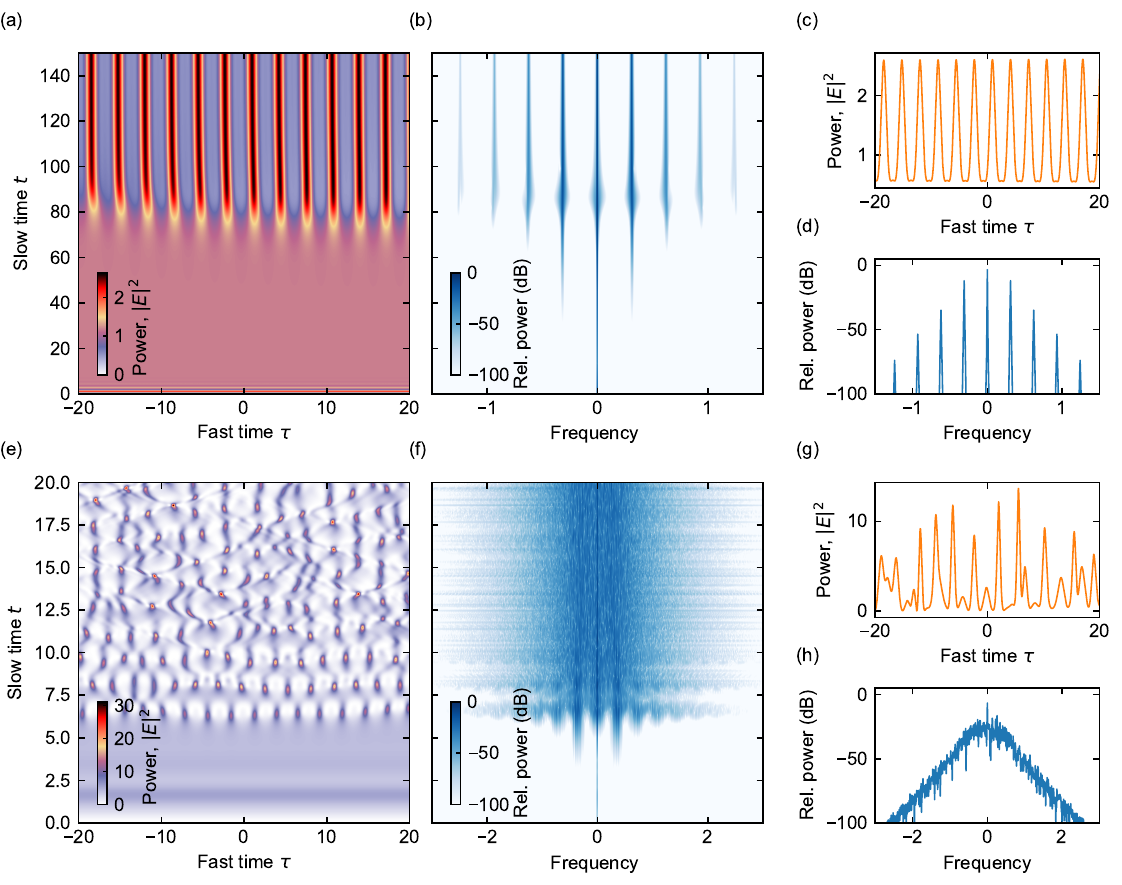}}
  \caption{Numerical simulation results, showing the dynamical evolution of modulation instability for two different detunings. (a)--(d) consider $\Delta = -1.5$, with panels (a) and (b) showing the evolution of the temporal and spectral intensity profiles, respectively. (c) and (d) show the final simulated temporal and spectral intensity profiles, respectively. (e)--(h) are as in (a)--(d) but for a larger detuning $\Delta = 3$. All simulations use normalized driving power $X = 10$ and assume an initial condition consisting of low-amplitude white noise.}
  \label{fig:MIdynamics}
\end{figure}


\subsection{From modulation instability to cavity solitons}
To understand how CSs arise from MI, it is useful to first consider the case of weak driving, such that the cw response of the cavity is mono-stable, and examine the MI states that emerge as the system  passes the threshold of instability (i.e., the point where the intracavity intensity $Y = 1$). In the immediate vicinity of the threshold, the pattern states that emerge can be written as
\begin{equation}
E(\tau) = E_\mathrm{s} + A_\mathrm{M} \cos(\Omega_\mathrm{M}\tau + \phi_\mathrm{M}),
\end{equation}
where $E_\mathrm{s}$ is the homogeneous steady-state at threshold ($|E_\mathrm{s}|^2 = Y \approx 1)$, $\Omega_\mathrm{M}$ is the frequency of maximal gain, and $A_\mathrm{M}$ and $\phi_\mathrm{M}$ are the amplitude and phase of the modulation, respectively. We note that, because of the system's time translation symmetry, $\phi_\mathrm{M}$ can take any value. 

In a seminal 1987 paper by Lugiato and Lefever~\cite{lugiato_spatial_1987}, the authors apply bifurcation theory to analytically study the patterned states close to the instability threshold. This analysis revealed that, in the anomalous dispersion regime, the modulation amplitude of patterned states near threshold satisfies the following condition:

\begin{equation}
|A_\mathrm{M}|^2 =\frac{72(\Delta-2)^2}{4(30\Delta-41)} (1-Y).
\label{eq:MIamp}
\end{equation}

Equation~\eqref{eq:MIamp} reveals that, depending on the detuning $\Delta$, the modulation amplitude, $|A_\mathrm{M}|^2$, can exhibit two qualitatively different behaviours as a function of the intracavity power $Y$. If $\Delta < 41/30$, the modulation amplitude increases with the intracavity power $Y$, whilst the opposite is true for $\Delta > 41/30$. These two behaviours correspond to two different types of bifurcations relating to the instability: the former corresponds to a \emph{supercritical} bifurcation whilst the latter corresponds to a \emph{sub-critical} bifurcation. 

\begin{figure}[b]
  \centerline{\includegraphics[width = \textwidth, clip=true]{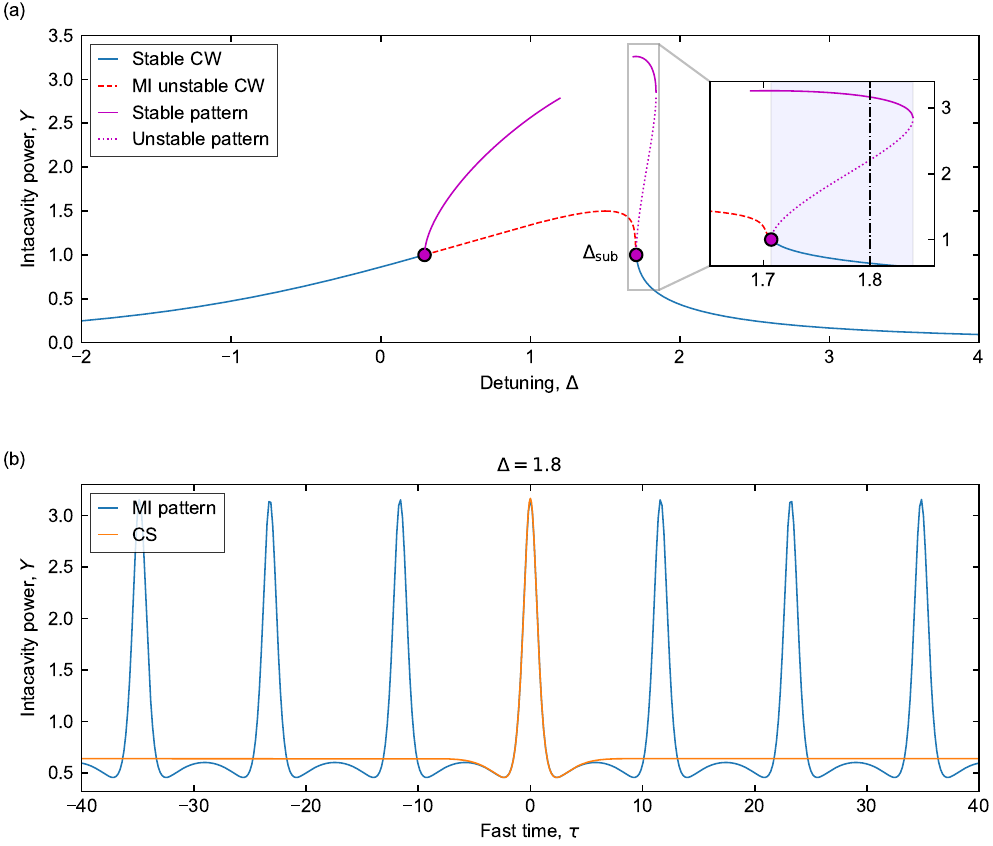}}
  \caption{(a) Magenta curves show the bifurcation characteristics of MI patterns for a normalized driving power $X = 1.5$, obtained using Newton continuation. The frequency of the patterns was set at the frequency of maximum gain at the two distinct MI thresholds. The curves correspond to the peak power of the patterned state. Inset shows the bifurcation characteristics of the pattern emerging sub-critically at $\Delta_\mathrm{sub}\approx 1.7 >41/30$. The solid blue and dashed red curves show the stable and modulationally unstable cw steady-states. While not shown for clarity, the Newton solutions agree with predictions from Eq.~\eqref{eq:MIamp} in the immediate vicinity of the bifurcation points. Lastly, note that only portions of the MI patterns are shown for clarity. (b) Blue and orange curves superimpose a stable steady-state pattern and a CS that exist at the same detuning of $\Delta = 1.8$.}
  \label{fig:MIsubcritical}
\end{figure}

At the sub-critical bifurcation (occurring for $\Delta>41/30$), the modulated pattern that emerges coexists with a stable cw background. This can be readily understood from Eq.~\eqref{eq:MIamp}: for $\Delta>41/30$, the requirement that $|A_\mathrm{M}|^2 \geq 0$ implies $Y < 1$. To quantitatively illustrate this point, the magenta curves in Fig.~\ref{fig:MIsubcritical}(a) depict the bifurcation characteristics of the periodic patterns that emerge at the MI thresholds for a normalized driving power $X = 1.5$. These curves were obtained by using Newton continuation to find those steady-state pattern solutions of the LLE whose frequency coincides with the most unstable frequency at the MI threshold (i.e., $\Omega_\mathrm{M}$). The mathematical formalism behind Newton continuation is presented in Section~\ref{sec:Newton}.

For the small driving power $X = 1.5$ considered, the cw response [shown as solid blue and dashed red curves in Fig.~\ref{fig:MIsubcritical}(a)] does not exhibit bistability. The single cw state is unstable for a range of detunings, with two MI thresholds occurring for detunings smaller and larger than the peak of the tilted resonance. For small detunings ($\Delta < 41/30$), the MI pattern emerges \emph{supercritically}, with the pattern branch emerging towards the region where the cw response is unstable. In contrast, at the MI threshold associated with a larger detuning ($\Delta > 41/30$), the bifurcation is \emph{sub-critical}, and the pattern emerges towards the region where the cw response is stable. Here, due to the sub-critical nature of the bifurcation, a region of detunings exists where the patterned state can coexist with a stable ($Y < 1$) cw background. 

Crucially, this analysis demonstrates that the coexistence between a patterned state and a stable cw background enables a hybrid field configuration, wherein the intracavity field exhibits a patterned state over one region of fast time and reverts to a cw state elsewhere. Bright cavity solitons (CSs) can be viewed in precisely in this manner -- as a single cycle of the modulation instability (MI) pattern embedded within a stable cw background [c.f. Fig.~\ref{fig:MIsubcritical}(b)].
We refer the reader to~\cite{parra-rivas_bifurcation_2018} for a detailed analysis of the bifurcation structure of Turing pattens and bright cavity solitons in this regime.

\subsection{CS existence and characteristics}
\label{sec:CSexist}
The analysis above focused on the case of weak driving, where the cw response is mono-stable. For larger driving powers (which are more relevant to temporal CS experiments), the cw response becomes bistable and the MI threshold in the $\Delta 
> 41/30$ regime can no longer be accessed by reducing the detuning. Indeed, for $X\approx 2$, the Kerr tilt is sufficiently strong such that the MI threshold in the $\Delta > 41/30$ regime falls within the homogeneously unstable middle branch of the cw response. Nonetheless, the qualitative origins of CSs remains the same: a patterned state bifurcates sub-critically and coexists with the stable, lower-state cw background.  

However, for higher driving powers ($X\gg1$, corresponding to all experimental configurations discussed in this article), the 
sub-critical pattern ceases to exist~\cite{parra-rivas_bifurcation_2018} and the requisite coexistence between a patterned state and a stable cw background is realized at detunings above the upswitching point $\Delta_\uparrow$ that marks the beginning of cw bistability; the stable cw background that surrounds the solitons then corresponds to the lower level of the cw bistability curve. The solitons exist up to a finite maximum detuning, which can be shown using soliton perturbation theory [see Section~\ref{sec:Lagragian}] to be approximately given by $\Delta_\mathrm{max} \approx \pi^2 X/8$, with the validity of the approximation improving as $X$ increases~\cite{barashenkov_existence_1996,herr_temporal_2014}.

Mathematically, just like MI patterns, CSs correspond to steady-state solutions of the LLE Eq.~\eqref{LLE}. The blue trace in Fig.~\ref{fig:CSbif}(a) shows the peak power of CS solutions of the LLE as a function of detuning $\Delta$, obtained for a constant driving power $X = 10$ using a Newton solver and numerical continuation~\cite{coen_universal_2013}. Also shown for reference are the (peak) power levels of the cw steady-state solutions, as well as of the MI pattern solutions that bifurcate supercritically from the cw background (and whose frequency coincides with the maximum MI gain at threshold). The results shown in Fig.~\ref{fig:CSbif}(a) corroborate the approximate CS existence limits described above ($\Delta_\mathrm{CS}\in[\Delta_\uparrow,\Delta_\mathrm{max}]$), and show how the MI and CS branches approximately meet at the up-switching point. The results also reveal the existence of two CS branches: a lower saddle branch (dotted trace)  that is always trivially unstable, and an upper branch that is stable at large detunings. As the detuning approaches the up-switching point $\Delta_\uparrow$, complex dynamical instabilities emerge due the presence of a Hopf bifurcation located at $\Delta_{\mathrm{Hopf}}$ (dashed trace)~\cite{leo_dynamics_2013,anderson_observations_2016}. These instabilities will be discussed in detail in Section~\ref{sec:CSinstabilities}.

Figures~\ref{fig:CSbif}(b) and (c) show illustrative temporal and spectral intensity profiles for a stable CS obtained at a cavity detuning of $\Delta=10$. This temporally localised solution is in stark contrast to the extended MI pattern solution obtained at $\Delta=-1.52$ and shown in Figs.~\ref{fig:CSbif}(d) and (e). Whilst there is no exact analytical expression for the CS solution, an approximation can be derived from perturbation theory for $\Delta\gg1$ ~\cite{wabnitz_suppression_1993,nozaki_chaotic_1985}, see Section~\ref{sec:Lagragian}. It reads
\begin{equation}
E(\tau) = \sqrt{2\Delta}\,\mathrm{sech}\left( \sqrt{\Delta}\,\tau \right) e^{i\phi} + E_\downarrow, \label{eq:CSsol}
\end{equation}
where $E_\downarrow$ is the lower-state cw field amplitude and satisfies Eq.~\eqref{eq:normss} while the phase~$\phi$ satisfies $\cos \phi = 2\sqrt{2\Delta}/(\pi S)$. It is worth noting that the maximum detuning of CS existence~$\Delta_\mathrm{max}$ corresponds to the situation where the soliton is in-phase with the driving field ($\cos\phi = 1$), representing the condition of maximum energy transfer from the drive to the soliton.

\begin{figure}
  \centerline{\includegraphics[width = \textwidth, clip=true]{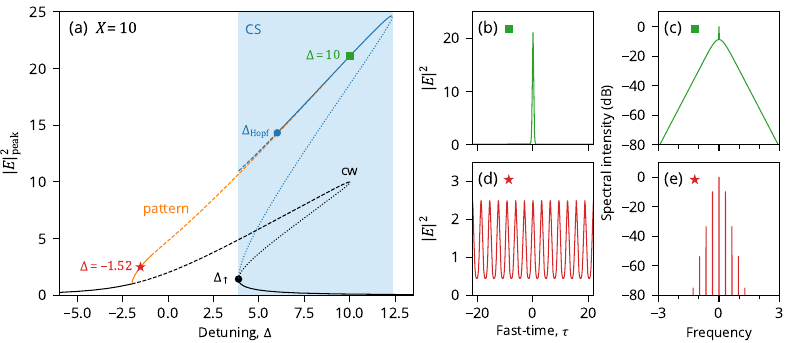}}
  \caption{(a) The peak power of solutions of the LLE as a function of detuning $\Delta$, obtained for a constant driving power $X = 10$. The CS, MI and cw responses are plotted as blue, orange and black traces respectively. Solid lines represent stable solutions, and dotted and dashed lines unstable solutions due to saddle and Hopf instabilities respectively. The location of the Hopf bifurcation is marked as $\Delta_{\mathrm{Hopf}}$. Panels (b) and (c) show the temporal and spectral profiles of a CS at $\Delta=10$. Panels (d) and (e) show the temporal and spectral profiles of an extended MI pattern at $\Delta=-1.52$. Adapted with permission from S. Coen \textit{et al.}, Opt. Lett. \textbf{38}, 1790-1792 (2013)~\cite{coen_universal_2013}. Copyright \textcopyright~2013 Optical Society of America. 
  \label{fig:CSbif}}
\end{figure}

The approximate solution given by Eq.~\eqref{eq:CSsol} immediately reveals that the soliton's duration (spectral width) decreases (increases) with an increase in detuning. Accordingly, the minimum normalised 3~dB duration (full width at half maximum) attainable for a given driving power is $\Delta\tau'_\mathrm{min}~=~1.763\,\sqrt{8/(\pi^2 X)}$, which in dimensional units becomes~\cite{coen_universal_2013}
\begin{equation}
    \Delta\tau_\mathrm{min} = 1.763\, \sqrt{\frac{4|\beta_2|}{\mathcal{F}^2\gamma\theta P_\mathrm{in}}}. \label{eq:mindur}
\end{equation}
This minimum duration naturally corresponds to the maximum attainable spectral width, given explicitly by $\Delta f_\mathrm{max} = 0.315/\Delta\tau_\mathrm{min}$. Whilst the scaling given by Eq.~\eqref{eq:mindur} may suggest that the soliton duration (bandwidth) can be reduced (increased) indefinitely by, e.g., increasing the driving power $P_\mathrm{in}$, in practice higher-order effects become increasingly important as the soliton bandwidth increases, leading to the eventual failure of the approximate soliton solution given by Eq.~\eqref{eq:CSsol}, and hence of the scaling given by Eq.~\eqref{eq:mindur}. Section~\ref{sec:DW_and_recoil} discusses the primary mechanism that limits the validity of this analysis, namely higher-order dispersion.

\begin{figure}
  \centerline{\includegraphics[width = 0.8\textwidth, clip=true]{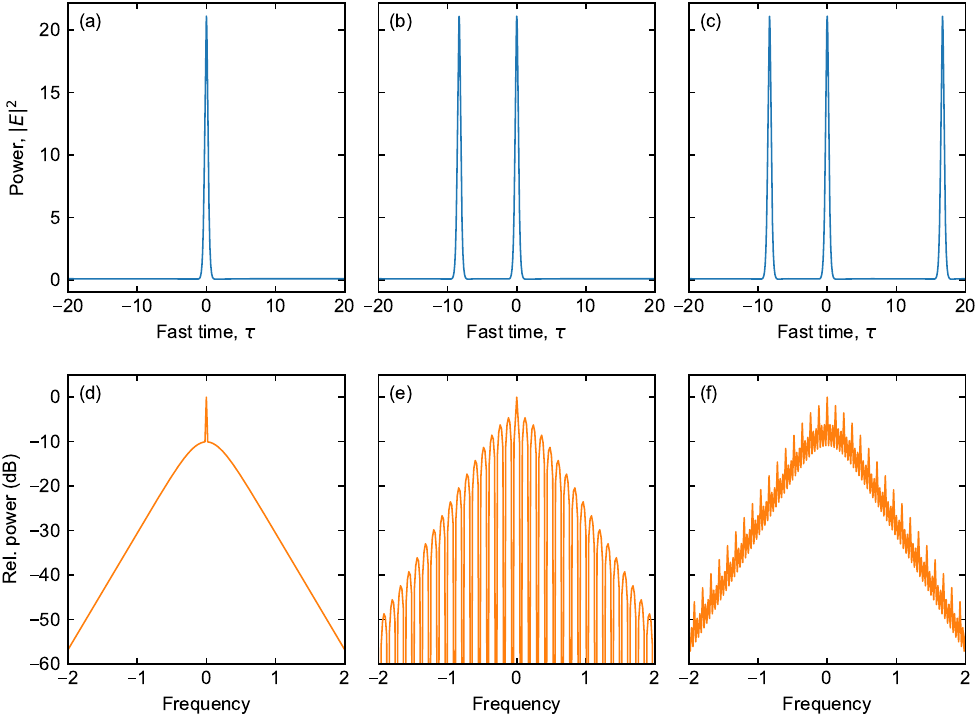}}
  \caption{(a)--(c) and (d)--(f) respectively show numerically simulated steady-state temporal and spectral intensity profiles of steady-state intracavity configurations consisting of different arrangements of CSs. All calculations use $X = 10$ and $\Delta = 10$.}
  \label{fig:multiCS}
\end{figure}

CSs of the LLE exhibit translational invariance. That is to say that they can occupy any position along the fast time $\tau$. Moreover, several CSs can exist simultaneously for the same set of parameters. As an example, Fig.~\ref{fig:multiCS} shows an assortment of steady-state solutions to the LLE for the same parameters ($X = 10$ and $\Delta = 10$), consisting of different configurations of multiple solitons. In the frequency domain, the coexistence of multiple, temporally-offset, solitons gives rise to a spectral interference pattern, which is a common experimental signature of multi-soliton states and that can be used to extract the solitons' relative temporal separations~\cite{brasch_photonic_2016,webb_experimental_2016}.

In the pure LLE model, coexisting solitons only interact via their exponentially decaying tails, which implies that the interactions become negligible for separations in excess of a few soliton widths. However, a number of higher-order perturbations can give rise to long-range interactions, which may only permit a discrete set of steady-state configurations. Such interactions will be discussed in Section~\ref{sec:CSbinding}.

\subsection{Controlled CS excitation and annihiliation}
\label{sec:controlled_excitation}

Because they coexist with a stable low-intensity cw state, CSs do not form spontaneously from an empty cavity; some form of temporally localized intracavity perturbation is required to excite (or ``write'') them. In this Section, we describe the main excitation schemes reported in the literature that allow for the deterministic excitation (and annihilation) of CSs. In the subsequent Section, we then describe a particularly popular means of non-deterministic soliton excitation that does not yield direct control over the excited soliton configurations, but benefits from ease of application.  

\begin{figure}[p]
  \centerline{\includegraphics[width = 0.72\textwidth, clip=true]{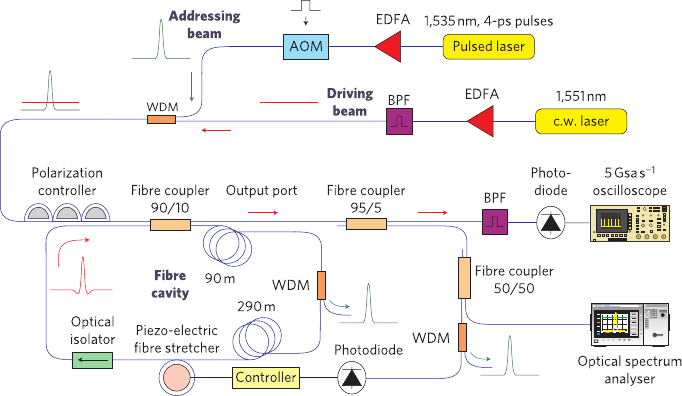}}
  \caption{Schematic illustration of the experimental setup used in 2010 for the first experimental observation of temporal CSs. A 380-m-long fibre resonator was driven with a cw laser at 1551~nm and individual solitons were excited by launching isolated pulses from a mode-locked laser at 1535~nm into the resonator. The detuning between the 1551~nm laser and a cavity resonance was stabilized using a piezo-electric fiber stretcher inside the resonator, and an optical isolator was used to prevent the build-up of stimulated Brillouin scattering. Reprinted with permission from F. Leo \textit{et al.}, Nat. Photonics. \textbf{4}, 471-476 (2010)~\cite{leo_temporal_2010}. Copyright \textcopyright~2015 Springer Nature.}
  \label{fig:fleo_experiment}
\bigskip
  \centering
  \includegraphics[width = 0.7\textwidth, clip=true]{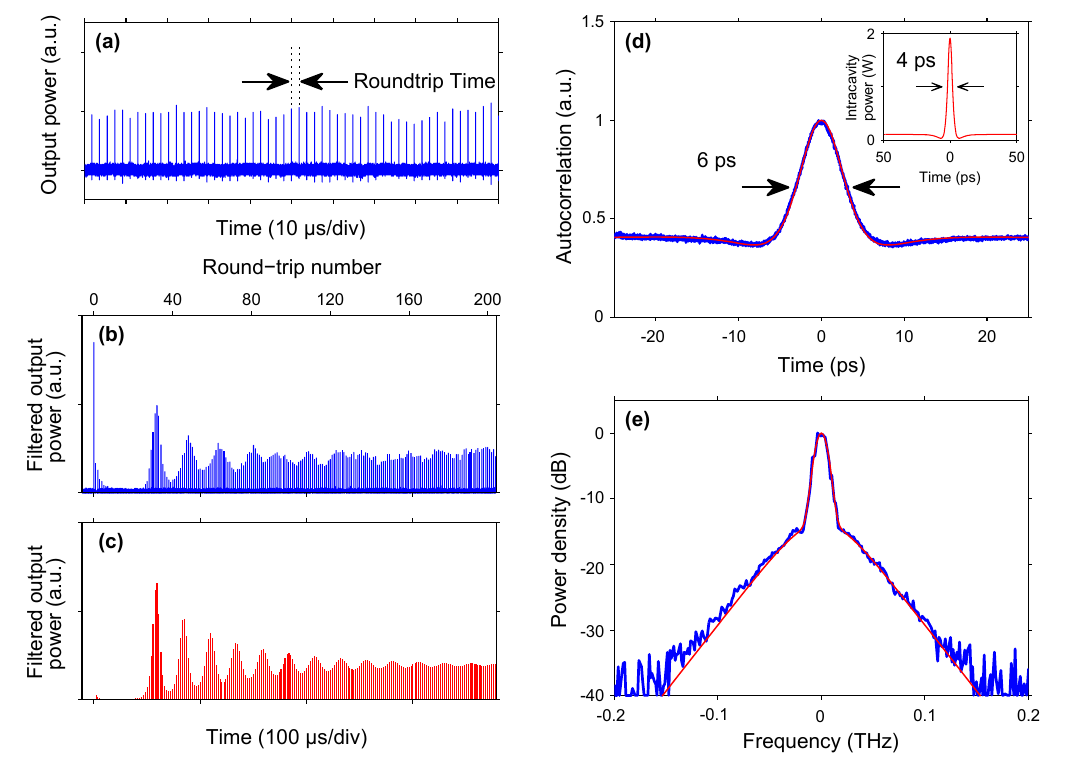}
  \caption{(a) Measured temporal signal exiting the cavity after a single CS has been excited. (b) Experimental measurement of the excitation dynamics using a bandpass filer to isolate the soliton signal from the driving field. (c) Numerical simulation of the excitation dynamics shown in (b) taking into account the full detection scheme. (d)~Experimental autocorrelation trace of the CS. The measured 6~ps duration corresponds to a CS with a 4~ps duration (see inset). (e) The measured spectrum at the output of the cavity showing the cw driving field superimposed on the hyperbolic secant profile of the soliton. Adapted with permission from F. Leo \textit{et al.}, Nat. Photonics. \textbf{4}, 471-476 (2010)~\cite{leo_temporal_2010}. Copyright \textcopyright~2015 Springer Nature.}
  \label{fig:CSexcitation1}
\end{figure}

\subsubsection{Incoherent cross-phase modulation}

The first experimental observation of temporal CSs was achieved in a macroscopic ring resonator made from a 380-m-long segment of single-mode optical fiber driven with a narrow-linewidth laser at 1551~nm~\cite{leo_temporal_2010} -- see Fig.~\ref{fig:fleo_experiment}. In that work, the solitons were excited by injecting into the resonator a single picosecond pulse from a mode-locked laser at 1535~nm. This mode-locked laser pulse, which only remains in the cavity for a fraction of a  roundtrip (before being extracted out with a wavelength-division-multiplier), imparts a temporally localized perturbation on the lower-state cw background via cross-phase-modulation (XPM). Under the correct operating conditions, this perturbation can then reshape into a CS. Figure~\ref{fig:CSexcitation1}(a) shows the experimentally measured optical intensity exiting the cavity after the CS has been excited. The measurement is made at the cavity's 95/5 output coupler using a high-speed photodiode and clearly shows a single ultrashort pulse circulating in the cavity. Figure~\ref{fig:CSexcitation1}(b) shows a more detailed measurement of the excitation process, obtained by applying a bandpass filter to isolate the soliton from the driving field. The large peak on the far-left signals the moment where the mode-locked laser pulse is injected into the resonator, and we see how the CS emerges and grows over a few tens of round trips and eventually reaches its steady-state. These excitation dynamics are in good agreement with numerical modeling that takes into account the full detection scheme [see Fig.~\ref{fig:CSexcitation1}(c)]. Once excited, the soliton persists in the cavity, displaying an autocorrelation trace with 6-ps-duration and a hyperbolic secant spectral profile that are in good agreement with numerical modeling [Figs.~\ref{fig:CSexcitation1}(d) and~(e)].

\subsubsection{Coherent phase and amplitude modulation}

Besides incoherent XPM perturbations, controlled CS excitation has also been achieved by applying localized phase~\cite{jang_writing_2015} and amplitude~\cite{wang_addressing_2018} modulations directly to the cavity driving field. These methods have also been shown to allow for the reverse process, i.e., the controlled annihilation (or ``erasure'') of an existing soliton. Figure~\ref{fig:Phasewriting}(a) shows results from numerical simulations from ref.~\cite{jang_writing_2015} illustrating how a localized phase perturbation [with profile $\Gamma(\tau)$ along the fast time axis] applied synchronously over a finite number of round trips [as determined by the rectangle function $A(t)$] allows one to deterministically excite a soliton [at round trip~100 in Fig.~\ref{fig:Phasewriting}(a)], and then to annihilate an existing soliton [at round trip~300 in Fig.~\ref{fig:Phasewriting}(a)]. Figure~\ref{fig:Phasewriting}(b) shows an experimental measurement that demonstrates such phase writing and erasure, applied here to a five-bit soliton sequence. Specifically, the pseudocolor plot in Fig.~\ref{fig:Phasewriting}(b) was constructed by recording oscilloscope traces of the cavity output at a frame rate of 1~frame/s, and vertically concatenating the resultant traces atop each other. During the recording of the data, carefully controlled phase perturbations were applied to the cavity driving field so as to systematically write and erase solitons. Note that the resonator used was made from a 100-m-long segment of single-mode optical fiber, such that the fast time span in Fig.~\ref{fig:Phasewriting}(b) only encompasses a small portion of the 0.5~$\mu$s round-trip time.

\begin{figure}[h]
  \centerline{\includegraphics[width = \textwidth, clip=true]{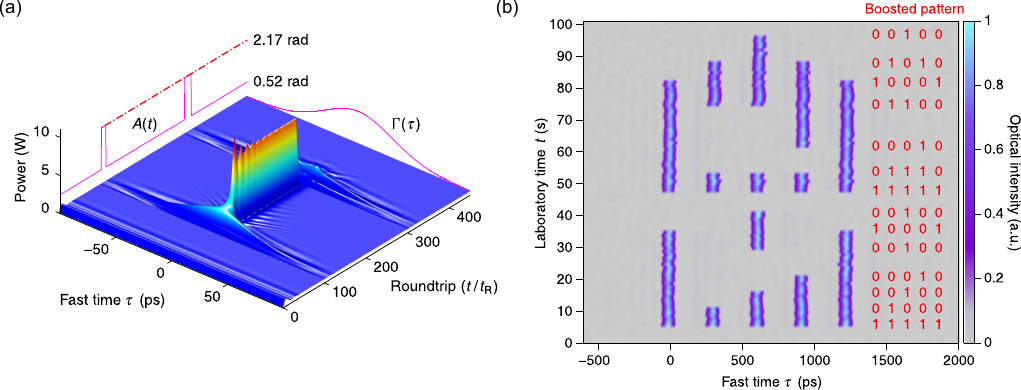}}
  \caption{(a) Numerical simulation results showing how abrupt phase perturbations to the cavity driving field allow the deterministic excitation and erasure of CSs. Here, the driving field is of the form $E_\mathrm{in} = P_\mathrm{in}^{1/2}\exp[i\phi(t,\tau)]$ where the phase modulation $\phi(t,\tau) = A(t)\Gamma(\tau)$ with $\Gamma(\tau)$ describing the temporal profile of the modulation and $A(t)$ a rectangle function modelling an abrupt boost to the modulation amplitude. (b)~Experimental results showing the deterministic excitation and erasure of CSs in a 100-m-long fibre resonator with $\Gamma(\tau)$ a $\sim50$~ps Gaussian pulse. The manipulations are applied on a 5-bit-CS sequence as indicated. Adapted with permission from J. Jang \textit{et al.}, Opt. Lett. \textbf{40}, 4755-4758 (2015)~\cite{jang_writing_2015}. Copyright \textcopyright~2015 Optical Society of America.}
  \label{fig:Phasewriting}
\end{figure}

\subsection{Soliton excitation via detuning scanning}
\label{sec:excitation_by_detuning}

The advantage of the excitation methods described above is that they allow for the systematic and deterministic excitation (and erasure) of desired CS patterns. However, they are largely restricted to macroscopic fiber resonators and cannot be easily applied to high-Q microresonators due to the large FSR and strong thermal nonlinearity of such systems.

Temporal CSs were first experimentally observed in a monolithic microresonator in 2014~\cite{herr_temporal_2014}. That pioneering work used a $35.2$-GHz-FSR resonator made from magnesium fluoride, and achieved soliton excitation by adiabatically scanning the detuning across a resonance from blue to red (i.e., from negative $\Delta$ to positive $\Delta$). Figure~\ref{fig:MicroresonatorCS} shows the spectrum of a single CS state generated in this resonator using this technique. The clean sech$^2$ spectral profile and low-noise RF spectrum (shown in the inset) demonstrate clearly that the system is operating in a CS state. Due to the ease with which it can be implemented, detuning scanning has subsequently become the dominant means through which CSs are excited, both in high-Q microresonators as well as in macroscopic fiber resonators. We note that, the excitation of CSs through detuning scanning is sometimes referred to as a `soft' excitation mechanism, as it does not require any external perturbation to the driving field beyond the detuning scan itself. This stands in contrast to the `hard' excitation mechanisms discussed in Section~\ref{sec:controlled_excitation}, which do require the direct external perturbation of the driving field~(see also~\cite{matsko_hard_2012}).

\begin{figure}[t]
  \centerline{\includegraphics[width = 0.65\textwidth, clip=true]{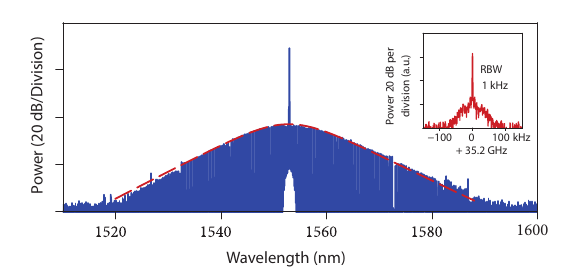}}
  \caption{Experimental measured spectrum of a single CS observed in a 35.2 GHz MgF$_2$ disk microresonator using detuning scanning (blue trace). Superimposed is the sech$^2$ spectral profile of a single CS (red trace). The inset shows the RF spectrum measured about the comb's fundamental 35.2 GHz beat note. This low noise trace provides further evidence the system is operating in a CS state. Adapted with permission from T. Herr \textit{et al.}, Nat. Photonics. \textbf{8}, 145-152 (2014)~\cite{herr_temporal_2014}. Copyright \textcopyright~2014 Springer Nature.}
  \label{fig:MicroresonatorCS}
\end{figure}

The fact that CSs can be excited by scanning the detuning across a resonance can be understood by considering the typical soliton existence (bifurcation) curve in Fig.~\ref{fig:CSbif}. Specifically, for low detunings, the only state available is the homogeneous cw state. As the detuning is increased, that state eventually becomes modulationally unstable and breaks into a pattern that is initially stable but eventually destabilizes. After the detuning increases above the up-switching point $\Delta_\uparrow$, the stable lower cw state comes into existence, thus permitting some of the initial MI fluctuations to act as perturbations that reshape into CSs. 

Figure~\ref{fig:CSscan} shows results from numerical simulations that illustrate the dynamics described above, depicting the evolution of the intracavity temporal [Fig.~\ref{fig:CSscan}(a)] and spectral [Fig.~\ref{fig:CSscan}(b)] intensity profiles as the detuning is adiabatically increased. The normalised drive power used is $X=10$. In this particular realisation of the simulation, Fig.~\ref{fig:CSscan}(a) clearly shows the formation of 4 distinct CSs which persist between detunings of $\Delta \sim 5$ and $\Delta \sim 12.3$. Also shown in Fig.~\ref{fig:CSscan}(c) is the corresponding evolution of the average intracavity energy, which shows a clear drop as the intracavity field enters the CS regime. This feature is often referred to as the ``soliton step''. This step is a characteristic signature of CS excitation when scanning the detuning~\cite{herr_temporal_2014}, and arises because much of the intracavity field drops to the lower branch cw state at this point. It is worth noting that, in the simulations shown in Fig.~\ref{fig:CSscan}, all the solitons decay at the same detuning $\Delta_\mathrm{max}\approx \pi^2 X/8$. This is in contrast with most experiments, where solitons are typically found to decay sequentially, giving rise to multiple ``soliton-step'' features. This discrepancy originates from the presence of higher-order effects (e.g., thermo refractive effect, mode interactions, broadband noise, etc.) that are not included in the simple LLE, but that can modify the instantaneous environment experienced by different solitons. 

Whilst detuning scanning is arguably the simplest method to excite CSs, it suffers from two important disadvantages on which we elaborate below together with mitigation strategies. 

\begin{figure}
  \centerline{\includegraphics[width = \textwidth, clip=true]{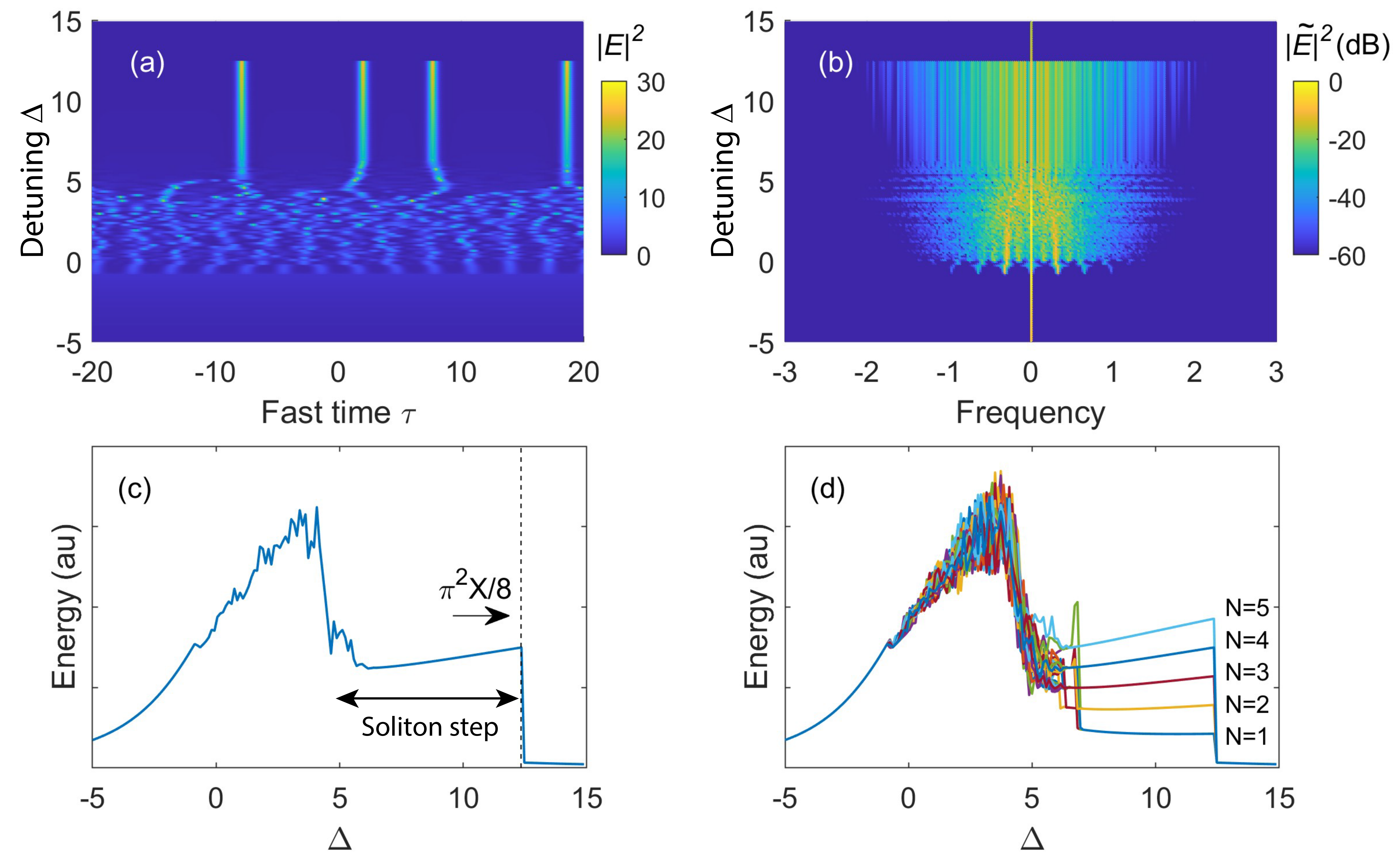}}
  \caption{LLE numerical simulation results illustrating the intracavity dynamics as the detuning~$\Delta$ is scanned from $-5$ to 15 at a normalized driving strength of $X=10$. (a)~Temporal evolution of the intracavity intensity and (b), (c) associated evolution of, respectively,  the spectral intensity and  the average energy in the cavity. (d) Average cavity energy evolution for 50~independent realizations, each with a different noise seed. The number of solitons~(N) formed in each realization can be deduced from the height of the step.}
  \label{fig:CSscan}
\end{figure}

\subsubsection{Deterministic excitation}
\label{sec:DeterministicEX}
Due to the stochastic origins of MI, the number and configuration of solitons that arise via detuning scanning are essentially random. This is in stark contrast to the controlled excitation schemes described in Section~\ref{sec:controlled_excitation}. As an example, Fig.~\ref{fig:CSscan}(d) shows numerically simulated average intracavity energy for 50 independent detuning scan realizations, obtained with different noise seeds, and we see clearly how the different realizations produce different soliton configurations. Several methods have been proposed and tested to overcome this issue and to achieve deterministic soliton generation via detuning scanning, some of which are summarised below:

\begin{itemize}
    \item \textbf{Forward/backward tuning.} Once a given soliton sequence has been non-deterministically excited via forward detuning scanning (i.e., from blue to red), it may be possible to systematically remove one soliton at a time by scanning the detuning backwards (i.e., from red to blue). This is enabled by the thermal nonlinearity described in Section~\ref{sec:thermal} and can be understood as follows. Let $P$ represent the average (normalized) power of a single soliton. The effective steady-state detuning for an $N$ soliton state is given by $\Delta(N) = \Delta_\mathrm{c} - \zeta' N P$, where $\Delta_\mathrm{c}$ is the normalized cold cavity detuning and $\zeta'$ describes the  normalized thermal detuning shift [see Eq.~\eqref{Eq:thermDet}]. As the detuning is scanned backwards (i.e., $\Delta_\mathrm{c}$ reduced), the effective detuning eventually reaches the up-switching point below which solitons cannot exist, $\Delta(N)\rightarrow\Delta_\uparrow$. However, since $\zeta'>0$, at that cold cavity detuning $\Delta(N-1)>\Delta_\uparrow$, implying that a $N-1$ soliton configuration can still exist. In this way, scanning the detuning backward allows to annihilate solitons one at a time, eventually bringing the intracavity field to the single-soliton ($N = 1$) state~\cite{guo_universal_2017}.


    \item \textbf{Active capture.} Because different numbers of solitons are associated with different average power levels, a servo-controller that stabilizes the frequency of the injected laser so as to lock the average power at a set level can be used to systematically obtain a state with a given number of solitons. In this scheme, the system effectively performs multiple detuning ramps until the desire average power (soliton number) is reached, at which point the servo stabilizes the system. Since the average power of a soliton depends on the effective cavity detuning, stabilizing the soliton also stabilizes this detuning --- suppressing fluctuations that would otherwise render solitons short-lived. This technique was first experimentally demonstrated using a silica wedge microresonator~\cite{yi_active_2016}.
        
    \item \textbf{Phase-modulated and pulsed driving.} As will be described in more detail in Section~\ref{sec:trapping}, phase or amplitude modulations applied to the external driving field can induce corresponding modulations in the cavity parameters. CSs naturally drift in response to such parameter gradients, enabling robust trapping at specific temporal positions. This mechanism can be successfully exploited to control the number and configuration of solitons that emerge from detuning scanning~\cite{obrzud_temporal_2017-1, xu_harmonic_2020}
  
    \item \textbf{MI control.} The fundamental reason why detuning scanning results in non-deterministic soliton configurations is the fact that the solitons are excited by the chaotic fluctuations of the MI state that precedes the solitons. By judiciously engineering the detuning scan rate and driving power, it is possible to reach the soliton state directly from the stable MI state, thus resulting in a deterministic soliton pattern~\cite{jaramillo2015deterministic,karpov_dynamics_2019}. In addition, mode shifts, arising from interactions with other modes or engineered resonator topology (e.g., coupled resonators or photonic crystal structures) can modify the MI gain spectrum so as to force the gain to peak at a 1~FSR frequency shift. In this case, the emergent MI pattern only exhibits a single cycle within the resonator round trip, leading to the deterministic generation of a single-soliton state. This scenario is closely linked to the formation of ``type 1'' combs reported in early microresonator frequency comb literature~\cite{ferdous_spectral_2011,torres-company_comparative_2014}.
   
\end{itemize}

\subsubsection{Thermal instability}
\label{sec:Thermalinstability}

The emergence of CSs during a detuning scan is typically marked by a pronounced drop in intracavity average power --- a direct consequence of their coexistence with the lower-power cw background that surrounds them [see Fig.~\ref{fig:CSscan}(c)]. Considering, as is most common, resonators with a positive $dn/dT$ coefficient, a sudden drop in average power leads to an abrupt cooling of the resonator, and a concomitant blue-shift of the driven resonance: i.e., the effective detuning $\Delta(P) = \Delta_\mathrm{c} - \zeta'P$ abruptly increases. Significantly, if the thermal effect is sufficiently strong, the effective detuning can abruptly increase beyond the maximum detuning of soliton existence, $\Delta(P_\mathrm{CS}) > \Delta_\mathrm{max}$, thus destroying any solitons that have just emerged. For resonators with a negative $dn/dT$ coefficient (e.g., monolithic crystalline microresonators fabricated from CaF$_2$), the situation is even worse. Here, adiabatically scanning from blue to red detunings, as required for detuning-scanning excitation, is typically not possible due to thermal run-away~\cite{kobatake_thermal_2016}.

In resonators with weak thermal effect (e.g., macroscopic fiber resonators), it is possible to stably reach the soliton stage with a simple detuning scan. However, in resonators with a strong thermal effect (e.g., many microresonators), the thermal transient must be mitigated. Below we summarise four notable techniques used to that end.

\begin{itemize}

    \item \textbf{High-speed scanning.} In this approach, the pump laser is scanned into the soliton resonance at a sufficiently high velocity that the resonator does not reach thermal equilbrium at any point. The scan is then halted upon reaching the detuning required for soliton formation. By varying the scan velocity, it is thus possible to control the temperature of the resonator at the end of the scan. Stable soliton generation can then be achieved by adjusting the scan speed such that the final temperature matches the resonator’s steady-state temperature in the soliton regime. This technique was used to achieve the first successful demonstration of stable cavity soliton (CS) formation in a microresonator system~\cite{herr_temporal_2014}.

    \item \textbf{Power kicking.} This approach involves slowly tuning the drive laser into resonance from the blue-detuned side, halting the scan just before reaching the up-switching point. The driving power is then rapidly decreased, cooling the resonator, and resulting in an increase of the effective detuning $\Delta(P) = \Delta_\mathrm{c}-\zeta' P$. If this effective detuning increases past the up-switching point [$\Delta(P)>\Delta_\uparrow$], soliton excitation can occur. Subsequently, the drive power can be increased to raise the average intracavity power in the soliton regime, counteracting any further cooling and thereby stabilizing the resonator for steady-state operation~\cite{yi_soliton_2015, brasch_photonic_2016, brasch2016bringing}. 
   
    \item \textbf{Auxiliary pumping.} Thermal transients that occur during soliton generation can be effectively suppressed by introducing an additional auxiliary laser to drive the resonator~\cite{zhang_sub-milliwatt-level_2019,lu_deterministic_2019,niu_repetition_2018}. This secondary field is tuned into resonance with a detuning that is thermally self-stabilized within the blue-detuned region, and is typically configured to excite a different mode family from the main pump to avoid nonlinear mixing. As the main pump's detuning is scanned into the soliton regime, the resonator cools, causing all resonances to blue-shift. This shift brings the auxiliary laser closer to its own resonance, increasing its intracavity power and counteracting the cooling effect. In other words, the auxiliary laser helps to maintain the resonator at constant temperature during soliton generation, thus surpressing any thermal shifts. More details on the experimental implementation of auxiliary pumping can be found in Section~\ref{Auxillary_pumping}. The nonlinear interactions that can arise when two laser fields excite the same mode family will be discussed in Section~\ref{sec:extension}. 

    \item \textbf{Coupled cavities.} As discussed in Section~\ref{avoided_crossings}, the resonant frequency of a cavity mode can be shifted via linear coupling with a nearby mode. Carefully coupling the main resonator to a second smaller auxiliary cavity can thus allow the pump mode to be shifted whilst leaving the remaining modes of the soliton comb unperturbed. This approach was demonstrated in Ref.~\cite{helgason2023surpassing}, and enabled soliton formation to occur even when the pump remained blue‑detuned relative to the pump mode. Consequently, no delicate thermal balancing was required, and the soliton state could be reached simply by tuning the pump to the appropriate detuning.
    
    
\end{itemize}

\subsection{Breathers and other soliton instabilities}
\label{sec:CSinstabilities}

As discussed in Section~\ref{sec:CSexist}, the CS solutions of the LLE are unstable for low detunings close to the up-switching point [see Fig.~\ref{fig:CSbif}(a)]. In this Section, we discuss the qualitatively different instability regimes that the CSs can exhibit.

The transition from stable to unstable solitons occurs via a Hopf bifurcation as the detuning is decreased leading to persistent oscillatory (or ``breathing'') behavior for detunings close to the instability point [denoted $\Delta_\mathrm{Hopf}$ in Fig.~\ref{fig:CSbif}(a)]. An illustrative example of simple period-1 breathing dynamics, obtained at a drive power $X=10$ and detuning $\Delta=4.2$, are shown in Figs.~\ref{fig:CSbreathing}(a) and~(b). Experimental signatures of breathing CSs were first observed in macroscopic fiber ring resonators in 2013~\cite{leo_dynamics_2013}, and subsequently in several microresonator platforms~\cite{bao_observation_2016,lucas_breathing_2017,yu_breather_2017}. 
Beyond the simple, single period oscillations shown in Fig.~\ref{fig:CSbreathing}(a) and~(b), increasingly complex nonlinear behaviour emerges at higher driving powers~\cite{nozaki_chaotic_1985,leo_dynamics_2013,godey_stability_2014}.
Figure~\ref{fig:CSbreathing}(c) maps the full phase space of these dynamics as a function of $S=\sqrt{X}$ and $\Delta$, delineating distinct regions of instability: region (I) encompasses cw homogeneous solutions; region (II), stable cavity solitons; region (III), breathing solitons: included in this region one finds simple period-1 breathing dynamics, as well as sub-regions of period$-2$, $-3$, $-4$, ... $-n$ oscillations, and even regions of chaotic oscillation; the region labeled (IV) corresponds to ``transient chaos,'' where initially chaotic oscillation of the solitons are followed by their spontaneous decay; in the region labeled (V), the solitons trigger an expanding domain of chaotic MI that invades the lower-state homogeneous background; finally, the region (VI) corresponds to regions of MI. We note that, by leveraging a backwards detuning scan in a macroscopic fiber resonator with negligible thermal nonlinearity, it is possible to adiabatically scan the detuning from the stable CS regime all the way to the upswitching point $\Delta_\uparrow$, thus sampling all these different instabilities in a single experiment~\cite{anderson_observations_2016}.


Finally, we discuss briefly the instability region labelled (V) in more detail. This regime is characterized by the bistability between a self-sustaining chaotic MI state and the lower-level cw background, which implies that both states can exist individually in quasi-steady-state. However, the former is found to exhibit metastability at the expense of the latter: that is, a sufficiently strong, temporally localized perturbation applied to an intracavity field in the lower-level cw state triggers an expanding domain of chaotic MI that eventually fills the entire cavity. Note that  if the cavity driving field (or the cavity detuning) is appropriately modulated (e.g., in amplitude), it is possible to arrest the expansion of the MI domain, thus allowing for the realization of a system where localized domains of spatiotemporal chaos can be individually addressed~\cite{nielsen_nonlinear_2021}. 

\begin{figure}[t]
  \centerline{\includegraphics[width = 0.7\textwidth, clip=true]{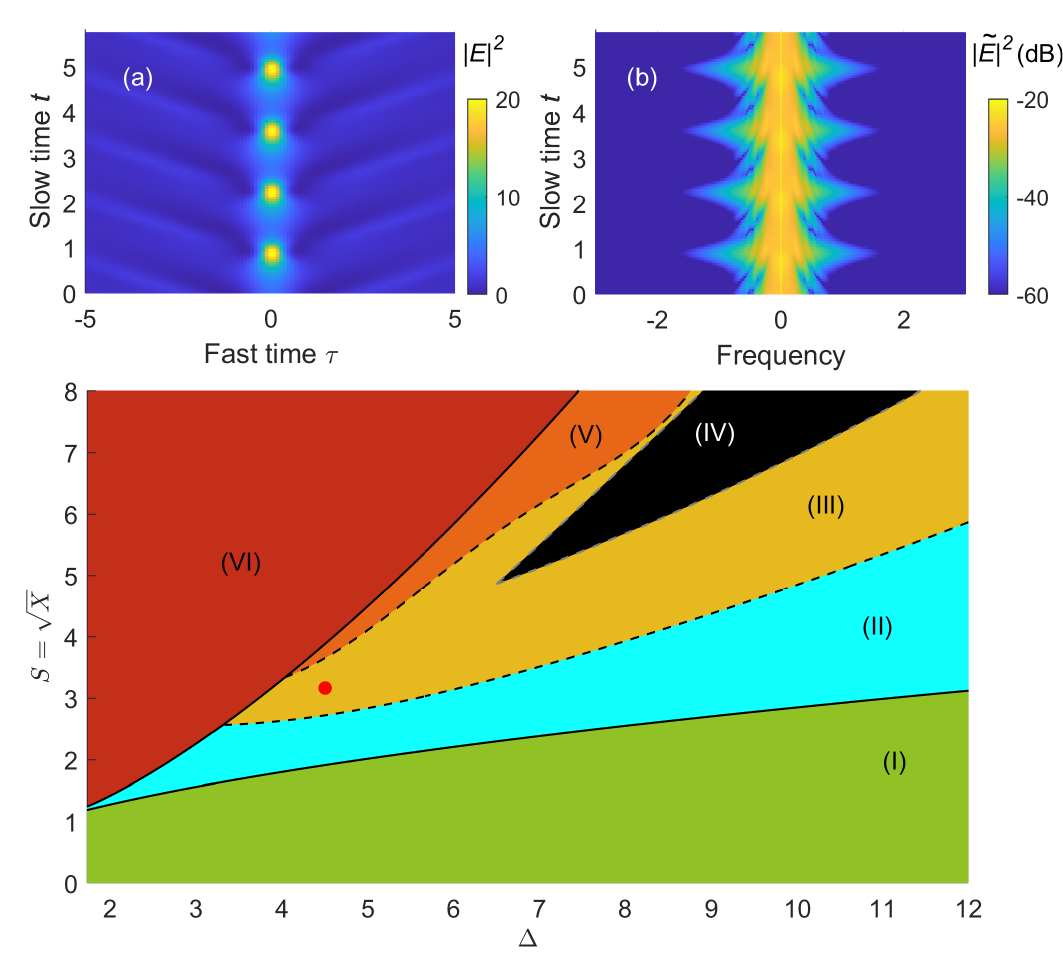}}
  \caption{The temporal and spectral evolution of a breathing CS are shown in panels (a) and (b). These results are obtained from numerical simulation of the normalised LLE at a driving strength and detuning of $X = 10$ and $\Delta = 4.2$ respectively. (c) Map of the phase space of the nonlinear dynamics of an anomalous dispersion Kerr cavity as a function of drive strength and detuning. The labeled regions correspond to: (I) cw homogeneous solutions, (II) stable cavity solitons, (III) breathing solitons, (IV) ``transient chaos'', (V) triggered domains of expanding chaotic MI, and (VI) MI. The red circle in panel (c) marks the location of the breather soliton shown in (a) and (b). Panel (c) adapted with permission from M. Anderson, ``Dynamics of temporal Kerr cavity solitons: Chaos, instability,
and multi-resonant solutions,'' MSc thesis, University of Auckland (2016)~\cite{anderson2016dynamics}. Copyright \textcopyright~2016 M. Anderson.}
  \label{fig:CSbreathing}
\end{figure}

\subsection{Trapping and synchronization}
\label{sec:trapping}

In their simplest form --- such as found in the cw driven LLE --- CSs form in an environment with full translational invariance. This symmetry allows CSs to reside freely at any location along the fast-time axis. However, when parameter inhomogeneities are introduced --- either deliberately or inadvertently --- such as through phase or amplitude modulations of the driving field, or fast-time-dependent fluctuations in cavity detuning, CSs can experience temporal drift toward a new equilibrium position. These inhomogeneities effectively behave as trapping potentials, confining the solitons to specific regions~\cite{jang_temporal_2015,hendry_spontaneous_2018,weng_spectral_2019}. This behavior, often described as ``plasticity,'' has been harnessed across numerous settings to facilitate control and manipulation of CS dynamics.

In what follows, we briefly discuss the different modalities that have been used to achieve trapping and manipulation of CSs. For a comprehensive review of the salient mathematical analyses, we refer to Ref.~\cite{erkintalo_phase_2022}.

\subsubsection{Phase modulation}
\label{sec:PM}

Of all the possible parameter inhomogeneities, phase modulation of the cavity driving field permits the most straightforward theoretical treatment. Consider the LLE with a phase modulated driving field $S\rightarrow S\exp[i\phi(\tau)]$. Anticipating that the intracavity field follows the phase of the drive, we substitute into Eq.~\eqref{LLE} (with anomalous dispersion, $\eta=-1$) the expression $E(t,\tau) = F(t,\tau)e^{i\phi(\tau)}$ and derive
\begin{equation}
\label{LLF}
    \frac{\partial F}{\partial t} = \left[-\alpha_\mathrm{e}(\tau)+i|F|^2-i\Delta_\mathrm{e}(\tau) -2\phi'(\tau)\frac{\partial}{\partial\tau} + i \frac{\partial^2}{\partial\tau^2} \right]F + S,
\end{equation}
where the effective loss and detuning coefficients are respectively defined by 
\begin{align}
    \alpha_\mathrm{e}(\tau) &= 1+\phi''(\tau), \\
    \Delta_\mathrm{e}(\tau) &= \Delta+[\phi'(\tau)]^2, 
\end{align}
with $\phi'(\tau)$ and $\phi''(\tau)$ designating the first and second fast time derivatives of the driving phase modulation~$\phi(\tau)$. For slow phase modulations that do not vary significantly over the soliton width, the modifications to the loss and detuning coefficients are small and can be ignored (such that $\alpha_\mathrm{e} = 1$ and $\Delta_\mathrm{e} = \Delta$). In this case, Eq.~\eqref{LLF} has the form of the standard LLE but with an additional drift term $-2\phi'(\tau)\partial F/\partial\tau$. This causes a CS located at fast time $\tau_\mathrm{cs}$ to acquire a temporal drift velocity of,
\begin{equation}
\label{eq:PMdrift}
    V_\mathrm{PM}(\tau_\mathrm{cs}) = \frac{d\tau_\mathrm{cs}}{dt} = 2\phi'(\tau_\mathrm{cs}).
\end{equation}

Equation~\eqref{eq:PMdrift} affords a simple physical interpretation. Phase modulation imposes a shift in the driving field’s instantaneous frequency, given by $\delta\omega = -\phi'(\tau)$. Since the CS center frequency is pinned to that of the driving field, this shift is inherited by the CS. Owing to dispersion, the resulting frequency shift leads directly to a modification of the CS’s group velocity. In dimensional units, the extra temporal delay per round trip can be written as: $\Delta\tau = -\beta_2 L \phi'(\tau)$, from which the drift velocity can be calculated to be $V_\mathrm{PM}=\Delta \tau/t_\mathrm{R}$ (which corresponds to the above after normalization)~\cite{jang_temporal_2015}.

Equation~\eqref{eq:PMdrift} reveals that CSs drift along phase gradients, with maxima and minima in the phase profile representing stable and unstable equilibria, respectively. Thus, phase modulation maxima of the driving field correspond to stable trapping points for CS. This behavior was experimentally verified in Ref.~\cite{jang_temporal_2015} using CSs generated in a 100-m-long fiber ring resonator. When a 110~ps Gaussian phase modulation was synchronously applied to the cavity driving field, CSs were observed to temporally lock to the peak of the modulation. Adiabatic variation of the peak’s position led the locked CSs to follow it, while abrupt shifts caused the CSs to drift toward the new peak location. This adiabatic `temporal tweezing' is shown in Fig.~\ref{fig:PMtrap}. Panel~(a) shows the location of three phase modulation peaks that are applied synchonrously to the driving field. Panel~(b) shows the optical signal of three CSs trapped to these peaks. As the position of the central phase modulation peak is varied, the center CS is seen to follow its position exactly. The drift velocity induced by phase modulation could be measured in realtime and compared to the predicted evolution of Eq.~\eqref{eq:PMdrift}. Figure~\ref{fig:PMtrap}(c) shows the experimental setup used to abruptly switch the position of a phase modulation peak by 45~ps. Panel~(d) shows the experimentally measured temporal position of a CS drifting in response to this shift (blue dots), overlaid with the numerically integrated trajectory from Eq.\eqref{eq:PMdrift} using experimental parameters (red trace). The agreement between this direct experimental measurement and the simple theory of Eq.~\eqref{eq:PMdrift} is excellent.

Phase modulation trapping has been experimentally used to suppress long-range (acoustic) interactions between solitons in macroscopic fiber resonators~\cite{jang_ultraweak_2013}, which has allowed for the experimental realization of a CS-based optical buffer operating at 10~Gb/s~\cite{jang_all-optical_2016}. It has also been used to systematically force two solitons to collide at a phase modulation peak, revealing merging and annihilation dynamics that underscore the solitons' dissipative character~\cite{jang_controlled_2016}. Beyond macroscopic resonators, phase-modulation trapping has also proven effective in microresonators, where it enables new functionalities, including the spectral purification of microwave signals~\cite{weng_spectral_2019}.

\begin{figure}[!t]
  \centering
  \includegraphics[width = 0.9\textwidth, clip=true]{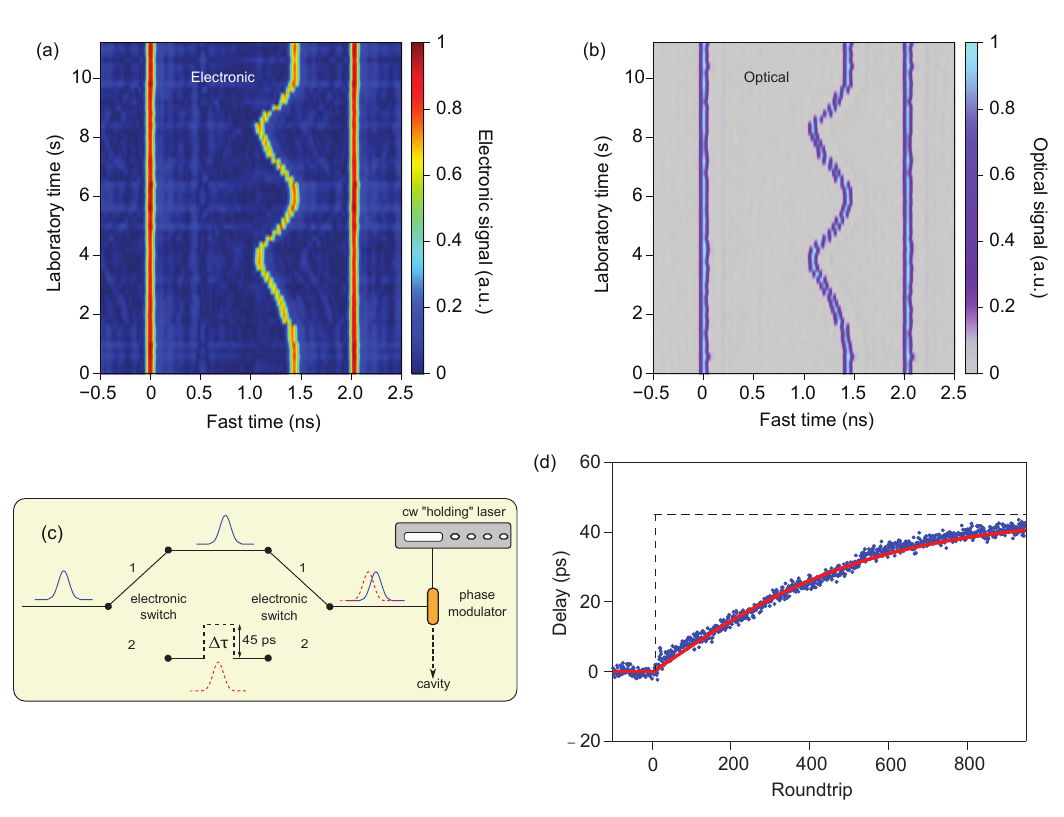}
  \caption{(a)~Measured position of three 110~ps phase modulation peaks that are applied synchonrously to the driving field. (b)~Measured positions of three CSs locked to these phase modulation peaks. (c)~Experimental setup used to impart an abrupt 45~ps shift to a phase modulation peak. (d)~Measured temporal evolution of the position of a CS following this abrupt shift (blue dots). The red trace shows the predicted trajectory obtained by integrating Eq.~\eqref{eq:PMdrift}. Adapted with permission from J. Jang \textit{et al.}, Nat. Commun. \textbf{6}, 7370 (2015)~\cite{jang_temporal_2015}. Copyright \textcopyright~2015 Springer Nature.
  }

  \label{fig:PMtrap}
\end{figure}

\subsubsection{Amplitude modulation}

Theoretical analysis of CS motion and trapping in the presence of driving field amplitude modulations (including pulsed driving) is not as straightforward as for phase modulations~\cite{hendry_spontaneous_2018,hendry_impact_2019,erkintalo_phase_2022,li_efficiency_2022}. However, it can be shown that a slow (relative to the soliton width) amplitude modulation $S(\tau)$ causes a CS at $\tau_\mathrm{cs}$ to undergo a time-domain drift with the rate
\begin{equation}
\label{eq:AMdrift}
    V_\mathrm{AM}(\tau_\mathrm{cs}) = \frac{d\tau_\mathrm{cs}}{dt} = a_\mathrm{AM}\left(S(\tau_\mathrm{cs}),\Delta\right) {\frac{dS}{d\tau}} \Big|_{\tau=\tau_\mathrm{cs}}^{}\,
\end{equation}
where $a_\mathrm{AM}(S,\Delta)$ is a drift coefficient that depends upon the driving amplitude (at the CS position) and detuning [see Fig.~\ref{fig:AMtrap}(a)]. This result is obtained by projecting the modulated driving onto the neutral mode of the CS ~\cite{hendry_spontaneous_2018,hendry_impact_2019,erkintalo_phase_2022} but can also be analyzed based on a reduced Lagrangian model~\cite{li_efficiency_2022}.

Equation~\eqref{eq:AMdrift} suggests that, as with phase inhomogeneities, CSs drift along amplitude gradients. However, the presence of the multiplicative drift coefficient $a_\mathrm{AM}(S,\Delta)$ qualitatively changes the trapping behavior. In particular, for each detuning $\Delta$, the drift coefficient is found to be monotonically decreasing with increasing driving strength~$S$ and to cross zero at a critical driving amplitude $S_\mathrm{c}$. This critical amplitude is shown as a black dashed line in Fig.~\ref{fig:AMtrap}(a). As a consequence, maxima of the amplitude modulation only correspond to stable equilibrium positions when $\text{max}[S(\tau)]\leq S_\mathrm{c}$. When this inequality does not hold (i.e., when $\text{max}[S(\tau)]>S_\mathrm{c}$), stable trapping positions are instead those positions along the edges of the driving field, where $S(\tau)=S_\mathrm{c}$. 

This behaviour is illustrated in Figs.~\ref{fig:AMtrap}(b) and~(c), which show numerical simulations of the equilibrium trapping position of a single CS driven by a Gaussian pulse, $S(\tau)~=~S_0~\exp(-\tau^2/\tau_0^2)$, for two distinct peak amplitudes~$S_0$. Both simulations use a detuning of $\Delta = 4$, for which~${S_c \sim 2}$.
In panel~(b), where $S_0 = 1.9 < S_c$, the soliton is trapped at the peak of the driving field. In contrast, panel~(c) shows that when $S_0 = 2.3 > S_c$, the soliton stabilises at an offset temporal position where $S(\tau) = S_c$. Panels (d) and~(e) illustrate the slow-time evolution of the soliton trajectories as they converge towards these stable trapping sites. Notably, the trapping position shown in Fig.~\ref{fig:AMtrap}(e) at $\tau \sim +10$ is only one of two possible stable equilibria in this regime; the driving field also satisfies $S(\tau) = S_c$ at $\tau \sim -10$. The final trapping location therefore depends on the soliton’s initial excitation position. In the case of spontaneous CS excitation, the trapping of CSs at the edges of (symmetric) amplitude-modulated driving fields can be considered as an example of spontaneous temporal symmetry breaking~\cite{xu_experimental_2014, hendry_spontaneous_2018}.

\begin{figure}[!t]
  \centering
  \includegraphics[width = 1.0\textwidth, clip=true]{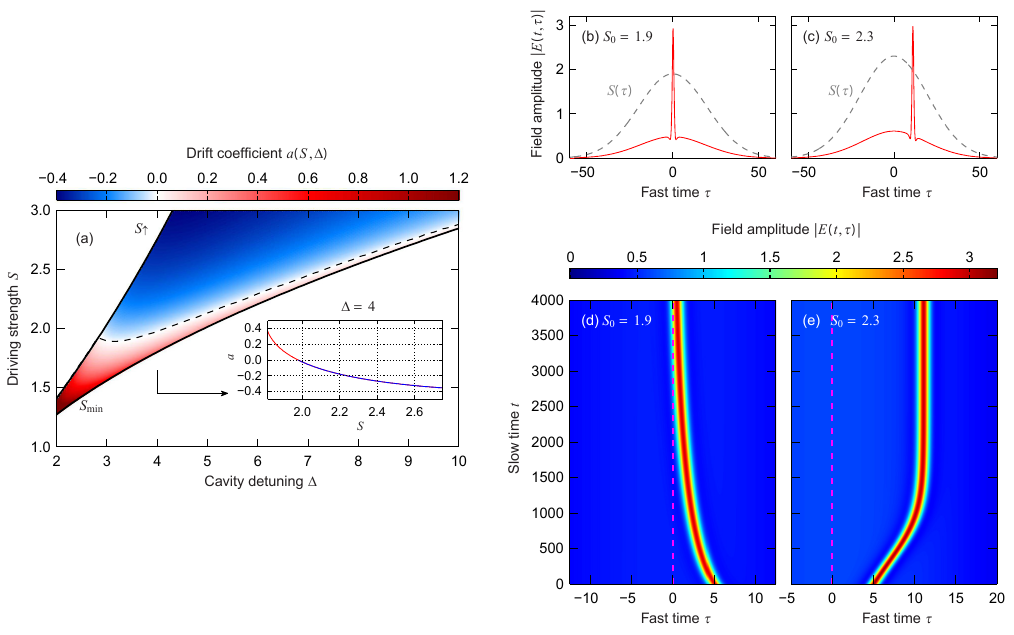}
  \caption{(a) Strength of the drift coefficient $a_{\mathrm{AM}}(S,\Delta)$ as a function of drive amplitude and detuning. The solid black lines labeled $S_{\mathrm{min}}$ and $S_\uparrow$ mark the minimum amplitude for CS existence and the up-switching point, respectively. The inset in panel (a) shows $a_{\mathrm{AM}}(S)$ at $\Delta = 4$, with the critical amplitude $S_c \sim 2$. (b, c) Numerically simulated final trapping locations of a CS driven by Gaussian pulses with peak amplitudes $S_0 = 1.9 < S_c$ and $S_0 = 2.3 > S_c$, respectively. (d, e) Slow-time evolution of a CS, initially excited at $\tau = 5$, toward the corresponding equilibrium trapping position in each case. Adapted with permission from I. Hendry \textit{et al.}, Phys. Rev. A \textbf{97}, 053834 (2018)~\cite{hendry_spontaneous_2018}. Copyright \textcopyright~2018 American Physical Society.}
  \label{fig:AMtrap}
\end{figure}

\subsubsection{Synchronization}
\label{sec:synchronization}

The descriptions above have focused on situations where the phase or amplitude inhomogeneities are perfectly synchronized with the round-trip time of the soliton. In the frequency domain, this implies that the Fourier components of the inhomogeneity coincide with the components of the soliton. Of course, in general the inhomogeneities may be applied desynchronously, such that the corresponding Fourier components differ from the free-running frequency comb lines of the soliton. However, provided that the desynchronization is sufficiently small, it is possible that the soliton nonetheless becomes trapped, effectively changing its group-velocity (comb line spacing) so as to synchronize with the inhomogeneity [see Fig.~\ref{fig:synchronization}]. In the frequency domain, this trapping can be seen as an Adler-type synchronization process, where Fourier components of the inhomogeneity capture the soliton comb lines closest to them.

\begin{figure}[!t]
  \centering
  \includegraphics[width = 1.0\textwidth, clip=true]{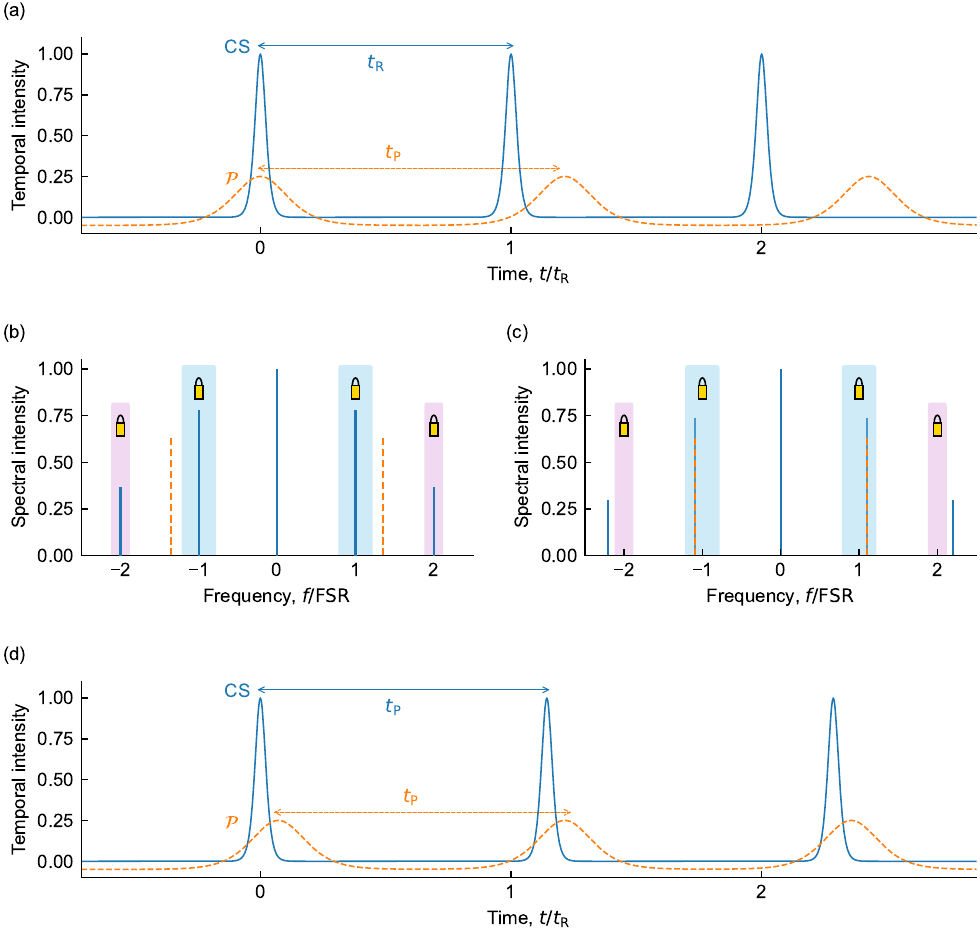}
  \caption{Schematic illustration of synchronization of a CS to an external perturbation. In (a) and (b), the perturbation (dashed orange curve) exhibits a large desynchronization with respect to the CS round trip time, such that the drift due to the inhomogeneity is insufficient to trap the soliton to the inhomogeneity. In the spectral domain (b), the Fourier components of the perturbation (dashed orange) are not sufficiently close to capture the soliton components. In (c) and (d), the Fourier components of the perturbation are sufficiently close to the soliton components to capture them, resulting in the soliton being trapped to the perturbation (d). The shaded regions in (b) and (c) illustrate ``locking windows'' within which Fourier components of the perturbation may capture the soliton components.
  }
  \label{fig:synchronization}
\end{figure}

As explained in Section~\ref{sec:pulsed}, the impact of desynchronization can be mathematically included in the analyses described above by considering a driving field that depends on both the fast and the slow time, $S(\tau)\rightarrow S(\tau+d \cdot t)$ [see Eq.~\eqref{pulsed_drive}], where $d$ represents a per-round-trip desynchronization~\cite{coen_convection_1999,hendry_impact_2019}. Here, we must note that technically $d$ represents the desynchronization between the driving field and the ``natural'' round-trip time of the cavity; the desynchronization with the soliton can be different as various perturbations (such as higher-order dispersion and stimulated Raman scattering) can perturb the soliton's group-velocity and hence round-trip time. Because the mathematical analyses considered in this subsection ignore such higher-order effects, the round-trip time of the soliton coincides with the natural round-trip time of the cavity.

When shifting into a reference frame where the driving is stationary ($\tau' = \tau + d\cdot t$), one obtains an LLE with an additional convective drift term proportional to $d$ [see Eq.~\eqref{GLLEpulsed}]. This adds to the drifts that are due to parameter inhomogeneities, such that the total time-domain drift rate experienced by the solitons becomes
\begin{equation}
    V_\mathrm{tot}(\tau_\mathrm{cs}) = \frac{d\tau_\mathrm{cs}}{dt} = d + V_\mathrm{m}(\tau_\mathrm{cs}),
\end{equation}
where $V_\mathrm{m}(\tau_\mathrm{cs})$ is the drift rate due to the parameter inhomogeneity [e.g., Eq.~\eqref{eq:PMdrift} for phase modulations and Eq.~\eqref{eq:AMdrift} for amplitude modulations]. 

Desynchronization changes the solitons' equilibrium positions from those where $V_\mathrm{m}(\tau_\mathrm{cs}) = 0$ to those where $V_\mathrm{m}(\tau_\mathrm{cs}) = -d$. In other words, desynchronization shifts the solitons to a position where drift due to the inhomogeneity exactly cancels the desynchronization. Of course, trapping can only be achieved over a finite range of desynchronizations $d\in [-\text{max}(V_\mathrm{m}(\tau)), -\text{min}(V_\mathrm{m})(\tau)]$, outside of which the drift due to desynchronization is too large to be compensated for by the inhomogeneity. It is important to reiterate that, when trapped in this manner, the solitons' group velocity changes such that they become synchronous with the inhomogeneity. In the spectral domain, this implies that the corresponding soliton frequency comb changes its line spacing such that the Fourier components of the inhomogeneity become part of the resultant frequency comb [i.e., the comb line spacing is locked to the frequency of the applied pertubation, see Fig.~\ref{fig:synchronization}]. 

The frequency domain synchronization interpretation is particularly evident in the trapping of CSs through the injection of a monochromatic auxiliary field into the resonator. Specifically, if the frequency of the injection laser is sufficiently close to a soliton comb line, the injection can capture the comb line, causing the two to have identical frequencies. This injection locking-type phenomenon has been referred to as ``Kerr-induced synchronization''~\cite{moille_kerr-induced_2023,wildi2023sideband,shandilya_all-optical_2025,moille_-chip_2025}, and can be understood in the (fast) time domain as the trapping of a CS to the phase- and amplitude modulated background that is due to the beating between the main and auxiliary driving fields. This latter interpretation can be readily understood by examining the mathematical form of a bichromatic driving field [see Eq.~\eqref{eq:bichromatic}], written in the following dimensional form for convenience: 
\begin{equation}
\label{eq:bichromatic2}
        E_\mathrm{in}(t,\tau) = E_{\mathrm{in}, 0} + E_{\mathrm{in}, 1}e^{-iqD_1\tau + i[\Delta\omega_1-\Delta\omega_0 - D_\mathrm{int}(q)]t},
\end{equation}
where $\Delta\omega_{0}$ ($\Delta\omega_{1}$) is the frequency detuning of the main, CS-generating driving field (the auxiliary injection field) with complex amplitude $E_{\mathrm{in}, 0}$ ($E_{\mathrm{in}, 1}$). Clearly, Eq.~\eqref{eq:bichromatic2} has the general form $E_\mathrm{in}(t,\tau) \equiv E_\mathrm{in}(\tau+d'\cdot t)$, where the dimensional desynchronization
\begin{equation}
d' = \frac{\Delta\tau}{t_\mathrm{R}} = -\frac{\Delta\omega_1 - \Delta\omega_0-D_\mathrm{int}(q)}{qD_1}.
\end{equation}
As described in Section~\ref{sec:polychromatic}, the numerator $\Delta\omega_1-\Delta\omega_0-D_\mathrm{int}$ describes the frequency deviation of the auxiliary injection field from the comb line with relative index $q$. Accordingly, synchronization is expected when the frequency of the auxiliary field is sufficiently close to a comb line (such that temporal desynchronization $d'$ is small), as indeed has been observed in recent experiments.

The concept of synchronization is also very evident in systems where a CS (or another nonlinear state) is generated in one resonator and injected to another CS-generating resonator~\cite{jang2018synchronization, zhao_all-optical_2024}. In this case, if the FSR difference between the resonators is sufficiently small, the soliton can be trapped to the inhomogeneity defined by the secondary state, adjusting its (free-running) comb line spacing to become synchronous with the injected perturbation.

\subsubsection{Detuning modulation}
\label{sec:CSdetmod}

Trapping and synchronization can also be performed through intracavity phase modulation, i.e., modulation of the cavity detuning. 
Although we note that slow modulation (compared to the FSR) of the detuning is widespread and can be implemented through thermal (see, e.g., \cite{wildi_phase-stabilised_2024} for a recent example) or piezoelectric control~\cite{liu_monolithic_2020}, we here focus on the case where the modulation is close to an integer multiple of the FSR as in the Sections above. 
It was only recently experimentally achieved because of the challenge of introducing fast intracavity phase modulation without degrading the finesse. Examples include fiber resonators (where the insertion loss can be compensated by an optical amplifier)~\cite{englebert_bloch_2023,englebert2026dynamics} and lithium niobate microresonators~\cite{he_high-speed_2023}.

Theoretically, the process is modeled by a fast-time-dependent detuning term in the LLE~\cite{englebert_bloch_2023,englebert2026dynamics}.
In the reference frame of the intracavity phase modulation, the LLE becomes:
\begin{equation}\label{eq:detuningLLE}
 \frac{\partial E(t,\tau)}{\partial t} = \left(-1-i\left[\Delta - V(\tau)\right]+i|E|^2 -d\dfrac{\partial}{\partial\tau}+i\dfrac{\partial^2}{\partial\tau^2} \right)E + S,
\end{equation}
where the potential $V(\tau)=\phi_\mathrm{INT}(\tau)/\alpha$ describes the intracavity phase modulation $\phi_\mathrm{INT}(\tau)$ and $d$ is the group velocity of light at the driving frequency $\omega_0$ in the modulation frame.
Unsurprisingly, Equation~\eqref{eq:detuningLLE} resembles Equation~\eqref{LLF}, which describes soliton propagation under a phase-modulated drive.
In the case of external phase modulation, the trapping range --- and the associated frequency shift --- is typically small compared to the soliton bandwidth, such that its impact on the soliton’s peak power and phase can be neglected~\cite{jang_temporal_2015}.
In contrast, intracavity phase modulation can induce frequency shifts on the order of the soliton bandwidth~\cite{englebert2026dynamics},
which significantly affects both the soliton peak power and duration.
As a result, the influence of such shifts cannot be neglected.
To capture these effects accurately, it is useful to analyze the case of detuning modulation using the Lagrangian method (see Section~\ref{sec:Lagragian}),
which enables to derive a reduced model describing the dynamics of relevant soliton parameters. 

Specifically, we use the ansatz,
\begin{equation}
E_s(t,\tau) = {B}(t)\operatorname{sech}\left[{B}(t)(\tau-\tau_\mathrm{cs}(t))/\sqrt{2}\right]\mathrm{e}^{(i\left[\phi(t)-\Omega_\mathrm{s}(t)(\tau-\tau_\mathrm{cs}(t))\right])}, 
\end{equation}
where ${B}$ is the soliton amplitude, $\Omega_\mathrm{s}$ is the frequency shift from the driving laser, $\phi$ is the phase, and $\tau_\mathrm{cs}$ is the CS position along the fast time. When applied to Eq.~\eqref{eq:detuningLLE},  following the steps described in Section \ref{sec:Lagragian}, we obtain the following reduced model for these four parameters, 
\begin{align}
    \frac{d{B}}{dt} &= -2{B}+\pi S\cos(\phi)\operatorname{sech}\left(\frac{\Omega_\mathrm{s} \pi}{\sqrt{2}{B}}\right),\\
    \frac{d\phi}{dt} &= \frac{{B}^2}{2}-\Delta+V(\tau_\mathrm{cs}) +\Omega_\mathrm{s}^2\\
    \frac{d\Omega_\mathrm{s}}{dt} &= -\frac{\Omega_\mathrm{s}}{{B}}\frac{d{B}}{dt} - 2\Omega_\mathrm{s} - \dfrac{dV(\tau_\mathrm{cs})}{d\tau_\mathrm{cs}},\\
    \frac{d\tau_\mathrm{cs}}{dt} &= -2\Omega_\mathrm{s} + d.
\end{align}
where we expanded the potential as
$V(\tau) = V(\tau_\mathrm{cs}) + (\tau-\tau_\mathrm{cs}) \frac{d V(\tau)}{d\tau}\bigg|_{\tau=\tau_\mathrm{cs}}$.

This system of equations illustrates how an intracavity phase potential influences the shape of the soliton. Specifically, the potential effectively changes the detuning~$\Delta$ and, through its slope, shifts the central frequency~$\Omega_\mathrm{s}$ of the soliton. In turn, this has an additional influence on the CS amplitude~$B$ and duration. A comprehensive description of the stationary solitons can be found in~\cite{englebert2026dynamics}.

In the reference frame of the modulation, synchronization is described by a stationary soliton ($d/dt \rightarrow 0$) for which $2\Omega_\mathrm{s} = d = -dV/d\tau_\mathrm{cs}$. Note that in real units, this can be cast as $\Omega_\mathrm{cs}=\delta\omega = -\mathcal{F}/(2\pi)\,\phi'_\mathrm{INT}$ which shows that the frequency shift is enhanced by a factor proportional to the finesse for \emph{intracavity} modulation compared with \emph{external} phase modulation of the driving (see Section~\ref{sec:PM}). If the modulation is synchronized to the FSR ($d=0$), the stationary solutions correspond to the phase modulation optima ($dV/d\tau_\mathrm{cs}=0$). As in the case of phase modulation of the driving (see section \ref{sec:PM}), the maximum is stable while the minimum is unstable because the dispersion is anomalous.
\begin{figure}[ht!]
\centering\includegraphics[width=14cm]{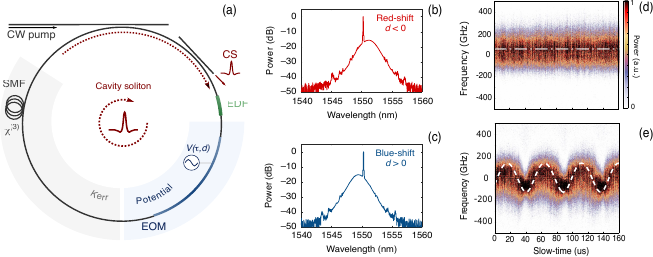}
\caption{(a) Fiber setup for intracavity detuning modulation. An Erbium-Doped Fiber Amplifier (EDFA) is included to compensate for the high insertion loss of the phase modulator but does not influence the system dynamics.
(b), (c) Soliton spectra corresponding to positive and negative desynchronization, respectively.
(d)~Instantaneous spectrum obtained via dispersive Fourier transform in the synchronization regime.
(e)~Same measurement in the regime of Bloch oscillations of cavity solitons. When the modulation slope is insufficient to counteract the soliton drift, the soliton moves in the modulation frame, resulting in periodic blue- and red-shifting of its spectrum. Panels (a--c) adapted with permission from N. Englebert \textit{et al.}, Light. Sci. \& Appl. \textbf{15} 117 (2026)~\cite{englebert2026dynamics}. Copyright \textcopyright~2026 Springer Nature. Panels (d,e) adapted with permission from N. Englebert \textit{et al.},  Nat. Phys. \textbf{19} 1014–1021 (2023) ~\cite{englebert_bloch_2023}. Copyright \textcopyright~2023 Springer Nature.}
\label{fig:detmod}
\end{figure}
However, if $d\neq0$, the soliton would initially drift in the modulation reference frame. As it propagates on the modulation slope, its frequency shifts and the soliton can eventually lock to the modulation at the point where the slope-induced frequency shift modifies the group velocity to match the drift ($d+dV/d\tau_\mathrm{cs}=0$). An example of an experimental spectrum in such regime is shown in Fig.~\ref{fig:detmod} where we see that the frequency shift in the synchronization regime can be comparable to the soliton bandwidth. The spectrum either redshifts or blue shifts depending on the sign of $d$.

This synchronization effect was also highlighted in a recent experiment of soliton microcomb generation in a $\text{LiNbO}_3$ microresonator~\cite{he_high-speed_2023}. By phase modulating the intracavity field, it was shown that the comb can be manipulated both at slow and high frequencies.
In the latter case (modulation close to the FSR) the comb can be disciplined to the external oscillator, a signature of soliton trapping~\cite{weng_spectral_2019}. 

Interestingly, if the slope of the modulation is not large enough to compensate for the drift, the soliton will continuously move in the modulation frame, which may lead to peculiar dynamics. For example, in the case of sinusoidal modulation, the soliton will be periodically blue shifted and red shifted [see Fig.~\ref{fig:detmod}(e)]. This behavior is reminiscent of the well-known Bloch oscillations in solid-state physics where a constant force applied on a particle in a periodic lattice leads to oscillatory behavior (see, e.g., \cite{leo_interband_1998}). 

Finally, we note that it was recently predicted that novel localized structures may emerge in the presence of a fast detuning modulation. In particular, parabolic modulation induces complex resonance structures, with secondary peaks, each corresponding to a higher-order Hermite-Gaussian-like localized structure\cite{sun_dynamics_2023}~.

\subsection{Higher-order dispersion}
\label{sec:DW_and_recoil}

The existence and characteristics of CSs discussed in Section~\ref{sec:CSexist} considered the canonical LLE. However, as described in Section~\ref{sec:norm}, the canonical LLE makes a number of simplifying approximations that are not always valid. One of the most significant approximations is the assumption that the group-velocity dispersion $\beta_2$ (or equivalently, $D_2$ --- see Section~\ref{sec:CM}) is constant and does not depend on frequency over the soliton bandwidth (i.e., $\beta_k = 0$ for $k>2$). Here we relax this assumption and consider how higher-order dispersion (HOD) perturbs CSs. 

\subsubsection{Dispersive waves and spectral recoil}
\label{sec:CS_DW}

The effect of HOD on temporal solitons has been extensively studied in the context of single-pass fiber propagation described by the (generalized) NLSE. It is well known that in this context HOD can cause the solitons to emit linear ``dispersive waves'' (DWs) at a frequency where the waves accumulate the same phase as the soliton with propagation~\cite{wai_nonlinear_1986,akhmediev_cherenkov_1995, agrawal_nonlinear_2013}. Referring to Eqs.~\eqref{GNLSE} and~\eqref{Dhat}, this phase-matching condition can to first order be written as
\begin{equation}
\label{eq:NLSE_PM}
    \frac{\gamma P_\mathrm{s}}{2} = \hat{D}(\Omega_\mathrm{DW}),
\end{equation}
where $P_\mathrm{s}$ is the soliton peak power, $\Omega_\mathrm{DW}$ is the angular frequency shift of the DW from the soliton, and the left-hand-side of the equation reflects the soliton's nonlinear phase.

In the absence of HOD, Eq.~\eqref{eq:NLSE_PM} simplifies to $\gamma P_\mathrm{s}/2 = \beta_2\Omega^2/2$, which has no roots because $\gamma$ and $\beta_2$ have opposite signs for bright solitons [Fig.~\ref{fig:DWPM}(a)]. It is in fact this phase-mismatch that underpins the stability of the NLSE soliton. HOD fundamentally changes the situation, permitting phase-matching to be satisfied [Fig.~\ref{fig:DWPM}(b)]. For example, in the simplest case, where dispersion is truncated to third-order ($\beta_k=0$ for $k>3$), phase-matching is approximately satisfied for $\Omega_\mathrm{DW}\approx -3\beta_2/\beta_3$, obtained by neglecting the soliton's nonlinear phase contribution (which is typically small). In this case, it is straightforward to show that $\Omega_\mathrm{DW}$ lies in the region of normal dispersion.

\begin{figure}[b]
  \centering
  \includegraphics[width = 0.7\textwidth, clip=true]{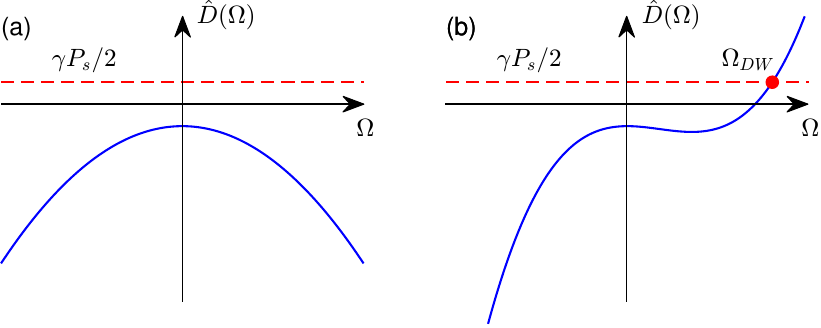}
  \caption{Schematic illustration of DW phase-matching in a conservative NLSE system with (a) second-order dispersion only --- no phase-matching possible --- and (b) higher-order dispersion, which enables phase-matching.
  }
  \label{fig:DWPM}
\end{figure}

In the NLSE soliton context, dispersive wave phase-matching results in the shedding of energy from the soliton to the dispersive wave, giving rise to a conspicuous spectral peak at $\Omega_\mathrm{DW}$. The efficiency of this process increases with the spectral overlap between the soliton and the phase-matched frequency, meaning substantial emission typically requires the soliton to be spectrally located close to the zero-dispersion wavelength~\cite{dudley_supercontinuum_2006}. Additionally, conservation of momentum dictates that the emitted radiation induces a spectral recoil of the soliton in the opposite direction. This recoil, mediated by group-velocity dispersion, also alters the soliton’s group velocity.

In the driven cavity context, the influence of higher-order dispersion on (cavity) solitons closely mirrors that observed in conservative systems. This is illustrated in Figs.~\ref{fig:DWemission}(a) and (b), which shows the simulated temporal and spectral profiles of a CS. The parameters used correspond to a 50~GHz FSR MgF$_2$ microresonator with $\beta_2 = -5~\mathrm{ps^2/km}$ and $\beta_3 = 1~\mathrm{ps^3/km}$ (for other parameters, see figure caption). This simulation reveals three key features. Firstly, the soliton spectrum develops a distinct dispersive wave peak at $\Omega_\mathrm{DW}\approx 3.8$ THz; secondly, the soliton center frequency has red-shifted due to spectral recoil; and thirdly, as a direct result of this offset in its center frequency, the soliton has acquired a constant drift in the original simulation reference frame. For the simulation parameters used in Fig.~\ref{fig:DWemission} this drift rate is $V=0.11$ fs/roundtrip.

\begin{figure}
  \centering
  \includegraphics[width = 0.8\textwidth, clip=true]{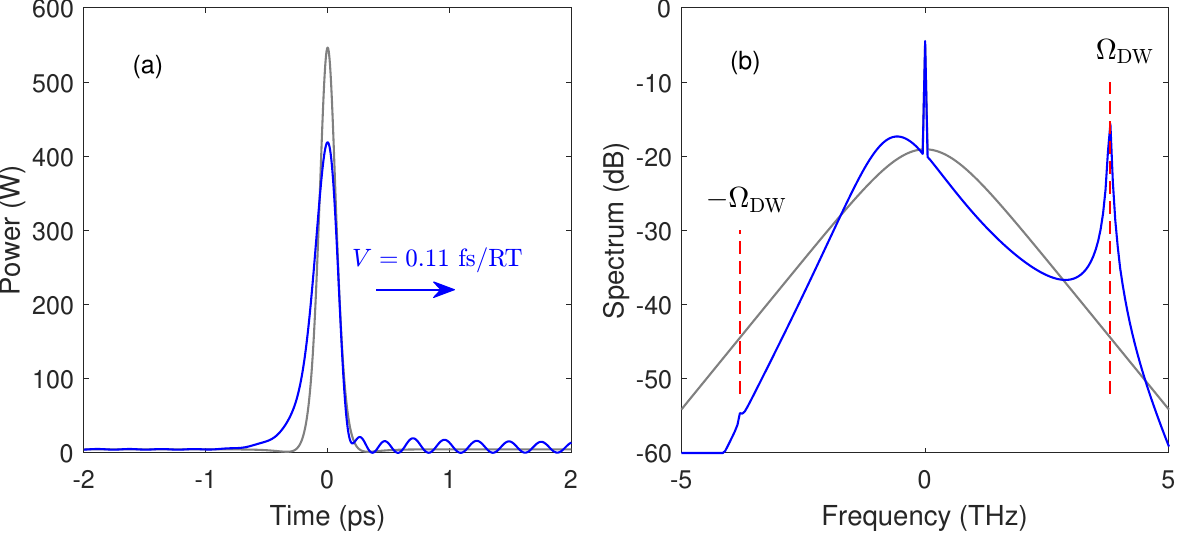}
  \caption{The temporal and spectral profiles of a CS subject to third-order dispersion are shown in panels (a) and (b) respectively (blue traces). These profiles are obtained from numerical simulation of the dimensional LLE using the following parameters typical of a cw-driven, critically-coupled MgF$_2$ microdisk resonator: $\mathrm{FSR} = 50$ GHz, $n=1.37$, Finesse$~=~30000$, $P = 50$ mW ($X=20$), $\gamma = 10^{-3}~\mathrm{W}^{-1}\mathrm{m}^{-1}$, $\beta_2 = -5~\mathrm{ps^2/km}$, $\beta_3 = 1~\mathrm{ps^3/km}$, and $\Delta = \delta_0/\alpha=10$. The superimposed gray traces show the CS profiles in the absence of HOD ($\beta_3=0$). The positions of the theoretically predicted DW frequencies $\pm\Omega_{\mathrm{DW}}$ as given by Eq.~\eqref{Eq:DWpm2} are plotted as red dashed lines, and show excellent agreement with the simulated DW peak positions.
  }
  \label{fig:DWemission}
\end{figure}

Despite their similarities, a key distinction exists in how conservative NLSE solitons and CSs respond to higher-order dispersion. Conservative solitons continuously emit radiation that drifts away in time, whereas CSs settle into a new steady-state featuring reshaped temporal and spectral profiles along with a modified group velocity. In the time domain, this steady-state takes the form of a modified soliton shape with the DW manifesting itself as an exponentially decaying, oscillatory tail locked to the side of the CS~\cite{coen_modeling_2013,jang_observation_2014,brasch_photonic_2016}. The presence of such a tail is clearly evident in Fig.~\ref{fig:DWemission}(a).

To study CSs in the presence of HOD, we consider the LLE written in a reference frame $\tau' = \tau - V t/t_\mathrm{R}$ co-moving with the (frequency-shifted) CSs, and in which the CSs are stationary. Here $V$ (which has units of fast time) represents the group-delay accumulated over one round trip by the temporal CSs with respect to the driving field. We get
\begin{equation}
t_\mathrm{R}\frac{\partial E(t,\tau')}{\partial t} = \left[-\alpha+i(\gamma L |E|^2-\delta_0) + {V \frac{\partial}{\partial\tau'}} + i L \hat{D}\left(i\frac{\partial}{\partial\tau'}\right) \right]E + \sqrt{\theta}E_\mathrm{in}.
\label{Eq:driftLLE}
\end{equation}

A phase-matching condition for the dispersive wave emission process can be derived in the cavity context by noting that, far away from the soliton, the field can be approximated as a superposition of the cw background (with complex amplitude $E_0$ and power $P_0 = |E_0|^2$) on top of which the soliton sits, and a small amplitude damped oscillation corresponding to the DW~\cite{jang_observation_2014}. Specifically, substituting the ansatz
\begin{equation}
    E(t,\tau') = E_0 + ae^{-iQ\tau'}+be^{iQ^\ast\tau'}
\end{equation}
into Eq.~\eqref{Eq:driftLLE} and linearizing the resulting equations with respect to the small amplitudes $a$ and $b$, one finds that the steady-state condition is equivalent with the following matrix equation
\begin{equation}\label{DW_matrix}
\begin{bmatrix}
K + i[L\hat{D}_\mathrm{o}(Q)-VQ] & i\gamma L E_0^2  \\
-i\gamma L (E_0^2)^\ast & K^\ast + i[L\hat{D}_\mathrm{o}(Q)-VQ]
\end{bmatrix}
\begin{bmatrix}
a  \\
b^\ast
\end{bmatrix} 
= 0,
\end{equation}
where $K = -\alpha -i\delta_0 + iL \hat{D}_\mathrm{e}(Q) + 2i\gamma L P_0$ with $\hat{D}_\mathrm{o}$ and $\hat{D}_\mathrm{e}$ denoting the odd and even parts of the dispersion operator (capturing $\beta_{2k+1}$ and $\beta_{2k}$ with $k\geq 1$, respectively). Notice how this result is a generalization (written in real units) of Eq.~\eqref{MI_matrix} derived for the study of intracavity MI. The above matrix equation has a non-trivial solution only when the determinant of the matrix is zero, which allows one to derive the following condition for the complex frequency $Q$ that determines the DW frequency (real part of $Q$) and decay rate (imaginary part of $Q$),
\begin{equation}
    \label{Eq:DWpm}
    -\alpha + i L \hat{D}_\mathrm{o}(Q) - iVQ \pm i \sqrt{[2\gamma L P_0 - \delta_0 + L\hat{D}_\mathrm{e}(Q)]^2 - (\gamma L P_0)^2} = 0.
\end{equation}

Equation~\eqref{Eq:DWpm} is the salient condition that determines the characteristics of DWs emitted by CSs. In contrast with the conservative NLSE case, Eq.~\eqref{Eq:DWpm} predicts that dispersive waves always come in pairs in the cavity context (associated with the plus and minus signs in the equation). It is straightforward to show that these pairs are related viz.\ $Q_+ = -Q_-^\ast$, signaling that the two DWs are symmetrically detuned in frequency from the driving field. A simple analysis shows that one of these DW signals is typically much stronger than the other one. This can clearly be seen in the numerically simulated spectrum shown in Fig.~\ref{fig:DWemission}(b), where in addition to a strong DW peak at $\Omega_{\mathrm{DW}}$, a much weaker peak is also visible at $-\Omega_{\mathrm{DW}}$. While many experiments often capture only the dominant DW, clear evidence of pair-wise DW generation has also been reported~\cite{jang_observation_2014}.

Equation~\eqref{Eq:DWpm} can be considerably simplified by noting that, because $P_0$ represents the power of the lower level of the bistable cw response (see, e.g., Fig.~\ref{fig:KerrTilt}), the term $(\gamma L P_0)^2\approx 0$ under typical conditions. For the $Q_+$ solution, this yields
\begin{equation}
\label{Eq:DWpm2}
   L\hat{D}(Q) - VQ + (2\gamma L P_0-\delta_0 + i\alpha) = 0.
\end{equation}
This simplified expression bears close similarity with the phase-matching condition given by Eq.~\eqref{eq:NLSE_PM}. In both cases, the phase-matching occurs at frequencies close to where $\hat{D}(\Omega_\mathrm{DW})\approx 0$, with the presence of nonlinear phase-shifts and pump-cavity detuning typically only slightly shifting the actual DW frequency. 

A useful simplification arises by neglecting both the nonlinear phase shift and propagation losses. Under this approximation, $Q=\Omega_\mathrm{DW}$ is real and the angular frequency shift of the dispersive wave (DW) satisfies the condition $L\hat{D}(\Omega_\mathrm{DW}) - V\Omega_\mathrm{DW} - \delta_0 = 0$. Utilizing Eq.~\eqref{Dint}, and writing the phase detuning $\delta_0$ in terms of the frequency detuning $\Delta\omega_0 = \delta_0/t_\mathrm{R}$, allows us to reformulate the phase-matching condition as:

\begin{equation}
\omega'_{\mu_\mathrm{DW}} = \omega_0+\mu_\mathrm{DW} D_1^\mathrm{(CS)},
\end{equation}
where $\omega'_{\mu_\mathrm{DW}}$ is the cavity resonance frequency for the DW (with mode index $\mu_\mathrm{DW}$), $\omega_0$ is the driving frequency, and $D_1^{(\mathrm{CS})} = D_1(1-V/t_\mathrm{R})$ is the comb line spacing (in angular frequency) of the CS. Note that, the CS comb line spacing differs from $D_1$ at the pump frequency due to the spectral recoil induced drift. This expression makes it clear that phase-matching for DWs can also be understood to occur when a soliton comb line coincides with, or lies near to, a linear cavity resonance.

From an applied point of view, dispersive waves offer an important mechanism to allow the substantial enhancement of selected spectral components within a CS spectrum. Indeed, a number of experiments have leveraged judicious dispersion management to design and realize resonators that allow for the generation of intense DWs at desired frequencies (see, e.g., \cite{brasch_photonic_2016,pfeiffer2017octave,li_stably_2017, yu2019tuning, liu_aluminum_2021}).

\subsubsection{Kelly sidebands}
\label{sec:Kelly}
As alluded to above, dispersive waves arise when the phase shift accumulated by the wave over one round trip matches that of the driving field at the coupler, enabling resonant energy transfer. More generally, constructive interference also occurs when this phase difference equals an integer multiple of $2\pi$. Neglecting losses and concentrating on the phasematched frequency shift, we can extend the phase-matching condition in Eq.~\eqref{Eq:DWpm2} to account for this possibility via,

\begin{equation}
\label{Eq:Kelly}
   L\hat{D}(\Omega_m) - V\Omega_m + 2\gamma L P_0-\delta_0 = 2\pi m,
\end{equation}
where $m$ is an integer. 

Dispersive waves (DWs) represent a specific instance of resonant radiation corresponding to the case $m = 0$. Radiations arising from $m \neq 0$ are commonly referred to as ``Kelly sidebands,'' as they are closely analogous to the characteristic sidebands observed in mode-locked soliton lasers~\cite{dennis_experimental_1994} whose origins were first elucidated in Refs.~\cite{kelly_characteristic_1992,gordon_dispersive_1992}. Figure~\ref{fig:Kelly} shows a CS spectrum obtained from numerical simulations with parameters corresponding to a 10-m-long fibre ring resonator with finesse $\mathcal{F} = 20$ driven at 1550~nm (see caption for additional parameters). Strong spectral sidebands are clearly visible at frequencies predicted by Eq.~\eqref{Eq:Kelly}, as highlighted by dashed vertical lines. 

\begin{figure}[b]
  \centering
  \includegraphics[width = 1\textwidth, clip=true]{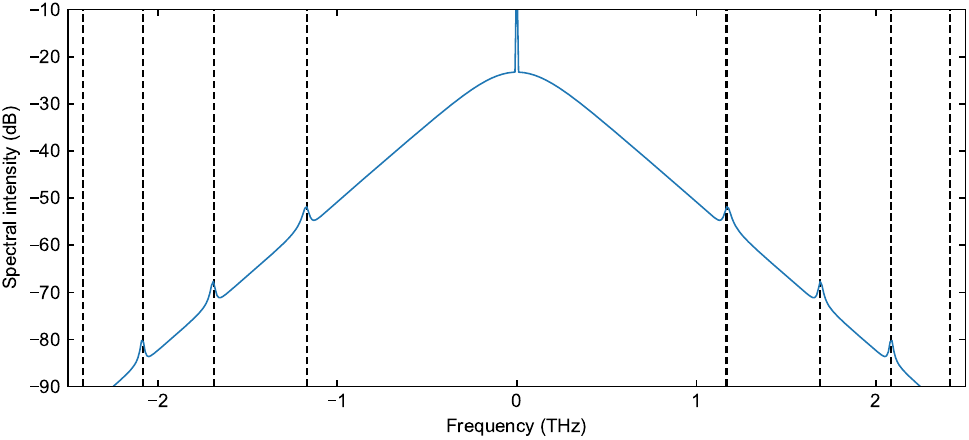}
  \caption{Numerically simulated spectral profile of a CS in a critically-coupled fibre ring resonator with $L = 10~\mathrm{m}$, Finesse $\mathcal{F} = 20$, $\gamma = 1.2\times10^{-3}~\mathrm{W}^{-1}\mathrm{m}^{-1}$, $\beta_2 = -21.4~\mathrm{ps^2/km}$, and $\Delta = \delta_0/\alpha=4$. The driving power $P_\mathrm{in} = 10~\mathrm{W}$, which can be readily achieved using nanosecond flat-top pulses. The dashed vertical lines correspond to positions of resonant Kelly-like sidebands as predicted by Eq.~\eqref{Eq:Kelly} for different integers $m$. The simulation results were obtained using the full Ikeda map described by Eqs.~\eqref{GNLSE} and~\eqref{map}.
  }
  \label{fig:Kelly}
\end{figure}

Whilst DWs manifest themselves both in (conservative) single-pass and cavity systems, Kelly sidebands typically only appear in the latter. This distinction stems from the difference in the mechanisms that underpin the radiations. DWs originate from a cascaded four-wave-mixing process, where phase-mismatches associated with individual steps of the cascade cancel out to phase-match the cascade as a whole~\cite{erkintalo_cascaded_2012}. Kelly sidebands, in contrast, require a periodic perturbation that acts upon the soliton so as to quasi-phase-match the cascade. Round-trip periodicity is intrinsically built into a cavity system, whereas in single-pass systems such periodicity typically requires the use of specialty fibers or waveguides~\cite{conforti2015parametric}. From a modeling perspective, capturing these features necessitates the use of the full Ikeda map described in Section~\ref{sec:Ikeda}; the mean-field LLE inherently averages over periodic perturbations from the cavity coupler and therefore cannot reproduce Kelly sidebands. Kelly sideband intensity increases with the strength of periodic perturbations, which explains why such sidebands in CS systems are most often observed in relatively low-finesse resonators. Moreover, because the sideband frequency scales inversely with resonator length, Kelly sidebands are also much more easily observed in macroscopic fiber resonators. 

The cavity coupler provides a minimum periodic perturbation that can give rise to Kelly sidebands. However, additional intracavity perturbations -- such as lossy components or dispersion modulations -- can magnify the sidebands and change their frequencies (if the period of the intracavity perturbation is a fraction of the round trip). These effects have been experimentally observed in dispersion managed systems ranging from macroscopic fiber resonators~\cite{wang_universal_2017, nielsen_invited_2018} to silicon nitride microresonators~\cite{anderson_dissipative_2023}. If their amplitude is sufficiently large, Kelly sidebands have also been observed to affect the existence of CSs, as demonstrated in Ref.~\cite{nielsen_invited_2018}.

\subsubsection{Avoided mode crossings}

As detailed in Section~\ref{avoided_crossings}, linear coupling between two mode families (or resonators) can lead to avoided mode crossings, where the resonance frequencies of the modes shift as they approach one another. When the interacting modes differ significantly in free spectral range (FSR), strong coupling tends to arise only within a narrow spectral window, resulting in frequency shifts confined to a handful of longitudinal modes. Because resonance frequencies are intrinsically linked to device dispersion [see Eq.~\eqref{Dint}], such linear mode coupling introduces strongly localized spectral perturbations to the dispersion profile. These perturbations can be modeled phenomenologically by assuming the resonance frequencies to be of the form~\cite{herr_mode_2014}
\begin{equation}
   \omega'_\mu = \omega'_0 + \mu D_1 + \frac{D_2}{2}\mu^2 + \frac{a/2}{\mu - b - 0.5}.
\end{equation}
with the parameter $a$ representing the amplitude of the frequency shift due to the mode coupling and $b$ the relative position of the avoided crossing, respectively. This yields a resulting integrated dispersion given by $D_\mathrm{int} = D_2\mu^2/2 + (a/2)/(\mu - b - 0.5)$.



Dispersion perturbations from avoided mode crossings can markedly influence a CS’s temporal and spectral characteristics~\cite{herr_mode_2014}. Although these modifications closely resemble those caused by DW emission, several qualitative distinctions arise. In particular, the narrow spectral localization of typical mode crossings gives rise to a sharp spectral peak, whereas smooth higher-order dispersion typically results in DWs that are spectrally broader. In the time domain, this implies that the decay of the oscillatory tail associated with mode crossings is much slower, which in microresonators with small round-trip time (large FSR) can result in the wrapping around of the oscillatory tail to surround the entire soliton. 

Besides affecting the CS temporal and spectral characteristics, mode-crossings can also have a significant impact on the solitons' existence and stability. In particular, mode crossings that occur close to the driving frequency or that are large in magnitude can prohibit solitons from existing~\cite{herr_mode_2014}. In general, it is desirable to have a resonator mode spectrum that is void of mode crossings to be able to access the entire range of soliton existence. Interestingly, however, mode-crossings can also be beneficial in some (other) contexts. For example, they can be leveraged to engineer quiet points~\cite{yi2017single}, increase the conversion efficiency in CS comb generation~\cite{helgason2023surpassing}, ensure spontaneous and deterministic soliton generation~\cite{yu_spontaneous_2021}, or to allow modulation instability to occur in the normal dispersion regime (see section~\ref{sec:SWexcitation}).

The above discussions have focused upon the case where the interacting modes have a large FSR difference, such that strong mode coupling only occurs around a narrow spectral region. However, it is also possible to engineer situations where the FSR difference is small, such that mode coupling occurs over an extended frequency range. This can particularly be achieved by fabricating on-chip devices consisting of resonators that are evanescently coupled to one another and that have very similar optical path lengths~\cite{soltani_enabling_2016, kim_dispersion_2017, yuan_soliton_2023}. 
In such devices, the mode coupling can be shown to add an anomalous dispersion contribution to one of the supermodes, and a normal dispersion contribution to the other. If sufficiently strong, this can compensate for the intrinsic dispersion of the uncoupled rings. This phenomenon has been successfully leveraged to realize windows of anomalous dispersion, and the subsequent generation of CSs in resonators that would exhibit strong normal dispersion without mode coupling~\cite{yuan_soliton_2023}.

\subsection{Spectral extension}
\label{sec:extension}
Dispersive wave emission can selectively amplify spectral components near the phase-matched frequency, making it valuable for applications that require enhanced signal strength at targeted frequencies. However, this technique is constrained by a key limitation: the DW phase-matched frequency is primarily set by the resonator’s dispersion and cannot easily be altered after fabrication. As such, this approach offers limited tunability and demands precise dispersion control during both the design and fabrication.

An alternate strategy to enhance targeted spectral components, or to extend the CS's spectral extent, involves injecting a second ``auxiliary'' driving field into the resonator at a frequency distinct from that of the primary pump responsible for CS generation~\cite{zhang_spectral_2020,zhang2022dark,moille_ultra-broadband_2021,qureshi_soliton_2022}. In this case, the CS driven by the primary pump can nonlinearly interact with the intracavity field associated with the auxiliary pump, giving rise to a secondary frequency comb centered around the auxiliary frequency. This secondary comb shares the same repetition rate as the main CS comb, but is frequency offset from the primary comb by a constant amount; that is, the two combs possess different carrier-envelope-offset frequencies.

The spectral extension described above can be interpreted in the time domain as originating from cross-phase modulation between the CS and the intracavity field at the auxiliary frequency. This nonlinear interaction induces a time-dependent phase shift that manifests as a localized spectral feature centered at the auxiliary frequency and group-velocity-matched to the CS. The resulting auxiliary comb thus represents the frequency-domain counterpart of this localized temporal structure. Alternately, in the frequency domain, the generation of the auxillary comb can be understood as arising from a phasematched cascade of individual four-wave-mixing (FWM) Bragg scattering processes~\cite{xu2013cascaded}. In this picture, CS comb lines mix with the auxiliary field at $\omega_\mathrm{a}$ to drive energy towards a new phasematched ``idler'' frequency $\omega_\mathrm{i}$ thus generating a new auxillary comb~\cite{qureshi_soliton_2022}. 

This FWM Bragg scattering cascade can in general result in the generation of new frequency components on either side (i.e., red- and blue-shifted) of the injected auxiliary field. However, due to phase-matching considerations, the process is typically biased towards one direction. In particular, under specific conditions, FWM Bragg scattering between a CS and an injected auxiliary laser can result in the generation of a sharp DW-like spectral feature (sometimes referred to as a ``synthetic DW''~\cite{moille_ultra-broadband_2021}). This process essentially ``translates'' the injected frequency to a new phasematched idler frequency. 

The phase-matching condition pertinent to the CS-auxiliary interaction can be found by looking for the idler frequency that accumulates the same phase as the field at the injected auxiliary frequency over one round trip~\cite{qureshi_soliton_2022}. This yields
\begin{equation}
\label{Eq:BraggPM1} 
    \hat{D}(\Omega_\mathrm{i})L = \hat{D}(\Omega_\mathrm{a})L + \delta_\mathrm{a}, 
\end{equation}

or equivalently,

\begin{equation}
\label{Eq:BraggPM2} 
    D_\mathrm{int}(\mu_\mathrm{i}) = D_\mathrm{int}(\mu_\mathrm{a}) - \Delta\omega_\mathrm{a}. 
\end{equation}

Here $\Omega_\mathrm{i,a} = \mu_\mathrm{i,a}D_1 \approx \omega_\mathrm{i,a}-\omega_0$ are the angular frequency shifts of the idler and auxiliary fields from the main CS-generating pump at $\omega_0$ (with corresponding relative mode numbers $\mu_\mathrm{i,a}$), respectively, and $\delta_\mathrm{a}$ ($\Delta\omega_\mathrm{a})$ is the phase (frequency) detuning of the auxiliary field from its closest cavity resonance. Equations~\eqref{Eq:BraggPM1} and~\eqref{Eq:BraggPM2} reveal that, for a fixed frequency CS-generating pump, the idler signal frequency $\omega_\mathrm{i}$ can be tuned by changing the mode driven by the auxillary pump -- modifying $D_\mathrm{int}(\mu_\mathrm{a})$ -- and/or by adjusting the auxiliary pump detuning $\Delta\omega_\mathrm{a}$. 

For large frequency separations between the soliton and auxiliary pumps, the dispersive terms in Eqs.~\eqref{Eq:BraggPM1} and \eqref{Eq:BraggPM2} generally dominate over detuning contributions. Ref.~\cite{moille_ultra-broadband_2021} demonstrated operation in this regime using a silicon nitride microresonator with a diameter of $46\,\mu\mathrm{m}$. Injecting a soliton pump at 1063 nm, and an auxiliary pump near 1560 nm (detuned by approximately $100$~THz) into this resonator, they were able to demonstrate the coherent spectral extension of the initial $\sim 100$ THz CS comb to over $\sim 250$ THz (1.6 octaves). Conversely, Ref.~\cite{qureshi_soliton_2022} working in the small frequency shift regime was able to demonstrate the effect of the phase detuning on new generated comb frequency. Figures~\ref{fig:spectral_extension}(a--c) show results obtained from a $\sim 1$~mm diameter magnesium fluoride microresonator generating a CS driven by a pump at 1550 nm, as the phase detuning ($\delta_\mathrm{a}$) of an auxiliary pump at 1582 nm is varied. The blue curves show the measured comb spectra at the output of the resonator for three different negative values of auxillary pump phase detuning. Here, changing the auxillary detuning can be seen to provide simple mechanism for control of the newly generated comb's frequency shift. Also plotted in Figs.~\ref{fig:spectral_extension}(d--f) is the respective RF spectrum measured in each case. The strong beat tone seen here is evidence of the constant (and tunable) frequency offset between the soliton and auxillary combs. 
 
\begin{figure*}[!t]
	\centering
	\includegraphics[width = 0.65\textwidth, clip=true]{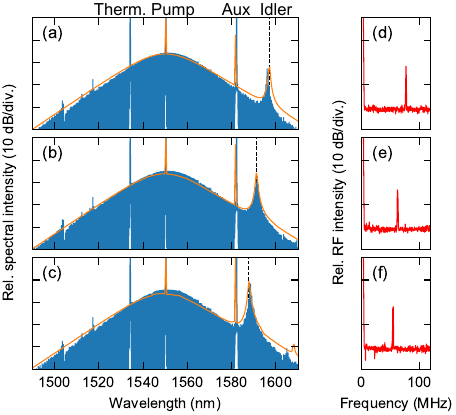}
	\caption{(a--c) Comb spectra measured at the resonator output for decreasing values of the auxillary detuning ($\Delta\omega_a/2\pi \sim 56, 48$ and $34$~MHz), resulting in a frequency tunable auxillary comb (blue curves). The orange curves show corresponding simulated spectra and the dashed black lines show the phasematched wavelength predicted by Eq.~\eqref{Eq:BraggPM1}. (d--f) Corresponding measured RF spectrum for each comb. The labels Therm, Pump, Aux and Idler indicate the locations of the thermal compensation field, the soliton pump, the auxilary field and the phasematched idler frequency respectively. Reprinted with permission from P. Qureshi \textit{et al.}, Commun. Phys. \textbf{5}, 123 (2022)~\cite{qureshi_soliton_2022}. Copyright \textcopyright~2022 Springer Nature.}
	\label{fig:spectral_extension}
\end{figure*}

\subsection{Raman scattering}\label{sec:Raman}

Raman scattering is a third-order nonlinear effect present in both glass and crystalline waveguides. It describes the inelastic scattering of an incident electromagnetic field from the vibrational modes (phonons) of the medium. This interaction leads to a direct exchange of energy between the field and the medium, resulting in a frequency shift of the scattered light. In silica, the peak of this shift corresponds to a phonon frequency of approximately $\sim13~\text{THz}$ (c.f. Fig.~\ref{fig:SRS}). Raman scattering enables efficient down-conversion of optical signals, and plays a significant role in shaping the propagation dynamics of short pulses~\cite{agrawal_nonlinear_2013}. In this section, we discuss its role in the formation of solitons in driven cavities, with an emphasis on silica resonators where the Raman response is well characterized. We analyze both the impact of the intra-pulse Raman effect on cavity solitons and the emergence of novel solitons facilitated by the Raman effect.

\subsubsection{Soliton self-frequency shift}
\label{sec:Raman_self_freq_shift}

When sufficiently short optical pulses propagate through a nonlinear medium with a broad Raman response, their low-frequency components can be amplified by the high-frequency components via stimulated Raman scattering (SRS). This phenomenon, known as intrapulse Raman scattering~\cite{dianov1985stimulated}, causes the pulse to undergo a progressive red-shift during propagation. It has been extensively studied in the context of single-pass soliton propagation, where it is commonly referred to as the soliton self-frequency shift~\cite{mitschke1986discovery,gordon1986theory}. Unsurprisingly, intrapulse Raman scattering also plays a significant role in the dynamics and formation of cavity solitons.
 
We begin by numerically continuing the normalised generalised LLE equation~\eqref{GLLEN} for a silica resonator, both with and without the Raman term. The Raman fraction, $f_R = 0.18$, and normalised fast time scale parameter, $\tau_s = \sqrt{|\beta_2|L/(2\alpha)}= 1.9$~ps, used correspond to that of a typical fiber cavity~\cite{wang_stimulated_2018}. The spectral profile of the Raman gain is obtained by using the multiple-
vibrational-mode model of Ref.~\cite{hollenbeck_multiple-vibrational-mode_2002}. The results are presented in Fig.~\ref{fig:Raman_continue} for a normalised drive power of $X=130$. Panel (a) illustrates the continuation of CS solutions in the presence (red) and absence (blue) of Raman scattering. The comparison reveals that Raman scattering significantly perturbs the CS branch, narrowing its existence range due to the emergence of a second Raman-induced Hopf bifurcation at detuning $\Delta_\mathrm{H2}$. Additionally, the spectra of individual soliton solutions exhibit a red-shift relative to the pump frequency. This spectral displacement is evident in panels (b) and (c), which display the temporal and spectral profiles of the CS solution at $\Delta = 62$. Unlike the conservative case --- where solitons continuously shift in frequency --- the dissipative nature of the cavity enforces a fixed stationary frequency shift to the CS spectrum. In experiments, the red-shift of CSs due to soliton self-frequency shift has been observed in both fiber and microresonator systems~\cite{anderson_observations_2016,wang_stimulated_2018,yi_soliton_2015,karpov_raman_2016,yi_theory_2016, webb_experimental_2016}. In Fig.~\ref{fig:Raman_continue}(d) we plot the experimentally measured spectrum of a Raman self-frequency-shifted CS in a 22 GHz silica wedge microresonator (blue trace), as reported in Ref.~\cite{yi_soliton_2015}. The frequency offset between the soliton center frequency from the pump field is clearly visible. The numerically simulated spectrum is superimposed (green trace) and shows excellent agreement.

\begin{figure*}[!t]
	\centering
	\includegraphics[width = 0.7\textwidth, clip=true]{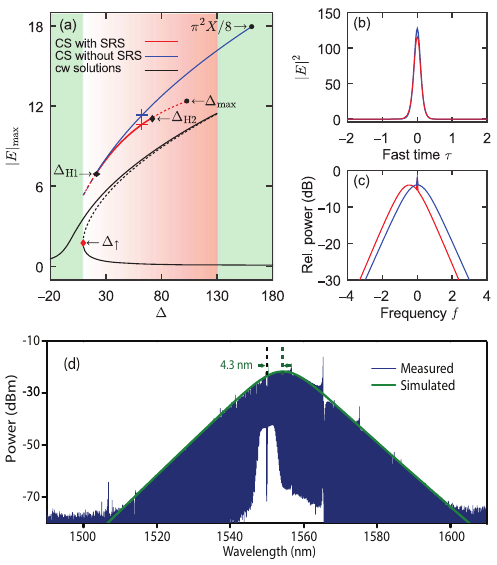}
	\caption{(a) Peak amplitude of soliton, $|E|_\mathrm{max}$, as a function of cavity detuning $\Delta$ at $X = 130$. The cw cavity solutions as shown as black curves, whilst the CS solutions with and without SRS, are drawn as red and blue curves. The dashed curves correspond to unstable solutions. (b, c) Temporal and spectral CS profiles obtained at $\Delta =62$. As in (a), red and blue curves correspond to CS solutions with and without SRS, respectively. The Raman fraction $f_R$ is set to $0.18$, and the fast time normalised time scale $\tau_s = 1.9$~ps. (d) Experimental measured (blue) and simulated (green) spectrum of a CS experiencing Raman SFS in a silica wedge microresonator (blue trace). Panels (a--c) adapted with permission from Y. Wang \textit{et al.}, Phys. Rev. Lett. \textbf{120} 053902 (2018)~\cite{wang_stimulated_2018}. Copyright \textcopyright~2018 American Physical Society. Panel (d) adapted with permission from Q. F. Yi \textit{et al.}, Optica \textbf{2}, 1078–1085 (2015)~\cite{yi_soliton_2015}. Copyright \textcopyright~2015 Optical Society of America.}
	\label{fig:Raman_continue}
\end{figure*}

To better understand and quantify the self-frequency shift of cavity solitons, we next apply the well known linear approximation of the Raman response which is valid as long as the soliton duration is much longer than the Raman response time~\cite{agrawal_nonlinear_2013}. Under that approximation, in normalised units, the Generalized LLE~\eqref{GLLEN} becomes
\begin{align}\label{eq:lle_raman}
	\frac{\partial E}{\partial t} = \left[-1-i\Delta + i \hat{D}\left(i\frac{\partial}{\partial\tau}\right) \right]E +i\left[|E|^2 -\tau_R \partial_\tau|E|^2\right]E + S. 
\end{align}
where $\tau_R = \tau'_R/\tau_\mathrm{s}$ and $\tau'_R=3~\mathrm{fs}$ for silica. 
This simpler model can be readily investigated with a Lagrangian analysis (see Section~\ref{sec:Lagragian}). The analysis presented here is a slightly extended version of that first presented in~\cite{yi_theory_2016}. Using the same ansatz as in section~\ref{sec:CSdetmod}, we recover a similar model as the case of detuning modulation but with the Raman effect as a source term for the frequency shift: 
\begin{align}\label{eq:Raman}
    \frac{d{B}}{dt} &= -2{B}+\pi S\cos(\phi)\operatorname{sech}\left(\frac{\Omega_s \pi}{\sqrt{2}{B}}\right),\\
    \frac{d\phi}{dt} &= \frac{{B}^2}{2}-\Omega_s^2-\Delta-\frac{d\tau_\mathrm{cs}}{dt}\Omega_s,\\
    \frac{d\Omega_s}{dt} &= -\frac{\Omega_s}{{B}}\frac{d{B}}{dt} - 2\Omega_s -\frac{4\tau_R}{15}{B}^4,\\
    \frac{d\tau_\mathrm{cs}}{dt} &= -2\Omega_s.
\end{align}
The self-frequency shift scaling (${B}^4$ or equivalently $1/\tau^4$) is unsurprisingly the same as for conservative solitons, but the added restoring force allows for stationary solitons ($d/dt \rightarrow 0$) with frequency shift and amplitude respectively given by 
\begin{align} 
\Omega_s &=-\frac{2\tau_R}{15}{B}^4,\\
{B} &= \sqrt{2(\Delta-\Omega_s^2)}.
\end{align}
By combining these expressions, we can directly link the cavity detuning and the self-frequency shift
\begin{equation}
    \Delta = \sqrt{\frac{15|\Omega_s
    |}{8\tau_R}}+|\Omega_s|^2.
\end{equation}

In Fig.~\ref{fig:Raman_Lagrange} we compare the soliton self frequency shift and peak amplitude predicted from the above Lagrangian results (blue traces) with those obtained by Newton continuation of the full LLE using the complete Raman gain profile provided by the multi-vibrational mode model (orange traces)~\cite{hollenbeck_multiple-vibrational-mode_2002}. The detuning range shown corresponds to the existence range of stable CS, and parameters used match those in Fig.~\ref{fig:Raman_continue}. The two approaches show an excellent agreement, with residual differences attributable primarily to the simplified linear Raman gain adopted in the Lagrangian analysis. As might be expected, the reduced model does not capture the Raman‑induced Hopf instability, for which a more complete ansatz is required to reproduce the soliton’s dynamical behaviour~\cite{nozaki_chaotic_1985}.

\begin{figure*}[!t]
	\centering
	\includegraphics[width = 0.7\textwidth, clip=true]{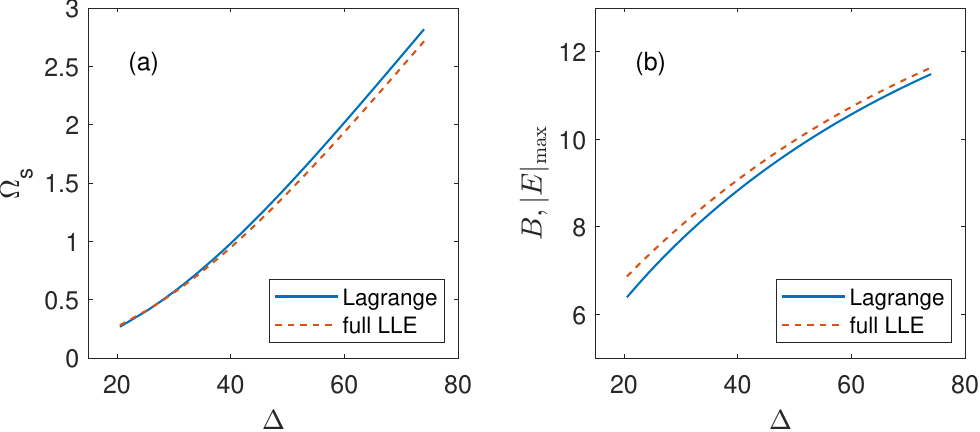}
	\caption{(a) Raman self frequency shift, $\Omega_s$, and (b) peak amplitude of the soliton, $B$ (or equivalently $|E|_\mathrm{max}$), versus detuning. The solid blue curves show the prediction of the reduced Lagrangian model using a linear Raman gain profile with $\tau'_R=3~\mathrm{fs}$; the dashed orange curves shows the prediction of a Newton continuation using the full LLE and a Raman gain profile provided by multi-vibrational mode model~\cite{hollenbeck_multiple-vibrational-mode_2002}. All other parameters match those in Fig.~\ref{fig:Raman_continue}.}
	\label{fig:Raman_Lagrange}
\end{figure*}

\subsubsection{Raman solitons}

In the previous subsection, Raman scattering was considered as a perturbative effect that influenced CS properties, such as center frequency and existence range. Researchers have also explored more extreme scenarios in which Raman scattering directly leads to the generation of new types of solitons. In 2017, Yang et al. experimentally demonstrated the existence of Raman `Stokes' solitons in a multimode silica microdisk resonator~\cite{yang_stokes_2017}. They observed that a standard CS could be excited in one mode family, and then, with increased drive power, a second, frequency down-shifted, Stokes soliton formed via stimulated Raman scattering into a separate mode family. Efficient generation of the Stokes soliton requires that it remains temporally overlapped with the primary CS during propagation. Thus, the group velocities of the primary and Stokes solitons must match, at a frequency that falls within the Raman gain spectrum. This condition can be written in terms of the FSRs of the two mode families as~\cite{yang_stokes_2017}:

\begin{equation}
 \mathrm{FSR}_{(\mathrm{Primary})}(\omega_p) = \mathrm{FSR}_{(\mathrm{Stokes})}(\omega_p+d\Omega),
\label{Eq:RS_groupvelocitymatch}
\end{equation}
where $\omega_p$ is the primary soliton's center frequency and $d\Omega$ is the frequency shift between the two solitons. Figure~\ref{fig:C3_VahalaRS}(a) shows the measured FSR as a function of wavelength for 4 different mode families of a $\sim 21.9$ GHz FSR silica microdisk resonator engineered to satify this condition at a pump wavelength of 1550 nm. This data reveals that the FSR of the primary soliton mode family at 1550 nm (green points) does indeed match to the FSR of the Stokes soliton mode family at $\sim 1593$ nm (red points, $d\Omega \sim -5.2$ THz). The measured output spectrum is plotted in Fig.~\ref{fig:C3_VahalaRS}(b). Both solitons are seen to exhibit clean sech$^2$ spectral profiles, with the center frequency of the Stokes soliton comb located at 1593 nm exactly as predicted by the group velocity matching condition. Further measurements of the spectral characteristics of these two solitons show that they possess identical comb line spacings, as demanded by Eq.~\eqref{Eq:RS_groupvelocitymatch}, but that the two combs are spectrally offset from each other [see inset to Fig.~\ref{fig:C3_VahalaRS}(b)]. This offset, similar to the frequency offset combs discussed in the spectral extension experiments of Section~\ref{sec:extension}, arises from the offset in resonant frequencies between the two mode families. Additional measurements also demonstrate that the Stokes soliton displayed a laser-like threshold behavior, with a linear increase in soliton energy observed once the drive power exceeded the threshold for Stokes soliton formation.

\begin{figure*}[!t]
	\centering
	\includegraphics[width = 0.65\textwidth, clip=true]{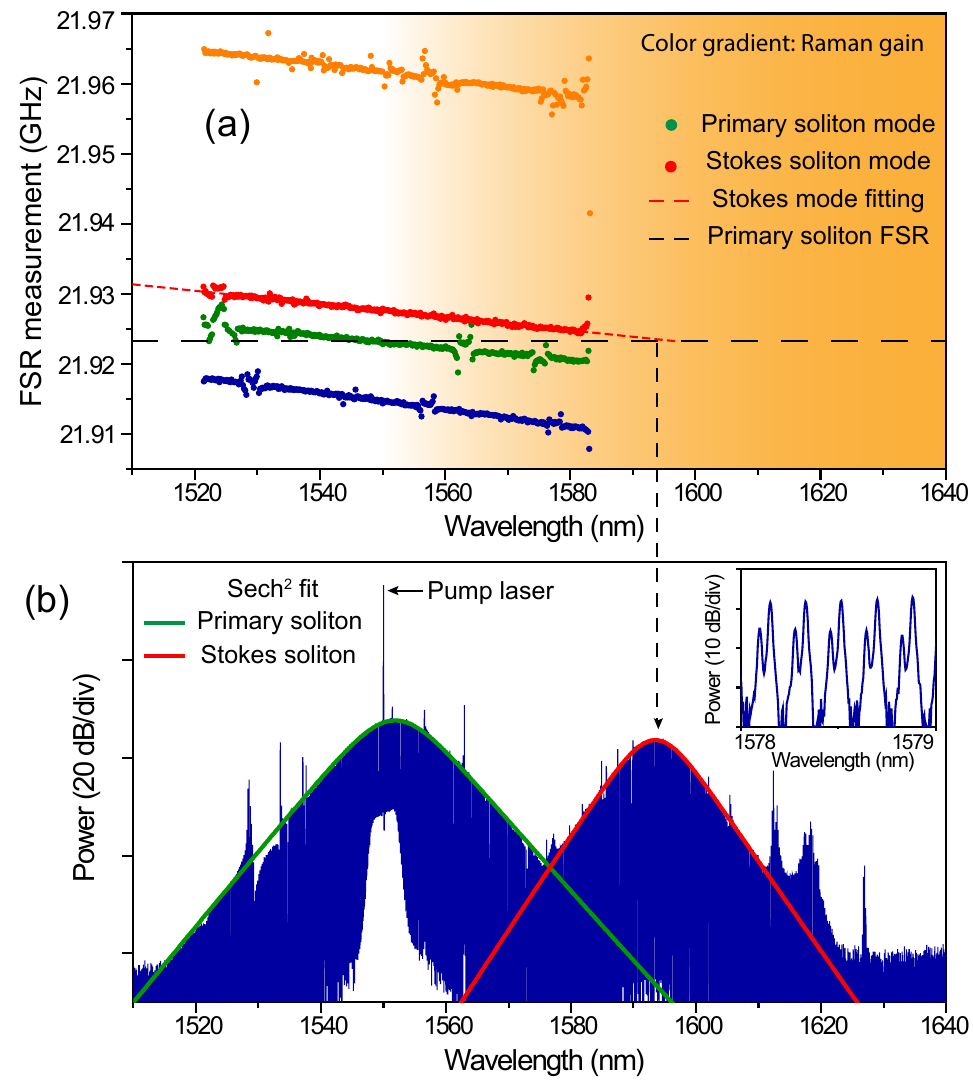}
	\caption{(a) Measured FSRs of 4 mode families in a $\sim 21.9$ GHz silica microdisk resonator. The black dashed line indicates the group velocity matching condition of Eq.~\eqref{Eq:RS_groupvelocitymatch}. (b) Measured spectrum of primary and Stokes solitons at the output of the cavity. Inset: Enlargement in the overlap region, revealing the two combs are offset from each other. Adapted with permission from Q.~F.~Yang \textit{et al.}, Nat. Phys. \textbf{12}, 53-57 (2017)~\cite{yang_stokes_2017}. Copyright \textcopyright~2017 Springer Nature.}
	\label{fig:C3_VahalaRS}
\end{figure*}

A second type of Raman soliton has also been observed in Kerr resonators. In this configuration, both the pump and soliton occupy the same mode family, and form an intracavity field composed of a single phase-coherent structure~\cite{li_ultrashort_2024}. This `phase-coherent' Raman soliton can be driven directly by an external pump without the need to generate a primary CS to excite the Stokes field. The existence of such solitons requires two separate conditions to be met. The first is a group velocity matching condition between the soliton and the pump equivalent to that of Eq.~\eqref{Eq:RS_groupvelocitymatch}. The second is a phase-matching condition that ensures the phases of the Raman signal and the pump evolve equally over one round trip. It is this second condition that enables the formation of a single phase-coherent spectral structure. Considering a pulsed drive field, these two conditions can be written as,
\begin{align}
    \label{eq:RamanSoliton1}
    &\frac{\Delta\tau}{L}+\hat{D}_1(d\Omega)=0\,,\\
    \label{eq:RamanSoliton2}
    &\frac{\Delta\tau}{L}d\Omega+\hat{D}(d\Omega)+q=0\,,
\end{align}
where $d\Omega$ is the frequency shift between the Raman soliton and the pump. The desynchronisation is set by $\Delta\tau = t_\mathrm{R}(\omega_p) - t_\mathrm{P}$ with $t_\mathrm{P}$ the period of the driving field, $\hat{D}$ is the dispersion operator defined in Eq.~\eqref{Dhat}, $q$ is the soliton's nonlinear phase shift and $\hat{D_1} = d\hat{D}/d\Omega$ is the group velocity mismatch relative to the pump. The use of desynchronised pulsed driving provides an additional degree of freedom with which to simultaneously satisfy both conditions. In Ref.~\cite{li_ultrashort_2024} the authors demonstrated that both conditions can be satisfied by operating in a standard dispersion shifted fiber, with the desynchronised pump located in the normal GVD regime exciting a Raman soliton whose center frequency is in the anomalous GVD regime. Figure~\ref{fig:C3_MaxRS}(a) shows the experimentally measured output using a $2.65$-m dispersion shifted fiber Fabry-Perot resonator. When the desynchronization~$\Delta\tau$ does not satisfy the two conditions above, the cavity output appears as a noisy narrow-band signal centered at the frequency that satisfies the group velocity matching condition (orange trace, $\Delta\tau = 356$~fs). However, when $\Delta\tau$ satisfies both the group velocity and phase matching conditions, a broadband Raman soliton state forms, with a spectrum spanning back to the pump frequency (blue trace, $\Delta\tau = 402$~fs). RF measurements confirm that, in this regime, the system is indeed operating in a low-noise soliton state. An Ikeda map simulation of output spectrum shows excellent agreement with the experimental measurement (red dashed trace). The temporal intensity envelope corresponding to the simulated spectrum is shown in  Fig.~\ref{fig:C3_MaxRS}(b), revealing a soliton with a temporal duration of $\sim 64$~fs. This is a remarkably short duration pulse for a meter-length oscillator constructed from commercially available fiber.

Further exploration of these phase coherent Raman solitons predict that they should also be observable under conditions of cw driving~\cite{li_continuous-wave_2024}. In this case, the group velocity matching condition is automatically satisfied and the frequency shift of the Raman soliton is set solely by Eq.~\eqref{eq:RamanSoliton2}. To date, this prediction is yet to be experimentally confirmed.  

\begin{figure*}[!t]
	\centering
	\includegraphics[width = 1 \textwidth, clip=true]{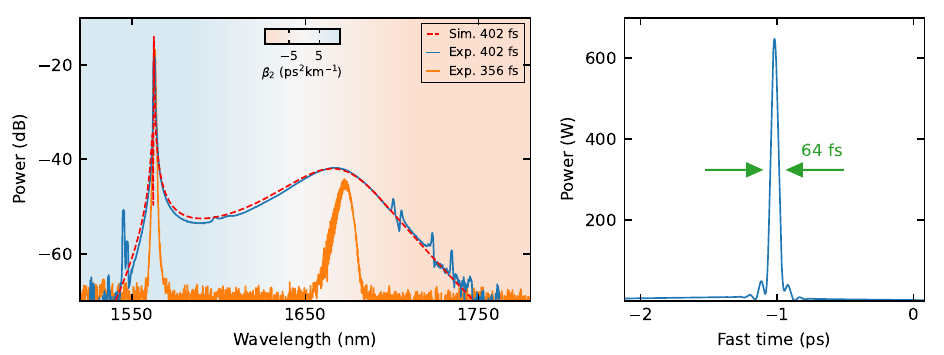}
	\caption{(a) Measured output spectrum of a 2.65 m dispersion shifted fiber cavity driven by a pulsed pump with a desynchronisation $\Delta\tau = 356$~fs (orange trace) and $\Delta\tau = 402$~fs (blue trace). The red dashed line shows the simulated spectrum at $\Delta\tau = 402$~fs. (b)~Simulated temporal intensity profile of the output soliton at $\Delta\tau = 402$~fs. Adapted with permission from Z. Li \textit{et al.}, Nat. Photonics \textbf{18}, 46-53 (2024)~\cite{li_ultrashort_2024}. Copyright \textcopyright~2024 Springer Nature.}
	\label{fig:C3_MaxRS}
\end{figure*}

\subsection{CS bound states}
\label{sec:CSbinding}

One of the more remarkable features of CSs is the ability for multiple solitons to co-exist simultaneously at different positions in the same cavity. Due to the translational invariance of the standard LLE, any perturbations to the positions of these solitons should in principle remain undamped. Thus, one might expect to observe a slow random walk in soliton position due to the unavoidable presence of noise in the system~\cite{firth_optical_1996-1,parra-rivas_interaction_2017}. However, experimental results consistently show the opposite: widely spaced solitons are observed to remain fixed relative to each other over extended periods~\cite{jang_ultraweak_2013,herr_temporal_2014,joshi_thermally_2016,brasch_photonic_2016,yi_active_2016,webb_experimental_2016,cole_soliton_2017,obrzud_temporal_2017-1}. This suggests the existence of long-range binding mechanisms between CS.

Reference~\cite{wang_universal_2017} proposed a `universal' mechanism to explain this locking behavior. When the standard LLE is perturbed by higher-order effects, such as the dispersive perturbation induced by a mode crossing~\cite{herr_mode_2014}, these disturbances can excite a spectral resonance within the soliton spectrum. This in turn leads to the development of an extended oscillatory tail in the temporal profile of the CS. Bound states are then observed to form at the minima of an effective interaction potential induced by this tail~\cite{malomed_bound_1991,cai_bound_1994}. The underlying physical origins of this trapping potential are naturally the intensity and phase trapping mechanisms discussed in section~\ref{sec:trapping}.

\begin{figure*}[!t]
	\centering
	\includegraphics[width = 0.65\textwidth, clip=true]{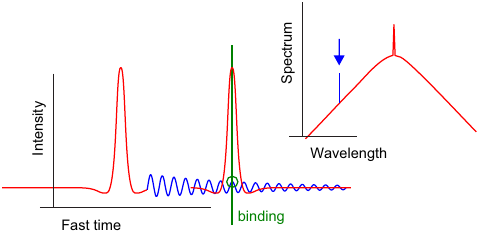}
	\caption{Schematic representation of CS binding mechanism. A perturbation to the LLE leads to a spectral resonance (blue arrow). As a result the temporal profile of the CS develops an extended oscillatory tail that enables the trapping of a neighbouring soliton. Reprinted with permission from Y. Wang \textit{et al.}, Optica \textbf{4}, 855-863 (2017)~\cite{wang_universal_2017}. Copyright~\textcopyright~2017 Optical Society of America.}
    \label{fig:C3_BSmechanism}
\end{figure*}

Figure~\ref{fig:C3_BSmechanism} schematically illustrates this mechanism. Here, a narrowband spectral feature results in an extended oscillatory tail, allowing multiple locking points between two solitons, each giving rise to a distinct bound state. Experiments in macroscopic fiber rings were conducted to validate this model~\cite{wang_universal_2017}. In fiber rings with significant third-order dispersion, CSs were observed to develop a substantial spectral peak at a dispersive wave resonance located 1.13 THz above the drive frequency. Experimentally, a pair of CS excited in this cavity were measured to reliably form into one of ten distinct available bound states. The final spacing between the two CS were measured to be between 4 and 11 ps, with each distinct state spaced by approximately 0.75 ps. This matched well with the predicted periodicity of the CS's oscillatory tail of 0.88 ps. Through the controlled injection of noise into the cavity, transitions between these bound states could be observed. Further experiments in fiber cavities were also able to observe CS bound states in systems where the higher-order perturbation was induced by the periodicity of the cavity (Kelly sidebands), or by the cavity's birefringence.

Soliton bound states are also commonly observed in optical microresonators, and in this context they are often referred to as soliton crystals\cite{cole_soliton_2017}. This terminology draws an analogy between the fixed temporal arrangement of CSs within the cavity and the periodic placement of atoms in a crystal~\cite{ashcroft_solid_1976}. In microresonators, soliton bound states often arise due to the presence of avoided mode-crossings, which induce localized perturbations to the dispersion of the soliton mode family~\cite{herr_mode_2014}. These perturbations lead to CSs with strong oscillatory tails that facilitate bound state trapping.

The temporal separation of a pair of bound CSs can be simply determined measuring the resultant spectral interference. For two solitons located at angular positions $\phi_1$ and $\phi_2$ within a cavity, the measured spectral intensity~$S_2$ of the total field can be written as
\begin{equation}\label{eq:TwoCSspectrum}
 S_2(\mu) \sim S(\mu) \left[ 1 + \mathrm{cos}((\phi_1-\phi_2)\mu)\right],
\end{equation}
where $S(\mu)$ is the spectral intensity of a single CS and $\mu$ is the mode index. Measuring the period of this spectral interference gives direct access to the angular separation of the two solitons in the cavity. This expression can also be generalised to include the case of $N$ solitons propagating simultaneously in the cavity. Here, there are $N(N-1)/2$ separate interference terms yielding a total optical spectrum,
\begin{equation}\label{eq:NCSspectrum}
 S_N(\mu) \sim S(\mu) \left[ N + 2\sum_{i=1}^{N}\sum_{j>i}^{N} \mathrm{cos}((\phi_i-\phi_j)\mu)\right]\,.
\end{equation}

Figure~\ref{fig:C3_boundstates} shows the spectra of a collection of different CS bound states for $N = 1, 2, 3$ and 4 solitons. The resultant soliton crystal spectra exhibit a highly complex structure, especially for large $N$. Of particular interest is the case, shown in the first column, where the solitons are spaced equi-distantly around the cavity. This situation is analogous to the equi-spaced pulses found in harmonically modelocked lasers. Here, the spectral interference fully suppresses all comb lines except those located at multiples of $N$-FSRs. Thus, by operating in one of these `perfect' soliton crystal states, it is possible to increase the line spacing of the output comb by a factor of $N$. An additional advantage of these perfect crystal states is that, with only every $N$th comb line present, the optical power in each comb line will increase by a factor $N^2$ when compared to the spectrum of a single CS~\cite{karpov_dynamics_2019}. The utility of these states is such that several groups have developed techniques that allow the deterministic excitation of $N$-state soliton crystals~\cite{karpov_dynamics_2019,he_perfect_2020,lu_synthesized_2021}. For example, in Ref.~\cite{lu_synthesized_2021}, a second counter-propagating auxiliary field was introduced to the resonator. Controlling the frequency of this second field allowed for the deterministic excitation of perfect soliton crystals with the number of equi-spaced solitons selectable between $N=1$ and $N=32$.

\begin{figure*}[!t]
	\centering
	\includegraphics[width = 0.75\textwidth, clip=true]{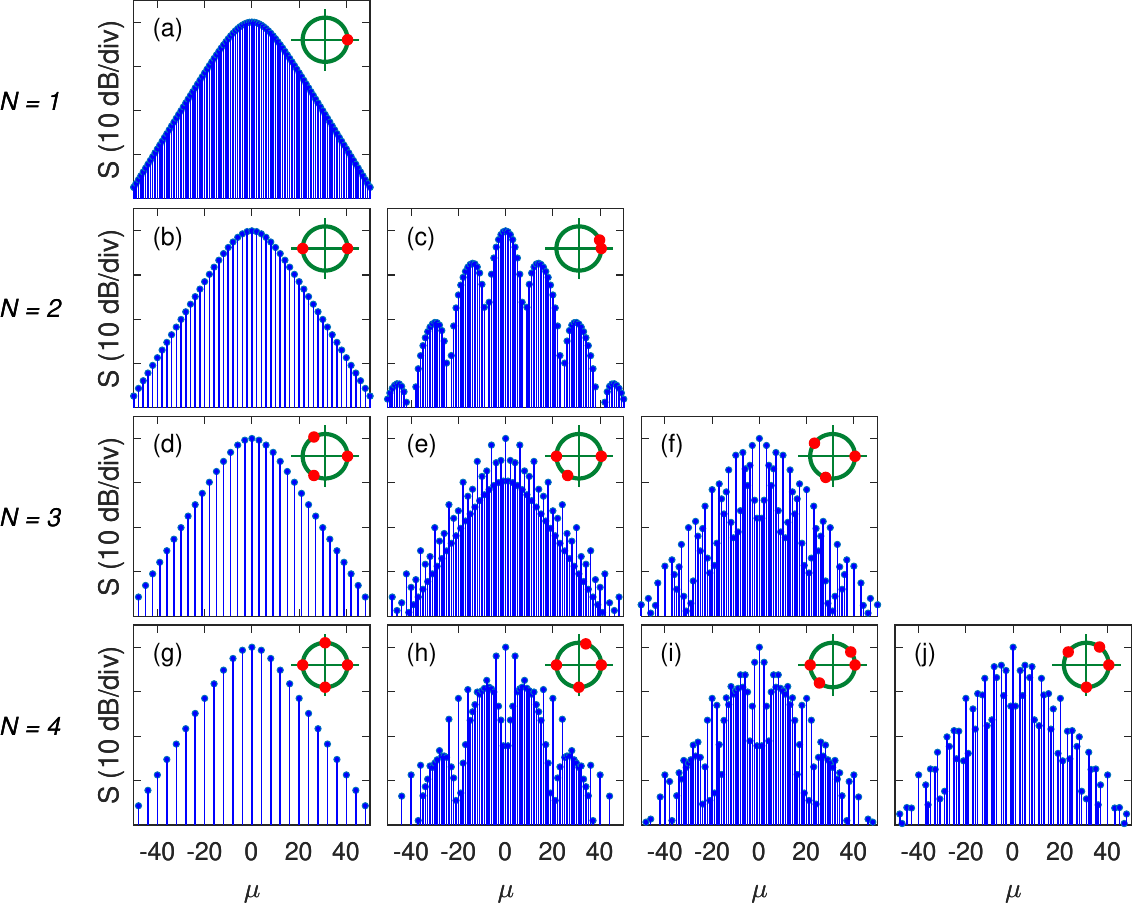}
	\caption{Spectra of solitons crystals for a variety of configurations with $N = 1,2,3$ and $4$ solitons in the cavity: (a) $\phi_i = 0$; (b) $\phi_i = 0, \pi$;  (c) $\phi_i = 0, \pi/8$; (d) $\phi_i = 0, 2\pi/3, 4\pi/3$; (e) $\phi_i = 0, \pi, 4\pi/3$; (f) $\phi_i = 0, 0.8\pi, 1.4\pi$; (g) $\phi_i = 0, \pi/2, \pi, 3\pi/2$; (h) $\phi_i = 0, 0.4\pi, \pi, 3\pi/2$; (i) $\phi_i = 0, 0.2\pi, \pi, 1.3\pi$; (j) $\phi_i = 0, 0.3\pi, 0.8\pi, 3\pi/2$.}
	\label{fig:C3_boundstates}
\end{figure*}

\subsection{CS timing jitter}

In the absence of any noise, an idealized CS pulse train output from a resonator would be perfectly periodic. However, in practice, noise is always present such that the amplitude, frequency, phase, and position (timing) of the soliton will fluctuate over time. In frequency space, the fluctuations result in the broadening of the optical linewidths of the corresponding frequency comb~\cite{kim2016ultralow}. 

There are many sources of noise and several ways their impact on the pulse train can be calculated and measured~\cite{matsko_timing_2013,drake_thermal_2020,lei2022optical}.
In this Section, we specifically discuss timing jitter induced by white noise. The analysis provided here is a simplified version of the work presented in~\cite{matsko_timing_2013}.
A fundamental source of white noise is quantum noise: the intracavity field interacts with vacuum fluctuations via both intrinsic and extrinsic loss channels, thereby perturbing the soliton. This white (i.e., temporally uncorrelated) noise can be incorporated into the dimensional Lugiato-Lefever equation as a Langevin force term $F(t,\tau)$~\cite{hasegawa_optical_2003,matsko_timing_2013}:
\begin{equation}
    t_\mathrm{R}\frac{\partial E(t,\tau)}{\partial t} =
        \left[ -\alpha-i\delta_0 + i L \beta_2 \frac{\partial^2}{\partial\tau^2} \right] E 
	+ i \gamma L |E|^2 E
        + \sqrt{\theta}\,E_\mathrm{in}+F(t,\tau).
    \label{eq:noiseLLE}
\end{equation}
The autocorrelation functions characterizing the noise source $F = F_r+ iF_i$ are,
\begin{align}
\langle F_r(t, \tau) F_r(t+t', \tau+\tau') \rangle &= A_F \, \delta(t') \, \delta(\tau'),  \\
\langle F_i(t, \tau) F_i(t+t', \tau+\tau') \rangle &= A_F \, \delta(t') \, \delta(\tau'),  \\
\langle F_r(t, \tau) F_i(t+t', \tau+\tau')\rangle &= 0.
\end{align}


In the case of quantum (shot) noise, the strength of the autocorrelation corresponds to one photon per mode coupled through loss channels, i.e., $A_F = t_\mathrm{R}\hbar\omega_0\alpha$~\cite{matsko_timing_2013}.
To investigate the impact of such noise on the soliton, we can use a Lagrangian analysis which corresponds to projecting the noise on each soliton parameter. We use the same soliton ansatz that is presented in more detail in Section~\ref{sec:Lagragian}. In dimensional units, it reads
\begin{equation}
    E_\mathrm{s}(t,\tau) = {B}\operatorname{sech}\left(\frac{{B}(\tau-\tau_\mathrm{cs})}{\sqrt{|\beta_2/\gamma|}}\right)e^{-i\Omega_\mathrm{s}(\tau-\tau_\mathrm{cs})}e^{i\phi}.
\end{equation}
The projection of the noise using this ansatz yields noise sources for $A$, $\tau_\mathrm{cs}$, $\Omega_\mathrm{s}$ and $\phi$.
As we are looking at jitter only, we consider a stationary amplitude and phase and focus on the slow time dynamics of the soliton frequency and position. The corresponding reduced model reads
\begin{align}
t_\mathrm{R} \, \frac{d\tau_\mathrm{cs}}{dt} &= \beta_2 L\Omega_\mathrm{s} + F_{\tau_\mathrm{cs}}\label{eq:jitterODEtime}\\
t_\mathrm{R}\frac{d\Omega_\mathrm{s}}{dt} &= -2\alpha\Omega_\mathrm{s} + F_{\Omega_\mathrm{s}}\label{eq:jitterODEfrequency}
\end{align}
with the noise sources
\begin{align}
F_{\tau_\mathrm{cs}}(t) &= \frac{1}{\mathrm{E}} \int_{-\infty}^{\infty} \tau \left[ E_\mathrm{s}^* F + F^* E_\mathrm{s} \right] \, d\tau  \\ 
F_{\Omega_\mathrm{s}}(t) &= \frac{i}{\mathrm{E}} \int_{-\infty}^{\infty} \left[ \frac{\partial E_\mathrm{s}^*}{\partial \tau} F - F^*\frac{\partial E_\mathrm{s}}{\partial \tau} \right] \, d\tau
\end{align}
where $\mathrm{E}=2A\sqrt{|\beta_2|/\gamma}$ is the soliton energy. The autocorrelation of the noise functions are given by
\begin{align}
\langle F_{\tau_\mathrm{cs}}(t) F_{\tau_\mathrm{cs}}(t+t') \rangle &= A_F\frac{\pi^2\mathrm{D}^2}{12\mathrm{E}} \, \delta(t') \\
\langle F_{\Omega_\mathrm{s}}(t) F_{\Omega_\mathrm{s}}(t+t') \rangle &= A_F\frac{1}{3\mathrm{ED}} \, \delta(t')
\end{align}
where $\mathrm{D}=(1/A) \sqrt{|\beta_2|/\gamma}$ is the soliton duration. For simplicity we neglect the cross correlation of the two noise sources.
The system of ODEs given in Eqs.~\eqref{eq:jitterODEtime} and~\eqref{eq:jitterODEfrequency} are linear and can be readily solved in the frequency domain.
Timing jitter is typically given as a power spectral density (PSD), i.e., the Fourier transform of the autocorrelation of the soliton position,
\begin{align}
S_{\tau_\mathrm{cs}}(\omega)&=\int^{\infty}_{-\infty}\langle\tau_\mathrm{cs}(t)\tau_\mathrm{cs}(t+t')\rangle e^{-i\omega t'}dt'\\
&=\int_{-\infty}^{\infty} \left( \lim_{T \to \infty} \frac{1}{T} \int_{-T/2}^{T/2} \tau_\mathrm{cs}(t) \tau_\mathrm{cs}(t + t') \, dt \right)e^{-i\omega t'}dt'\\
&\approx\frac{\hbar\omega_0\kappa}{\mathrm{E}\omega^2}\left[ \frac{\pi^2\mathrm{D}^2}{12} + \frac{|\beta_2|^2v_\mathrm{g}^2}{3\mathrm{D}^2(\omega^2+\kappa^2)} \right]
\label{Matsko_eqn}
\end{align}
where $v_g = L/t_R$ is the group velocity and the cavity linewidth $\kappa = 2\alpha/t_R$.

The first term corresponds to direct timing jitter. Since the soliton is free to move in time, we recover the standard $\omega^{-2}$ dependence characteristic of integrating a random process. The second term is the well-know Gordon Haus jitter, first introduced in the context of soliton transmission lines~\cite{gordon_random_1986}. 
The soliton center frequency is perturbed by noise and coupled to timing through the group velocity dispersion.
The key difference in the CS case, as compared to transmission lines, is that the center-frequency drift is constrained because of the continuous reinjection of the coherent driving field. This introduces an additional frequency dependence and leads to a roll-off near the cavity linewidth.

\begin{figure}[ht!]
\centering\includegraphics[width=10cm]{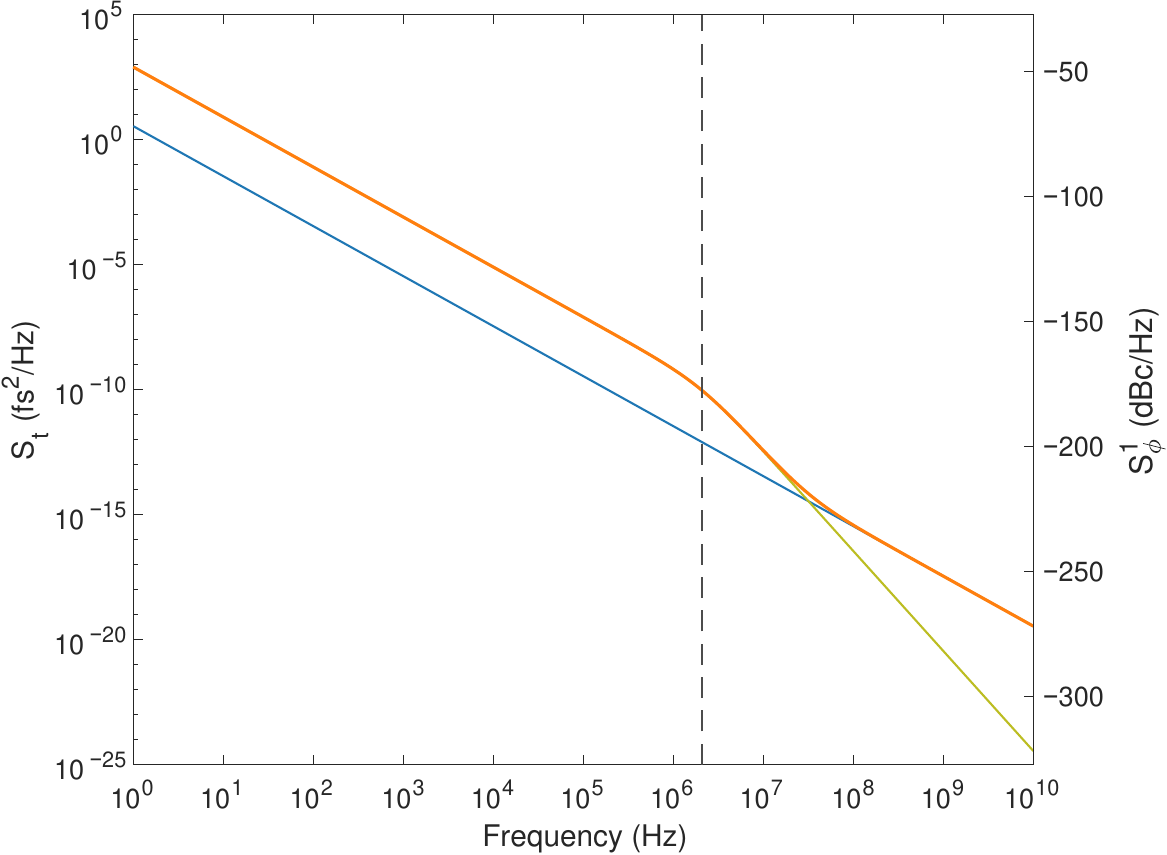}
\caption{Theoretical timing jitter and corresponding phase noise of a soliton propagating in a $9.4$~mm long silica microresonator. Orange: power spectral density of the timing jitter (left axis) and phase noise of the first beatnote at $21.9$~GHz (right axis). Green line: contribution from Gordon haus jitter. Blue line: contribution from direct timing jitter. The black dashed line indicated the cavity linewidth $\kappa = 2\pi \times 2.1~\text{MHz}$}
\label{fig:jitter}
\end{figure}

Both terms of Eq.~\eqref{Matsko_eqn} and the total spectral density of noise are plotted in Fig.~\ref{fig:jitter}. The physical parameters correspond to a $9.4$ mm-long silica resonator, yielding a free spectral range (FSR) of $21.9$~GHz~\cite{bao_quantum_2021}. The linewidth is $\kappa = 2\pi \times 2.1~\text{MHz}$ and dispersion
$\beta_2 = -22~\text{ps}^2/\text{km}$.
The soliton parameters are
$D = 120~\text{fs}$ and 
$E = 140~\text{pJ}$. The noise source is quantum noise as defined above.

Figure~\ref{fig:jitter} shows that Gordon-Haus jitter is dominant at low frequencies, with the contribution of direct jitter only becoming significant at high frequencies ($\omega>\kappa$).
Overall, such a soliton pulse train displays very low jitter, which can be harnessed for low noise microwave generation. 
The phase noise PSD of the $n^{\mathrm{th}}$ order RF beatnote, defined as
\begin{equation}
    S^{n}_{\phi}(\omega)=\int_{-\infty}^{\infty} \left( \lim_{T \to \infty} \frac{1}{T} \int_{-T/2}^{T/2} \phi_n(t) \phi_n(t + t') \, dt \right)e^{-i\omega t'}dt'
\end{equation}
 can be calculated considering the simple connection: $\phi_n(t)=2\pi nf_r\tau_\mathrm{cs}(t)$ where $f_r$ is the repetition rate. For example, the conversion of jitter to phase noise for the first harmonic is shown in Fig.~\ref{fig:jitter}. It highlights that very high-quality microwave signal  can in principle be generated from microresonator soliton combs. 

Although such low RF noise has not yet been experimentally demonstrated, due to the dominance of other noise sources~\cite{lei2022optical}, the potential of solitons as low-noise sources was underscored by a measurement of quantum-limited soliton diffusion in a silica resonator (with parameters as previously described). To isolate quantum noise, two counter-propagating solitons were excited, and their temporal separation was monitored. The results agree with the theoretical predictions presented in this section across the kHz to 100 kHz spectral band, where Gordon–Haus noise dominates~\cite{bao_quantum_2021}.

Finally, a recent study has revealed that the timing jitter of a soliton pulse train can be reduced by self-trapping induced by dispersive waves~\cite{jin_self-suppressed_2024}. This interaction introduces an additional decay term in Eq.~\eqref{eq:jitterODEfrequency}, effectively enhancing the damping of the Gordon Haus contribution. Using this approach, a reduction of up to 18~dB in phase noise was observed at a 10~kHz offset from the first beatnote in an $11.4$~GHz free spectral range (FSR) silica microresonator~\cite{jin_self-suppressed_2024}.

\subsection{CS conversion efficiency}

The efficiency with which energy from the input driving field is converted to a CS is an important figure of merit for many applications, and has accordingly been extensively examined~\cite{bao2014nonlinear,yi_soliton_2015,yi_theory_2016, zhang_advances_2024,yang_efficient_2024}. In what follows, we derive the CS conversion efficiency in a simple, cw-driven cavity and review strategies to enhance it.

The starting point of the analysis is to assume the validity of the approximate CS solution given by Eq.~\eqref{eq:CSsol}. In dimensional units, the soliton part of the solution reads
\begin{equation}
    E(\tau) = \sqrt{\frac{2\delta_0}{\gamma L}} \,\operatorname{sech}\left(\sqrt{\frac{2\delta_0}{|\beta_2|L}}\,\tau\right).
\end{equation}
Assuming a single soliton in the cavity, the integrated CS energy is
\begin{equation}
E_\mathrm{CS} = \sqrt{\frac{8\delta_0 |\beta_2|}{\gamma^2 L}}.
\end{equation}
The conversion efficiency $\eta$ is defined as the ratio of the out-coupled soliton energy ($\theta E_\mathrm{CS}$) and the energy in the input driving field integrated over one round trip ($E_\mathrm{d} = P_\mathrm{in}t_\mathrm{R}$), yielding
\begin{equation}
\label{Eq:CS_CE}
    \eta = \frac{\theta E_\mathrm{CS}}{E_\mathrm{d}} = \sqrt{\frac{8\theta^2\delta_0 |\beta_2|}{P_\mathrm{in}^2t_\mathrm{R}^2\gamma^2 L}}.
\end{equation}
Equation~\eqref{Eq:CS_CE} shows that the conversion efficiency increases with the detuning~$\delta_0$. As discussed in Section~\ref{sec:CSexist}, solitons only exist within a finite range of detunings, with the maximum detuning given in dimensional units as $\delta_\mathrm{max} = \pi^2\gamma P_\mathrm{in} L \theta / (8\alpha^2)$. This yields the maximum attainable CS conversion efficiency 
\begin{equation}
\eta_\mathrm{max} = \sqrt{\frac{\pi^2\theta^3|\beta_2|}{\alpha^2 P_\mathrm{in}t_\mathrm{R}^2\gamma}}.
\label{eq:etamax1}
\end{equation}

To gain more insights, it is useful to express the optimum conversion efficiency $\eta_\mathrm{max}$ in terms of the soliton duration $\Delta\tau_\mathrm{min}$ given by Eq.~\eqref{eq:mindur}:
\begin{equation}
    \eta_\mathrm{max} = {\frac{1}{1.763}} \frac{\pi^2}{2}\frac{\theta^2}{\alpha^2}\frac{\Delta\tau_\mathrm{min}}{t_\mathrm{R}} = {\frac{2\pi^2}{1.763}} \frac{Q_\mathrm{tot}^2}{Q_\mathrm{ext}^2}\frac{\Delta\tau_\mathrm{min}}{t_\mathrm{R}},
    \label{eq:etamax2}
\end{equation}
where $Q_\mathrm{ext} = \omega_0t_\mathrm{R}/\theta$ and $Q_\mathrm{tot} = \omega_0t_\mathrm{R}/(2\alpha)$ are the extrinsic and total resonator quality factors, respectively. Expressed in this form, and noting that $\text{max}[\theta/\alpha] = 2$ (see Section~\ref{sec:Ikeda}), we see that the CS conversion efficiency is fundamentally determined by the ratio of the soliton duration and the resonator round-trip time. Alternatively, considered from a frequency domain perspective, Eq.~\eqref{eq:etamax2} shows that the conversion efficiency is inversely proportional to the number of generated comb lines. For typical parameters, $\tau_\mathrm{min} \ll t_\mathrm{R}$, highlighting that the optimum conversion efficiency for CSs is usually small.

Below we give illustrative values for optimum efficiencies corresponding to parameters typical of different Kerr resonator paradigms (all assuming critical coupling, $\alpha = \theta$ for simplicity):
\begin{itemize}
    \item A 100-m-long resonator (2~MHz FSR) made from standard single-mode optical fiber with $\beta_2~\approx~-21~\mathrm{ps^2/km}$, $\gamma \approx 1.2~\mathrm{W^{-1}\,km^{-1}}$, and finesse $\mathcal{F} = \pi/\alpha \approx 22$ driven with 1~W of cw power (as in Ref.~\citenum{jang_ultraweak_2013}) would yield $\eta_\mathrm{max}=1\times 10^{-5}$. 

    \item A magnesium fluoride resonator with 35~GHz FSR, $\beta_2\approx -9.4~\mathrm{ps^2/km}$, $\gamma \approx 0.4~\mathrm{W^{-1}\,km^{-1}}$, and finesse $\mathcal{F} \approx 78\times 10^4$ driven with 20~mW of cw power (as in Ref.~\citenum{herr_temporal_2014}) would yield $\eta_\mathrm{max}=7.5\times 10^{-3}$.

    \item A silicon nitride resonator with 1~THz FSR, $\beta_2\approx -5.5~\mathrm{ps^2/km}$, $\gamma \approx 1000~\mathrm{W^{-1}\,km^{-1}}$, and finesse $\mathcal{F} \approx 1000$ driven with 240~mW of cw power (as in Ref.~\citenum{pfeiffer2017octave}) would yield $\eta_\mathrm{max}=3\times 10^{-2}$.

\end{itemize}

The examples above demonstrate that, as expected, the CS conversion efficiency generally increases for resonators with larger FSRs. This can be intuitively understood by the fact that large-FSR resonators have short round-trip times $t_\mathrm{R}$, thus resulting in larger attainable duty cycles $\Delta \tau_\mathrm{min}/t_\mathrm{R}$. 

\subsubsection{Strategies to improve CS conversion efficiency}

The examples above demonstrate that CS conversion efficiency is typically small, restricted to a few percent even for very small resonators with large FSRs. To overcome this deficiency, several strategies have been developed and demonstrated:
\begin{itemize}
\item \textbf{Pulsed driving.} Because they are underpinned by the instantaneous Kerr nonlinearity, CSs only interact with (and extract energy from) the region of the cavity background field that they temporally overlap with. For cw driving (which the preceding conversion efficiency analysis assumed), energy contained in the background field outside of the solitons' temporal extent is effectively wasted. 
This issue can be mitigated by employing pulsed (rather than cw) driving. Assuming a rectangular driving pulse profile with peak power $P_\mathrm{in}$ and duration $\Delta \tau_\mathrm{d}$, it is straightforward to show that the maximum conversion efficiency is still given by Eqs.~\eqref{eq:etamax1} and~\eqref{eq:etamax2}, but with the round-trip time $t_\mathrm{R}$ replaced with the driving pulse duration $\Delta\tau_\mathrm{d}$. This replacement highlights how the improved conversion efficiency can be traced to reducing the energy outside of the solitons' extent. Of course, in practice, the total conversion efficiency of a pulse driven system must also include a consideration of the efficiency with which the driving pulses are generated. This, in general, will greatly reduce the actual conversion efficiency of pulse driven Kerr combs.

Pulsed driving has been extensively used in CS experiments performed in macroscopic optical fiber ring resonators to achieve (peak) power levels sufficient to exploring complex CS dynamics (see, e.g., Refs.~\cite{anderson_observations_2016,anderson_coexistence_2017}). In the context of conversion efficiency, the benefits of pulsed driving were first highlighted by Obrzud et al., who used 2.1-ps-long pulses derived from an electro-optic comb generator to excite CSs in a fiber-based Fabry–Pérot microresonator with an FSR of 9.77~GHz~\cite{obrzud_temporal_2017-1}. Several other studies have subsequently demonstrated CSs and associated frequency combs in resonators (both macroscopic and microscopic) driven with short (picosecond-scale) pulses~\cite{xu_harmonic_2020,jia_photonic_2020,li_ultrashort_2024,anderson_photonic_2021}.

\item \textbf{Multi-soliton states.} The conversion efficiency analysis above assumed a single soliton in the cavity. A multi-soliton state with $N$ solitons circulating the cavity will naturally be associated with $N$ times larger energy, and hence $N$ times better conversion efficiency.~\cite{cole_soliton_2017,karpov_dynamics_2019,he_perfect_2020,lu_synthesized_2021} The caveat of this approach is, however, that the interference between the solitons can result in a complex spectral envelope, or an integer reduction of the line spacing of the corresponding frequency comb (if operating in a ``soliton crystal'' state). Such spectral modifications may not be suitable for some applications. 

\item \textbf{Coherent combining.} As described in section~\ref{sec:synchronization}, CSs generated in two separate resonators that are linearly coupled can synchronize their repetiton rate (comb line spacing). If driven with the same laser source, the solitons will be in-phase and can be coherently combined. Such coherent combining can overcome the power limitations of conventional CS systems~\cite{kim2023coherent}.    

\item \textbf{Coupled cavities.} CS conversion efficiency can be increased in ``photonic molecule'' systems comprised of two linearly coupled cavities. There are two schemes that have been demonstrated. First, in a pump recirculation scheme described in \cite{xue_super-efficient_2019}], a nonlinear soliton generating cavity (with anomalous dispersion) is coupled to a linear pump cavity (with normal dispersion) that is externally driven. In this scheme, the soliton cavity is internally driven by the field circulating in the pump cavity. Thanks to the resonant enhancement provided by the pump cavity, solitons with specific characteristics can be generated with external driving power levels considerably lower than in the case where a single resonator is directly driven. In this way, the pump recycling improves the CS conversion efficiency by reducing the denominator in Eq.~\eqref{Eq:CS_CE}. The scheme was experimentally realized in Ref.~\cite{xue_super-efficient_2019} using mutually coupled macroscopic fiber ring cavities; it was shown that a CS that would require $2.47$~W of external driving power in a single, directly driven cavity, could be sustained with just $0.27$~W when leveraging coupled cavities. The absolute conversion efficiency was still small due to the fiber resonators used, but numerical simulations signalled that conversion efficiencies close to unity could be attainable in ideal resonators with no intrinsic losses.

The second scheme leverages directly the resonance shift (avoided mode crossing) that is due to the coupling of the two cavities. Here, an auxiliary resonator is coupled to an externally-driven, soliton-generating cavity so as to engineer a desired resonance shift at the driven mode (and only the driven mode). In this case, the detuning of the pump light becomes decoupled from the soliton, allowing the pump to be on-resonance whilst the rest of the modes experience red detuning (as required for soliton existence). As in the previous scheme, this allows solitons to be sustained with significantly smaller external pump power levels, thus permitting improved conversion efficiencies close to unity. The scheme was experimentally realized by Helgason et al.~\cite{helgason2023surpassing} using photonic integrated silicon nitride resonators, achieving single-soliton conversion efficiencies as high as 54~\%.

Finally, we note that if the coupled resonator system contains elements that violate the mean-field assumptions that underpin a (coupled) LLE model (such as e.g. large internal loss associated with the resonator coupling or dispersion modulations along either ring), then the intracavity dynamics are better described with a coupled Ikeda map, which tracks the field evolution of each ring on a roundtrip by roundtrip basis. A full description of the equations required to model coupled Ikeda systems can be found in Refs.~\cite{xue_super-efficient_2019,helgason2023surpassing}.

\item \textbf{Laser cavity-soliton combs}. An alternative strategy has been demonstrated in Ref.~\cite{bao_laser_2019}, where the microresonator is integrated directly into a fiber ring laser rather than being driven by an external pump. This configuration produces background‑free soliton combs with markedly higher conversion efficiency than conventional CS combs.

\item \textbf{Normal dispersion}. Localized structures can also be realized in resonators with normal dispersion. These structures are qualitatively different from the bright CSs discussed in this Chapter, and can be associated with significantly improved conversion efficiencies. This topic will be discussed in detail in Section~\ref{sec:normal_conversioneff}.   

\end{itemize}


\subsection{CSs in self-injection locked systems}
\label{sec:SIL_CS}

In a conventionally driven CS system, the pump laser must be optically isolated from the Kerr cavity to prevent unwanted feedback interfering with the laser's operation. Moreover, deterministic CS excitation typically relies on complex excitation schemes, as detailed earlier in section~\ref{sec:DeterministicEX}. These dual requirements pose significant barriers to the development of integrated CS sources suitable for low-cost mass production. A promising solution to these problems comes in the form of the self-injection-locked (SIL) Kerr cavity.  In this scheme, a pump laser diode is directly connected to a Kerr cavity without optical isolation. The backscattered light from the resonator enables self-injection locking of the pump laser frequency, leading to a substantial reduction in its linewidth~\cite{dahmani_frequency_1987,liang_whispering-gallery-mode-resonator-based_2010,liang_ultralow_2015,kondratiev_self-injection_2017}.  At high drive powers, the nonlinear properties of the Kerr resonator become more important and exert a strong influence on the optical feedback from the cavity. In this regime, it has been demonstrated both theoretically and experimentally, that the optical properties of the SIL pump laser can be manipulated such that it possesses both sufficient optical power, and the required detuning from the cavity resonance to enable stable CS generation. Furthermore, with precise tuning of system parameters, it is possible to operate in a regime that allows for the deterministic formation of single CSs.

To analyse the SIL Kerr cavity system, we follow the approach adopted in Ref.~\cite{shen_integrated_2020}. We note that, an alternative examination of the system’s nonlinear dynamics presented in Ref.~\cite{voloshin2021dynamics} also offers complementary insights. The analysis of Ref.~\cite{shen_integrated_2020} divides the system into three parts, the forward propagating resonator soliton field $E_\mathrm{s}$, the counterpropagating backscattered resonator field $E_\mathrm{b}$, and the driving laser field $E_\mathrm{L}$. To simplify the analysis it is explicitly assumed that the gain bandwidth of the laser diode used is smaller than the FSR of the resonator. As a result, only the driven mode of the resonator will contribute to the locking process. This allows us to neglect the fast time dependence from the backscattered and laser fields. Under these assumptions, the forward propagating resonator field $E_\mathrm{s}$ can be written in the form of a standard LLE equation, Eq.~\eqref{SILEqn_Es}, with additional terms describing the cross-phase modulation and linear coupling of the counterpropagating field $E_\mathrm{b}$. The evolution of the backscattered field is then governed by an ordinary differential equation, Eq.~\eqref{SILEqn_Eb}, that contains terms that describe the effect of cross-phase modulation and linear coupling with the forward propagating field (both time-averaged over one roundtrip). Finally, a third equation, Eq.~\eqref{SILEqn_EL}, described the dynamics of the driving laser field $E_\mathrm{L}$ under the influence of the backscattered field~$E_\mathrm{b}$.
\begingroup
\begin{align}
t_\mathrm{R}\frac{\partial E_\mathrm{s}(t,\tau)}{\partial t} &=
    \left[ -\alpha_\mathrm{R}-i\delta_\mathrm{R} - i\beta_2 L \frac{\partial^2}{\partial\tau^2} 
	+ i \gamma L (|E_\mathrm{s}|^2 + 2|E_\mathrm{b}|^2) \right] E_\mathrm{s} \notag\\
    &\qquad + i\beta\alpha_\mathrm{R} E_\mathrm{b}
    - \sqrt{\theta_\mathrm{R} \theta_\mathrm{L}}\, e^{i\varphi}\, E_\mathrm{L},\label{SILEqn_Es}\\
t_\mathrm{R}\frac{dE_\mathrm{b}(t)}{dt} &=
    \left[ -\alpha_\mathrm{R}-i\delta_\mathrm{R} 
	+ i \gamma L (|E_\mathrm{b}|^2 + 2\langle{|E_\mathrm{s}|^2}\rangle) \right] E_\mathrm{b} +i\beta\alpha_\mathrm{R}\langle{E_\mathrm{s}}\rangle,\label{SILEqn_Eb}\\
t_\mathrm{R}\frac{dE_\mathrm{L}(t)}{dt} &=
    \left[ -\alpha_\mathrm{L}-i(\delta_\mathrm{R}-\delta_\mathrm{L}) 
	+ \frac{g(1+i\alpha_\mathrm{H})}{2} \frac{1}{1+|E_\mathrm{L}|^2/|E_\mathrm{sat}|^2} \right] E_\mathrm{L}\notag\\
    &\qquad - \sqrt{\theta_\mathrm{R}\theta_\mathrm{L}}\,e^{i\varphi}\,E_\mathrm{b}.\label{SILEqn_EL}
\end{align}
\endgroup

Eq.~\eqref{SILEqn_Es} is written in the standard LLE form presented in Section~\ref{sec:mean_field_model} with the resonator's roundtrip time, length, dispersion, nonlinear coefficient, loss coefficient, and detuning defined as $t_\mathrm{R}$, $L$, $\beta_2$, $\gamma$, $\alpha_\mathrm{R}$ and $\delta_\mathrm{R}$ respectively. $\beta$ is the dimensionless backscattering coefficient that couples $E_\mathrm{s}$ and $E_\mathrm{b}$. For Eq.~\eqref{SILEqn_EL}, the laser's gain, saturation power, amplitude to phase coupling coefficient, loss coefficient, and phase detuning from the laser's cavity resonance are $g$, $|E_\mathrm{sat}|^2$, $\alpha_\mathrm{H}$, $\alpha_\mathrm{L}$ and $\delta_\mathrm{L}$, respectively. Finally the driving terms of Eqs.~\eqref{SILEqn_Es} and~\eqref{SILEqn_EL} are written in terms of the cavity and laser intensity coupling coefficients $\theta_\mathrm{R}$ and~$\theta_\mathrm{L}$, and $\varphi$ the propagation phase shift accumulated between the laser and the resonator.

To proceed, one makes the further assumption that the timescales of the evolution of the laser field are much faster than those of the resonator, such that the laser power and phase can be assumed to adiabatically track the feedback yielding an Adler type equation for the pump phase variable $z=-e^{i\varphi}e^{i\phi_\mathrm{L}(t)}$,
\begin{align}
t_\mathrm{R}\frac{1}{iz}\frac{dz}{dt} =
    \left[ \delta_\mathrm{L}-\delta_\mathrm{R}+\alpha_\mathrm{L}\alpha_\mathrm{H}+K\, \mathrm{Im}\left(e^{i\phi}\frac{E_\mathrm{b}}{i\beta z F} \right)\right],\label{SILEqn_z}
\end{align}
with $\phi_\mathrm{L}$ the phase of the laser field, $\phi=2\varphi-\mathrm{tan}^{-1}(\alpha_\mathrm{H})+\mathrm{arg}(\beta)+\pi/2$, $F=\sqrt{\theta_\mathrm{L}\theta_\mathrm{R}|E_\mathrm{L}|^2}$ the effective driving strength, and $K=|\beta|\,\theta_\mathrm{L}\theta_\mathrm{R}\,\sqrt{1+\alpha_\mathrm{H}^2}$ the locking strength. 

\subsubsection{cw response of SIL Kerr cavity}

\begin{figure*}[b]
	\centering
	\includegraphics[width = 0.5\textwidth, clip=true]{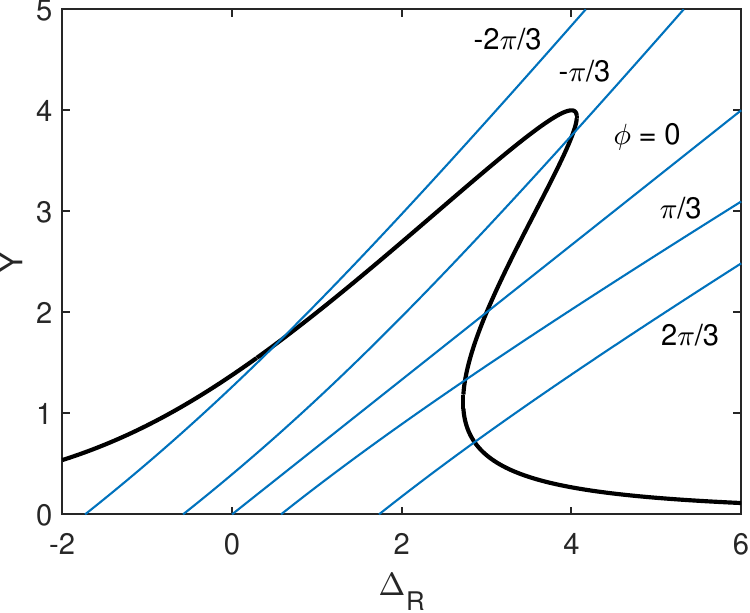}
	\caption{cw response of a self-injection-locked Kerr cavity at a normalised drive power $|S|^2 = 4$. The black trace shows the solution to the cavity amplitude condition, Eq.~\eqref{SILEqn_CW1}, and the blue traces the solution to the phase condition, Eq.~\eqref{SILEqn_CW2}, for feedback phases~$\phi$ as indicated.}
	\label{fig:C3_SIL_CW}
\end{figure*}

It is instructive to first consider the steady-state homogeneous response of the above system of equations. Setting the fast- and slow-time derivatives to zero, and assuming a weak background field, yields two normalized equations that must be simultaneously satisfied,
\begin{align}
  &Y^3 - 2\Delta_\mathrm{R} Y^2+(1+\Delta_\mathrm{R}^2) Y + S^2 = 0,
  \label{SILEqn_CW1}\\
  &\Delta_\mathrm{R} \approx \frac{3}{2}Y - \frac{1}{\tan(\phi)} + \frac{\sqrt{4+Y^2\sin^2(\phi)}}{2\sin(\phi)},
  \label{SILEqn_CW2}
\end{align}
with $\Delta_\mathrm{R}=\delta_\mathrm{R}/\alpha_\mathrm{R}$, $Y=\gamma L|E_\mathrm{s}|^2/\alpha_\mathrm{R}$ and $S^2=\theta_\mathrm{L}\theta_\mathrm{R}\gamma L|E_\mathrm{L}|^2/\alpha_\mathrm{R}^3$. Equation~\eqref{SILEqn_CW1} is the usual mean-field cubic Kerr bistable response [cf. Eq.~\eqref{eq:cwcubic_norm}] while Eq.~\eqref{SILEqn_CW2} gives the approximate cavity detuning derived under the assumption of a large locking strength~$K$~\cite{shen_integrated_2020}. These two equations can be solved graphically with the cw solution at the intersection of the two curves. Figure~\ref{fig:C3_SIL_CW} illustrates the system’s behavior at a normalized drive strength of $S^2=4$. Critically, it shows that~$\varphi$, the linear phase shift imparted by the waveguide section between the two cavities, acts as a free parameter that allows operation across a broad range of cavity detunings. Crucially, it gives access to operating points on the otherwise unstable intermediate branch of the tilted resonance. Since the nonlinear structures of the LLE emerge from a cw background when the detuning is correctly set, this cw analysis strongly suggests that such a system could be applied to access soliton comb states. 

\subsubsection{Soliton formation in SIL Kerr cavity}

\begin{figure*}[b]
	\centering
	\includegraphics[width = 0.75\textwidth, clip=true]{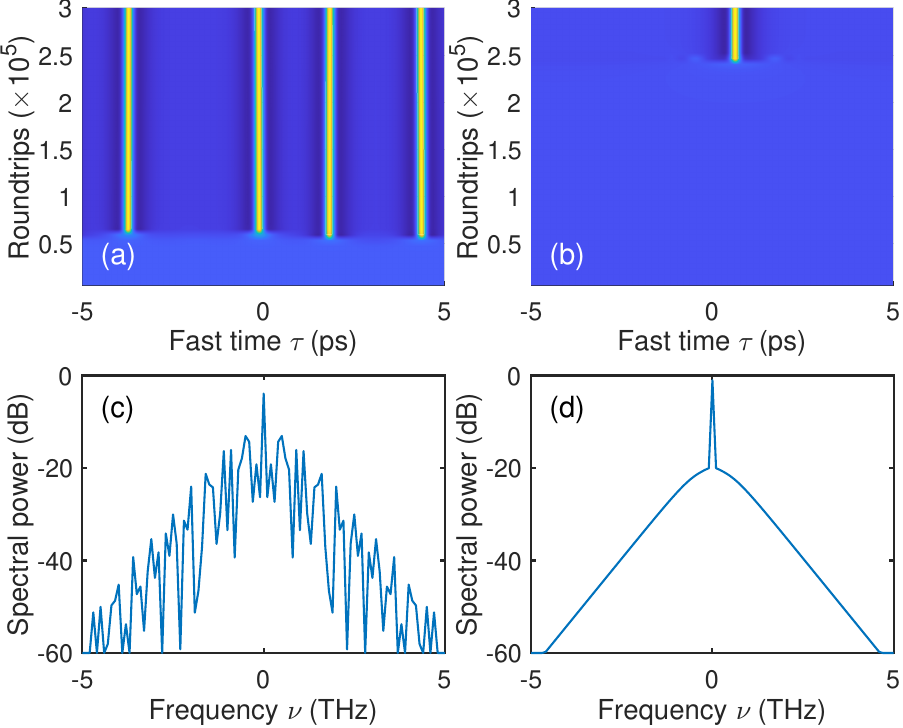}
	\caption{Slow time evolution of the intracavity power $|E_\mathrm{s}|^2$ of a SIL cavity at phase detunings of (a) $\phi=0.15$ and (b) $\phi=0.28$. The steady-state spectra of (a) and (b) are shown in panels (c) and (d) respectively. Simulation parameters are stated in the text.}
	\label{fig:C3_SIL_LLEsim}
\end{figure*}

To reveal the full dynamics of this SIL system we consider the numerical integration of Eqs.~\eqref{SILEqn_Es}--\eqref{SILEqn_EL}. For concreteness, we consider the specific example of a critically coupled Kerr resonator with an FSR of $100$~GHz, a finesse of $2000$, and waveguide parameters $\gamma = 1~\mathrm{W^{-1}\,m^{-1}}$ and $\beta_2 = -60~\mathrm{ps^2/km}$. The laser cavity parameters are set to $2\alpha_\mathrm{L}=\theta_\mathrm{L}=0.5$, a drive power of 76~mW and a laser cavity detuning of $\Delta_\mathrm{L} = \delta_\mathrm{L}/\alpha_\mathrm{R}=5$. These parameters are chosen to match the normalized values used in the simulations of Ref.~\cite{shen_integrated_2020}. 

In Fig.~\ref{fig:C3_SIL_LLEsim}(a) we plot the slow time temporal evolution of the power intracavity field $|E_\mathrm{s}|^2$ for a feedback phase of $\phi=0.15$. This shows the familiar CS excitation dynamics with the generation, in this case, of four stable CS located at random positions around the cavity. Increasing $\phi$ results in the generation of fewer solitons within the cavity. At a phase detuning of $\phi=0.28$, simulations show the deterministic generation of a single CS within the cavity [Fig.~\ref{fig:C3_SIL_LLEsim}(b)]. Importantly, we note that, in this configuration, the cavity is self-starting with a single stable CS forming whenever the laser pump field is applied to the cold Kerr cavity, with no detuning sweep, or special excitation procedure required. In this sense, the system can truely be described as turn-key. Figures~\ref{fig:C3_SIL_LLEsim}(c) and (d) show the steady state optical spectrum obtained at $\phi=0.15$ and $\phi=0.28$, respectively. They show, as expected, the characteristic spectral interference of a multi-soliton field and the sech profile of a single CS. Further simulations reveal that this system is also able to support other nonlinear structures typically observed in standard Kerr cavities such as stable and unstable modulation instability, breathing CS, and perfect soliton crystals.

\subsubsection{Experimental realisation of SIL cavity solitons}

\begin{figure*}[b]
	\centering
	\includegraphics[width = 0.75\textwidth, clip=true]{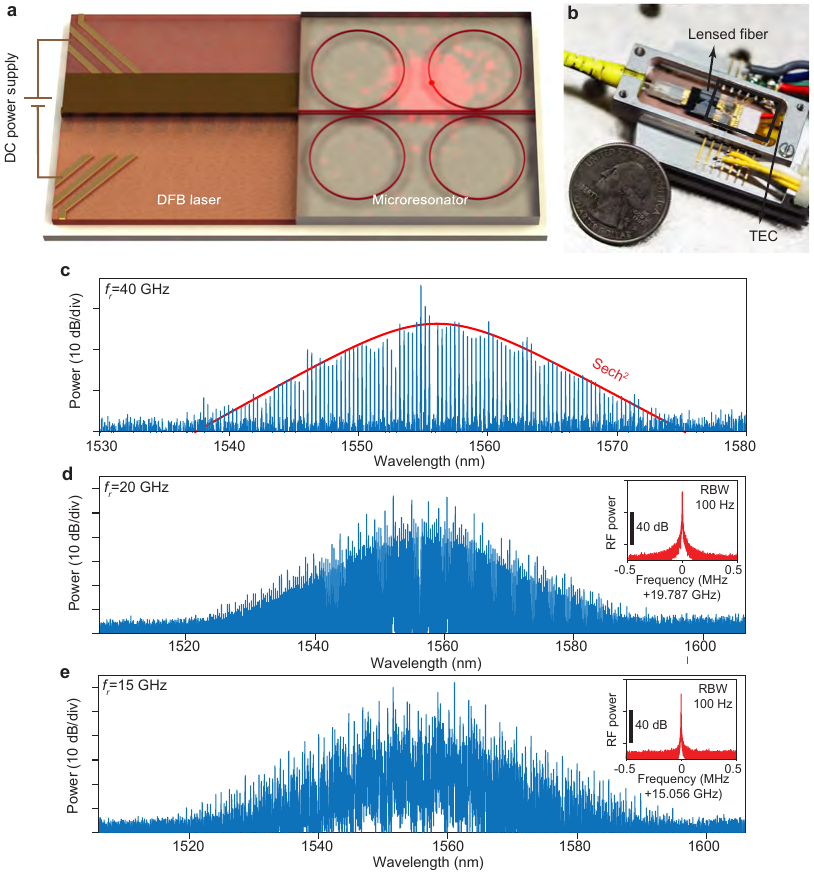}
	\caption{(a) Image of DFB laser chip butt-coupled to a silicon nitride microresonator chip. (b) Image of entire SIL setup integrated into compact butterfly package. (c) Single CS spectrum measured at the output of a SIL Kerr cavity with a FSR of 40~GHz. (d,e) Multi-soliton SIL comb states with repetition rates of 20 and 15~GHz. Adapted with permission from B. Shen \textit{et al.}, Nature \textbf{582}, 365-369 (2020)~\cite{shen_integrated_2020}. Copyright \textcopyright~2020 Springer Nature.}
	\label{fig:C3_SIL_expt}
\end{figure*}

The use of self-injection locking to lock the frequency of a pump laser diode to a Kerr cavity and realise a comb output dates back to the early dates of microresonator comb research~\cite{savchenkov_tunable_2008}. Following the experimental demonstration of CSs in optical microresonators~\cite{herr_temporal_2014}, reports of SIL Kerr CSs quickly followed in both monolithic crystalline~\cite{liang_high_2015,pavlov_narrow-linewidth_2018} and on-chip microresonators~\cite{stern_battery-operated_2018,voloshin2021dynamics,raja_electrically_2019,shen_integrated_2020}.

Figure~\ref{fig:C3_SIL_expt} summarizes the experimental results reported in Ref.~\cite{shen_integrated_2020}. Fig.~\ref{fig:C3_SIL_expt}(a) shows the experimental configuration with a DFB laser chip directly butt-coupled to one of the four ring resonators integrated onto a silicon nitride chip. The entire setup could then be packaged in a compact butterfly package, as shown in Fig.~\ref{fig:C3_SIL_expt}(b). Figure~\ref{fig:C3_SIL_expt}(c) shows the measured optical spectrum of a single CS state observed in a SIL silicon nitride microring with an FSR of 40~ GHz. The superimposed red trace shows the expected $\operatorname{sech}^2$ profile of an isolated CS. The deviation of the experimental spectrum from this curve is attributed to the influence of mode-crossings and the dispersion of the waveguide-resonator coupling parameter. Figures~\ref{fig:C3_SIL_expt}(d,e) show the measured spectra of multi-soliton states with repetition rates of 20 and 15~GHz respectively. Insets to these figures show the corresponding RF spectra, which exhibit narrowband electrical beatnotes at the comb repetition rate, confirming the coherence of these optical structures.

More recent work has been able to demonstrate SIL Kerr combs where both the central optical frequency and repetition rate can be locked to external references, yielding a fully phase-stabilised CS Kerr comb~\cite{wildi_phase-stabilised_2024}, and SIL Kerr combs in a photonic crystal cavity where intracavity Bragg reflectors can be harnessed to provide significantly more control over the feedback field~\cite{ulanov_synthetic_2024}. In addition, SIL comb operation has also been demonstrated in normal dispersion Kerr cavities~\cite{lihachev_platicon_2022}. These normal dispersion structures will be discussed in more detail in Chapter 4.

\section{Localized structures in the normal dispersion regime}
\label{sec:Normal}

In the preceding Chapter, we explored the nonlinear dynamics of a Kerr resonator with anomalous chromatic dispersion. Such systems were shown to permit the excitation of bright, localized, nonlinear dissipative structures known as temporal cavity solitons (CSs). We now shift our attention to Kerr cavities with a self-focusing nonlinearity ($\gamma>0$) and normal chromatic dispersion. In this scenario, the sign of the second-order GVD coefficient is reversed ($\beta_2 > 0$ or $D_2 < 0$, or in normalized units $d_2 = \eta = +1$). This change breaks the balance between dispersion and nonlinearity critical to the formation of CSs. Furthermore, under conditions of cw driving, the upper branch of the homogeneous steady-state solution of the intracavity field is unconditionally stable, preventing the use of the soft-excitation mechanisms previously exploited in the anomalous dispersion regime (c.f.\ Section~\ref{sec:excitation_by_detuning}). In this Chapter we show that, despite these differences, the normal dispersion regime is home to an equally-important class of localized nonlinear dissipative structures that we will now explore. Finally, we note that the analysis presented here can also be used to describe the mathematically equivalent case of Kerr cavities with a self-defocusing nonlinearity ($\gamma<0$) and anomalous dispersion.

\subsection{Switching waves}
\label{sec:SW}

The canonical form of localized dissipative structure present in normal dispersion Kerr resonators is the switching wave (SW)~\cite{coen_convection_1999, parra-rivas_origin_2016}. These structures arise in nonlinear distributed systems that exhibit a multi-valued response, and are formally defined as an inhomogeneous solution, static or traveling, that provides a localized connection --- or interface --- between two different homogeneous solutions \cite{nicolis_self-organization_1977, vasilev_autowave_1979, rozanov_transverse_1982}. As such, they constitute heteroclinic orbits of the underlying dynamical system. SW structures have been reported in many fields such as in chemical kinetics, including flame propagation~\cite{zeldovich_mathematical_1985, frank-kamenetskii_diffusion_2015}, plasma physics and semiconductors~\cite{volkov1969physical, grigoryants_switching_1987}, or in hydrodynamics, where they are often referred to as fronts~\cite{pomeau_front_1986}. In biological systems, they are found, e.g., in population dynamics~\cite{kanel_wave_2006} and are also associated with waves of advances of epidemics or genes~\cite{fisher_wave_1937} or with the action potential in nerves~\cite{hodgkin_quantitative_1952}

In Kerr cavities, SWs were first studied in the context of transverse effects in diffractive Fabry-Perot cavities by Rozanov et~al~\cite{rozanov_transverse_1982}. In the analysis that follows, we restrict our attention to the properties of temporal SWs in dispersive Kerr cavities only. In the simplest scenario, we consider a cw-driven normal dispersion resonator operated at a detuning for which the cavity exhibits a Kerr-induced bistable response (see Section~\ref{Kerrtilt}). For practical purposes, we will only consider SWs between stable cw states, namely the upper and lower branch solutions~\cite{parra-rivas_origin_2016}.

%
%

Figure~\ref{fig:C4_SWintro}(a) shows the temporal intensity profile~$Y=|E(\tau)|^2$ of a typical SW of the canonical LLE, Eq.~\eqref{LLE} (i.e., with GVD restricted to second order normal-dispersion $\eta=+1$, and without SRS), for normalized driving parameters $X = 5$ and~$\Delta = 4$. The black curve in Fig.~\ref{fig:C4_SWintro}(b) highlights the corresponding tilted resonance with the red markers and dotted red lines indicating the respective levels of the homogeneous solutions associated with the upper and lower branches. This clearly shows the intrinsic character of a SW as a localized connection between these two homogeneous states. 

\begin{figure}
    \centering
    \includegraphics{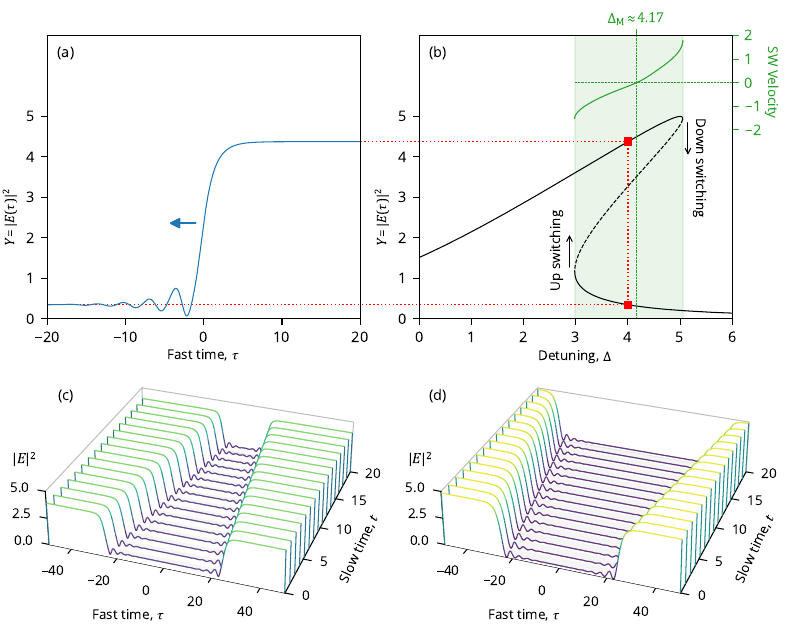}
    \caption{Basics of SWs illustrated for a normalized driving power~$X=5$. (a) Temporal intracavity intensity profile $Y=|E(\tau)|^2$ of a SW for detuning~$\Delta=4$. (b) The black curve shows the associated tilted resonance of the homogeneous steady-state solution (dashes indicate unstable states), with the red dotted lines highlighting the relation to the SW plotted in (a). In the bistable region, we also plot (green curve) how SW velocities ($d\tau/dt$) varies with detuning. (c), (d) Slow-time evolution of a pair of SWs, respectively, for $\Delta = 3.4 < \Delta_\mathrm{M}$, where the upper state invades the lower state, and $\Delta = 4.7 > \Delta_\mathrm{M}$, where the lower state invades the upper state. $\Delta_\mathrm{M} \approx 4.17$ for~$X=5$.}
    \label{fig:C4_SWintro}
\end{figure}

It is important to note that SWs are typically not stationary and \emph{drift} with respect to the driving. In fact, dynamical simulations would reveal that the SW shown in Fig.~\ref{fig:C4_SWintro}(a) travels along the fast-time axis~$\tau$ (in the direction indicated by the blue arrow in the figure), which would signal that it propagates at a different group velocity than the driving field. SW velocities depend both on normalized driving power~$X$ and detuning~$\Delta$~\cite{rosanov_diffractive_1990, coulibaly_universal_2014, parra-rivas_origin_2016, garbin_experimental_2017}. For $X=5$, those velocities are plotted as the green curve in Fig.~\ref{fig:C4_SWintro}(b). As can be seen, they vary monotonically with detuning, which is a general trend. For detunings below a critical value ($\Delta < \Delta_\mathrm{M}$), SWs drift such that the upper state will progressively invade the lower state [as in Fig.~\ref{fig:C4_SWintro}(a)]. Conversely, when $\Delta > \Delta_\mathrm{M}$, the drift is reversed and the lower state will invade the upper state. Only at the critical detuning ${\Delta = \Delta_\mathrm{M}}$, referred to as the Maxwell point (which varies with driving power), is the drift velocity of SWs zero [indicated by the green dashed lines in Fig.~\ref{fig:C4_SWintro}(b)]. For an intuitive understanding of this behavior, it can be observed that, for detunings below the Maxwell point, the system is closer to its up-switching point, so that the upper state is energetically favored, and consequently will invade the lower state~\cite{parra-rivas_origin_2016}. The same argument can be applied when operating above the Maxwell point with the system now closer to its down-switching point, favoring the lower branch. To illustrate this key point, results of numerical simulations performed for a normalized driving power~$X=5$ (for which the Maxwell point is located at ${\Delta_\mathrm{M} \approx 4.17}$) are presented in Figs.~\ref{fig:C4_SWintro}(c) and~(d), and explicitly demonstrate this behavior. Specifically, for detunings (c) below ($\Delta=3.4$) and (d) above ($\Delta=4.7$) the Maxwell point, we show the slow-time evolution of a pair of SWs, which takes the intracavity field from the upper branch solution to the lower branch solution and then back again (because of periodic boundary conditions, SWs must always come in pairs with opposite symmetry). We note that the same behavior is also observed when varying the driving power~$X$ in a cavity operated at a fixed detuning~$\Delta$. Here, likewise, we find that the SW solutions will drift such that for drive powers $X < X_\mathrm{M}$ (respectively, $X>X_\mathrm{M}$) the lower (upper) branch solution will invade the upper (lower) branch.

The above analysis might suggest that the operation of stable SW structures is thus only possible when operating exactly at the Maxwell point --- a fact that would render such structures of little practical use. Indeed, one could think that the relative drift of a SW will continue until either the upper state has fully invaded the lower state ($\Delta < \Delta_\mathrm{M}$), or the lower state has fully invaded the upper state ($\Delta > \Delta_\mathrm{M}$). SWs exhibit however oscillatory tails~\cite{ackemann_chapter_2009} as can be seen in Fig.~\ref{fig:C4_SWintro}(a). The asymptotic profile where the SWs connect to the homogenous cw states can in fact be calculated analytically based on the same analysis as for dispersive waves attached to CSs as presented in Section~\ref{sec:CS_DW}~\cite{jang_observation_2014, malaguti_dispersive_2014, macnaughtan_temporal_2023, bunel_broadband_2024}. In the case of SWs with pure second-order GVD, Eq.~\eqref{Eq:DWpm} (with $V$ now interpreted as the velocity of the SW) predicts that oscillations are only present where the SW field approaches the lower branch solution. In contrast, the SW profile is monotonic where it connects to the upper branch. This is clearly visible in Fig.~\ref{fig:C4_SWintro}(a).

The presence of these oscillatory tails allow SWs to interact and bind to each other in close analogy with the binding of CSs discussed in the previous chapter (c.f. Section~\ref{sec:CSbinding}). As a direct consequence of this interaction, one finds there exists a range of detunings around the Maxwell point within which two SWs can arrest their relative motion and form stable bound states~\cite{parra-rivas_origin_2016,parra-rivas_dark_2016}. Because the profile of the SW is monotonic as it leaves the upper branch, with an oscillatory tail only present as the field approaches the lower branch, the binding between SWs only occurs through interaction of these lower branch tails. As a result, the two SWs will lock together to form a localized `dark' pulse -- a low-intensity structure that is embedded within a high-intensity background~\cite{parra-rivas_origin_2016,parra-rivas_dark_2016}. The multiple oscillations present in the SW tail permit binding at multiple separations between the SWs giving rise to a set of discrete localized structures. In the context of normal dispersion Kerr cavities, such structures are also referred to in the literature as platicons~\cite{lobanov_frequency_2015, lobanov_generation_2015, lobanov_thermally_2021, lihachev_platicon_2022, liu_stimulated_2022}, dark CSs or dark pulse Kerr frequency combs~\cite{coillet_azimuthal_2013,liang_generation_2014,xue_mode-locked_2015,xue_normal-dispersion_2015, parra-rivas_origin_2016}. 

\subsection{Bifurcation analysis of switching waves}

References ~\cite{parra-rivas_origin_2016,parra-rivas_dark_2016} used numerical continuation to reveal the full bifurcation structure of SWs in a normal dispersion Kerr cavity. These results are summarised in Figs.~\ref{fig:C4_SWbifurcation_structure} and ~\ref{fig:C4_SWbifurcation_profiles}. Figure~\ref{fig:C4_SWbifurcation_structure} shows the numerical continuation of SWs solutions at a detuning of $\Delta = 4$ as a function of normalised drive amplitude $u_0=\sqrt X$. Here, the homogeneous solutions of the stable lower ($u_\mathrm{b}$) and upper ($u_\mathrm{t}$) branches, and the unstable middle ($u_\mathrm{m}$) branch, are shown as solid and dashed blue lines respectively. The saddle node bifurcations between these homogeneous solutions are labelled $\mathrm{SN_{hom1,2}}$, and the region of the lower branch that is modulationally unstable indicated by the dashed line between the points MI and $\mathrm{SN_{hom1}}$. The continued SWs solutions are located at a pump amplitudes close to the Maxwell point ($u_\mathrm{M} \sim 2.2$) and plotted in red. The zoomed in right-hand panel of Fig.~\ref{fig:C4_SWbifurcation_structure} shows that these SW solutions are initially unstable (dashed red line) as they depart from the upper branch solution at $\mathrm{SN_{hom2}}$. They subsequently approach the first of a series of saddle-nodes, the first seven of which are labeled $\mathrm{SN_1}$ to $\mathrm{SN_7}$. Stable and unstable SW solutions are then found to exist between successive pairs of these saddle-nodes (shown in Fig.~\ref{fig:C4_SWbifurcation_structure} as solid and dashed red lines). The resultant temporal profile of the first four stable SW solutions are shown in Fig.~\ref{fig:C4_SWbifurcation_profiles} and correspond to locations labelled (a)--(d) in Fig.~\ref{fig:C4_SWbifurcation_structure}. These stable SW solutions can clearly be seen to comprise of a pair of SWs locked together through their overlapping tails, with each new solution possessing an additional oscillation period at the center of the dark structure, and an associated increased temporal separation. Moreover, each successive SW solution is confined to a progressively narrower range of driving amplitudes. This behavior is referred to as collapsed snaking and can be understood as a consequence of the exponential decay of the oscillations of SW tails, with these oscillations quickly becoming too weak to lock the SWs position against their inherent drift velocity~\cite{parra-rivas_origin_2016}.

\begin{figure*}[!t]
	\centering
	\includegraphics[width = 0.7\textwidth, clip=true]{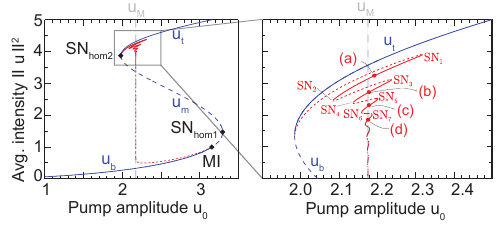}
	\caption{Bifurcation diagram of SW solutions at a detuning of $\Delta = 4$ plotted as a function of normalized drive amplitude $u_0=\sqrt X$. The right-hand panel shows the zoomed in area where stable dark pulse solutions reside. Stable and unstable structures are plotted as solid and dashed lines respectively. Reprinted with permission from P.~Parras-Rivas \textit{et al.}, Opt. Lett. \textbf{41}, 2402-2405 (2016)~\cite{parra-rivas_origin_2016}. Copyright \textcopyright~2016~Optical Society of America.}
	\label{fig:C4_SWbifurcation_structure}
\end{figure*}

\begin{figure*}[!t]
	\centering
	\includegraphics[width = 0.7\textwidth, clip=true]{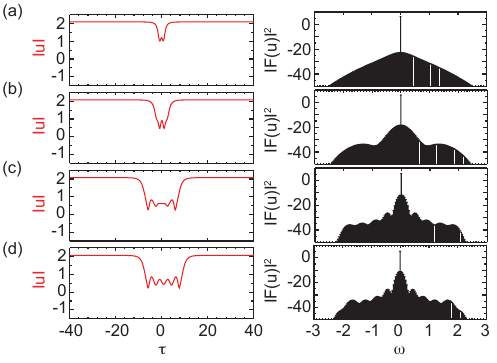}
	\caption{The temporal profile of the magnitude of the normalised intracavity field amplitude $|u|$ (left hand panels), and associated spectrum plotted in dB (right hand panels), obtained at locations labelled (a)-(d) in Fig.~\ref{fig:C4_SWbifurcation_structure}. Reprinted with permission from P.~Parras-Rivas \textit{et al.}, Opt. Lett. \textbf{41}, 2402-2405 (2016)~\cite{parra-rivas_origin_2016}. Copyright \textcopyright~2016~Optical Society of America.}
	\label{fig:C4_SWbifurcation_profiles}
\end{figure*}

The above analysis also reveals the striking fact that, in contrast to cavity solitons, modulation instability does not play a role in the existence of SWs. In the normal dispersion regime the upper branch of the homogeneous cavity solution is unconditionally stable, whilst the lower branch does permit MI, but only around a small range of detunings close to the up-switching point (as illustrated in Fig.~\ref{fig:C4_SWbifurcation_structure}). The SW structures in Fig.~\ref{fig:C4_SWbifurcation_structure} are clearly seen to exist in a region of the parameter space where both the upper and lower states are modulationally-stable. Reference~\cite{parra-rivas_dark_2016} also includes a more detailed analysis of the bifurcation structure of SWs at larger values of $X$ and $\Delta$ that reveals the emergence of both oscillatory and chaotic SW solutions in this region. Additionally, they are able to identify `2-soliton' solutions that comprise pairs of bound SW structures, and show that these structures also exhibit stable and unstable behaviors. Finally, we note that other studies have demonstrated that in the presence of higher-order dispersion, or nonlocal coupling, SWs do exhibit oscillatory tails as they approach the upper branch solution~\cite{gelens_nonlocality-induced_2010, fernandez-oto_strong_2013, colet_formation_2014, gelens_formation_2014, tlidi_localized_2015}. In these situations, is it thus possible to realize the formation of `bright' SW structures through the same binding mechanism described above~\cite{fernandez-oto_strong_2013, tlidi_localized_2015}.

\subsection{Switching wave excitation}
\label{sec:SWexcitation}
We now shift our attention to the various techniques that have been developed for the excitation of SWs. As highlighted at the beginning of this chapter, the fact that the upper branch of a Kerr cavity's homogeneous steady-state response is modulationally-stable precluded the use of the MI-induced, soft-excitation methods exploited for CS excitation in the anomalous dispersion regime. Rather, in a cavity with purely normal dispersion, scanning the pump detuning across a cavity resonance (in the positive $\Delta$ direction) will not excite any localized structures, and the intracavity field will simply follow the trajectory of the upper branch of the tilted Kerr resonance. Consequently, the excitation of SWs, and their associated normal dispersion comb structures, requires more complicated methods. These methods can generally be divided into one of two different approaches. Firstly, engineered local dispersion perturbations to enable phasematched MI even in the presence of normal dispersion~\cite{savchenkov_kerr_2012,liu_investigation_2014,lobanov_frequency_2015,xue_mode-locked_2015,xue_normal-dispersion_2015,soltani_enabling_2016,kim_dispersion_2017,kim_turn-key_2019,helgason_dissipative_2021,spektor_photonic_2024}, and secondly through so-called `hard-excitation' mechanisms that involve direct perturbation of either the driving, or intracavity field~\cite{lobanov_generation_2015, garbin_experimental_2017, lottes_excitation_2021, xu_frequency_2021, liu_stimulated_2022, macnaughtan_temporal_2023, li_experimental_2023, anderson_dissipative_2023, bunel_broadband_2024}.

\subsubsection{Linear mode coupling}

As detailed in Section \ref{avoided_crossings}, the interaction between distinct spatial modes modifies the eigenstates, resulting in a shift in resonance frequencies when compared to scenarios without mode coupling. Specifically, within the (avoided) mode crossing regions, these resonance shifts are notably pronounced. This phenomenon can enable the fulfillment of phase matching conditions for MI, even when the overall dispersion remains normal. The methodologies that leverage mode coupling to facilitate comb generation can be categorized into two subtly distinct scenarios. The first scenario is characterized by a strong coupling condition, wherein the FSRs of the two interacting modes are closely aligned, and the coupling coefficient is substantial. Within the (avoided) mode crossing region, a considerable number of resonances are engaged, spanning a broad spectral range. Employing the same notation as introduced in Section \ref{avoided_crossings}, the natural resonance frequencies of the two cavities can be written as
\begin{align}
    \omega_{01}^{\prime} = \omega_0 + \mu D_{1,1}, \\
    \omega_{02}^{\prime} = \omega_0 + \mu D_{1,2},
\end{align}
where $D_{1,1}=2\pi \mathrm{FSR}_{1}$ and $D_{1,2}=2\pi \mathrm{FSR}_{2}$. Here, we have disregarded the intrinsic dispersion inherent to each cavity. By substituting $\omega_{01}^{\prime}$ and $\omega_{02}^{\prime}$ into Eq. (\ref{ompn}), we can then determine the resonance frequencies of the two super modes as follows
\begin{align}
    \omega_{\pm}^{\prime} &= \omega_0 + \mu D_1 \pm \sqrt{\frac{\mu^2 \Delta D_1^2}{4}+|\zeta|^2 } \\
    &\approx \omega_0 + \mu D_1 \pm |\zeta| \pm \frac{\mu^2 \Delta D_1^2}{8|\zeta|}
\end{align}
where $D_1=(D_{1,1}+D_{1,2})⁄2$ and $\Delta D_1=D_{1,1}-D_{1,2}$. The final approximation is derived by expanding the square root term into a Taylor series, assuming strong coupling (large $|\zeta|$), and retaining only the first two terms. The integrated dispersion is then
\begin{equation}
  D_{\mathrm{int},\pm} = \omega_{\pm}^{\prime} - (\omega_0+\mu D_1 \pm |\zeta|) = \pm \frac{\mu^2 \Delta D_1^2}{8|\zeta|}
\end{equation}
It is evident that the mode coupling effect has a propensity to direct one super mode ($\omega_+^{\prime}$) towards anomalous dispersion, while steering the other super mode ($\omega_-^{\prime}$) towards normal dispersion. Figure \ref{Mode_coupling_dispersion} provides a visual representation of the resonance frequencies with and without the influence of mode coupling. In the case of $\omega_+^{\prime}$, the anomalous dispersion induced by mode coupling may outweigh the intrinsic normal dispersion, resulting in a region characterized by overall anomalous dispersion. When this mode is driven, the anticipated behavior is akin to that of an ordinary anomalous dispersion cavity, potentially leading to the formation of bright solitons~\cite{soltani_enabling_2016,kim_dispersion_2017,yuan_soliton_2023}. 
\begin{figure*}[!t]
	\centering
	\includegraphics[width = 8cm, clip=true]{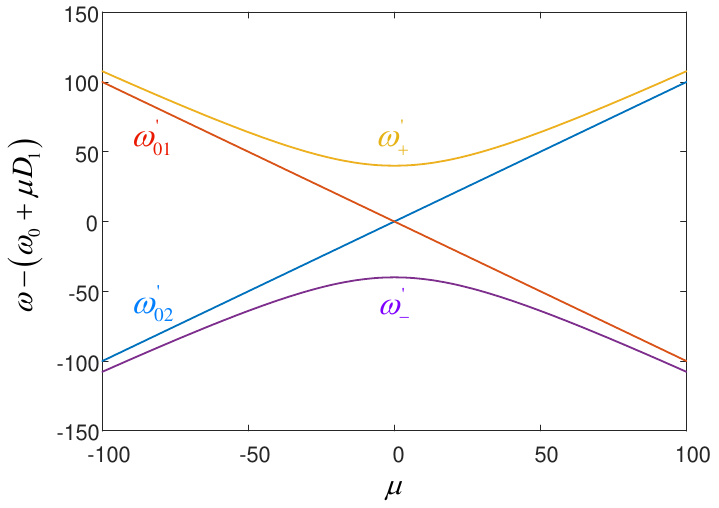}
	\caption{Dispersion engineering with avoided mode crossings} \label{Mode_coupling_dispersion}
\end{figure*}

The second category of mode-coupling-assisted comb generation is associated with a weak coupling scenario. In this case, the two modes possess significantly distinct FSRs. Consequently, only a limited number of resonances from the two modes align and are influenced by the mode coupling. For the sake of simplicity, and without loss of generality, we presume that only one resonance of the mode under consideration is altered as a result of mode coupling. This could be either the resonance that is being pumped or the one where a sideband comb line is generated. The examination of MI entails a three-wave mixing process: the pump mode, the resonance-shifted mode, and its counterpart. Drawing parallels with the analysis presented in Section \ref{sec_MI}, and assuming that the resonance of the lower sideband at frequency $-\Omega$ is the one that is shifted, the matrix of Eq. (\ref{MI_matrix}) is subsequently adjusted as follows:
\begin{equation}
\text{M} = \begin{bmatrix}
-1+i\kappa+i\Delta_\mathrm{a} & iE_\mathrm{s}^2  \\
-i(E_\mathrm{s}^2)^\ast & -1-i\kappa
\end{bmatrix},
\end{equation}
where $\Delta_\mathrm{a}$ represents an additional phase shift related to the real resonance frequency shift $\Delta \omega$ by
\begin{equation}
\Delta_\mathrm{a} = -\frac{\Delta \omega \,t_\mathrm{R}}{\alpha}
\end{equation}
The eigenvalues are
\begin{equation}
  \lambda_{\pm} = -1+i \frac{\Delta_\mathrm{a}}{2} \pm \sqrt{Y^2 - \left(\Delta-\eta \Omega^2 -2Y-\frac{\Delta_\mathrm{a}}{2}\right)^2}\,.
\end{equation}
This expression shows that frequency comb lines will grow up ($\lambda_+$ with positive real part) around~$\pm \Omega$ provided that~$Y>1$. The phase shift $\Delta_\mathrm{a}$ required to achieve the maximum MI gain is then
\begin{equation}
  \Delta_\mathrm{a} = 2(\Delta-\eta \Omega^2)-4Y\,.
\end{equation}
Figures \ref{MI_dark_pulse}(a)-(c) depict the numerical simulation results when $\Delta_\mathrm{a}$ is applied to the $-1\mathrm{st}$ mode to facilitate phase-matched MI. The time-domain waveform resembles a Turing roll, characterized by a narrow spectrum that encompasses only a few comb lines. The gradual temporal drift of this pattern, stemming from the imaginary component of the eigenvalue, suggests a minor deviation in its repetition rate relative to the resonator's FSR. This Turing roll can then act as a catalyst for the excitation of dark pulses. As elucidated in the preceding section, the presence of dark pulses is not contingent upon any MI gain. Consequently, the mode coupling effect can be effectively ``deactivated'' once the Turing roll has been established. Figure~\ref{MI_dark_pulse}(d) illustrates the subsequent evolution of the intracavity field when~$\Delta_\mathrm{a}$ is set to zero. A pair of switching waves emerge, converge, and ultimately coalesce to form a stable dark pulse. The spectrum of the dark pulse significantly broader in comparison to the MI-induced Turing roll. 
 In practical experiments, the mode coupling effect can be deliberately introduced and precisely controlled using a dual-coupled-cavity configuration~\cite{xue_normal-dispersion_2015,kim_turn-key_2019,helgason_dissipative_2021}. The system consists of a primary cavity, which is pumped for frequency comb generation, and an auxiliary cavity with a slightly different free spectral range (FSR). To achieve resonance matching, a tuning mechanism --- such as a microheater for thermal tuning --- is employed to adjust the auxiliary cavity's resonance frequency. The strength of the mode coupling effect can be modulated by tuning the auxiliary cavity's resonance closer to or farther from that of the primary cavity, thereby enabling dynamic control over the coupling strength.

\begin{figure*}[!t]
	\centering
	\includegraphics[width = 12cm, clip=true]{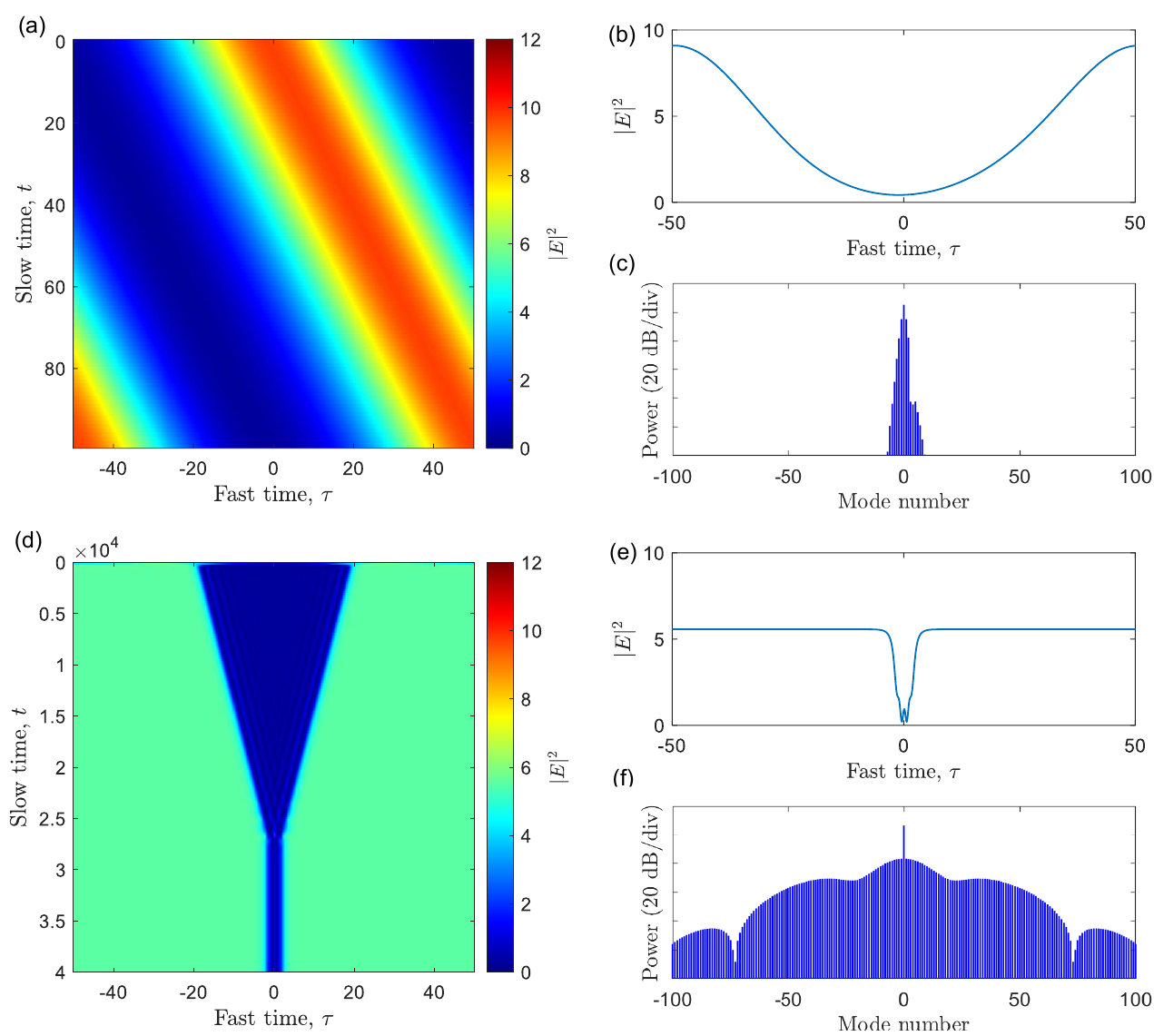}
	\caption{Mode coupling assisted dark pulse generation. The normalized cavity roundtrip time is 100. The pumping parameters are $X=7$ and $\Delta=5$. The initial state is the homogeneous upper-branch solution. (a) Evolution of the intracavity field when a proper amount of phase shift induced by mode coupling ($\Delta_{\mathrm{a}}=-12.1$) is applied to the -1st mode. (b),(c) Waveform and spectrum of the generated Turing roll. (d) Evolution of the intracavity field when the phase shift is turned off ($\Delta_{\mathrm{a}}=0$). (e),(f) Waveform and spectrum of the generated stable dark pulse. } \label{MI_dark_pulse}
\end{figure*}

Applying the resonance shift to the $+\Omega$ mode yields analogous conclusions, with the eigenvalues computed as follows
\begin{equation}
  \lambda_{\pm} = -1-i \frac{\Delta_\mathrm{a}}{2} \pm \sqrt{Y^2 - \left(\Delta-\eta \Omega^2 -2Y-\frac{\Delta_\mathrm{a}}{2}\right)^2}\,.
\end{equation}

When the phase shift $\Delta_\mathrm{a}$ is applied to the pump mode, it is equivalent to applying an opposite shift $-\Delta_\mathrm{a}$  to both $-\Omega$ and $+\Omega$. The eigenvalues are then
\begin{equation}
  \lambda_{\pm} = -1 \pm \sqrt{Y^2 -  \left(\Delta+\Delta_\mathrm{a}-\eta \Omega^2-2Y\right)^2}
\end{equation}
Here, it should be noted that $\pm \Omega$ represent all the sideband pairs, not just one specific pair as when the phase shift is applied to the sideband mode. The condition for the onset of MI (at some $\pm \Omega$) is given by
\begin{equation} \label{MI_region}
  \Delta > 2Y-\sqrt{Y^2-1} - \Delta_\mathrm{a} \quad  \text{and}  \quad Y>1
\end{equation}
It is expedient to visualize the MI region on the Kerr tilt curve, as depicted in Fig. \ref{MI_platicon}(a). The region delineated by Eq. (\ref{MI_region}) is shaded. In the absence of a mode coupling-induced phase shift ($\Delta_\mathrm{a}=0$), the MI region (gray)  predominantly covers the middle branch. With active mode coupling and a positive $\Delta_\mathrm{a}$, the MI region (light blue) shifts to the left, now encompassing a substantial portion of the upper branch. Consequently, frequency combs can be generated by simply scanning the pump into resonance. Figures \ref{MI_platicon}(b)-(d) present a simulated example of this phenomenon. Notably, the bound SW pair exhibits a significantly larger proportion of lower intensity, giving the structure the appearance of a ``bright'' pulse rather than a ``dark'' pulse. Such bright pulses have been termed ``platicons'' by Lobanov et al.~\cite{lobanov_frequency_2015}. The platicon pulse width can be adjusted by varying the pump detuning. It is important to note that the SWs here cannot be directly associated with the homogeneous solutions of the pumped mode. This is because the additional phase shift applied to the pump introduces a global interaction between the intracavity pump and the SW pair. Consequently, a modified effective pump intensity must be used to relate platicons to the dark pulse solution of the ordinary LLE (see supplementary materials of~\cite{chenghao_quantum_2023}). 
\begin{figure*}[!t]
	\centering
	\includegraphics[width = 12cm, clip=true]{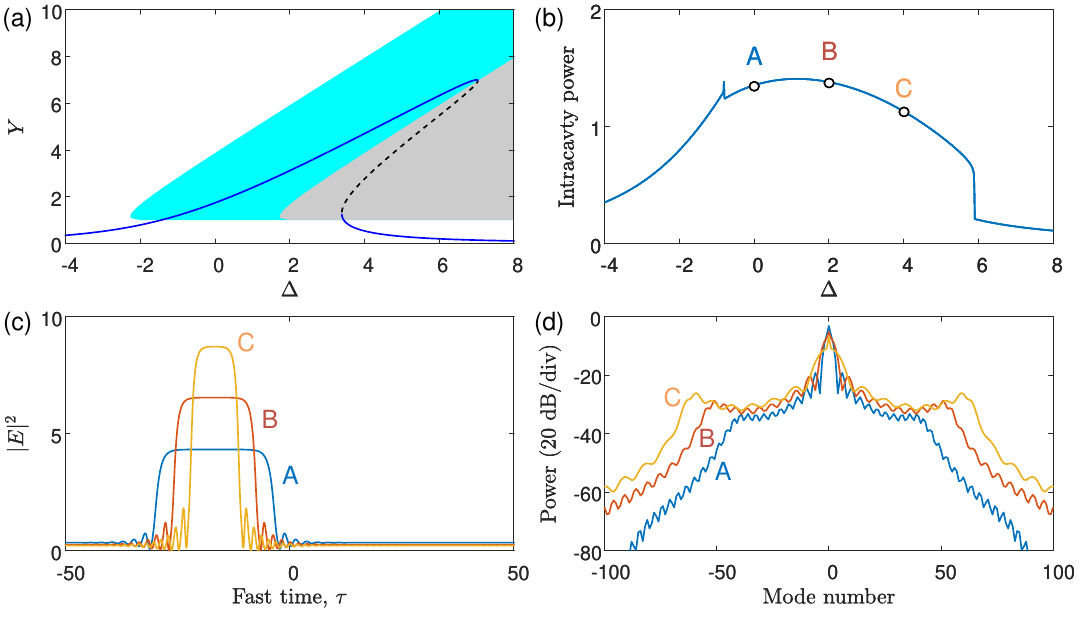}
	\caption{Mode coupling assisted platicon generation. The normalized cavity roundtrip time is 100. The pump power is $X=7$. (a) MI region visualized on the Kerr tilt curve. The results with ($\Delta_{\mathrm{a}}=4$) and without ($\Delta_{\mathrm{a}}=0$) mode-coupling-induced phase shift are indicated by light blue and gray respectively. (b) Intracavity power transition trace when $\Delta_{\mathrm{a}}=4$ and the pump scans across the resonance. The transient waveforms and spectra at the locations denoted by A-C are shown in (c) and (d), respectively.} \label{MI_platicon}
\end{figure*}

Both dark pulse and platicon combs have been successfully demonstrated in experiments, with some exemplary results presented in Fig. \ref{Dark_platicon}~\cite{xue_mode-locked_2015,helgason_dissipative_2021}. The oscillatory tails of the SWs, as predicted by simulations, are clearly visible. Reference~\cite{xue_mode-locked_2015} further characterized the spectral and temporal phase profiles using spectral line-by-line shaping and waveform reconstruction. As shown in Fig.~\ref{Dark_platicon}(a), the spectral phase of the SW combs is not flat but exhibits a complex structure. This complexity arises from the frequency chirp inherent to the fronts and oscillatory tails, which are represented by different colors in Fig. \ref{Dark_platicon}(b). A distinct advantage of dark pulse and platicon combs is the significantly reduced susceptibility to thermal instability when compared to the generation of bright solitons in the anomalous dispersion regime. This is attributed to the substantially less pronounced intracavity power transitions. Consequently, there is no necessity for intricate methodologies such as power kicking or sophisticated pump scanning techniques.
\begin{figure*}[!t]
	\centering
	\includegraphics[width = 12cm, clip=true]{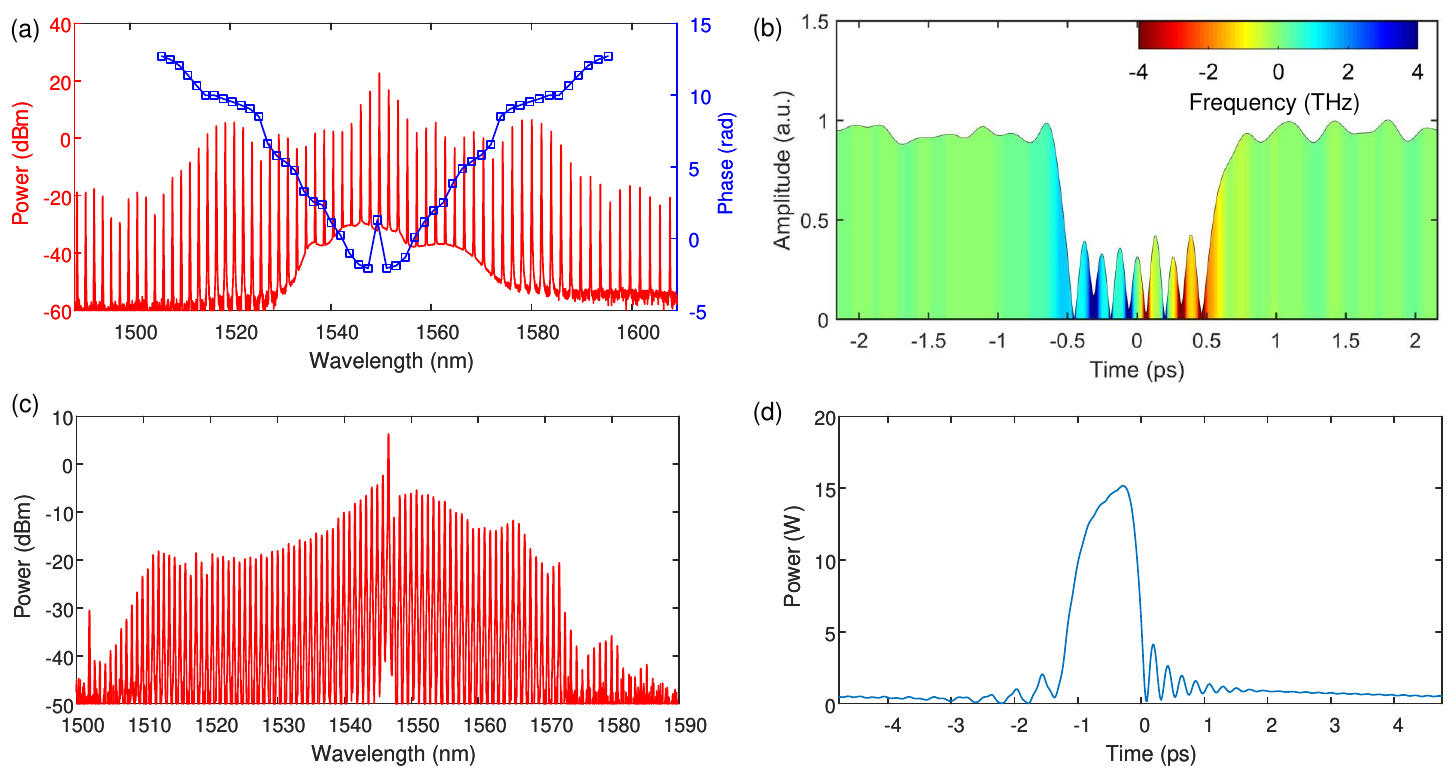}
	\caption{Experimental demonstrations of dark pulse (a,b) and platicon (c,d) combs. (a,c) Spectrum. (b,d) Waveform. The spectral (temporal) phase (frequency) profiles are also shown in (a,b). Panels (a,b) adapted with permission from X. Xue \textit{et al.}, Nat. Photonics \textbf{9}, 594-600 (2015)~\cite{xue_mode-locked_2015}. Copyright \textcopyright~2015 Springer Nature. Panels (c,d) adapted with permission from O. B. Helgason \textit{et al.}, Nat. Photonics \textbf{15}, 305-310 (2021)~\cite{helgason_dissipative_2021}. Copyright \textcopyright~2021 Springer Nature.} \label{Dark_platicon}
\end{figure*}

\subsubsection{Nonlinear mode coupling}

Nonlinear mode coupling can also be harnessed for the excitation of frequency combs in the normal dispersion regime. One form of nonlinear mode coupling arises from the interaction between the fundamental wave and its second harmonic. While Kerr comb generation typically relies on third-order nonlinearity ($\chi^{(3)}$), there are microresonator platforms capable of simultaneously supporting second-order nonlinearity ($\chi^{(2)}$). Notable examples include aluminum nitride, lithium niobate, AlGaAs, and so on. Intriguingly, even amorphous materials lacking intrinsic second-order nonlinearity, such as silicon nitride, can exhibit second-order nonlinearity due to surface effects~\cite{lettieri_second-harmonic_2002,ning_strong_2012, levy_harmonic_2011} or the photogalvanic effect~\cite{lu_efficient_2021,nitiss_optically_2022}. In systems where both $\chi^{(2)}$ and $\chi^{(3)}$ nonlinearities are present, the nonlinear dynamics can be modeled using the following set of coupled equations,
\begin{align}
  t_\mathrm{R} \frac{\partial E_1}{\partial t} & = \left( -\alpha_1-i\delta_0-i\frac{\beta_{21}L}{2}\frac{\partial^2}{\partial\tau^2} + i\gamma_1 L |E_1|^2 + i2\gamma_{12}L|E_2|^2 \right)E_1 + i\kappa L E_2 E_1 + \sqrt{\theta_1} E_\mathrm{in} \\
  t_\mathrm{R} \frac{\partial E_2}{\partial t} & = \left( -\alpha_2-i2\delta_0 -i\Delta k L -\Delta k^{\prime}L\frac{\partial}{\partial \tau} -i\frac{\beta_{22}L}{2}\frac{\partial^2}{\partial\tau^2} + i\gamma_2 L |E_2|^2 + i2\gamma_{21}L|E_1|^2 \right)E_2 + i\kappa^* L E_1^2
\end{align}
where $E_{1,2}$ represents the complex amplitude of the fundamental and second-harmonic waves; $t_R$ is the round-trip time measured at the fundamental frequency; $\alpha_{1,2}$, $\beta_{21,22}$, $\gamma_{1,2}$ are loss, second-order dispersion, and Kerr nonlinearity coefficients of the fundamental and second-harmonic waves, respectively; $\gamma_{12}$ and $\gamma_{21}$ are cross-phase modulation coefficients; $\kappa$ is the second-order coupling coefficient; $L$ is the round-trip length; $\Delta k = 2\beta_{01}-\beta_{02}$ is the wave vector mismatch; $\Delta k^{\prime} = \beta_{12}-\beta_{11}$ is the group velocity mismatch; $\theta_1$ is the power coupling ratio for the fundamental wave; and $E_\mathrm{in}$ is the pump.

When the resonator is driven at the fundamental frequency, the sidebands at the fundamental frequency may become strongly coupled to their counterparts in the second harmonic through sum- and difference-frequency mixing with the fundamental pump, as illustrated in Fig. \ref{SH_comb} (a). This energy exchange can lead to a shift in the resonance of the sidebands, mirroring the avoided mode crossing effect induced by linear mode coupling. Consequently, the resonance shift can promote phase-matched MI. The concept of second-harmonic assisted four-wave mixing and the generation of dark pulses in the normal dispersion regime have been experimentally demonstrated ~\cite{xue_second-harmonic-assisted_2016}. The measured comb spectra in the fundamental and second-harmonic wavelength regions are shown in Figs. \ref{SH_comb} (b) and (c) respectively.
\begin{figure*}[!t]
	\centering
	\includegraphics[width = 12cm, clip=true]{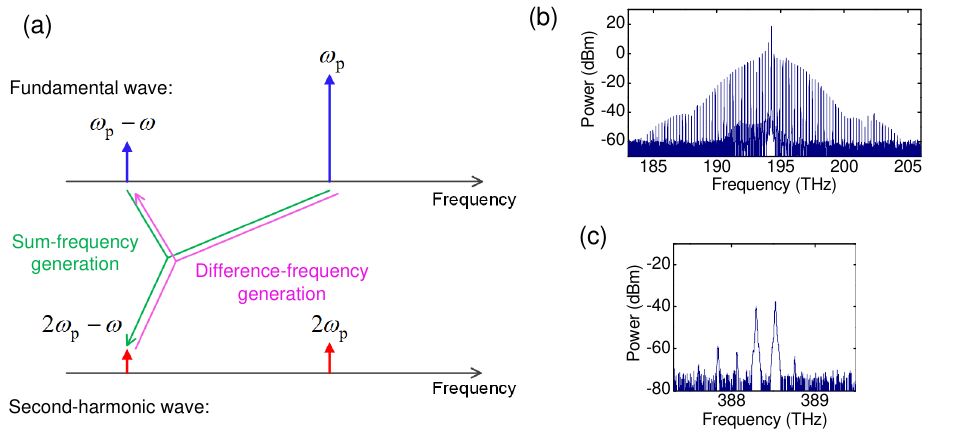}
	\caption{Dark pulse comb generation via second-harmonic assisted four-wave mixing. (a) Illustration of the nonlinear mode coupling through sum- and difference-frequency conversion. (b) Comb spectrum measured in the fundamental wavelength region. (c) Spectrum in the second-harmonic wavelength region. Adapted with permission from X. Xue \textit{et al.}, Light. Sci. \& Appl. \textbf{6}, e16253 (2016)~\cite{xue_second-harmonic-assisted_2016}. Copyright \textcopyright~2016~Springer Nature.} \label{SH_comb}
\end{figure*}

An alternative mechanism for initiating nonlinear mode coupling in comb formation in the normal dispersion regime is cross-phase modulation (XPM). The XPM effect can be induced between distinct spatial modes traveling in the same direction or between modes propagating in clockwise (CW) and counter-clockwise (CCW) directions when the resonator is excited bidirectionally. In the context of a bidirectional pumping scheme, the CW and CCW fields can be characterized by the following set of normalized equations:
\begin{align}
  \frac{\partial E_\mathrm{CW}}{\partial t} & = \left[ -1 -i\Delta_\mathrm{CW} -i\eta\frac{\partial^2}{\partial \tau^2} +\left( |E_\mathrm{CW}|^2 + 2\left< |E_\mathrm{CCW}|^2 \right>\right) \right] E_\mathrm{CW} +S_\mathrm{CW} \\
  \frac{\partial E_\mathrm{CCW}}{\partial t} & = \left[ -1 -i\Delta_\mathrm{CCW} -i\eta\frac{\partial^2}{\partial \tau^2} +\left( |E_\mathrm{CCW}|^2 + 2\left< |E_\mathrm{CW}|^2 \right>\right) \right] E_\mathrm{CCW} +S_\mathrm{CCW} 
\end{align}
Here, $\left<|E_\mathrm{CW}|^2\right>$ and $\left<|E_\mathrm{CCW}|^2\right>$ are roundtrip-averaged intracavity powers given by
\begin{equation}
  \left< |E_\mathrm{CW,CCW}|^2\right> = \frac{1}{T_\mathrm{R}}\int_{0}^{T_\mathrm{R}} |E_\mathrm{CW,CCW}|^2 \mathrm{d}\tau
\end{equation}
where $T_\mathrm{R}=t_\mathrm{R}⁄\tau_\mathrm{s}$  is the normalized round-trip time. The equations for the homogeneous solutions are
\begin{align}
  Y_\mathrm{CW} \left[ 1 + \left[ \Delta_\mathrm{CW} - (Y_\mathrm{CW}+2Y_\mathrm{CCW})\right]^2 \right]  & = X\\
  Y_\mathrm{CCW} \left[ 1 + \left[ \Delta_\mathrm{CW} +\Delta_\mathrm{d} - (Y_\mathrm{CCW}+2Y_\mathrm{CW})\right]^2 \right ]  & = rX
\end{align}
where $\Delta_\mathrm{d}=\Delta_\mathrm{CCW}-\Delta_\mathrm{CW}$ is the pump detuning difference and $r=|S_\mathrm{CCW} |^2⁄|S_\mathrm{CW} |^2$ is the pump power imbalance. The scenario with symmetric pumping (i.e., $r=1$ and $\Delta_\mathrm{d}=0$) has been discussed in Section \ref{sec_ssb}. Here we focus on the condition with asymmetric pumping. The plot with asymmetric pump power and symmetric detuning ($r=0.6$, $\Delta_\mathrm{d}=0$) is shown in Fig.~\ref{Bi_pump}(a). The cw pump power is higher in this example, resulting in a higher intracavity power in the CW direction. If we define an effective detuning for the CW field as
\begin{equation}
\Delta_\mathrm{eff,CW} = \Delta_\mathrm{CW} - 2Y_\mathrm{CCW},
\end{equation}
we can convert the plot to represent the canonical bistable cavity response with a unidirectional pump, as shown in Fig.~\ref{Bi_pump}(b). Subsequently, we can directly apply the findings from the MI analysis conducted in the previous section. Notably, the MI region, which predominantly encompasses the middle branch in the bistable cavity response, now transitions to encompass a significant portion of the upper branch. Frequency combs can then be accessed by scanning the pump into resonance. Similar results can be obtained when the bidirectional pump detuning is asymmetric and the pump power is symmetric (i.e., $r=1$, $\Delta_\mathrm{d}\neq 0$). The generation of dark pulse combs using bidirectional pumping has been successfully demonstrated by Yang et al.~\cite{yang_cross-phase_2024}.
\begin{figure*}[!t]
	\centering
	\includegraphics[width = 9cm, clip=true]{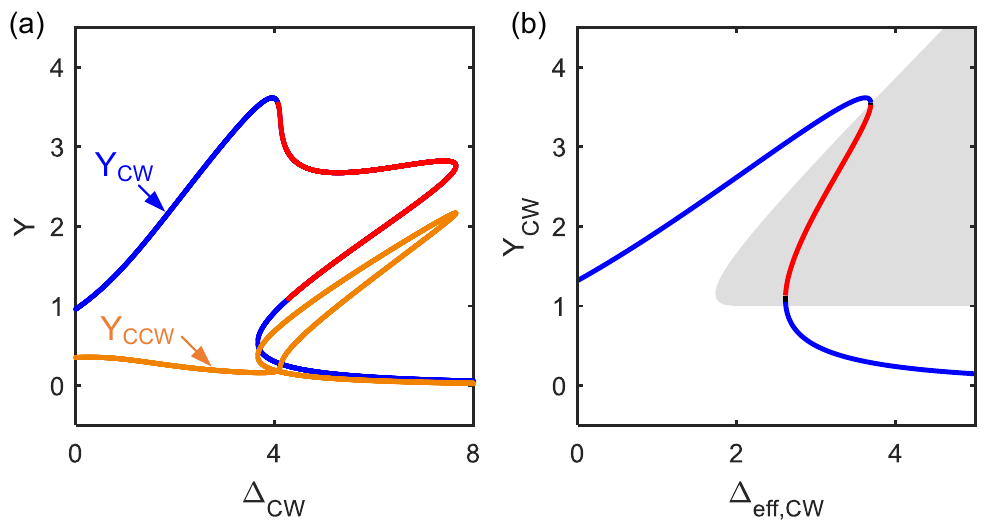}
	\caption{(a) Homogeneous solutions for the bidirectional pumping scheme with $r=0.6$, $\Delta_\mathrm{d}=0$, and $X=3.6$. (b) Re-plotting of the CW trace in (a) with respect to the effective detuning $\Delta_{\mathrm{eff,CW}}$. The middle branch is marked in red, and the MI region is shaded in gray. Adapted with permission from B. Yang \textit{et al.}, Opt. Lett. \textbf{49}, 4425-4428 (2024)~\cite{yang_cross-phase_2024}. Copyright \textcopyright~2024~Optica Publishing Group.} \label{Bi_pump}
\end{figure*}

\subsubsection{Direct writing} 

Switching waves can also be excited in purely normal dispersion cavities through hard-excitation mechanisms whereby perturbations are applied directly to driving field~\cite{garbin_experimental_2017 , lottes_excitation_2021}. In Ref.~\cite{garbin_experimental_2017} the direct writing of SWs into a 90 m fiber cavity with a net normal dispersion was reported. To achieve this, an electro-optic intensity modulator was used to imprint a localized $\sim 200$ ps dip into the center of an $\sim1$~ns quasi-cw driving field for 10 cavity round trips. After this time, the drive power was returned to its original level. As a result of this perturbation, a pair of SWs were excited at the two edges of the dip. This can be seen directly in the experimentally measured intracavity intensity plotted as a function of round-trip number in Fig.~\ref{fig:C4_SWexpt_drift}. The perturbation to the driving field is applied at round trip $\sim 250$, at which point the SW pair is seen to form and begin to drift according to the dynamics outlined in the previous Section. These measurements were made at a normalized drive power of $X = 19.5$ for which the Maxwell point is found at $\Delta_\mathrm{M} \sim 10.4$. Fig.~\ref{fig:C4_SWexpt_drift} shows the SW evolution with (a) the cavity detuning set at $\Delta=7.5$ below the Maxwell point (with the upper state invading the lower state, and (b) the cavity detuning set at $\Delta =11.1$ above the Maxwell point, with the lower state invading the upper state. In addition, Ref.~\cite{garbin_experimental_2017} was also able to repeat this measurement at a detuning close to the Maxwell point ($\Delta = 9.9$) and directly observe the formation of stable locked SW structures. Two separate realizations of this experiment are shown in panels (a) and (b) of Fig.~\ref{fig:C4_SWexpt_boundstate} with the final stable locked state evident from round trip $\sim 1000$ onwards. Due to slightly different driving conditions, the final dark pulse structure has a different temporal duration in each case. We also note that the temporal resolution of the real time oscilloscope used to make this measurement was insufficient to directly resolve the oscillatory tails of the SWs responsible for this locking. However, later experiments were able to directly resolve these temporal features~\cite{macnaughtan_temporal_2023, bunel_broadband_2024}.

\begin{figure*}[!t]
	\centering
	\includegraphics[width = 0.7\textwidth, clip=true]{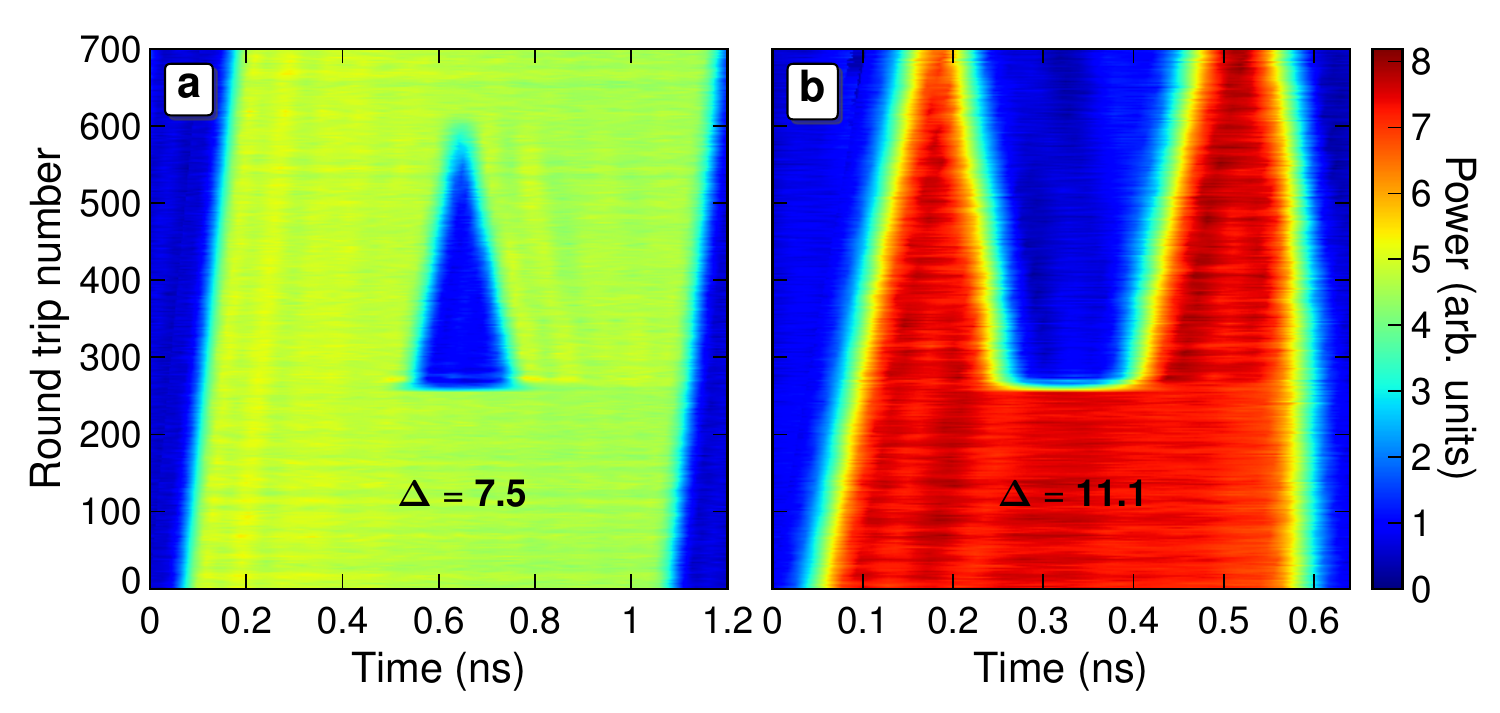}
	\caption{Experimental measurement of the temporal evolution of the intracavity intensity over 700~round trips: (a) $\Delta = 7.5 < \Delta_\mathrm{M}$, (b) $\Delta = 11.1 > \Delta_\mathrm{M}$. The SW is excited by through the addition of a 200 ps dip to the center of $\sim1$ ns drive pulse for 10 round trips starting at round trip 200. Adapted with permission from B. Garbin \textit{et al.}, Eur. Phys. J. D \textbf{71}, 240 (2017)~\cite{garbin_experimental_2017}. Copyright \textcopyright~2017~Springer Nature.}
	\label{fig:C4_SWexpt_drift}
\end{figure*}

\begin{figure*}[!t]
	\centering
	\includegraphics[width = 0.7\textwidth, clip=true]{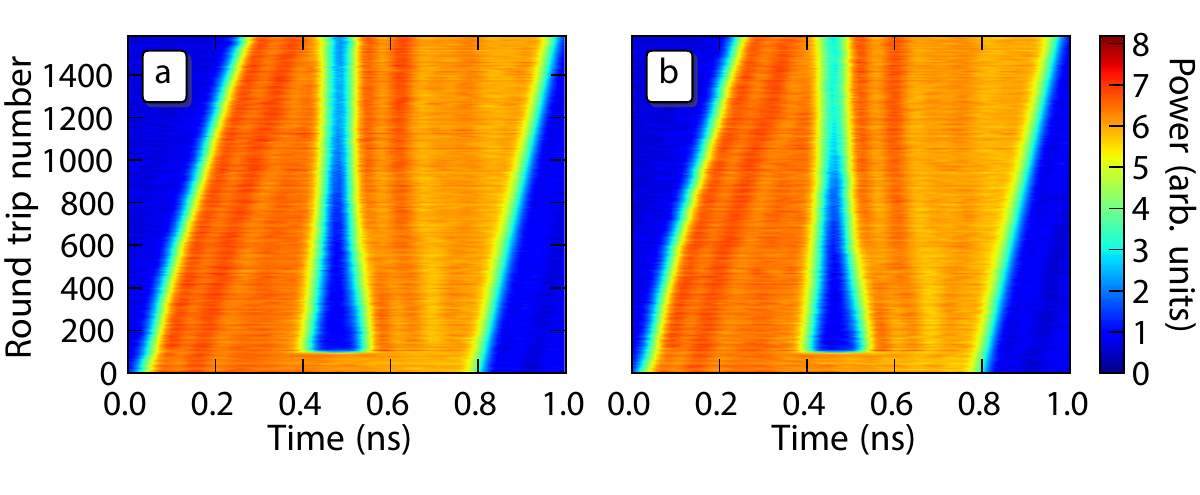}
	\caption{Experimental measurement of stable dark pulse formation when operating at $\Delta = 9.9$, close to the Maxwell point $\Delta_\mathrm{M} \sim 10.4$. Panels (a) and (b) show different realizations of the same experiment, with different duration dark pulse structures obtained in each case. Adapted with permission from B. Garbin \textit{et al.}, Eur. Phys. J. D \textbf{71}, 240 (2017)~\cite{garbin_experimental_2017}. Copyright \textcopyright~2017~Springer Nature.}
	\label{fig:C4_SWexpt_boundstate}
\end{figure*}

\subsubsection{Pulsed driving}

Switching waves can also form in normal dispersion cavities that are pulse driven~\cite{lobanov_generation_2015, xu_frequency_2021, liu_stimulated_2022, li_experimental_2023, anderson_dissipative_2023,  macnaughtan_temporal_2023, bunel_broadband_2024}. The formation dynamics in this case can be simply described in light of the analysis presented above. We first consider the case of a perfectly synchronized driving field for which $\Delta\tau = t_\mathrm{R} - t_\mathrm{P} = 0$, where $t_\mathrm{R}$ is the cavity round-trip time of the cavity and $t_\mathrm{P}$ is the period of the pulsed driving beam (see also Section~\ref{sec:pulsed}). Below, we also refer to $X_\mathrm{max}$ as the normalized  peak power of the driving field, with the driving power varying from 0 to $X_\mathrm{max}$ across the pump pulses. When operating at a fixed detuning $\Delta$, there are two possibilities: 

\begin{itemize}
    \item $X_\mathrm{max} < X_\mathrm{M}$ --- The lower state will invade the upper state across the whole driving pulse. Therefore, the entire intracavity field will occupy the lower branch and no SW will form.
    \item $X_\mathrm{max} > X_\mathrm{M}$ --- The lower state will prevail as above in the wings of the pulse, specifically at fast-time locations where the local driving power is below the Maxwell point, $X(\tau) < X_\mathrm{M}$. In contrast, where $X_\mathrm{max} \ge X(\tau) > X_\mathrm{M}$, the upper state invades the lower state, and at those locations the cavity can stably remain on the upper branch.
\end{itemize}

\begin{figure*}[!t]
	\centering
	\includegraphics[width = \textwidth, clip=true]{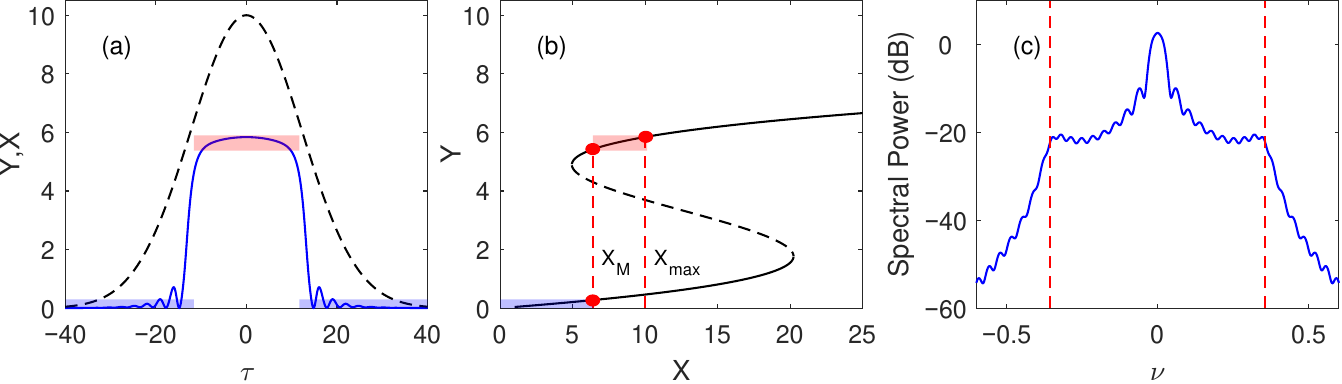}
	\caption{Numerical simulations of the role of SWs in a normal GVD resonator driven with Gaussian pulses, with peak power  $X_\mathrm{max}=10$ and detuning $\Delta=5$. (a)~Temporal intensity profiles of the driving field~$X(\tau)$ (dashed black line) and intracavity field~$Y=|E(\tau)|^2$ (solid blue line). (b)~Associated bistability curve of the homogeneous (cw) steady-state solutions. The red (blue) color bars, matching between (a) and~(b), delineate ranges of intracavity powers where $X>X_\mathrm{M}$ ($X>X_\mathrm{M}$), respectively. (c) Spectral intensity of the intracavity field. All units are normalized.}
	\label{fig:C4_SWpulsed}
\end{figure*}

In the second case, SWs will form between the regions occupied by the lower and upper states, i.e., at fast time locations that correspond to $X(\tau) = X_\mathrm{M}$. This behavior is illustrated by the LLE simulation results shown in Fig.~\ref{fig:C4_SWpulsed}(a). Here, we have considered a normal dispersion Kerr resonator operating at a detuning $\Delta = 5$ and synchronously driven by Gaussian pulses with a normalized peak power $X_\mathrm{max} = 10$. An example of temporal intensity profiles of the driving and intracavity fields are shown as black dashed and blue solid lines, respectively. The presence of a pair of SWs, symmetrically located on either side of the intracavity pulse, is plainly evident. Fig.~\ref{fig:C4_SWpulsed}(b) shows the cw bistability response at the same detuning. On that plot, we have delineated with solid red and blue rectangles the ranges of intracavity powers~$Y$ for which the driving power~$X$ is respectively higher and lower than the Maxwell point~$X_\mathrm{M}$ (for $\Delta=5$, $X_\mathrm{M} \sim 6.4$). The same ranges of intracavity powers are shown in~(a) with matching color bars, which demonstrates the role of the Maxwell point in setting the location of the two SWs. Finally, Fig.~\ref{fig:C4_SWpulsed}(c) shows the optical spectrum of the intracavity field. Prominent peaks in the wings of the spectrum can be associated with the oscillatory tails of the SWs. As discussed in Section~\ref{sec:SW}, the frequency of those peaks can be predicted analytically based on Eq.~\eqref{Eq:DWpm}. This prediction is marked by red dashed lines in the plot and found to be in excellent agreement with simulation results. In normalized units, this frequency can be approximated (in the limit of a large detuning~$\Delta$) by $\nu_\mathrm{tail} \simeq \pm\sqrt{\Delta}/2\pi$, which can be used to provide a useful estimate of the spectral extent of the SWs and associated frequency combs.

\begin{figure*}[!t]
	\centering
	\includegraphics[width = 0.5\textwidth, clip=true]{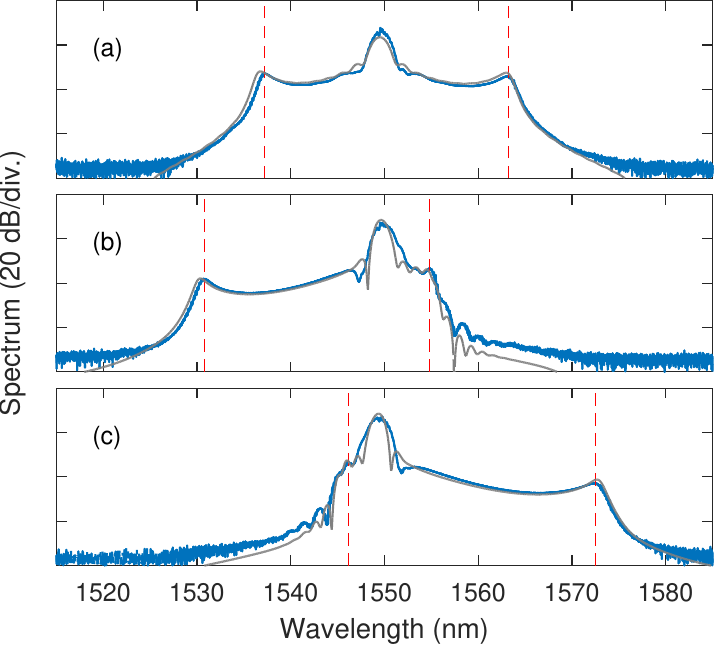}
	\caption{Pulsed driven SW spectra measured in a normally dispersion fiber cavity (blue traces), and associated LLE simulations (gray traces), at pump-cavity desynchronisations of (a) $\Delta \tau = 0$ fs, (b) $\Delta \tau = -14.3$ fs and (c) $\Delta \tau = +17$ fs. Adapted with permission from Y. Xu \textit{et al.}, Opt. Lett. \textbf{46}, 512-515 (2021)~\cite{xu_frequency_2021}. Copyright \textcopyright~2021~Optica Publishing Group.}
	\label{fig:C4_SWdesynch}
\end{figure*}

Experimental realizations of pulsed driven SW combs have been reported in both microresonators~\cite{liu_stimulated_2022, li_experimental_2023, anderson_dissipative_2023} and fiber cavities~\cite{xu_frequency_2021, macnaughtan_temporal_2023, bunel_broadband_2024}. In both cases, the pump source used is derived from an electro-optic comb to enable accurate control of both the detuning and the repetition rate. An interesting extension to the generation of pulsed-driven SW-based combs occurs when we consider desynchronized driving fields such that $\Delta \tau  \neq 0$. In this case, and in order for the SWs to remain stationary with respect to the driving field, the SWs' own natural velocities must balance the convective drift induced by the desynchronization, $V + \Delta\tau/\alpha = 0$ [using normalized units, see Eq.~\eqref{d_coefficient_normalized}]. This leads to an asymmetric temporal intensity profile for the intracavity pulse, and to an asymmetric arrangement of the SW spectral peaks, now found at approximately $\nu_\mathrm{tail} \simeq (-\Delta\tau/\alpha \pm \sqrt{\Delta})/(2\pi)$. This enables the spectral extent of the output spectrum to be controlled by varying the repetition rate of the driving field. To illustrate this point, Fig.~\ref{fig:C4_SWdesynch} shows simulated and experimentally measured pulsed-driven SW-based comb spectra reported in Ref.~\cite{xu_frequency_2021} at desynchronisations of (a) $\Delta \tau = 0$ fs, (b) $\Delta \tau = -14.3$ fs and (c) $\Delta \tau = +17$ fs. The phasematched locations predicted by Eq.~\eqref{Eq:DWpm} are shown as red dashed lines and again seen to provide an accurate estimate of the spectral extent of the resultant comb. In particular, the spectra appear blue-shifted (respectively, red-shifted) for $\Delta\tau <0$ ($\Delta\tau>0$), as expected.

\subsection{Conversion efficiency}
\label{sec:normal_conversioneff}

The conversion efficiency of SW combs, including dark pulses and platicons, is generally much higher than that of bright solitons in the anomalous-dispersion regime even when no special strategies are employed. This can be readily observed by noting that SW combs typically exhibit a much smaller contrast between the residual pump line and the other comb lines outside the cavity. For instance, conversion efficiencies exceeding 30\,\% have been demonstrated for dark pulses in an ordinary single microcavity driven by a cw pump~\cite{xue_microresonator_2017}. Unlike bright solitons, deriving a closed-form analytical expression for the efficiency of dark pulse Kerr combs is challenging. However, the conclusion derived for bright solitons in Eq.~\eqref{eq:etamax2} is also qualitatively applicable to dark pulses, i.e., their conversion efficiency is related to their duty cycle. Specifically, a larger duty cycle results in higher conversion efficiency. Theoretically, the optimum conversion efficiency for dark pulses in an ordinary single microcavity is around 50\,\%, assuming the absence of intrinsic loss~\cite{xue_microresonator_2017}. In microcavities with avoided mode crossings for the pump, even higher efficiencies (nearly 100\,\%) are possible, as the mode coupling disrupts the localization of the dark pulses/platicons~\cite{lobanov_frequency_2015}. Experimentally, conversion efficiencies as high as 41\,\% and 49\,\% have been achieved using dual-coupled microcavities by Kim et al.~\cite{kim_turn-key_2019} and Helgason et al.~\cite{helgason_dissipative_2021}, respectively.

\subsection{Zero-dispersion solitons}

We now consider the localized nonlinear structures that arise in Kerr cavities where the normalised second order dispersion coefficient  ($d_2$, as defined in section~\ref{sec:norm}) is close to zero. In this scenario, the influence of higher-order dispersion, particularly third-order dispersion (TOD), must be taken into account, revealing a new range of multi-peak structures~\cite{li_experimental_2020,li_observations_2021, anderson_zero_2022, xiao_near-zero-dispersion_2023}. This evolution is illustrated schematically in Fig.~\ref{fig:C4_ZW_schematic} which shows, clockwise from the left, the nonlinear dissipative structures found in a Kerr cavity with, initially, purely anomalous dispersion ($d_2 = -1, d_3 = 0$), progressing to a small second-order dispersion with a significant contribution from TOD ($d_2 \sim 0, d_3 \sim 1$), then finally to a cavity with pure normal dispersion ($d _ 2 = +1, d _ 3 = 0$). Following this path, the structures present are seen to evolve from conventional temporal CSs, to CSs with large dispersive-wave tails, to a new class of multi-peak structures in the region where TOD dominates, before transforming into SW-based structures with oscillating tails on both the upper and lower levels, and finally standard normal dispersion SW structures~\cite{anderson_zero_2022}. 

\begin{figure*}[!t]
	\centering
	\includegraphics[width = 0.9\textwidth, clip=true]{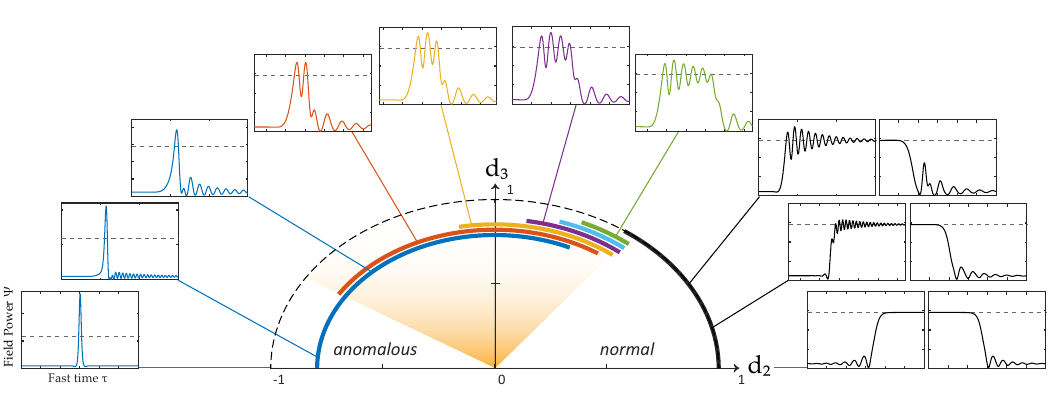}
	\caption{Evolution of nonlinear localized structures in a Kerr cavity as dispersion is swept from (clockwise from left): pure anomalous dispersion ($d_2 = -1, d_3 = 0$), to small second-order dispersion with TOD dominant ($d_2 \sim 0, d_3 \sim 1$), to pure normal dispersion ($d_2 = +1, d_3 = 0$). Reprinted with permission from M.~H.~Anderson \textit{et al.}, Nat. Commun. \textbf{13}, 4764 (2022)~\cite{anderson_zero_2022}. Copyright \textcopyright~2022 Springer Nature.}
	\label{fig:C4_ZW_schematic}
\end{figure*}

An analysis of the bifurcations of the localized structures present in the regime of dominant TOD confirms the existence of a wide range of both bright and dark multi-peaked structures that follow a collapsed-snaking bifurcation sequence~\cite{li_experimental_2020}. Consequently, for a given set of drive parameters, multiple stable localized states are realizable. This was confirmed in experiments carried out in a dispersion shifted fiber ring in Ref.~\cite{li_experimental_2020}. At a pump wavelength of 1563~nm, that ring was measured to possess weakly normal second-order dispersion ($\beta_2 = 0.25 ~\mathrm{ps^2/km}$), and TOD of $\beta_3 = 0.135~\mathrm{ps^3/km}$. Using quasi-cw driving, the authors were able to excite multiple stable structures when operating the ring at a fixed drive power and detuning. The experimentally measured spectra of these states are shown in Figs.~\ref{fig:C4_Expt_ZDCS}(a--d) in blue. Plotted in orange are the spectral and temporal traces recovered from the associated LLE simulations. The measured spectra show a characteristic strong dispersive wave peak on the high frequency side the spectrum. The measured position of this peak agrees almost exactly with the predicted DW position [c.f. Eq.~\eqref{Eq:DWpm2}] shown as a red dashed line. In addition, the low frequency side of the spectrum exhibited a number of distinct spectral dips. LLE simulations reveal that spectra with $n$ dips on the low-frequency side of the spectrum correspond to temporal structures with $n+1$ peaks. These zero-dispersion structures have also been reported in microresonator systems with both pulsed~\cite{anderson_zero_2022, xiao_near-zero-dispersion_2023} and cw driving~\cite{zhang_microresonator_2023, ji_engineered_2023}.

\begin{figure*}[!t]
	\centering
	\includegraphics[width = \textwidth, clip=true]{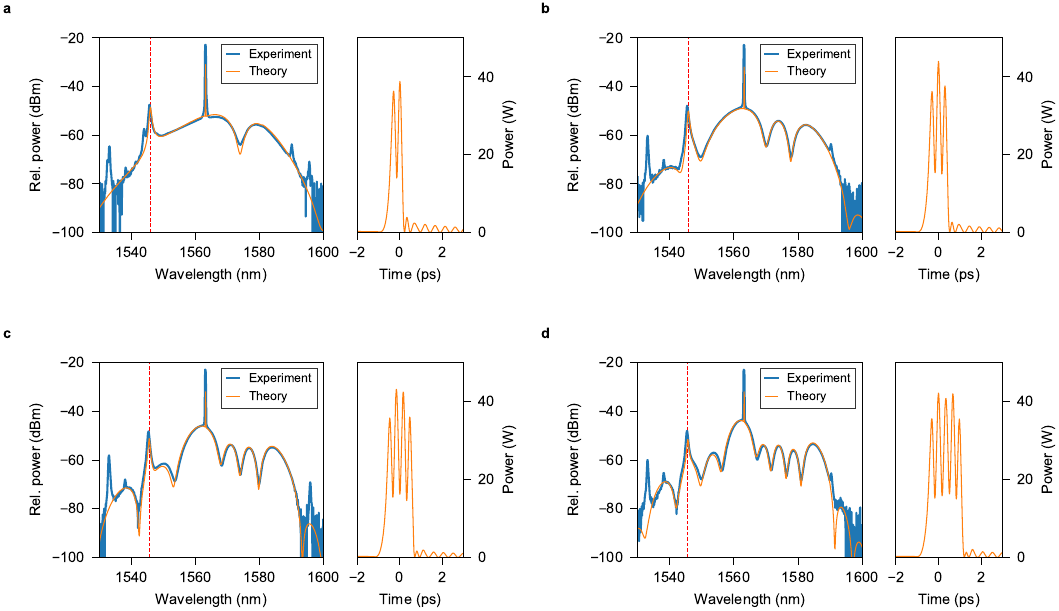}
	\caption{Measured bright structures with weak normal dispersion driving. Panels (a)–(d) display experimentally measured (blue solid curves) and numerically simulated (orange solid curves) optical spectra on the left. The corresponding temporal profiles extracted from numerical simulations are shown on the right. The dashed red vertical lines indicate the theoretically predicted dispersive wave wavelengths. Reprinted with permission from Z. Li \textit{et al.}, Optica. \textbf{7}, 1195-1203 (2020)~\cite{li_experimental_2020}. Copyright \textcopyright~2020~Optical Society of America.}
	\label{fig:C4_Expt_ZDCS}
\end{figure*}

\subsection{Other localized states in the normal-dispersion regime}

In addition to the SW structures discussed above, several studies have reported the emergence of other distinct localized states when the cavity dispersion is either all-normal or partially normal. These findings are summarized as follows:

\begin{itemize}
\item \textbf{Self-injection-locked laser SWs.} Self-injection locking of the laser diode with the microcavity represents a promising approach to realizing ultracompact microcomb sources. Self-injection-locked turnkey soliton generation has been successfully demonstrated in the anomalous-dispersion regime (c.f. Section~\ref{sec:SIL_CS}). Notably, the self-injection locking scheme also provides an alternative pathway for generating SWs in the normal-dispersion regime~\cite{jin_hertz-linewidth_2021,lihachev_platicon_2022}. The interaction between the laser diode and the nonlinear microcavity allows for an operating point in the middle branch of the bistable cavity, where modulational instability (MI) is present. Consequently, SWs can be excited without the need for avoided mode crossings. Furthermore, the separation between the SWs (i.e., the dark pulse width) can be flexibly controlled by adjusting the feedback phase~\cite{wang_self-regulating_2022}.

\item \textbf{Filter-driven solitons.} It is well established in the mode-locked laser research community that stable pulses can be generated in the normal-dispersion regime when a bandpass filter is inserted into the cavity~\cite{grelu_dissipative_2012}. These filter-driven pulses are typically highly chirped~\cite{renninger_dissipative_2008,kieu_sub-100_2009}. One significant advantage of filter-driven solitons is their ability to achieve pulse energies much higher than those of standard anomalous dispersion NLSE solitons. Inspired by mode-locked lasers, filter-driven solitons have also been investigated in coherently driven passive cavities. Spiess et al.\ demonstrated the generation of chirped solitons in a pulse-driven fiber cavity with large net-normal dispersion and a spectral filter~\cite{spiess_chirped_2021}. The pulse extracted at the cavity output could be compressed using grating pairs, resulting in a minimum duration of~$1.08$~ps. A notable feature of chirped solitons highlighted by the authors is their ability to exist in a low-Q cavity with roundtrip loss exceeding 90\,\%, which may offer new opportunities for nonlinear pattern formation. Filter-driven solitons were also demonstrated in ~\cite{xue_dispersion-less_2023}. Therein, unlike for the fiber cavity used by \cite{spiess_chirped_2021} which had a large net-normal dispersion, a programmable spectral shaper is inserted into the cavity to precisely compensate for the fiber dispersion, resulting in both close-to-zero second-order and higher-order dispersion. Nearly chirp-free pulses, termed “dispersion-less solitons,” were generated. The absence of all orders of dispersion clearly indicates that pulse formation is primarily dominated by the filtering effect and the Kerr nonlinearity. With a rectangular gate filter, Nyquist-pulse-like solitons with ultra-flat spectra were observed.

\item \textbf{Dispersion-managed solitons.} The concept of dispersion-managed solitons was initially developed for mode-locked lasers~\cite{chen_dispersion-managed_1999}. The laser cavity comprises elements with alternating normal and anomalous dispersion, causing the pulses to stretch and compress periodically within the cavity. This mechanism is also known as stretched-pulse mode locking. Stretched-pulse mode-locked lasers can generate pulses that are much shorter than those produced by traditional soliton lasers in the anomalous-dispersion regime. Bao and Yang first numerically explored the possibility of generating stretched solitons in coherently driven dispersion-managed cavities~\cite{bao_stretched_2015}. Dong et al.\ and Li et al.\ later experimentally demonstrated stretched solitons in dispersion-managed fiber cavities~\cite{dong_stretched-pulse_2020} and microcavities~\cite{li_real-time_2020}, respectively. In~\cite{dong_stretched-pulse_2020}, the fiber cavity is constructed by splicing different types of fiber with opposite dispersion. In~\cite{li_real-time_2020}, dispersion management was achieved by adiabatically varying the width of the microcavity waveguide. Unlike conventional anomalous-dispersion solitons, which have hyperbolic secant profiles, dispersion-managed cavity solitons exhibit Gaussian profiles both in the time domain and the spectral domain. This characteristic enables the generation of frequency combs with better spectral flatness than conventional solitons.
\end{itemize}

\section{Experimental  techniques}
In this section, we highlight some of the key experimental questions that must be considered to successfully generate and sustain CSs in macroscopic fiber cavities and in microresonators. As we outline below, each of these platforms presents it own set of distinct experimental challenges.

\subsection{Macroscopic fiber cavities}

Macroscopic fiber cavities typically range from one to hundreds of meters in length~\cite{leo_temporal_2010,xu_frequency_2021}. Smaller fiber cavities, approximately centimeters in length or less, have also been reported, but their properties more closely resemble those of the microresonator platforms discussed in the next section~\cite{obrzud_temporal_2017-1, li_experimental_2023}. Macroscopic optical fiber cavities can be constructed using either a ring or Fabry-Perot geometry. Fiber rings can be easily fabricated by splicing a length of fiber between two ports of a fused fiber coupler. Fabry-Perot cavities, on the other hand, require dielectric mirrors at both ends of a length of fiber. These mirrors can be either butt-coupled~\cite{li_ultrashort_2024} or directly deposited onto the fiber ends~\cite{obrzud_microphotonic_2019}. Despite the fact that optical fiber is an extremely low-loss optical waveguide, macroscopic fiber cavities typically only exhibit modest finesse in the range of $\sim 50$--$500$ due to the relatively large excess loss introduced at the fiber coupler or the mirror-fiber interface.

Despite this modest finesse, the long length of macroscopic cavities can result in very narrow cavity linewidths (hence very large quality factors~$Q$). Considering, for illustrative purposes, a $L=100$~m fiber ring cavity with a finesse~$\mathcal{F}$ of 100, we find the cavity resonance width [cf.~Eq.~\eqref{eq:finesse}] to be only $\Delta f = \mathrm{FSR}/\mathcal{F} = c/(nL\mathcal{F}) \sim 20~$kHz. At 1550~nm wavelength, this corresponds to a Q~factor of $10^{10}$ [cf.\ Eq.~\eqref{eq:Qfactor}]. For this reason, the linewidth of the driving laser used must be carefully considered. Fiber DFB lasers, which typically exhibit sub-kHz linewidths, are often chosen to drive Kerr fiber cavities for this reason~\cite{coen_continuous-wave_2001, leo_temporal_2010}. Additionally, the resonance frequencies of a macroscopic fiber cavities are highly sensitive to environmental perturbations, since both the length and the refractive index of an optical fiber change with temperature. For silica fibers, the change in refractive index [$dn/dT \sim 10^{-5}~^{\circ}\mathrm{C}^{-1}$] is approximately 20~times larger than the relative change in length [$(dL/dT)/L\sim 0.5\times10^{-6}~^{\circ}\mathrm{C}^{-1}]$, and so is the dominant thermal effect~\cite{haynes_crc_2014}. The change in cavity resonance frequency for a change of temperature~$\Delta T$ can thus be expressed as, 
\begin{equation}
    \label{eq:Temperature_stability}
    \frac{\Delta\omega_0'}{2\pi} \sim -\frac{mc}{n^2L}\frac{dn}{dT}\Delta T\,,
\end{equation}
where~$m$ is the integer order of the cavity resonance ($\omega_0'=2\pi\, m\, \mathrm{FSR}$). Considering the same optical fiber ring as above, driven at a pump wavelength of 1550~nm (for which $m=10^8$), we find that a change in temperature of only $\Delta T\sim 15\ \mu\mathrm{K}$ is required to shift the cavity resonance frequency by a full cavity linewidth ($20$~kHz). Consequently, active cavity stabilization is invariably required in macroscopic fiber cavities to maintain the drive laser at a fixed detuning relative to the resonances of the cavity.

\subsubsection{Active stabilisation}

When considering the active stabilization of a macroscopic fiber cavity, two options immediately present themselves. Either the laser frequency can be controlled to follow the shifting cavity resonances~\cite{jang_ultraweak_2013}, or a dynamic fiber stretcher can be added to the cavity to lock the cavity mode to the frequency of the input laser~\cite{coen_continuous-wave_2001, leo_temporal_2010}. Both techniques have been successfully reported in experiments, though the former is typically easier to implement as commercially available  fiber DFB lasers offer continuous tunability over $\sim$ GHz frequency ranges through the addition of integrated piezoelectric stretchers.

Instead of directly locking the drive laser to a resonance of the fiber cavity, additional flexibility can be achieved by introducing a second weak control beam to the cavity. This beam, derived from the same laser as the pump, and frequency-shifted using an acousto-optic modulator (AOM), is then locked to a cavity resonance via standard side-locking or Pound-Drever-Hall (PDH) techniques~\cite{li_ultrashort_2024}. Once the control beam is stably locked, the detuning of the pump laser can be simply and accurately controlled by adjusting the RF drive frequency of the AOM. This method allows the pump laser's detuning to be stably tuned over a full FSR of the cavity. To avoid interference between the drive and control fields, the control laser can be operated with an orthogonal polarization to the pump, or injected in the counter-propagating direction in a fiber ring.

\subsubsection{Suppression of stimulated Brillouin scattering}

In macroscopic fiber rings, it is also necessary to suppress the competing nonlinear effect of stimulated Brillouin scattering (SBS). SBS arises in optical fibers when a strong cw pump induces a co-propagating acoustic wave through electrostriction. This acoustic wave in turn modulates the fiber’s refractive index, creating a moving Bragg grating that causes the incident pump to both frequency shift and backscatter, resulting in a counter-propagating red-shifted Stokes wave~\cite{agrawal_nonlinear_2013}. In a step-index silica fiber, the frequency shift of the counter-propagating SBS field from the pump frequency is $\sim -10$~GHz. As a result of this stimulated scattering process, once the Brillouin threshold is exceeded, the forward-propagating pump power saturates, preventing the observation of weaker Kerr nonlinear effects. The situation is further exacerbated in a fiber cavity where the backscattered SBS field can also resonate, causing the cavity to function as a Brillouin laser~\cite{stokes_all-fiber_1982,braje_brillouin-enhanced_2009}. The ratio of the threshold powers for parametric oscillation (MI) and Brillouin lasing can be written as,
\begin{equation}
    \label{eq:Threshold_power_ratio}
    \frac{P_{\mathrm{T,parametric}}}{P_{\mathrm{T,Brillouin}}}= \frac{1}{2\gamma}\frac{g_B}{A_{\mathrm{eff}}}\,,
\end{equation}
where $g_B/A_{\mathrm{eff}}$ is the Brillouin gain coefficient of the fiber. For a standard SMF-28 single mode fiber operating at a wavelength of 1550 nm, the Brillouin gain coefficient is typically $g_B/A_{\mathrm{eff}} \sim 0.15~\mathrm{W^{-1}\,m^{-1}}$, while the Kerr nonlinear coefficient $\gamma \approx 2.5 \times 10^{-3}\ \mathrm{W^{-1}\,m^{-1}}$. Equation~\eqref{eq:Threshold_power_ratio} thus predicts the driving power threshold for SBS in a fiber ring to be 30-times lower than that of parametric oscillation~\cite{jang_strong_2012}. This implies that additional Brillouin suppression techniques are required to enable the operation of a macroscopic fiber ring as a pure Kerr cavity. Below, we outline some common methods used to suppress SBS in optical fiber rings.

\begin{itemize}

\item \textbf{Intracavity isolation} --- Perhaps the simplest method to suppress the backscattered Brillouin field is to add an optical isolator to the fiber loop. With isolation factors in excess of $\sim 50$~dB readily achievable, this provides an extremely efficient method of SBS suppression and allows the operation of cw driven fiber rings with negligible Brillouin response~\cite{coen_continuous-wave_2001,leo_temporal_2010}. The main downside to this technique is the large intracavity loss imposed by the isolator. This typically restricts the finesse of such cavities to $\sim 30$--$60$, which necessitates the use of high drive powers to achieve a sufficient driving strength to excite and maintain localized structures. We also note that, even with an isolator in place, forward-scattered acoustic waves~\cite{shelby_resolved_1985,shelby_guided_1985} are still present and can sometimes induce unexpected ultra-weak long-range interactions between CSs~\cite{erkintalo_bunching_2015}.

\item \textbf{Pump phase modulation} --- Phase modulation of the input pump field broadens its effective bandwidth and thus reduces SBS gain. It has long been used in single-pass fiber experiments to suppress Brillouin scattering~\cite{agrawal_nonlinear_2013}. In a fiber ring, the modulation frequencies applied must be matched to multiples of the cavity FSR in order to efficiently drive the cavity~\cite{coen_continuous-wave_2001}. Although pump phase modulation offers some level of SBS suppression in fiber rings, it is generally insufficient to prevent Brillouin lasing by itself and must be used in conjunction with other methods discussed in this Section.

\item \textbf{Pulsed driving} --- Synchronously driving a macroscopic fiber ring with either nanosecond or picosecond pulses provides very efficient SBS suppression~\cite{anderson_observations_2016, anderson_coexistence_2017, wang_universal_2017, luo_resonant_2015, xu_frequency_2021}. As with phase modulation, this stems from the increased bandwdith of the driving field. When using nanosecond driving pulses, the driving field can often be regarded as quasi-continuous wave (quasi-cw) in relation to the approximately picosecond-duration localized structures that typically arise. An additional advantage of nanosecond pulsed driving is that it enables experimenters to achieve significantly higher drive powers compared to cw driving. While picosecond driving pulses offer even stronger SBS suppression, the temporal duration of the pump pulse and the localized structures are often commensurate, such that their evolution cannot be considered independently. This leads to the more complex pulsed-driving dynamics discussed in Chapters~\ref{sec:CSanomalous} and~\ref{sec:Normal} with pulse desynchronization playing an important role~\cite{obrzud_microphotonic_2019, hendry_impact_2019, xu_frequency_2021}.

\item \textbf{Short cavities} --- There exists an intermediate length scale between macroscopic fiber resonators and optical microresonators where passive SBS suppression is possible through judicious choice of the resonator length. The gain bandwidth of SBS in a step-index silica optical fiber is approximately 100~MHz. When the FSR of the fiber cavity exceeds this bandwidth, it becomes possible to select a fiber length such that the SBS field is no longer resonant with the cavity. The shorter the cavity length, the larger the FSR, and thus the more effective the suppression. Experiments performed in short-length optical fiber rings show that for $L \sim 30$~cm over 30~dB of SBS suppression is possible~\cite{jang_strong_2012}. We also note that, in these short cavities, SBS can also sometimes be harnessed to improve the optical characteristics of the output soliton comb. For example, in Ref.~\cite{jia_photonic_2020}, a 10 cm fiber Fabry-Perot cavity was reported in which an external pump produced an internal pump field with significantly improved spectral linewidth through SBS. This internal pump was then used to drive the formation of a very low-noise soliton comb.

\end{itemize}

\subsubsection{A generic macroscopic fiber ring experiment}

Finally, we conclude this Section by presenting a schematic diagram of an experiment designed to generate localized dissipative structures in a macroscopic fiber ring. We note that this is included as an illustrative example only, designed to highlight how the various parts of such an experiment fit together, and that many other variants have been reported in the literature. The schematic diagram, shown in Fig.~\ref{fig:C5_generic_fiber_ring}, highlights the three key sections of such an experiment: the Kerr cavity, the driving laser section, and the active stabilization of the cavity detuning.

The Kerr cavity, implemented here as a fiber ring, contains both an input coupler to inject the driving field and a drop-port coupler to directly sample the intracavity field. Both the driving field and the probe field used for active stabilization are derived from the same laser. In this example, 95\,\% of the laser output is directed to the driving field path. Here, this light passes through an amplitude modulator to carve nanosecond pulses into the drive field at a repetition rate synchronous with that of the fiber cavity, ensuring effective SBS suppression. The output of the modulator is then amplified and filtered to remove unwanted amplified spontaneous emission before entering the cavity at the input coupler and circulating in an anti-clockwise direction. A polarization controller is included before the input coupler to align the pump with one of the two principal axes of the cavity.

The weak probe field is sent through an acousto-optic modulator (AOM) to provide a controllable frequency offset from the drive field. It then passes through a phase modulator that forms part of a Pound-Drever-Hall (PDH) locking circuit~\cite{black_introduction_2001}, before entering the fiber ring at the input coupler and circulating in the opposite (clockwise) direction to the drive field. When the PDH locking is activated, the probe frequency is locked to the peak of a low-power linear resonance of the cavity. The drive field can then be independently adjusted to any desired cavity detuning via the RF frequency applied to the AOM.

\begin{figure*}[!t]
	\centering
	\includegraphics[width = 0.9\textwidth, clip=true]{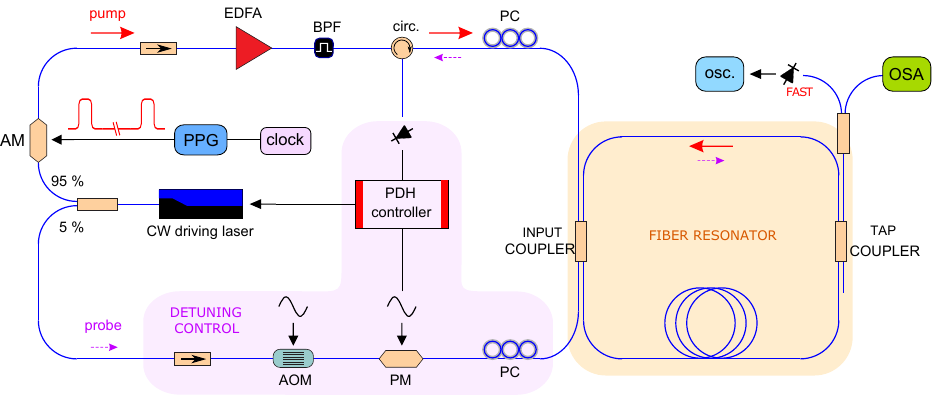}
	\caption{Schematic diagram of a generic experimental realization of a coherently driven Kerr fiber ring experiment. Components: AM -- amplitude modulator, AOM -- acousto optic modulator, BPF -- optical band-pass filter, circ -- optical circulator, EDFA -- Erbium doped fiber amplifier, osc -- oscilloscope, OSA -- optical spectrum analyzer, PC -- polarisation controller, PPG -- electronic pulse pattern generator. Reprinted with permission from Z. Li \textit{et al.}, Optica. \textbf{7}, 1195-1203 (2020)~\cite{li_experimental_2020}. Copyright \textcopyright~2020~Optical Society of America.}
	\label{fig:C5_generic_fiber_ring}
\end{figure*}

\subsection{Optical microresonators}

The realization of localized dissipative structures in optical microresonators present a different set of experimental challenges. Though no clear definition exists, for the purposes of this discussion we use the term microresonator to include any optical cavity with a round-trip path of the order of centimeters or smaller. The large resultant FSRs ($\sim$ GHz to THz) mean that acoustic effects such as SBS are almost never resonant unless the cavity is explicitly designed for this purpose~\cite{grudinin2009brill,lin_nonlinear_2017}. The small footprint also means that optical microresonators are far less sensitive to environmental perturbations than macroscopic fiber rings. Rather, microresonators often exhibit a large thermal nonlinearity that brings with it a new set of experimental complications.

\subsubsection{Passive locking via thermal nonlinearity}

Considering a pump laser that is being tuned into a resonance, one observes an increase in the  intracavity optical power, and this is associated by an increase in the resonator’s temperature. Since the refractive index of the resonator depends on temperature, this temperature rise also shifts the resonance frequency one is trying to tune into. In an optical microresonator, with its small footprint and high finesse, this thermal effect can be quite pronounced and has important consequences which we outline below. 

A mathematical description of the thermal nonlinearity is presented in Section~\ref{sec:thermal} and we now study its experimental implications. We consider first a resonator material with a positive~$dn/dT$ coefficient. In this case, when slowly scanning the laser into resonance from blue to red (i.e., from negative detuning to positive detuning, as required for CS excitation), one observes the resonance position continuously shifting to longer wavelengths, i.e., running away from the pump laser, as the power in the resonator (and hence the temperature) increases [cf.\  Eq.~\eqref{eq:Temperature_stability}]. As a result, the recorded resonance profile appears, not as the expected Lorentzian profile of a `cold' cavity, but rather as a `thermal triangle' whose spectral width extends over a much wider range of optical frequencies~\cite{carmon_dynamical_2004}. This behavior is illustrated schematically in Fig.~\ref{fig:C5_thermal_response}. 

\begin{figure*}[!t]
	\centering
	\includegraphics[width = 0.55\textwidth, clip=true]{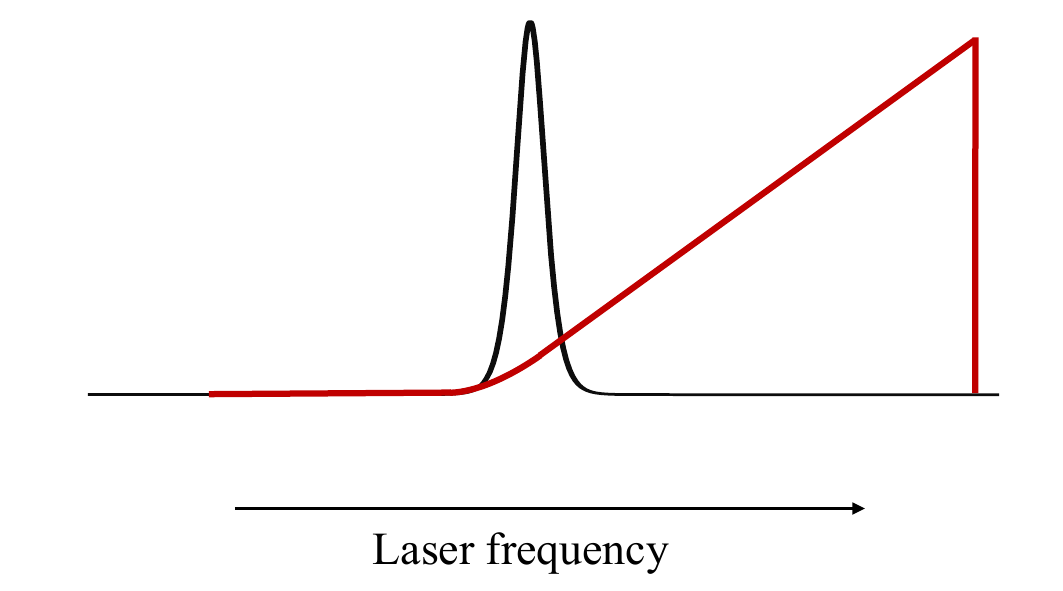}
	\caption{The Lorentzian resonance of a `cold' cavity (black trace), and the thermal triangle observed for a microresonator resonance with a positive $dn/dT$ coefficient (red trace).}
	\label{fig:C5_thermal_response}
\end{figure*}

One fortuitous consequence of this behavior is the ability to passively lock the detuning of the driving laser from the resonant mode. Consider a laser tuned to a frequency within the positive slope region of the thermal triangle. Here, if the laser frequency decreases, so too does the intracavity power and thus the resonance frequency. Conversely, if the laser frequency increases, the intracavity power and the resonance frequency both increase. Thus, the positive slope of the thermal triangle enables a passive thermal locking whereby the cavity mode's resonant frequency essentially tracks that of the driving laser~\cite{carmon_dynamical_2004}. Thanks to this passive lock, many microresonator experiments do not need to implement any active frequency stabilization. In contrast, in resonator materials with a negative $dn/dT$ (such as $\mathrm{CaF}_2$ crystals), the direction of the thermal triangle is reversed and it is not possible to stably scan into a resonance from blue to red as required for CS formation. This has considerably complicated attempts to observe CSs in such platforms~\cite{kobatake_thermal_2016}.

\subsubsection{Accessing cavity soliton states using an auxiliary field}
\label{Auxillary_pumping}

Despite the positive effect of thermal locking as discussed above, the generation of CSs out of chaotic MI patterns in microresonators typically leads to sudden changes in intracavity average power levels, with associated strong thermal shifts of resonances. These strong thermal nonlinearities can complicate attempts to scan directly into a CS state. A variety of techniques have been developed to address this challenge, including variable scan rates, power kicks, auxiliary pumps, and coupled cavities. The theoretical underpinnings of these methods have been discussed in detail in Section~\ref{sec:Thermalinstability}. In the following, we focus on the experimental considerations involved in implementing one of the most widely adopted approaches: the use of an auxiliary field for thermal compensation.

Recall that when attempting to access the CS regime by directly scanning the pump laser from negative (blue) detuning to a detuning that supports CSs, the intracavity field first traverses the nonlinear regimes of stable and then unstable modulation instability. Indeed, it is the unstable MI field that seeds CS formation. Unfortunately, the transition from unstable MI to CSs is often accompanied by a substantial drop in average intracavity power (c.f. Fig.\ref{fig:CSscan}). If uncompensated, this drop will lead to a rapid cooling of the cavity, on a time scale set by the thermal time constant of the resonator, and to an associated blue-shift in the cavity resonance. In microresonator systems, this shift is often large enough to drag the driven resonator mode completely out of resonance with the drive field, preventing direct access to the CS state via detuning scanning.

The use of an auxiliary laser to mitigate this thermal effect was independently proposed by three research groups~\cite{zhang_sub-milliwatt-level_2019,lu_deterministic_2019,niu_repetition_2018}. The principle is illustrated in Fig.~\ref{fig:C5_thermal_auxilary}. Initially, an auxiliary laser is tuned into a second resonance, typically from a different mode family than the soliton resonance, resulting in heating and subsequent red-shifting of both resonances (panel~a). The pump laser is then tuned into resonance. As the pump enters resonance, the increase in intracavity power further red-shifts both resonances, reducing the detuning of the auxiliary laser (panel~b). Finally, continuing to detune the pump across the MI-CS boundary threshold results in a sudden drop in intracavity power. This blue-shifts both resonances, however, the blue-shift is immediately counteracted by a red-shift as the auxiliary pump is forced back into resonance (panel~c). When the power and initial detuning of the auxiliary laser are correctly set, it is thus possible to traverse the MI-CS boundary with only a minimal change in the total intracavity power. This significantly reduces the induced thermal shift and can allow direct access to the CS state through manual tuning of the pump laser.

\begin{figure*}[!t]
	\centering
	\includegraphics[width = 0.6\textwidth, clip=true]{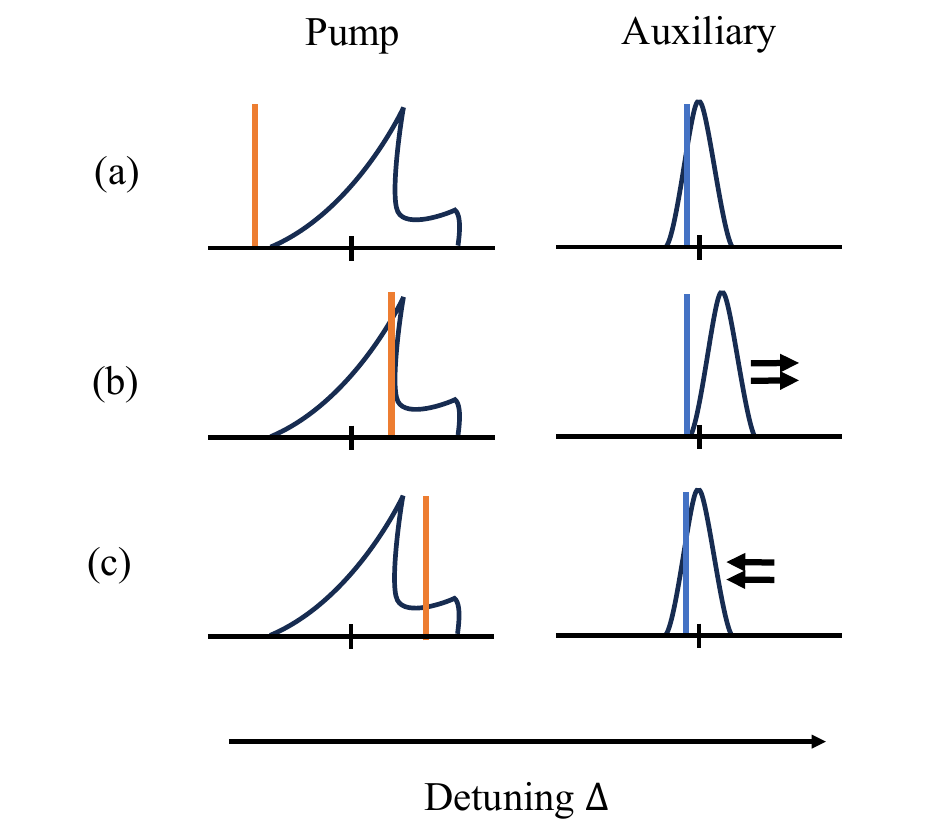}
	\caption{Illustration of thermal compensation procedure: (a) pump laser blue detuned from CS resonance, auxiliary laser tuned into auxiliary resonance, (b) pump laser detuned to peak of unstable MI regime, auxiliary laser forced off resonance by thermal shift. (c) Drop in intracavity power as pump detuning reaches CS regime offset as auxiliary laser forced back into resonance. Adapted with permission from P. Qureshi, ``Multimode Microresonator Optical Frequency Combs,'' PhD thesis, University of Auckland (2024)~\cite{qureshi_multimode_2024}. Copyright \textcopyright~2024 P. Qureshi.}
	\label{fig:C5_thermal_auxilary}
\end{figure*}

The experimental implementation of this auxiliary technique requires careful optimization of the auxiliary field. Ideally, to avoid any corruption of the soliton comb, there should be no interaction between the soliton and auxiliary fields other than through their associated thermal shifts. In microrings, strong isolation between the two fields can be achieved by injecting the auxiliary laser in the counter-propagating direction to that of the soliton pump. Similarly, operating the auxiliary field in a different mode family, or at a frequency far-detuned from the pump, has been demonstrated with good effect. In Refs.~\cite{wildi_thermally_2019,nishimoto_thermal_2022}, a thermal auxiliary field derived from a frequency-shifted copy of the pump laser has also been successfully demonstrated. This eliminates the need for an additional independent laser for the auxiliary field, with the frequency-shifted field actually driving the same mode as the pump. 

Optimizing the power and initial detuning of the auxiliary laser is key for the successful implementation of this technique. This can be achieved by repeatedly scanning the pump laser across the soliton resonance while maintaining the auxiliary laser at a fixed frequency. An example of this is shown in Fig.~\ref{fig:C5_Experimental_compensation} for a soliton resonance in a monolithic MgF$_2$ crystalline resonator. When the auxiliary laser is out of resonance, as shown in Fig.~\ref{fig:C5_Experimental_compensation}(a), a thermally elongated MI stage followed by a short soliton step is observed. The power and detuning of the auxiliary laser are then adjusted so as to extend the frequency range of the soliton step as far as possible. Notice, as shown in Fig.~\ref{fig:C5_Experimental_compensation}(b), that this also compresses the MI stage as the auxiliary laser also compensates the pump resonance’s thermal triangle. Once fully optimized, the pump laser scanning can be halted, and the pump manually tuned into the soliton resonance. A more detailed, step-by-step procedure to optimize the thermal compensation of the auxiliary laser is provided in the supplementary material of Ref.~\cite{zhang_spectral_2020}.

\begin{figure*}[!t]
	\centering
	\includegraphics[width = 0.6\textwidth, clip=true]{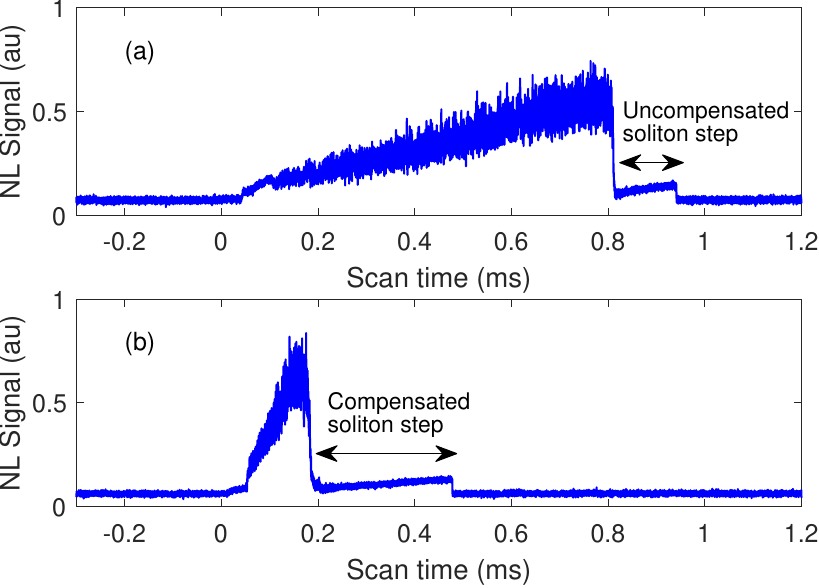}
	\caption{Measured nonlinear signal from a 56 GHz FSR MgF$_2$ resonator as the pump frequency is scanned across the soliton resonance (scan rate $\sim 20$~Hz). The nonlinear signal is obtained by removing the pump frequency from the microresonator output using a band-stop filter, then measuring the remaining comb signal on a slow-photodiode (bandwidth $\sim 10$~MHz). Trace (a) is measured with the auxiliary laser off-resonance, trace (b) is measured with the auxiliary detuning optimised for maximum thermal compensation. Adapted with permission from A. Su, ``Experimental Observations of Cavity Solitons in a Magnesium Fluoride Microdisk,'' MSc thesis, University of Auckland (2020)~\cite{su2020cavitysoliton}. Copyright \textcopyright~2020 A. Su.}
	\label{fig:C5_Experimental_compensation}
\end{figure*}

\subsection{Experimental characterization of localized dissipative structures}

In this section we outline a range of experimental instruments and methods that have been used to identify and characterize localized dissipative structures (LDS) in both macroscopic fiber rings and optical microresonators.

\begin{itemize}

\item \textbf{Optical spectrum analysis} --- Direct measurement of the optical spectrum output from a resonator is a key tool for the characterization of Kerr frequency combs. For instance, as shown in Fig.~\ref{fig:C5_Combined_measurement}(a), a single CS propagating in the absence of higher-order dispersion, can be readily identified by its clean sech$^2$ spectral profile. Similarly, as discussed in section~\ref{sec:CSbinding}, the presence (and temporal separation) of multiple CSs can be inferred from the interference patterns observed in the output spectrum as shown in Fig.~\ref{fig:C5_Combined_measurement}(b). Both standard grating-based optical spectrum analyzers and Fourier transform spectrometers are suitable for measuring Kerr resonator spectra. Note that the acquisition time of these instruments is generally far slower than the rapid nonlinear dynamics characteristic of nonstationary Kerr states (e.g. the unstable modulation‑instability regime). As a result, the recorded spectra appear artificially smoothed, and the underlying system dynamics must be inferred using alternative diagnostic techniques.

\item \textbf{Electronic spectrum analysis} --- The photocurrent generated by direct photodetection of the optical output from a Kerr resonator can be analyzed using an RF electronic spectrum analyzer (ESA) to gain insights into the stability and coherence of the output comb. The ESA profile of a stable localized dissipative structure is characterized by a series of sharp spectral peaks at multiples of the cavity’s FSR upon a low-noise background. If the LDS undergoes periodic breathing, additional sharp spectral peaks will appear at lower frequencies. Conversely, an incoherent modulation instability (MI) state will exhibit low-frequency broadband RF noise. Notably, when attempting to excite a cavity soliton (CS) by scanning the cavity detuning, the sharp transition of the ESA spectrum from a broadband noise trace to a low-noise signal is an important experimental signature that the resonator has enter a CS state. An example of this behavior can be seen in Fig.~\ref{fig:C5_Combined_measurement}(c)--(h) which shows the experimentally measured optical and RF spectra obtained from a 65~GHz FSR MgF$_2$ resonator as it is directly scanned into a CS state.

\item \textbf{Low speed photodiode measurements} --- The use of low-speed photodiodes to measure the output of a Kerr cavity also provides considerable information about the nonlinear state of the system. If the bandwidth of the photodiode is set to be much faster than the pump scan rate, but much lower than the cavity FSR, the output photocurrent provides a reliable measure of the average intracavity power of the resonator. The addition of an optical bandpass filter to block the pump and transmit a portion of the newly generated comb frequencies will sometimes provide a clearer picture of the nonlinear evolution of the cavity. For example, in Fig.~\ref{fig:C5_Combined_measurement}(i) and~(j), we plot the received photocurrent output from a crystalline MgF$_2$ resonator with an FSR of $\sim56$~GHz. The bandwidth of the photodiode is 10~MHz, whilst the pump laser is scanned across the resonance at a rate of $\sim20$~Hz. The received signal is thus a faithful representation of the average intracavity power. Figure~\ref{fig:C5_Combined_measurement}(i) shows the photocurrent detected when the total optical field exiting the cavity is measured. This shows the presence of a clean soliton step at the end of the resonance and implies that, with the application of suitable thermal compensation, direct access to stable soliton states should be possible. Figure~\ref{fig:C5_Combined_measurement}(j) shows the same signal, but measured after a bandpass filter that transmits only the high-frequency components of the output comb, while fully blocking the pump frequency. On this trace, it is possible to discern not only the soliton step (shaded in red), but also the transition between the stable MI (green shading) and unstable MI (blue shading) regions.

\begin{figure*}[!t]
	\centering
	\includegraphics[width = 0.9\textwidth, clip=true]{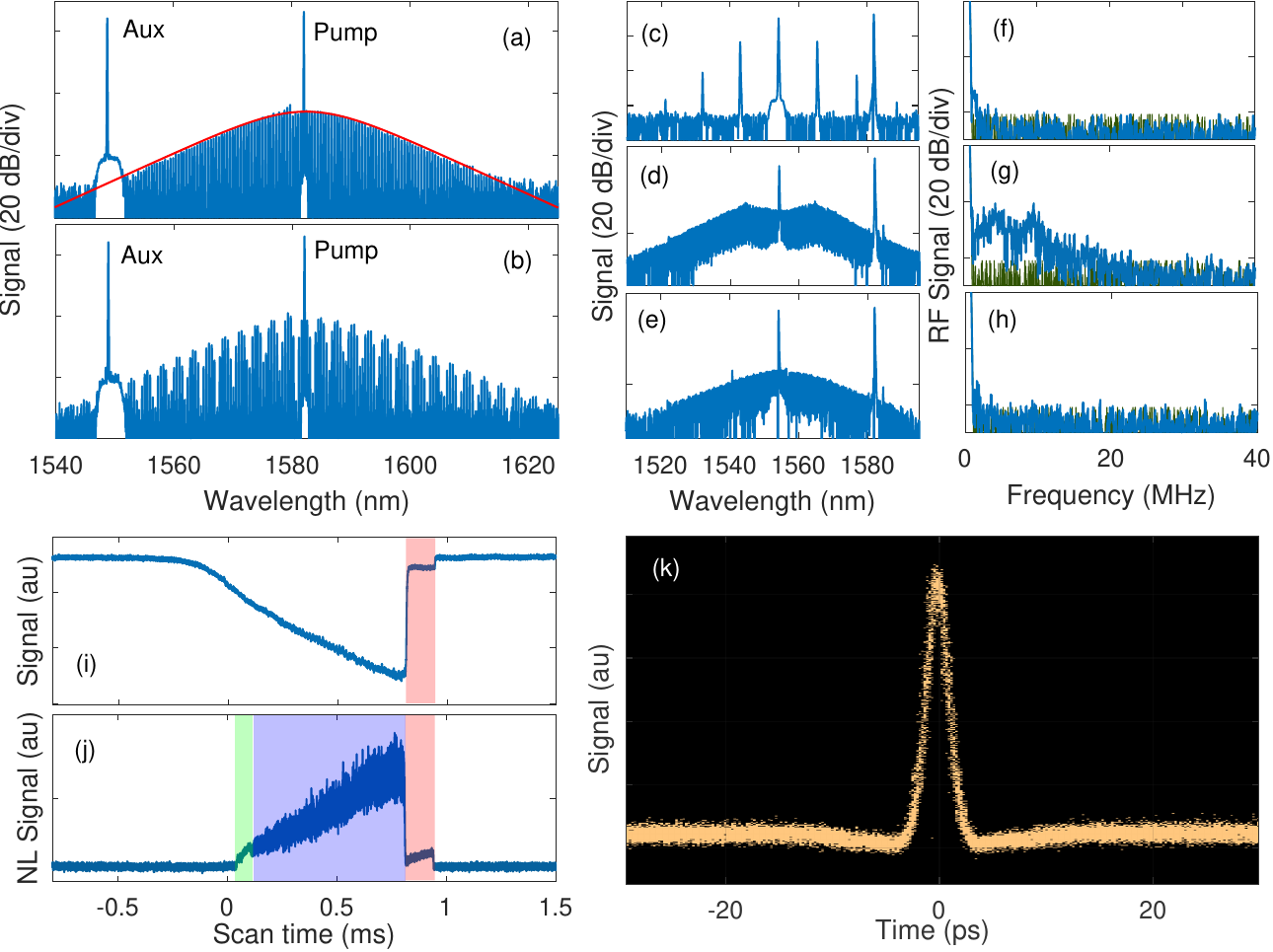}
	\caption{Illustrative experimental measurements. (a) Measured optical spectrum of a single CS generated in a $56$~GHz FSR MgF$_2$ resonator. The red trace shows the excellent agreement of a superimposed sech$^2$ fit to this spectrum. (b) A two-CS crystal state in the same resonator. The measured spectral interference period implies a temporal separation of $3.06$~ps between the two CSs. (c)--(e) Measured optical spectra from the same resonator at detunings showing stable MI, unstable MI, and a single CS state, respectively. (f)--(h) Corresponding RF spectra. The large low-frequency noise of the unstable MI state, and the low-noise of the CS state are clearly visible. (i) Slow photodiode trace as the pump laser is scanned across the soliton resonance of this cavity. (j) The same trace recorded after the addition of a high-pass optical filter to block the pump frequency. (k) Temporal intensity profile of CS in a 100~m optical fiber cavity measured using a picosecond resolution optical sampling scope. Panels (a--j) adapted with permission from A. Su, ``Experimental Observations of Cavity Solitons in a Magnesium Fluoride Microdisk,'' MSc thesis, University of Auckland (2020)~\cite{su2020cavitysoliton}. Copyright \textcopyright~2020 A. Su. Panel (k) adapted with permission from J.~K.~Jang \textit{et al.}, Opt. Lett. \textbf{41}, 4526-4529 (2016)~\cite{jang_all-optical_2016}. Copyright \textcopyright~2016~Optical Society of America.}
	\label{fig:C5_Combined_measurement}
\end{figure*}

\item \textbf{Realtime oscilloscopes} --- The timescales of macroscopic fiber cavities are such that it is possible to directly record the round trip-by-round trip evolution of the intracavity field using a fast-photodiode and a high-sample-rate realtime oscilloscope. Even for a fiber cavity, this measurement is typically not able to resolve the detailed picosecond time-scale structures of typical LDS in a fiber ring. However, a high-speed realtime measurement is still capable of providing useful insight into the dynamics of the cavity. For example, Fig.~\ref{fig:C4_SWexpt_drift} shows the realtime measurement evolution of SW fronts as a function of cavity round trip. This measurement used a 12.5 GHz bandwidth photodiode and a 40 GS/s realtime oscilloscope yielding a temporal resolution of $\sim 30$~ps. Although this resolution is insufficient to resolve the $\sim$~ps duration structure of individual SWs, important properties, such as their drift velocities, can be directly measured from such a recording~\cite{garbin_experimental_2017}.

\item \textbf{Optical sampling oscilloscope} --- In macroscopic cavities, it is possible to obtain direct picosecond resolution measurements of the temporal profiles of CSs using commercially available optical sampling oscilloscopes. Figure~\ref{fig:C5_Combined_measurement}(k) shows the measured temporal intensity profile of a cavity soliton (CS) circulating in a 100-meter macroscopic fiber ring~\cite{jang_all-optical_2016}. The characteristic CS temporal profile, featuring a sech-shaped peak riding on top of a continuous wave background with dips on either side of the central peak, can be clearly observed here.

\item \textbf{Electro-optic (EO) comb sampling} --- In optical microresonators, a dual comb measurement can be employed to recover the high-speed dynamics of the intracavity field. By interfering the cavity output with an external electro-optic (EO) comb, an interferogram is produced that can be processed to reveal the evolution of the temporal profile of the intracavity field~\cite{yi_imaging_2018}. The EO comb used typically has a smaller bandwidth than the Kerr comb under investigation, so the detailed temporal profile of the LDS under investigation cannot be resolved. However, similar to the real-time oscilloscope measurement described above, this measurement still provides highly useful insights into the real-time dynamics of the cavity.

\item \textbf{Electro-optic downconversion} --- The FSR of a typical microresonator can easily exceed $100$~GHz, placing it well beyond the range of direct electronic detection. In such cases, external electro‑optic modulators can be used to imprint additional EO sidebands onto the original microcomb, enabling the recovery of a down‑converted representation of the comb’s FSR. An example of this measurement approach is presented in Ref.~\cite{del2012hybrid}.

\item \textbf{Measurement of spectral phase} --- Significant additional insights into the behavior and properties of LDS in Kerr cavities can be obtained by measuring both the amplitude and phase of the output field. The well-known nonlinear technique of frequency-resolved optical gating (FROG) has been employed to make such measurements in both macroscopic fiber rings and microresonators~\cite{jang_ultraweak_2013,herr_temporal_2014}. In both cases, the low peak power of the cavity output complicates the measurement, often necessitating external optical amplification. In this case, care must be taken to ensure the additional amplification stage does not distort the measurement. Linear techniques to measure the full spectral field have also been reported~\cite{twayana_differential_2022}. One such technique, known as the stepped-heterodyne measurement, involves stepping a local oscillator between successive adjacent comb lines and measuring their phase difference~\cite{reid_stepped-heterodyne_2010}. This allows for a full recovery of the spectral and temporal field of the output field. As a linear technique, such methods typically do not require additional amplification of the output comb signal.

\item \textbf{Vector network analysis} --- Conventional vector network analyzers (VNAs) have been deployed to measure the pump detuning in a Kerr cavity operating in a soliton state~\cite{guo_universal_2017}. The input pump is passed through a phase modulator driven by the VNA’s output (port 1). The cavity’s output is then measured with a high-speed photodiode and fed back into the VNA (port 2). The pump is tuned into a CS state using the techniques described earlier. Measuring the VNA's S21 parameter then reveals peaks corresponding to the detuning between the pump field and the linear cavity resonance (the C-peak) and between the soliton field and the nonlinear cavity resonance (the S-peak). Since this measurement is performed in-situ, it is immune to any thermal shifts of the cavity resonance and provides a direct measurement of the pump-cavity detuning~\cite{guo_universal_2017}.

\end{itemize}

\section{Numerical/theoretical techniques}

In this section, we outline the use of some of the key analytical and numerical methods required to analyze the properties and evolution of localized dissipative structures in Kerr cavities.

\subsection{Numerical simulation of the Ikeda map and LLE}

Direct numerical simulation of the partial differential equations (PDE) that underlie the Ikeda map and the Lugiato-Lefever equation can be easily carried out on a standard desktop computer. Below, we present the key steps required to successfully implement these simulations.

\subsubsection{Numerical simulation of the Ikeda map}

Turning our attention first to the numerical simulation of the Ikeda map, we note that this is explicitly a two-step algorithm requiring 1) the numerical integration of the GNLSE~\eqref{GNLSE}, to determine the field evolution over one round trip, and 2) the application of the input coupler boundary condition, Eq.~\eqref{map}. This procedure is then repeated to track the evolution of the field over the time scale of interest. The numerical integration of the GNLSE itself is an extremely well known problem in nonlinear optics. Numerous numerical methods have been proposed and successfully implemented, but by far the most popular approach is the split-step Fourier technique~\cite{agrawal_nonlinear_2013}. Here, the PDE is split into its nonlinear and dispersive parts, each of which is solved separately over a propagation step. Provided the step-size is sufficiently small, this method provides a very good approximation to the solution of the full PDE. In addition, the split-step method is very computationally efficient as the dispersive step is implemented directly in the frequency domain using the fast Fourier transform (FFT). Example GNLSE codes are readily available~\cite{dudley_supercontinuum_2006,redman_gnlse-python_2021}, and can be simply adapted for the numerical simulation of the Ikeda map. 
\subsubsection{Numerical simulation of the LLE}

As demonstrated in Section 2 of this Article, under a carefully defined set of approximations, the Ikeda map can be reformulated as a single mean-field propagation equation: the Lugiato-Lefever equation (LLE). The LLE is also amenable to numerical simulation via the split-step Fourier technique. Below, we outline the numerical solution of the LLE in more detail. To that end, we recast the LLE Eq.~\eqref{GLLE} or~\eqref{GLLEpulsed} into its dispersive, nonlinear, and driving components,
\begin{equation}
    \label{GLLEsplitstep}
    t_\mathrm{R} \frac{\partial E}{\partial t} = \left[ iL\hat{D}\left(i\frac{\partial}{\partial\tau}\right) + \hat{N}\right] E + \sqrt{\theta}\, E_\mathrm{in},
\end{equation}
where $\hat{D}$ is the linear dispersion operator, Eq.~\eqref{Dhat}, potentially also including desynchronization, and $\hat{N}$ is a nonlinear operator, 
\begin{equation}
    \hat{N} = -\alpha-i\delta_0 + i \gamma L \left[ (1-f_\mathrm{R})|E|^2 + f_\mathrm{R} h_\mathrm{R}(\tau')\ast|E|^2 \right].
    \label{eq:Nhat}
\end{equation}
Note that this formulation includes the linear loss and detuning in the nonlinear operator, but they can also be easily incorporated into the dispersive operator if desired, e.g., to include wavelength-dependent losses. These operators can also be considered in normalized units.

As is traditional with the split-step Fourier algorithm, over each (slow) time step, from $t_0$ to $t_0+dt$, the dispersive operator is considered separately from the other terms in the equation, and integrated in the Fourier domain as,
\begin{equation}
    \label{Dhat_solution}
    E(t_0+dt,\tau) = \mathrm{FT}^{-1} \left[ \exp\left( iL \frac{dt}{t_\mathrm{R}} \hat{D}(\Omega) \right)\, \mathrm{FT}\left[ E(t_0,\tau)\right] \right]
\end{equation}
where $\mathrm{FT}$ and $\mathrm{FT}^{-1}$ denote, respectively, the forward and inverse Fourier transforms and are implemented in practice using the fast Fourier transform (FFT) algorithm. Note, care must be taken to respect the sign conventions used in the respective Fourier operations. In this Article, as in most of nonlinear optics, the forward transform is defined with a positive exponent [Eq.~\eqref{FFT}]. This is opposite to the definition used by most common numerical implementations of the FFT. This issue, once identified, can be simply remedied by calling the numerical \texttt{IFFT()} command to implement the forward transform and the \texttt{FFT()} command to implement the inverse transform.

Likewise, the nonlinear operator, Eq.~\eqref{eq:Nhat}, can be integrated together with the driving term directly in the time domain. Because the driving term is inhomogeneous, the procedure is a bit different than for integration of the GNLSE, Eq.~\eqref{GNLSE}. We find
\begin{equation}
    \label{Nhat_solution}
    E(t_0+dt,\tau) = \left[ E(t_0,\tau) + \frac{\sqrt{\theta}\,E_\mathrm{in}}{\hat{N}} \right] \exp\left(\frac{dt}{t_\mathrm{R}}\hat{N}\right) - \frac{\sqrt{\theta}\,E_\mathrm{in}}{\hat{N}}.
\end{equation}

As with the NLSE, more complex multi-step numerical methods can be used to evaluate this nonlinear step. Such methods result in an improvement in simulation accuracy, allowing the use of large step sizes, but come at the expense of increased complexity. For more details, the interested reader is referred to~\cite{agrawal_nonlinear_2013}.

\subsubsection{Step size and temporal and spectral windows}

Accurate simulation of both the LLE and NLSE require carefully consideration of the simulation's temporal and spectral windows, as well as the numerical step size used. Specifically, the temporal profile of the intracavity field is stored in an array length $N$ that spans a fast-time duration~$\tau_\mathrm{span}$. This sets the spectral extent of the fast-time frequency window between $-(N/2)/\tau_\mathrm{span}$ to $[(N/2)-1]/\tau_\mathrm{span}$. For microresonators, it is often convenient to set the duration of the fast-time temporal window to $t_\mathrm{R}$, the resonator round-trip time. This choice has two immediate advantages. Firstly, it automatically encodes the periodic boundary condition of the temporal intracavity field, thus correctly accounting for any field components that overflow this window. Secondly, the array containing the spectral field is spaced by exact multiples of the cavity FSR, considerably simplifying the calculation of the output comb spectrum. In this situation, it is necessary only to adjust the number of points in the temporal array such that the spectral window contains sufficient bandwidth to fully contain the spectral field. Care must be taken to monitor the evolution of the spectral field during a simulation to ensure it does not overflow these limits. For longer cavities, it becomes impractical to store the temporal field in an array that spans a whole round trip. In this case, it is necessary to carefully select the spans of both the temporal and spectral windows such that neither the temporal nor spectral field overflow their limits. 

Accurate split-step simulations also require the use of a sufficiently small step size. A common criterion used for NLSE simulations is to set the step size to be much smaller than the natural nonlinear and dispersive lengths of the waveguide~\cite{agrawal_nonlinear_2013}. In practice, the step size of a simulation can be optimized by repeatedly running the same simulation at different step sizes and observing the point at which further reduction in step size can no longer be detected in the evolution of the intracavity field. In this context, a common signature of too large a step size is the appearance of spurious high-frequency components in the field's spectrum~\cite{severing_spurious_2023}. Example codes that implement the full numerical algorithm required to solve the LLE are readily available~\cite{moille_pylle_2019}.

\subsubsection{Choice of LLE or Ikeda map?}

The Ikeda map offers a more comprehensive description of the nonlinear evolution of a Kerr cavity, as it is not constrained by the approximations required to derive the mean field Lugiato-Lefever Equation. Nonlinear phenomena such as Kelly sidebands~\cite{nielsen_invited_2018}, period-2 modulation instability~\cite{bessin2019real}, and super cavity solitons~\cite{anderson_coexistence_2017} explicitly violate these approximations and can only be accurately modeled using an Ikeda map. However, an Ikeda simulation necessitates the explicit evaluation of the boundary condition, Eq.~\eqref{map}, at every round trip. This imposes a maximum slow-time step size equal to the cavity roundtrip time~$t_\mathrm{R}$, which can be computationally expensive, particularly when simulating high-finesse resonators, whose photon lifetime~$t_\mathrm{ph}$ can be much larger than~$t_\mathrm{R}$ [cf.\ Eq.~\eqref{eq:photon_lifetime}]. By contrast, the LLE is not subject to this restriction, allowing for step sizes much larger than~$t_\mathrm{R}$.
 
\subsubsection{Newton method}
\label{sec:Newton}

The numerical techniques described above explicitly simulate the temporal dynamics of the intracavity field over subsequent round trips, mimicking what actually occurs physically. This is useful to investigate soliton excitation protocols, breathing states, and other dynamical behaviors. However, if one is simply interested in steady state solutions, these only emerge in dynamical simulations after all transients have died off, which can sometimes take an appreciable amount of computation time. Some solutions (including unstable ones) might also be difficult to obtain in this way. Instead, it can be much more efficient to directly seek steady state solutions of the underlying equations using a Newton solver. Such a solver is more directly applicable to mean-field-type models (scalar or coupled) [Eqs.~\eqref{GLLE}, \eqref{GLLE2}, \eqref{GLLEpulsed}, or~\eqref{Eq:SSB2}], which is what we describe in more detail below, but can in principle also be applied to the Ikeda map with some extra effort (e.g., using quasi-Newton methods, like secant or Broyden's method \cite{press_numerical_2007, wang_universal_2017}). 

\paragraph{General considerations.}

In a generic form, mean field models are partial differential equations that can be written as $\partial E(t,\tau)/\partial t = f(E)$. Note that our notation here is meant to implicitly include multi-modal or multi-cavity configurations, described by systems of coupled partial differential equations. In these cases, $f$~would be a multi-dimensional function while $E$~would be a vector with several components. Steady-state solutions are independent of slow-time~$t$, for which the equation reduces to $f\bigl( E(\tau) \bigr)=0$, and can thus be found by looking for the roots of the function~$f$. By discretizing the equation over a fast-time grid~$\tau_i$ [leading to $N$~equations for the $N$~complex 
unknowns $E(\tau_i)$], this problem can be tackled by generalized (multi-dimensional) root finding algorithms, and the Newton method (also known as the relaxation method) is particularly suited to this task \cite{press_numerical_2007}. 

Before proceeding, it is important to note that the Newton method, in its simplest form, can only converge to isolated roots. cw-driven Kerr resonators however typically exhibit translational invariance, which implies that every steady-state localized structure solution is associated with a continuum of infinitely many other solutions, simply translated along the fast-time axis with respect to each other. This prevents the Newton method to converge. To make the problem well-posed, it must be augmented with a pinning condition that will force the sought structure to be uniquely positioned in fast-time. Such pinning conditions can in essence be chosen arbitrarily and will typically differ for different types of localized structures. For bright CSs, a simple condition is to impose that the derivative of the intensity~$|E(\tau)|^2$ (or, for multi-modal problems, the intensity of one of the component) with respect to fast-time~$\tau$ be zero in the middle of the simulated temporal window. This forces the peak of the structure to be located at that position. For front-like solutions, such as SWs or domain walls \cite{coen_convection_1999, garbin_experimental_2017, garbin_dissipative_2021}, one can impose instead that the intensity in the center be half-way between the intensities at the two boundaries. The introduction of a pinning condition, however, makes the system of equations overdetermined: we now have $N+1$ equations for $N$~unknowns, which brings us to the next problem. 

On top of the pinning condition, it is important to consider that even localized structures that are ``stationary,'' in the sense of maintaining their shape and amplitude during propagation, will often drift along the numerical fast-time axis over subsequent round trips, due to their group velocity being different from that of the driving. This can be caused, e.g., by DW emission and spectral recoil due to high-order dispersion or through stimulated Raman effects as was already mentioned in Sections~\ref{sec:DW_and_recoil} and~\ref{sec:Raman_self_freq_shift}. SW solutions also drift for most parameters, except at the Maxwell point (see Section~\ref{sec:SW}). To take this into account, a drift term [a first-order fast-time derivative of the field with coefficient~$d_1$, see, e.g., Eq.~\eqref{GLLEN}] must be included into the partial differential equation to compensate for the natural drift velocity of the sought structure. In essence, this amounts to introduce an artifical desynchronization such that the structure remains stationary, i.e., does not move along the fast-time axis. The coefficient~$d_1$ is treated as an additional unknown variable, which together with the pinning condition makes the system of equations to solve consistent ($N+1$~equations for $N+1$~unknowns). The drift velocity of the solutions, which are sometimes difficult to evaluate accurately with direct integration, is therefore provided for free by the Newton solver, which can be seen as an advantage of the method. Note that in the case of pulsed driving, no pinning condition and no extra drift term are necessary when using the Newton method because the driving pulse will effectively pin the solution along the fast-time axis. A drift term may still be included in the equation to model actual desynchronization effects, but the corresponding coefficient would be set as an extra parameter, rather than being treated as an unknown.

\paragraph{Iterative step.}

To avoid a number of pitfalls, the Newton method is better implemented based on real unknowns. Each (complex) sample of the electric field  $E(\tau_i)$ must therefore be split into a pair, $\mathrm{Re}(E(\tau_i)), \mathrm{Im}(E(\tau_i))$. In a practical implementation, all these unknowns can be packed into a single (real valued) column vector, including the drift coefficient if relevant. To simplify the notation, below we still refer to this vector as $E$, i.e., $E=\bigl[ (\mathrm{Re}(E(\tau_i)), \mathrm{Im}(E(\tau_i)) )\ [i=1\ldots N], d_1 \bigr]^T$ (for a total of up to $2N+1$ elements). Similarly, the function $f(E)$ must be split into its real and imaginary parts, and for consistency we assume that it includes the pinning condition as an additional component. Thus, below, $f(E)$ represents a $2N+1$-dimensional set of real-valued functions in the $2N+1$ real components of~$E$.

Starting from an initial guess~$E_0$ for the vector of unknowns (the choice of which will be discussed below), the Newton method advances towards an (ideally) improved approximate~$E_1$ of the root of~$f$ by applying the well known formula,
\begin{equation}
    \label{eq:newtonstep}
    E_1 = E_0 - \lambda \left.J\right|_{E_0}^{-1}\,f\left( E_0 \right) \qquad \text{where} \qquad
    \left.J\right|_{E_0} = \left. \frac{\partial f}{\partial E} \right|_{E_0}\,.
\end{equation}
Here $J$ is the Jacobian matrix of the (discretized) problem ($2N+1$ by $2N+1$) taken with respect to all the (real) unknowns and evaluated for~$E_0$. If $E_1$ is not a sufficiently good approximation of the solution [i.e., if $f(E_1)$ is not sufficiently close to zero], the above formula is then iterated, with $E_1 \rightarrow E_0$. Note that the above equations constitute a direct generalization to higher dimensions of the famous one-dimensional Newton-Raphson algorithm (for which $\lambda=1$) that evaluates successive x-intercepts of the tangent to the function~$f$ to approach a root.

In high dimensional implementations of the Newton algorithm, if the guess~$E_0$ is not sufficiently close to the solution, a full Newton step ($\lambda=1$) of the iteration above, Eq.~\eqref{eq:newtonstep}, may not necessarily lead to an improved solution. In fact, it may take it wildly away from the actual solution. In that case, one may start with a reduced value of~$\lambda$ or, better, implement a globally convergent Newton method~\cite{press_numerical_2007}, which will guarantee some progress towards the solution at each step, by adapting the step size~$\lambda$ automatically. The optimal value of~$\lambda$ will typically progressively increase towards the optimal theoretical value of~1, after which quick final quadratic convergence is then observed. Progress may still become stalled however, which can be detected by observing ever decreasing adaptive values of $\lambda$. This should be monitored and remedied by starting from a different guess solution~$E_0$.

It is also important to realize that the Jacobian matrix~$J$ does not need to be actually inverted to implement the iterative step of Eq.~\eqref{eq:newtonstep}. Rather, one can solve the linear system $J x = f(E_0)$ for $x$ using standard linear algebra routines available in most numerical packages, and replace the iterate with $E_1 = E_0 - \lambda x$. This provides much improved speed and accuracy. Note that $f(E_0)$ has typically been calculated already to test the quality of the previous step, so should not have to be re-computed again.

\paragraph{Choice of the initial guess.} The choice of the initial guess~$E_0$ is critical to ensure fast convergence of the Newton method, or in fact to ensure convergence at all. Unfortunately, there is no simple rule to select a good initial guess, and trial and error may be inevitable. When seeking cavity soliton solutions, a good starting choice is to use the approximate analytical estimate provided by Eq.~\eqref{eq:CSsol} (or a solution of the same form with slightly different parameters), with the lower state background obtained by solving Eqs.~\eqref{eq:normss}--\eqref{eq:cwcubic_norm}. For SWs or domain wall solutions, one can use a $\tanh$ connection of appropriate width between the two adjacent homogeneous states. In the case of calculations involving the generalized LLE with several extra terms (high order dispersion, stimulated Raman scattering, ...), the solution may be sufficiently distorted that these simple analytical estimates may not suffice to ensure convergence. In that case, the result of a split-step Fourier simulation may be used as a starting point. Even if obtained for different parameters, these can then be changed at a later stage, e.g., using continuation (see below). Once some solutions are obtained, a good strategy is also to fit those solutions to models based on sech or tanh functions, for which only the fitting parameters need to be stored, to be reused as starting points at a later stage. 

\paragraph{Determination of the Jacobian matrix~$J$.} In the case of the generalized LLE [below referring to Eq.~\eqref{GLLEN} in normalized form], the Jacobian matrix can be calculated exactly at each Newton step, which is better and faster than taking numerical derivatives. Ignoring stimulated Raman scattering for now ($f_\mathrm{R}=0$), we first note that the loss, detuning, and nonlinearity are purely local terms and only contribute $2\times2$ block diagonal submatrices to the Jacobian for each grid point~$\tau_i$. These are straightforward to obtain as
\begin{equation}
    \begin{bmatrix}
        -1 - 2\, \mathrm{Re}( E(\tau_i))\, \mathrm{Im}(E(\tau_i)) & -\left(\mathrm{Re} (E(\tau_i))\right)^2 - 3 \left(\mathrm{Im} (E(\tau_i))\right)^2 + \Delta\\
        3\left(\mathrm{Re} (E(\tau_i))\right)^2 +  \left(\mathrm{Im} (E(\tau_i))\right)^2 - \Delta & -1 + 2\, \mathrm{Re}(E(\tau_i))\, \mathrm{Im}(E(\tau_i))
    \end{bmatrix}
    \label{eq:J_NL}
\end{equation}
where the two rows (respectively, columns) correspond to (derivatives with respect to) the real and imaginary parts of the LLE (the electric field). For ease of coding, these can be computed in the complex domain then split into real and imaginary components.

With regard to the chromatic dispersion term, $i \hat{D} ( i\partial/\partial\tau) E(\tau)$, there are two options. The first one is to use the well known finite difference formulas to calculate derivatives of successive orders in the polynomial expansion of~$\hat{D}$ [Eq.~\eqref{eq:Dhat_normalized}]. The advantage is that the resulting Jacobian matrix is sparse, speeding up the calculations. With extra effort, this also enables the use of non-periodic boundary conditions, which may be more practical when looking for certain types of solutions like switching waves. However, for broadband combs such as those generated in microresonators, relying on finite differences can produce significant distortions even when taking into account a large number of dispersion orders. E.g., the location of dispersive wave peaks can be tens of nanometers away from their expected locations in the spectrum. In that case, calculating derivatives based  on (fast) Fourier transforms provides much enhanced accuracy. One then writes the dispersion term as $i\, \mathrm{FT}^{-1} \hat{D}(\Omega)\, \mathrm{FT}\, E(\tau)$. We note that this is a linear operation and therefore, in the context of the implementation of the Newton method on a discretized space, where $E$ is a column vector, $\mathcal{D}=i\, \mathrm{FT}^{-1} \hat{D}(\Omega)\, \mathrm{FT}$ is a matrix. Here $\hat{D}(\Omega)$ has to be interpreted as a diagonal matrix and $\mathrm{FT}$ is the matrix representing the discrete Fourier transform. The matrix $\mathcal{D}$ is the contribution of the dispersion term to the Jacobian matrix. Specifically, using the same arrangement of rows and columns as in Eq.~\eqref{eq:J_NL}, each element $\mathcal{D}_{ij}$ can be split into a $2\times2$ block as
\begin{equation}
    \begin{bmatrix}
        \mathrm{Re}(\mathcal{D}_{ij}) & -\mathrm{Im}(\mathcal{D}_{ij})\\
        \mathrm{Im}(\mathcal{D}_{ij}) & \mathrm{Re}(\mathcal{D}_{ij})
    \end{bmatrix}
\end{equation}
Note that in comparison to using finite difference derivatives, this method also has the advantage that $\hat{D}(\Omega)$ is not restricted to being a polynomial expansion. Instead, one can consider an arbitrary dispersion profile, including, e.g., contributions from avoided mode crossings. 

To save time, the matrix~$\mathcal{D}$ may be computed only once at the beginning at the calculation, and then re-used when building up the Jacobian at each Newton step. In fact, given the periodic boundary conditions inherent to the use of the fast Fourier transform, only one column need to be pre-calculated and stored, the other ones being obtained by cyclic permutations. For example, the first column of $\mathcal{D}$ can be obtained by calculating the dispersion term for the first base vector, i.e., an $E$ column-vector of the form $\left[ 1, 0, 0, 0, \ldots \right]^T$. Finally we note that the drift term, $-d_1 \partial E/\partial \tau$, must typically be treated separately when using a pinning condition because $d_1$ is then an extra unknown, but the same principle can be applied. Note that the row of the Jacobian corresponding to the pinning condition must also be calculated separately, but this is usually straightforward. 

With regard to the Raman contribution, which involves a convolution of the Raman susceptibility with the temporal intensity of the field, and despite this term being nonlinear, one can still note that the convolution itself is a linear operation. That operation can thus be stored in a pre-calculated matrix (or in fact a single column vector) using the same method as for the dispersion operator and re-used when setting up the Jacobian.  

\paragraph{Linear stability analysis.} It is important to realize that the Jacobian matrix that needs to be calculated for the Newton iteration, Eq.~\eqref{eq:newtonstep}, is the same as that involved in the evolution of small (linear) perturbations~$\delta E$ to stationary solutions~$E$ of the LLE (provided one excludes the extra row and column associated with the pinning condition),
\begin{equation}
    \frac{d \delta E}{dt} = J\,\delta E\,.
\end{equation}
As the Newton method progresses and converges towards a solution, the Jacobian of the LLE for that solution is thus readily available at the end of the computation. Establishing whether that solution is linearly stable or unstable thus simply requires the extra step of calculating the eigenvalues of the Jacobian. If none of the eigenvalues have positive real parts, the solution is linearly stable, and unstable otherwise. The nature of the unstable eigenvalues can also help in identifying certain types of instabilities. For example, a Hopf bifurcation (which signals breathing CSs) corresponds to a pairs of complex conjugates  eigenvalues. Note that because of the translation symmetry of the problem, the Jacobian will always have a trivial zero eigenvalue (the Goldstone mode), which can be ignored.  

\paragraph{Continuation.} Once a Newton solver has been implemented, it is only a small extra step to implement a continuation procedure. That technique enables to very quickly calculate multiple solutions for different parameter values by letting the solver vary by itself a selected parameter. Using, e.g., the pseudo arc-length continuation method, the solver is then able to automatically follow solutions around fold points, i.e., through unstable branches of hysteresis cycles, a feat which cannot be achieved using direct integration methods. See for example Chap.~10 of~\cite{kuznetsov_elements_2004} for more detail.

\subsection{Lagrangian analysis and reduced analytical model}
\label{sec:Lagragian}

In Sections \ref{sec:CSdetmod} and~\ref{sec:Raman_self_freq_shift}, we discussed a number of results based on a reduced analytical model that describes the (slow) temporal evolution of key temporal cavity soliton parameters (amplitude, position, phase, and frequency). Here, we present the theoretical background behind that model as well as a step-by-step derivation based on the perturbed variational (Lagrangian) method. This method is part of a broader class of perturbative methods that have been used extensively in the past half-century to study soliton solutions of various nonlinear equations (see, e.g., \cite{whitham_linear_1974, anderson_variational_1978, kaup_solitons_1978, anderson_variational_1983, kivshar_dynamics_1989, hasegawa_optical_2003} and references therein). All of these methods start from either an exact soliton solution of an unperturbed equation or an approximate ansatz and aim to derive ordinary differential equations for the evolution of a limited set of key soliton parameters in the presence of conservative and/or dissipative perturbations. 

The derivation below is based on the Lagrangian formalism as presented in~\cite{hasegawa_soliton-based_2000}, but the same model can also be retrieved via the inverse scattering method~\cite{nozaki_chaotic_1985} or the moments method~\cite{ankiewicz_comparison_2008}. Specifically, we follow the procedure detailed in~\cite{hasegawa_optical_2003} and apply it to the generalized LLE, taking into account three dissipative perturbations: driving, losses, and stimulated Raman scattering. Additionally, we also show how the reduced model can be applied to deduce the expression for the maximum detuning at which a CS can exist (c.f.\ Section~\ref{sec:CSexist}).

Note that for this Article we restrict ourselves to the case of pure second-order anomalous GVD and we use the linear approximation of the Raman gain, i.e., we base our developments on Eq.~\eqref{eq:lle_raman}. Also, to match with previously published results based on the perturbed NLSE, we use a different scaling of the fast time, $T = \tau/\sqrt{2}$ (with corresponding angular frequency $\kappa = \sqrt{2}\,\Omega$) and characteristics Raman gain slope $T_\mathrm{R} = \tau_\mathrm{R}/\sqrt{2}$. With these new variables, the normalized generalized LLE can be cast as
\begin{equation}
  \label{eq:nlse_hasegawa}
  i\frac{\partial E}{\partial t} + \frac{1}{2} \frac{\partial^2 E}{\partial T^2} + |E|^2 E - \Delta E = iR\,,
\end{equation}
with
\begin{equation}
    R = -E -i\, T_\mathrm{R} E \frac{\partial |E|^2}{\partial T}  + S\,.
    \label{eq:R_hasegawa}
\end{equation}
In this form, the left-hand side of Eq.~\eqref{eq:nlse_hasegawa} only contains conservative terms while all the dissipative terms are grouped in~$R$. When $R=0$, Eq.~\eqref{eq:nlse_hasegawa} matches with the canonical self-focusing NLSE with the additional detuning term.

\subsubsection{Lagrangian density}

We recall that the Lagrangian formalism represents a dynamical system as a variational problem based on a single real function called the Lagrangian, or Lagrangian density,~$\mathcal{L}$, which encodes all the system's properties. The evolution of the system is obtained by applying the principle of stationary action, which states that a physical trajectory makes the action integral --- the integral of the Lagrangian over the trajectory --- stationary under small variations. This leads to the Euler–Lagrange equation(s), which determine the equation(s) of motion.

In the case of Eq.~\eqref{eq:nlse_hasegawa}, complications stem from the right-hand side, $iR$, because dissipative terms are non-variational: this means that no Lagrangian exists that can describe the whole system. Instead, we will treat~$R$ as a perturbation and only the left-hand side (containing the canonical NLSE terms and the detuning) is described by a Lagrangian density, which is well known and reads~\cite{anderson_variational_1978, anderson_variational_1983, hasegawa_optical_2003}
\begin{align}
  \mathcal{L} &= \frac{i}{2} \left( E^* \frac{\partial E}{\partial t} - E \frac{\partial E^*}{\partial t} \right) + \frac{1}{2}\left( |E|^4- \left| \frac{\partial E}{\partial T} \right|^2\right) - \Delta|E|^2,\notag\\
  &= \frac{i}{2} \Bigl[ E^* (\partial_t E) - E (\partial_t E^*) \Bigr] + \frac{1}{2}\left[ E^2 {E^*}^2- (\partial_T E)(\partial_T E^*) \right] - \Delta E E^*\,.\label{eq:lagragiandensity}
\end{align}
As is standard practice, instead of separating the complex envelope of the electric field~$E(t,T)$ into its real and imaginary parts, we treat $E$ and~$E^*$ (and their derivatives with respect to $t$ and~$T$) as independent variables when varying the action. Hence, formally, the Lagrangian density above $\mathcal{L} = \mathcal{L}(t,T,E,E^*,\partial_t E, \partial_T E, \partial_t E^*, \partial_T E^*)$. The second line of Eq.~\eqref{eq:lagragiandensity} is written to make that aspect explicit. The action integral is
\begin{equation}
    \label{eq:action}
    S = \iint \mathcal{L}(t,T,E,E^*,\ldots)\,dt\,dT\,.
\end{equation}
Requesting the action to be stationary, $\delta S=0$, with respect to variations of~$E$ and~$E^*$, leads to the well-known Euler-Lagrange equations (a classic result we will not derive here). Specifically, for variations in $E^*$, we have,
\begin{equation}
    \frac{\delta\mathcal{L}}{\delta E^*} = \frac{\partial\mathcal{L}}{\partial E^*} - \frac{\partial}{\partial t}\frac{\partial\mathcal{L}}{\partial (\partial_t E^*)}-\frac{\partial}{\partial T}\frac{\partial\mathcal{L}}{\partial (\partial_T E^*)} = 0\,.
    \label{eq:EulerLagrange}
\end{equation}
Here we use the symbol~$\delta$ to denote functional derivatives, i.e., derivatives with respect to changes in \emph{functions} (also referred to as calculus of variations). Applying Eq.~\eqref{eq:EulerLagrange} to the Lagrangian density Eq.~\eqref{eq:lagragiandensity} is straightforward and leads to Eq.~\eqref{eq:nlse_hasegawa} excluding the dissipative terms, i.e., with $R=0$. Eq.~\eqref{eq:nlse_hasegawa} can therefore be formally written as
\begin{equation}
  \label{eq:pertrubednlse}
  \frac{\delta\mathcal{L}(t,T,E,E^*,\ldots)}{\delta E^*} = iR\,. 
\end{equation}
This form makes clear that the dissipative perturbations make the action non stationary. For that reason, $R$~is referred to as the residual. Also note that because the Lagrangian density~$\mathcal{L}$ is a real-valued function, the corresponding equation for variations of~$E$, $\delta\mathcal{L}/\delta E$, is simply the complex conjugate of the above and does not need to be considered separately. 

\subsubsection{Perturbed variational method}

We now discuss how to take into account the non-variational perturbations contained in~$R$ using the technique initially described in \cite{anderson_variational_1978}. We start by postulating an ansatz for the solution that depends on a finite number of real parameters. For the study of cavity solitons, we will fully specify the shape along the fast-time~$T$, while the parameters will only depend on the slow time~$t$, 
\begin{equation}
  E(t,T) \rightarrow E(T, q_1(t), q_2(t), \ldots )
  \label{eq:ansatz_formal}
\end{equation}
The parameters $\{q_j\} = q_1(t), q_2(t), \ldots$ are typically known as collective coordinates in reference to traditional Lagrangian mechanics. By reformulating the original problem into a variational principle for the parameters, the corresponding Euler-Lagrange equations will yield ordinary differential equations governing the dynamics of the parameters. In this way, the infinite-dimensional PDE, Eq.~\eqref{eq:nlse_hasegawa}, is approximated by a low-dimensional dynamical system that captures the essential behavior of the solution. We note that the ansatz does not a priori contain the exact solution to the problem. However, because the variational method is based on a minimization principle (least action), we can expect to find a best fit solution given the constraints.

We proceed by considering the variation of the action~\eqref{eq:action} with respect to $E$ and~$E^*$,
\begin{equation}
    \delta S = \iint
    \left( \frac{\delta\mathcal{L}}{\delta E^*} \delta E^* + \frac{\delta\mathcal{L}}{\delta E} \delta E \right) \,dt\,dT\,.
\end{equation}
In this expression, we then substitute $\delta\mathcal{L}/\delta E^*$ (and its conjugate) using Eq.~\eqref{eq:pertrubednlse}. We also take into account that $\delta E$ (and $\delta E^*$) are now constrained by the ansatz, Eq.~\eqref{eq:ansatz_formal}, and that the admissible variations must lie in the tangent space of the ansatz, i.e.,
\begin{equation}
  \delta E = \sum_j \frac{\partial E}{\partial q_j} \delta q_j(t)\,.
\end{equation}
Overall, this yields
\begin{equation}
    \delta S = \iint \sum_j\,
    i \left( R \frac{\partial E^*}{\partial q_j} - R^* \frac{\partial E}{\partial q_j} \right) \delta q_j(t) \,dt\,dT\,.
    \label{eq:deltaS_Rprojection}
\end{equation}
Independently, we can write the action as
\begin{equation}
    S = \iint \mathcal{L}(t,T,E,E^*,\ldots)\,dt\,dT = \int L_\mathrm{eff} ( t, \{q_j\}, \{\dot{q}_j\} )\,dt\,,
    \label{eq:action_qj}
\end{equation}
with the effective, or reduced, Lagrangian $L_\mathrm{eff}$ obtained by inserting the ansatz directly into the Lagrangian density and carrying out the fast-time integral (with dots denoting slow-time derivatives),
\begin{equation}
  L_\mathrm{eff}(t, \{q_j\}, \{\dot{q}_j\}) = \int \mathcal{L} \left[ t,T, E(T, \{q_j(t)\}), E^*(\ldots), \ldots \right]\, dT\,.
  \label{eq:Leff}
\end{equation}
The second equality in Eq.~\eqref{eq:action_qj} represents a variational problem for the effective Lagrangian~$L_\mathrm{eff}$, based on the finite set of ansatz parameters~$\{ q_j(t) \}$. Using the Euler-Lagrange theorem for variations of these parameters, the variation of the action can be expressed as,
\begin{equation}
    \delta S = \int \sum_j\,\frac{\delta L_\mathrm{eff}}{\delta q_j}\, \delta q_j(t)\,dt = \int \sum_j \left[ \frac{\partial L_\mathrm{eff}}{\partial q_j}-\frac{d}{dt}\left( \frac{\partial L_\mathrm{eff}}{\partial \dot{q}_j}\right)\right]\,\delta q_j(t)\,dt\,.
    \label{eq:deltaS_Leff}
\end{equation}
The two expressions for $\delta S$, Eqs.~\eqref{eq:deltaS_Rprojection} and~\eqref{eq:deltaS_Leff}, must be equal for all possible variations $\{\delta q_j(t)\}$. Because these are arbitrary functions of time, their coefficients must be equal pointwise in~$t$ for all values of~$j$ (fundamental lemma of the calculus of variations), leading to
\begin{equation}\label{eq:eulerlagrange}
    \frac{\partial L_\mathrm{eff}}{\partial q_j}-\frac{d}{dt}\left(\frac{\partial L_\mathrm{eff}}{\partial \dot{q}_j}\right)  = \int i\left(R\frac{\partial E^*}{\partial q_j}-R^*\frac{\partial E}{\partial q_j}\right)\, dT := I_{q_j}.
\end{equation}
with the symbol $I_{q_j}$ introduced to refer to the integrals on the right. The above expressions lead to dynamical evolution equations for the ansatz parameters under the influence of the non-conservative perturbation~$R$. Note that they can be interpreted as generalized Euler-Lagrange equations, with the residual projected onto the tangent space of the ansatz.

\subsubsection{Derivation of the reduced model of cavity solitons}

We can now apply the result of the above section, Eqs.~\eqref{eq:eulerlagrange}, to the generalized LLE, Eqs.~\eqref{eq:nlse_hasegawa}--\eqref{eq:R_hasegawa}. We assume the following ansatz for a cavity soliton,
\begin{equation}
  \label{eq:lagragianansatz}
  E(t,T) \rightarrow B\, \mathrm{sech} \left[ B (T-T_\mathrm{s}) \right] \exp{\left( -i \kappa_s(T-T_\mathrm{s}) + i\phi \right)}\,.
\end{equation}
with the set of parameters $\{q_j(t)\} = \{ B, T_\mathrm{s}, \kappa_\mathrm{s}, \phi \}$ representing amplitude, position, frequency, and phase of the cavity soliton, respectively, and all of them dependent on slow time~$t$. We make the choice to neglect the continuous-wave background to facilitate the calculations. This effectively limits the range of validity of the reduced model to high detunings, but is sufficient for the example calculation we present here. For a more accurate analysis, we refer the reader to~\cite{matsko_timing_2013, li_efficiency_2022}.

The first step is to calculate the effective Lagrangian, Eq.~\eqref{eq:Leff}, i.e., the fast-time~($T$) integral of the Lagrangian density~\eqref{eq:lagragiandensity} for the ansatz~\eqref{eq:lagragianansatz}. As our ansatz essentially matches an NLSE soliton, we can draw on results obtained in the context of the NLSE~\cite{hasegawa_soliton-based_2000, hasegawa_optical_2003}. We simply have to consider the additional contribution of the detuning, i.e., the $-\Delta |E|^2$ term in the Lagrangian density, 
\begin{equation}
    \int^{\infty}_{-\infty}(-\Delta |E|^2)\, dT
    = -\Delta \int^{\infty}_{-\infty} |E|^2\, dT
    = - B^2 \Delta  \int^{\infty}_{-\infty} \operatorname{sech}^2\left[ B (T-T_\mathrm{s}) \right] dT = -2 B \Delta\,,
    \label{eq:int_Eabs2}
\end{equation}
using $\int_{-\infty}^\infty \operatorname{sech}^2 x\, dx = 2$. Combining this result with the other known terms of the effective Lagrangian of the canonical NLSE~\cite{hasegawa_soliton-based_2000, hasegawa_optical_2003}, we find
\begin{equation}
  \label{eq:Leff_ansatz}
  L_\mathrm{eff} = -2{B}\left(\kappa_s\frac{dT_s}{dt}+\frac{d\phi}{dt}\right) + \frac{1}{3}{B}^3 - {B} (\kappa_s^2+2\Delta)\,.
\end{equation}
We now have to insert the above expression for $L_\mathrm{eff}$ into Eqs.~\eqref{eq:eulerlagrange}. For the left-hand sides, we have
\begin{align}
  \frac{\partial L_\mathrm{eff}}{\partial q_j}-\frac{d}{dt}\left(\frac{\partial L_\mathrm{eff}}{\partial \dot{q}_j}\right)
  &= 2 \frac{dB}{dt}\,, & (q_j &= \phi) \label{eq:EL_LHS1}\\
  &= -2B \left( \frac{dT_\mathrm{s}}{dt} + \kappa_\mathrm{s} \right)\,, & (q_j &= \kappa_\mathrm{s})\\
  &= 2 \frac{dB}{dt} \kappa_\mathrm{s} + 2 B \frac{d\kappa_\mathrm{s}}{dt}\,, & (q_j &= T_\mathrm{s})\\
  &= -2 \left(\kappa_s\frac{dT_s}{dt}+\frac{d\phi}{dt}\right) + B^2 - (\kappa_s^2+2\Delta)\,, & (q_j &= B)
  \label{eq:EL_LHS4}
\end{align}
while the derivatives of the ansatz with respect to each parameter, which appear in the right-hand sides, can be written as (using $d \operatorname{sech} x/dx = -\tanh x \operatorname{sech}x$)
\begin{align}
    \frac{\partial E}{\partial \phi} &= iE\,, &
    \frac{\partial E}{\partial T_\mathrm{s}} &= \left(B \tanh\left[ B (T-T_\mathrm{s}) \right] + i\kappa_s\right) E\,,\notag\\
    \frac{\partial E}{\partial \kappa_\mathrm{s}} &= -i(T-T_\mathrm{s}) E\,, &
    \frac{\partial E}{\partial B} &= \left( \frac{1}{B} - (T-T_\mathrm{s}) \tanh\left[ B (T-T_\mathrm{s}) \right] \right) E\,.\notag
\end{align}
Proceeding with the integrals~$I_{q_j}$ in Eq.~\eqref{eq:eulerlagrange}, we first consider $q_j = \phi$ and $\kappa_\mathrm{s}$. Using the expressions for the derivative as above, these can be written as
\begin{align}
    I_\phi &= 2\,\text{Real}\left[ \int_{-\infty}^\infty R^* E \, dT\right] &
    I_{\kappa_\mathrm{s}} &= -2\,\text{Real}\left[ \int_{-\infty}^\infty (T-T_\mathrm{s}) R^* E \, dT \right]
    \label{eq:IphiITsReal}
\end{align}
These integrals both involve $R^* E = -|E|^2 + S E + i\, T_\mathrm{R} |E|^2 (\partial |E|^2/\partial T)$ (respective contributions from loss, driving, and  stimulated Raman scattering). We note that the Raman term in this expression is purely imaginary, hence does not contribute to the two integrals above. Also, because $|E|^2$ is an even function of $T-T_\mathrm{s}$, the loss term does not contribute to $I_{\kappa_\mathrm{s}}$. We are left with
\begin{align}
    I_\phi &= -2 \int_{-\infty}^\infty |E|^2 \, dT + 
     2S\,\text{Real}\left[ \int_{-\infty}^\infty E \, dT\right] &
    I_{\kappa_\mathrm{s}} &= -2S\,\text{Real}\left[ \int_{-\infty}^\infty (T-T_\mathrm{s}) E \, dT \right]
\end{align}
We proceed as in Eq.~\eqref{eq:int_Eabs2}, also using $\int_{-\infty}^\infty \operatorname{sech} x \exp(-i \kappa x) dx = \pi \operatorname{sech}(\pi\kappa/2)$ (Fourier transform of the sech function), and the derivative $i \partial/\partial \kappa$ of the latter (Fourier transform of $x \operatorname{sech} x$). This yields
\begin{align}
    I_\phi &= -4B +2\pi S \cos(\phi) \operatorname{sech}\left(\frac{\pi\kappa_\mathrm{s}}{2B}\right) &
    I_{\kappa_\mathrm{s}} &= -\frac{\pi^2}{B} S \sin(\phi) \operatorname{sech}\left(\frac{\pi\kappa_\mathrm{s}}{2B}\right) \tanh\left(\frac{\pi\kappa_\mathrm{s}}{2B}\right)
    \label{eq:Iphi_and_Ikappas}
\end{align}

Next, we consider $I_{T_\mathrm{s}}$. We first note that the driving term ($S$) in the residual~$R$ does not contribute to the integral because $\int_{-\infty}^\infty (\partial E/\partial T_\mathrm{s})\, dT= \int_{-\infty}^\infty (-\partial E/\partial T)\,dT=0$, which comes from our CS ansatz approaching zero at infinity. Using the form of $\partial E/\partial T_\mathrm{s}$ derived above, we can then write $I_{T_\mathrm{s}}$ in a similar form as for $I_\phi$ and~$I_{\kappa_\mathrm{s}}$ in Eqs.~\eqref{eq:IphiITsReal}, obtaining
\begin{equation}   
   I_{T_\mathrm{s}} = 2\kappa_\mathrm{s}\,\text{Real}\left[ \int_{-\infty}^\infty R^* E \, dT\right] + 2\,\text{Imag}\left[ \int_{-\infty}^\infty B \tanh\left[ B (T-T_\mathrm{s}) \right] R^* E\,dT\right]\,.
\end{equation}
In $R^*E$, we now only have to consider the loss ($-|E|^2$) and Raman scattering terms. These terms being respectively purely real and purely imaginary, they contribute separately to the two integrals above, leading to,
\begin{align}
   I_{T_\mathrm{s}} &= -2\kappa_\mathrm{s} \int_{-\infty}^\infty |E|^2 \, dT + 2 T_\mathrm{R} \int_{-\infty}^\infty B \tanh\left[ B (T-T_\mathrm{s}) \right]\, |E|^2 \,\frac{\partial |E|^2}{\partial T}\,dT \\
   &= -4 \kappa_\mathrm{s} B - 4 T_\mathrm{R} B^6 \int_{-\infty}^\infty \tanh^2\left[ B (T-T_\mathrm{s}) \right] \operatorname{sech}^4\left[ B (T-T_\mathrm{s}) \right]\,dT = -4 \kappa_\mathrm{s} B - \frac{16}{15} T_\mathrm{R} B^5\,.
   \label{eq:ITs}
\end{align}
where we have used $d \operatorname{sech}^2 x/dx = -2\tanh x\operatorname{sech}^2 x$, and where the final integral is solved by change of variable, with $u=\tanh\left[ B (T-T_\mathrm{s}) \right].$ 

Finally, for $I_B$, using the form of $\partial E/\partial B$ derived earlier, we get
\begin{equation}
    I_B = 2\,\text{Imag}\left[ \int_{-\infty}^\infty \left( \frac{1}{B} - (T-T_\mathrm{s}) \tanh\left[ B (T-T_\mathrm{s}) \right] \right) R^* E\,dT \right]\,.
\end{equation}
Here, out of the three terms of $R^*E$, the loss term, purely real, does not contribute, while the Raman term vanishes because $|E|^2 (\partial |E|^2/\partial T)$ is an odd function of $T-T_\mathrm{s}$. Only the driving term remains. The final integration involves calculating an integral of the form $\int_{-\infty}^\infty x \tanh x \operatorname{sech} x \exp(-i \kappa x) dx$ which can be obtained from results quoted above by substituting $\tanh x \operatorname{sech} x = - d \operatorname{sech} x/dx$ and integrating by part. The final result is
\begin{equation}
    I_B = \kappa_\mathrm{s} \frac{\pi^2}{B^2} S \sin(\phi) \operatorname{sech}\left(\frac{\pi\kappa_\mathrm{s}}{2B}\right) \tanh\left(\frac{\pi\kappa_\mathrm{s}}{2B}\right)\,.
    \label{eq:IB}
\end{equation}

\subsubsection{Final form of the reduced model}

Putting all the results of the previous Section together, specifically Eqs.~\eqref{eq:EL_LHS1}--\eqref{eq:EL_LHS4} as well as Eqs.~\eqref{eq:Iphi_and_Ikappas}, \eqref{eq:ITs}, and \eqref{eq:IB}, we obtain the reduced model of the LLE with linear Raman gain [c.f.\ Eq.~\eqref{eq:lle_raman}],
\begin{align}
  \frac{dB}{dt} &= -2B + \pi S \cos(\phi) \operatorname{sech}\left(\frac{\pi\kappa_\mathrm{s}}{2B}\right)\,,\\
  \frac{dT_\mathrm{s}}{dt} &= -\kappa_\mathrm{s} + \frac{\pi^2}{2B^2} S \sin(\phi) \operatorname{sech}\left(\frac{\pi\kappa_\mathrm{s}}{2B}\right) \tanh\left(\frac{\pi\kappa_\mathrm{s}}{2B}\right)\,,\\
  \frac{d\kappa_\mathrm{s}}{dt} &= -\frac{\kappa_\mathrm{s}}{B} \frac{dB}{dt} -2 \kappa_\mathrm{s} - \frac{8}{15} T_\mathrm{R} B^4\,,\\
  \frac{d\phi}{dt} &= -\kappa_s\frac{dT_s}{dt} + \frac{B^2 - \kappa_s^2}{2}-\Delta - \kappa_\mathrm{s} \frac{\pi^2}{2B^2} S \sin(\phi) \operatorname{sech}\left(\frac{\pi\kappa_\mathrm{s}}{2B}\right) \tanh\left(\frac{\pi\kappa_\mathrm{s}}{2B}\right) \,.
\end{align}
The terms proportional to $1/B^2$ are often neglected as they decrease rapidly with the power of the cavity soliton, leading to
\begin{align}
  \frac{dB}{dt} &= -2B + \pi S \cos(\phi) \operatorname{sech}\left(\frac{\pi\kappa_\mathrm{s}}{2B}\right)\,, &
  \frac{dT_\mathrm{s}}{dt} &= -\kappa_\mathrm{s}\,,\\
  \frac{d\kappa_\mathrm{s}}{dt} &= -\frac{\kappa_\mathrm{s}}{B} \frac{dB}{dt} -2 \kappa_\mathrm{s} - \frac{8}{15} T_\mathrm{R} B^4
  &  \frac{d\phi}{dt} &= -\kappa_s\frac{dT_s}{dt} + \frac{B^2 - \kappa_s^2}{2}-\Delta\\ 
  &= - \frac{\kappa_\mathrm{s}}{B} \pi S \cos(\phi) \operatorname{sech}\left(\frac{\pi\kappa_\mathrm{s}}{2B}\right) - \frac{8}{15} T_\mathrm{R} B^4\, &
 &= \frac{B^2 + \kappa_s^2}{2} - \Delta\,.
\end{align}
This model has been derived previously, e.g., in~\cite{yi_theory_2016}. Note that the second equality in the bottom equations are obtained by  substituting the derivatives in the right-hand side with the top two equations. With the changes of variables $T=\tau/\sqrt{2}$ and $\kappa=\sqrt{2}\,\Omega$ introduced at the top of this Section, the model matches with that discussed in Section~\ref{sec:Raman_self_freq_shift}, where we also  present a comparison between the analytical soliton amplitude and frequency shift with results obtained from the generalized LLE (see Figure~\ref{fig:Raman_Lagrange}).

The reduced model used in Section~\ref{sec:CSdetmod} additionally includes desynchronization and a linearized trapping potential. This can be added simply to the derivation presented here, starting with an additional term $i d/(2\sqrt{2}) [E^* (\partial_T E)-E (\partial_T E^*)] + V(T) |E|^2$ in the Lagrangian density, Eq.~\eqref{eq:lagragiandensity}. In the effective Lagrangian, Eq.~\eqref{eq:Leff_ansatz}, these translates into terms $\sqrt{2}\, B \kappa_\mathrm{s} d + 2B V(T_\mathrm{s})$. The rest follows simply.

\subsubsection{Raman-free solution and maximum detuning}
\label{sec:sqrt2Delta}

In the absence of stimulated Raman scattering ($T_\mathrm{R}=0$), there is no source of frequency shift, hence $\kappa_\mathrm{s}=0$. The model reduces to
\begin{align}
    \frac{d{B}}{dt} &= -2B + \pi S \cos(\phi)\\
    \frac{d\phi}{dt} &= \frac{B^2}{2}-\Delta\,.
\end{align}
Looking for the stationary solutions, we get the well known equation for the cavity soliton amplitude, $B=\sqrt{2\Delta}$. We also find that the phase of the soliton (with respect to the driving beam) satisfies $\cos(\phi) = 2 \sqrt{2\Delta}/(\pi S)$. As $\cos(\phi)$ cannot exceed a maximum value of~1, this relation immediately leads to the maximal detuning for CS existence, $\Delta_\mathrm{max} = \pi^2 S^2/8$. At that maximum detuning, the cavity soliton is in phase with the driving beam ($\phi=0$) and its amplitude is $B_\mathrm{max} = \pi S/2$. We note that, although this is not an exact solution (in particular, we have neglected the cw background), these expressions provide a highly accurate approximation of the cavity soliton power and duration in most practical scenarios.

\section{Conclusions and Outlook}

The temporal cavity soliton is the canonical bright localised dissipative structure of a passively driven Kerr resonator. Since their first prediction in 1993~\cite{wabnitz_suppression_1993}, demonstration in macroscopic fiber resonators in 2010~\cite{leo_temporal_2010}, and realization in optical microresonators in 2014~\cite{herr_temporal_2014}, the study of CSs has rapidly evolved from a niche scientific curiosity into an exciting new frontier in nonlinear photonics. Today, the application of these remarkable nonlinear structures, and their associated normal dispersion counterparts, has been reported in areas as diverse as precision metrology~\cite{del2016phase}, ultrafast ranging~\cite{trocha_ultrafast_2018}, optical communications~\cite{marin-palomo_microresonator-based_2017-1,rizzo2023massively} and computing~\cite{xu2020photonic,xu202111}, and frequency synthesis across both the optical~\cite{spencer_optical-frequency_2018} and microwave domains~\cite{he_high-speed_2023,kudelin2024photonic}.

In this article, we present a comprehensive analysis of the governing equations that describe the underlying physics of coherently driven passive Kerr resonators. These equations reveal the fundamental nonlinear properties that underpin the system’s behaviour. These insights, in turn, provide the basis for understanding temporal cavity solitons --~self‑localized pulses of light that persist within the resonator thanks to a delicate double-balance between its anomalous dispersion and nonlinearity, and dissipation and driving. The manuscript then proceeds to a detailed exposition of the key properties of CSs including discussion of, procedures for their excitation and erasure, trapping and synchronisation, CS nonlinear dynamics, the influence of higher-order dispersion and Raman scattering, CS bound states and solitons crystals, timing jitter and conversion efficiency, and the operation of CSs in self-injection-locked systems. We then extend our analysis to the normal‑dispersion regime, and consider in detail the properties of optical switching waves: nonlinear fronts that connect coexisting homogeneous states of the intracavity field. We show these fronts can propagate, lock, and even form stationary localized domains depending on system's parameters. This reveals their key role in the formation of normal dispersion dark‑pulse structures. 

As research into this rapidly expanding field progresses, new insights continue to emerge. Notably, recent advances have unveiled new operating regimes, including parametrically driven~\cite{englebert_parametrically_2021,moille2024parametrically} and active cavities~\cite{englebert_temporal_2021}, self-injection-locked combs~\cite{shen_integrated_2020, voloshin2021dynamics} and disciplined CS combs locked to external references~\cite{weng_spectral_2019,kudelin2024photonic, moille_kerr-induced_2023,moille_-chip_2025}. Looking ahead, theoretical advances, combined with improved fabrication techniques and material engineering, can be expected to result in continued improvements in all aspects of comb performance. Such developments promise to unlock new opportunities in integrated photonics and quantum technologies, and map an exciting trajectory for future research and applications in the field.

\begin{backmatter}

\bmsection{FUNDING}

\medskip
\noindent S.~G.~M., S.~C., and M.~E. acknowledge support from the Marsden fund of the Royal Society Te Ap\={a}rangi of New Zealand and Te Whai Ao -- Dodd-Walls Centre for Photonic and Quantum Technologies, S.~C. and F.~L. acknowledge support from the Fonds de la Recherche Scientifique (F.R.S.-FNRS), F.~L. acknowledges support from the European Research Council (ERC) under the Horizon Europe research and innovation programme (Grant Agreement No. 101113343), and X.~X. acknowledges support from the Tsinghua-Toyota Joint Research Fund.
\medskip

\bmsection{DISCLOSURES}

\medskip
\noindent The authors declare no conflicts of interest.
\medskip

\bmsection{DATA AVAILABILITY}

\medskip
\noindent Data underlying the results presented in this paper are not publicly available at this time but may be obtained from the authors upon reasonable request.
\medskip

\bmsection{BIOGRAPHIES}

\medskip

\noindent
\begin{minipage}[t]{0.2\textwidth}
    \vspace{0pt}
    \includegraphics[width=2.5cm]{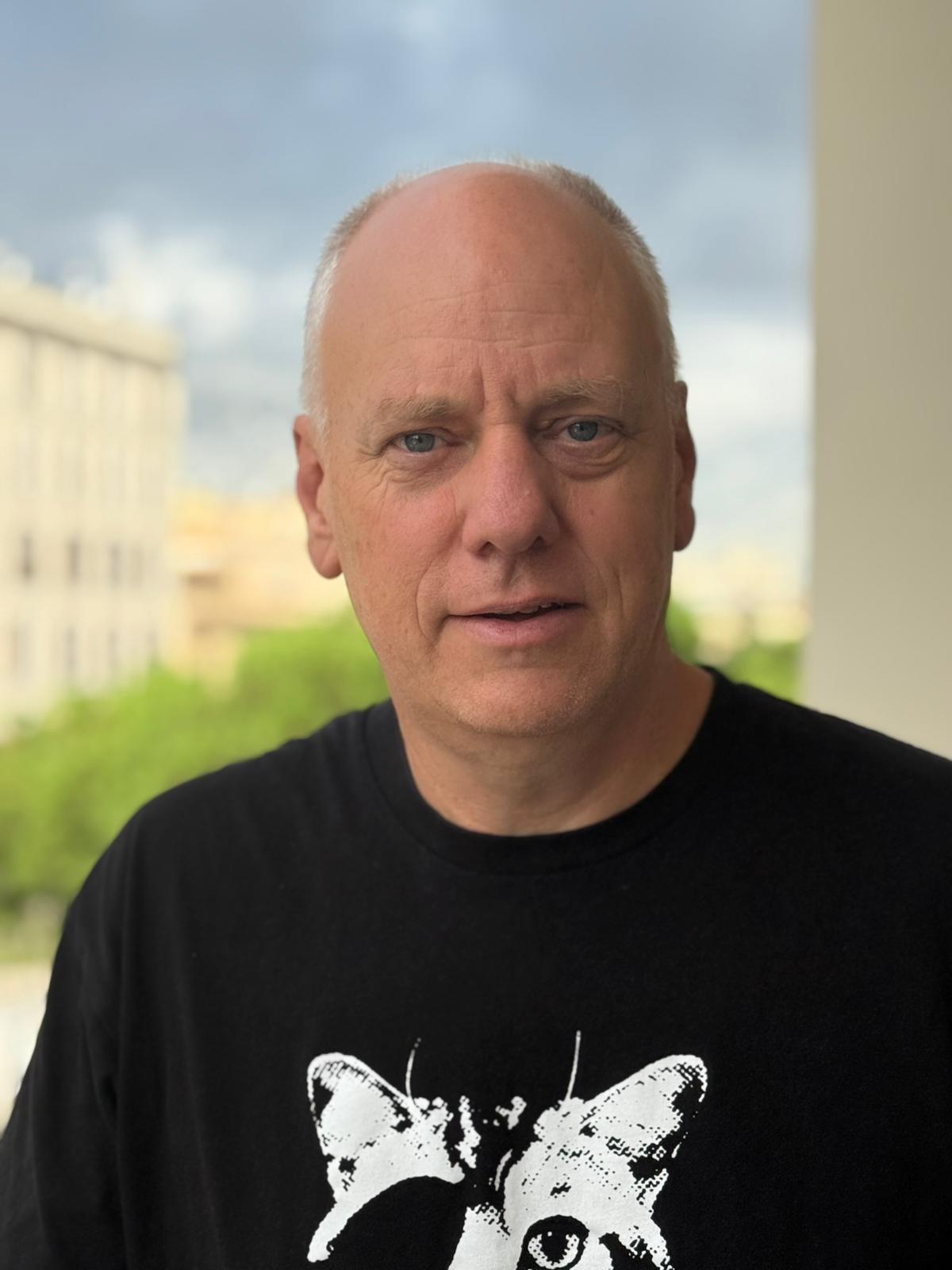}
\end{minipage}
\hfill
\begin{minipage}[t]{0.79\textwidth}
    \vspace{0pt}
    Stuart Murdoch is an Associate Professor in the Department of Physics at the University of Auckland (New Zealand), where he has been a member of academic staff since 2003. He received his PhD from the University of Auckland in 1997. From 1998 to 2000 he was a postdoctoral fellow at the Institut d’Optique, Orsay (France), working on Bose–Einstein condensation. He then spent three years at Photon Kinetics, Chandlers Ford (United Kingdom), developing instrumentation for optical fiber test and measurement. He runs an experimental optics laboratory at Auckland, with current interests in nonlinear fiber optics, optical microresonators, and Kerr‑comb generation.
\end{minipage}

\medskip

\noindent
\begin{minipage}[t]{0.2\textwidth}
    \vspace{0pt}
  \includegraphics[width=2.5cm]{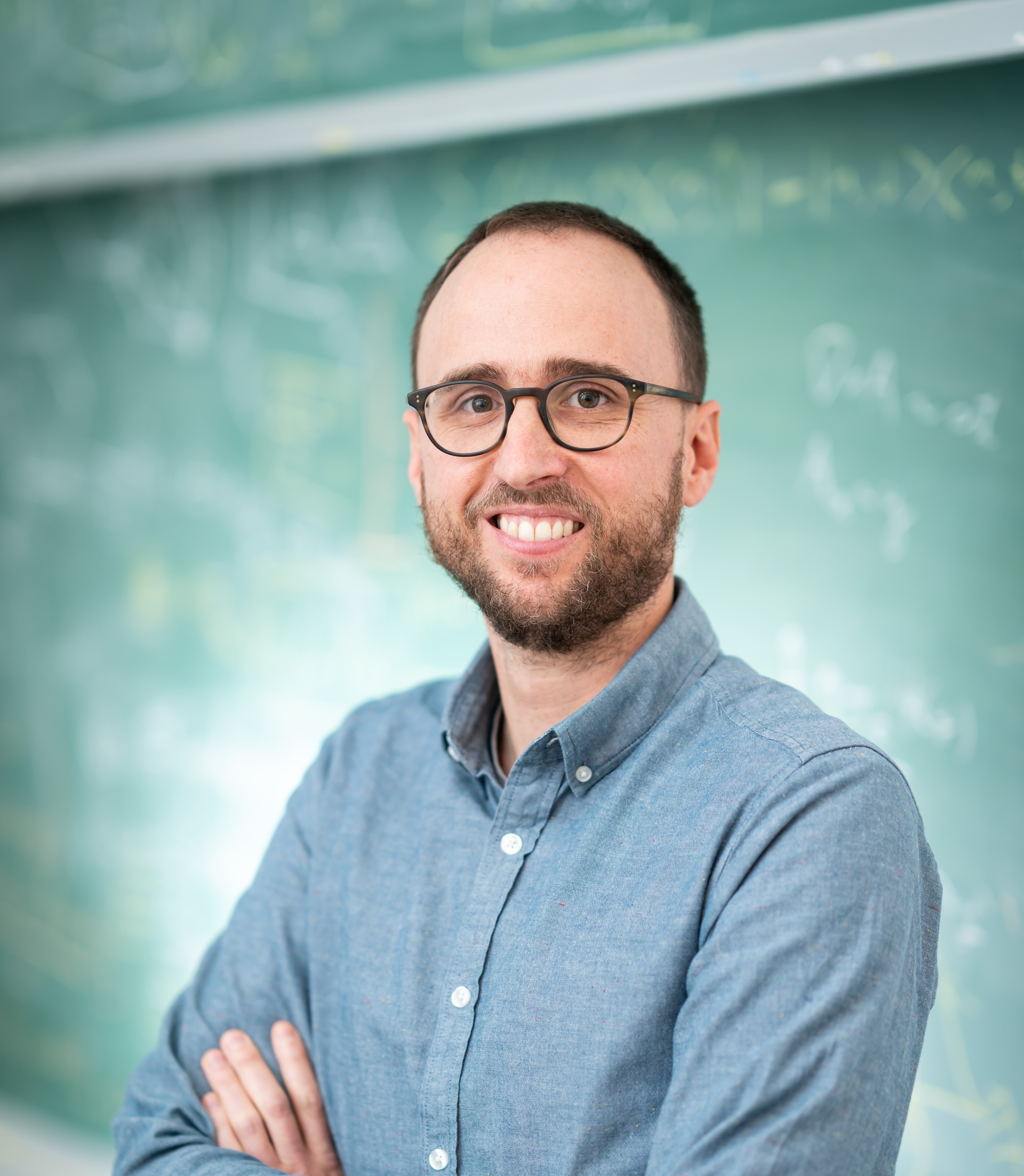}
\end{minipage}
\hfill
\begin{minipage}[t]{0.79\textwidth}
    \vspace{0pt}
  François Leo is a research associate at the University of Brussels. He received his PhD in 2010 from the University of Brussels, with a thesis entitled “Experimental and theoretical study of dissipative structures in optical resonators.” From 2011 to 2014, he worked at Ghent University on nonlinear silicon photonics, before moving to Auckland in 2015 to study nonlinear fiber resonators. Since 2017, he has been leading a research team focused on understanding and harnessing the dynamics of nonlinear resonators. His work combines experimental and theoretical approaches, with interests spanning both fundamental aspects and applications
\end{minipage}

\medskip

\noindent
\begin{minipage}[t]{0.2\textwidth}
    \vspace{0pt}
    \includegraphics[width=2.5cm]{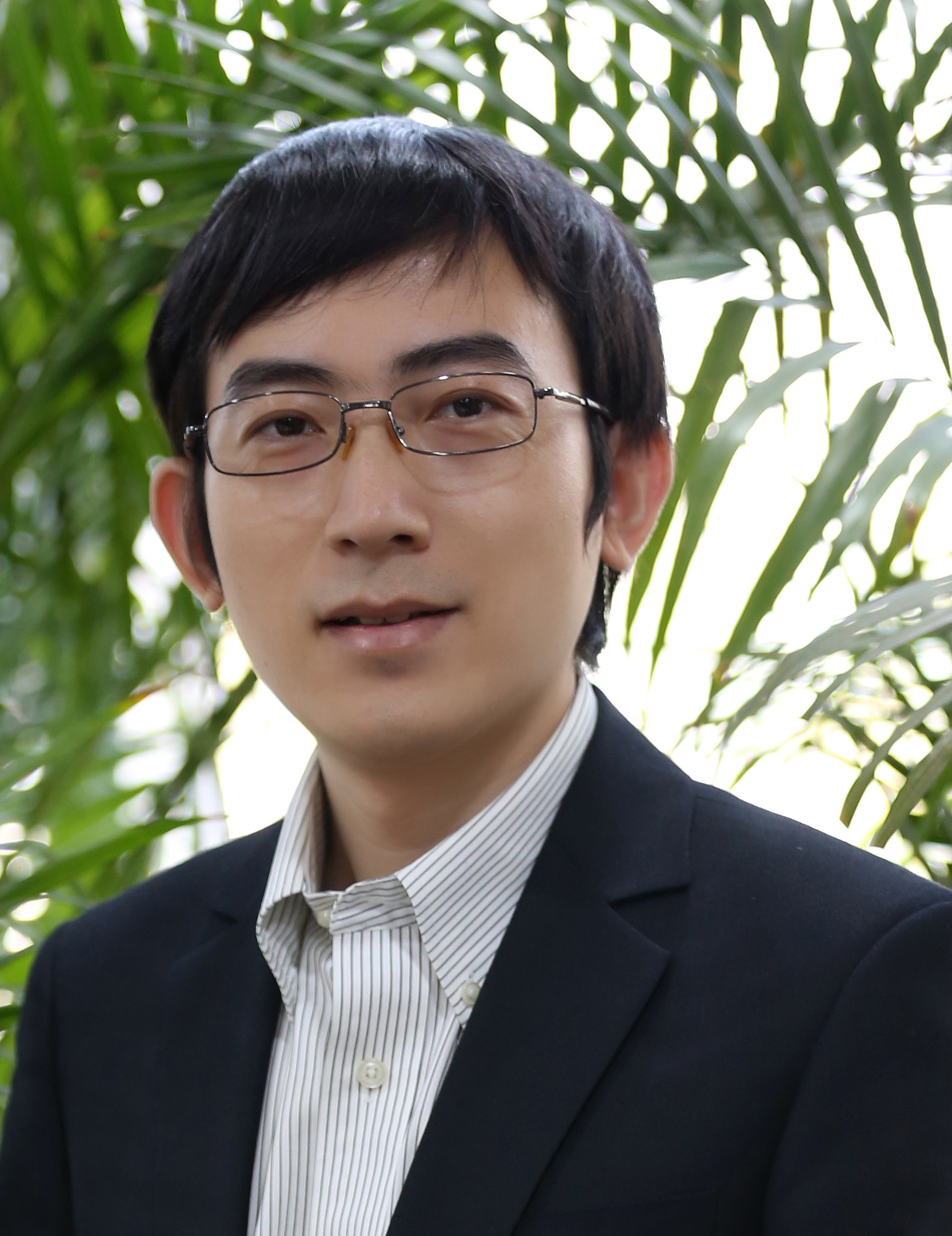}
\end{minipage}
\hfill
\begin{minipage}[t]{0.79\textwidth}
    \vspace{0pt}
    Xiaoxiao Xue received the B.S. and Ph.D. degrees (both with highest honors) in electronic engineering from Tsinghua University, Beijing, China, in 2007 and 2012, respectively. From 2013 to 2015, he was a Postdoctoral Researcher with the Ultrafast Optics and Optical Fiber Communications Laboratory at Purdue University. He joined the Department of Electronic Engineering at Tsinghua University in 2016, where he is currently an Associate Professor. He has authored or co-authored over 90 journal papers. His research interests encompass optical frequency comb generation, microwave photonic signal processing, radio-over-fiber, and phased array antennas. He was the recipient of the  Wang Daheng Optical Award (2012) and the Rao Yutai Basic Optical Award (2019) from the Chinese Optical Society for his contributions to microwave photonics and high-efficiency microcomb generation.
\end{minipage}

\medskip

\noindent
\begin{minipage}[t]{0.2\textwidth}
    \vspace{0pt}
   \includegraphics[width=2.5cm]{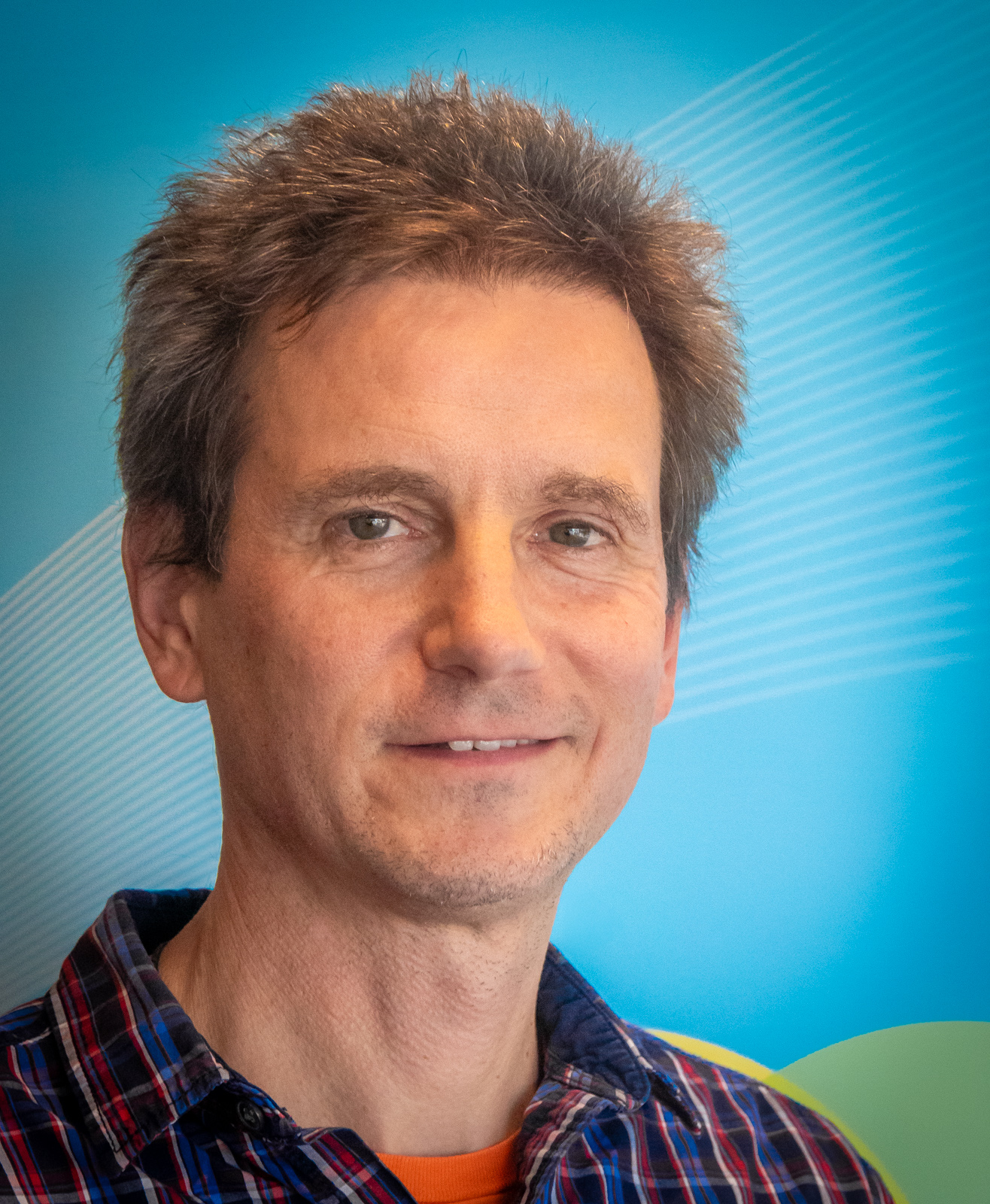}
\end{minipage}
\hfill
\begin{minipage}[t]{0.79\textwidth}
    \vspace{0pt}
    St\'ephane Coen received an Engineering and PhD in Engineering degrees from the Université Libre de Bruxelles (ULB) in 1996 and 1999, respectively. In 2000--2001, he was a postdoctoral fellow in the Physics Department of The University of Auckland (New Zealand), before becoming a permanent faculty member of that institution at the end of~2003. He was promoted to Professor in~2025. His research mainly focus on nonlinear photonics, in particular supercontinuum generation, Kerr resonators, and temporal cavity solitons. He has authored or co-authored 130 journal publications. In 2016, he was awarded the Hector medal of the Royal Society Te Ap\={a}rangi of New Zealand. He is a Fellow of Optica and of the Royal Society Te Ap\={a}rangi of New Zealand.
\end{minipage}

\medskip

\noindent
\begin{minipage}[t]{0.2\textwidth}
    \vspace{0pt}
    \includegraphics[width=2.5cm]{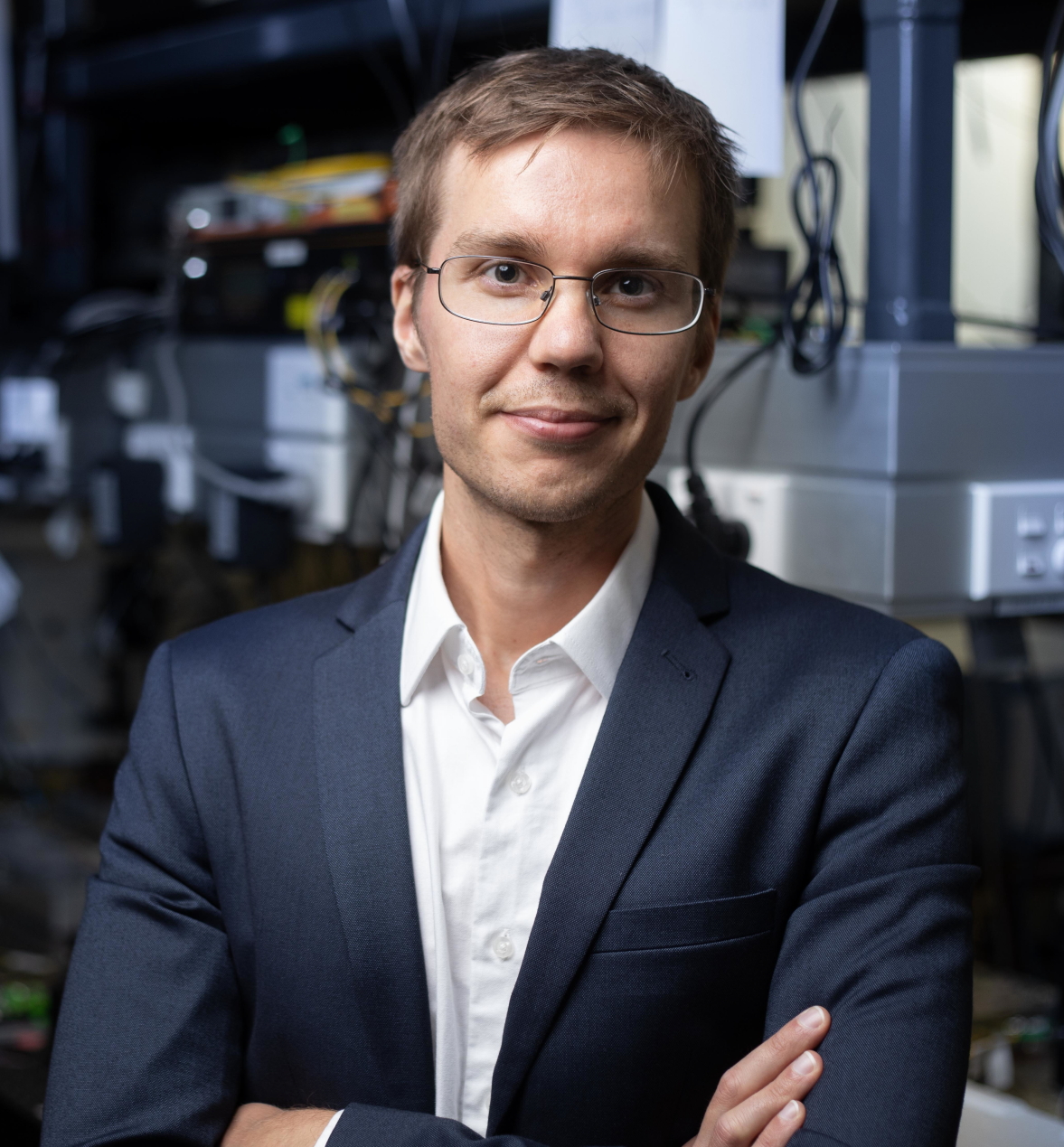}
\end{minipage}
\hfill
\begin{minipage}[t]{0.79\textwidth}
    \vspace{0pt}
     Miro Erkintalo received the BSc, MSc, and DSc (PhD equivalent) degrees from Tampere University of Technology, Finland, in March 2009, October 2009, and February 2012, respectively. He joined the University of Auckland in New Zealand as a research fellow in 2012 and subsequently became a member of permanent faculty. In 2025, he was promoted to full professor. His research interests focus on nonlinear photonics in waveguide and resonator geometries, including the dynamics of temporal (conservative and dissipative) solitons and optical frequency combs. He has authored or co-authored more than 100 peer-reviewed journal publications, and has received multiple awards for his research, including the 2019 New Zealand Prime Minister's Emerging Scientist Prize.
\end{minipage}
\end{backmatter}

\bibliography{refs_combined,refs_extra,refs_extra2}

\end{document}